\documentclass[preprint,review,12pt]{elsarticle}

\usepackage{tikz}
\usetikzlibrary{arrows.meta, positioning, calc, shapes.geometric}
\definecolor{stableblue}{RGB}{25,90,155}
\definecolor{gatewayorange}{RGB}{198,93,52}
\definecolor{inactivegrey}{RGB}{150,150,150}
\DeclareRobustCommand{\circled}[1]{\tikz[baseline=(char.base)]{ \node[shape=circle,draw,inner sep=1.4pt, line width=0.5pt] (char) {\small #1};}}
\usepackage{amsmath}
\usepackage{amssymb}
\usepackage{amsthm}
\usepackage{xcolor}
\usepackage{hyperref}
\usepackage{enumitem}

\theoremstyle{plain}
\newtheorem{theorem}{Theorem}
\newtheorem{proposition}[theorem]{Proposition}
\newtheorem{heuristic}[theorem]{Heuristic Criterion}
\theoremstyle{definition}
\newtheorem{definition}{Definition}

\journal{Journal of the Mechanics and Physics of Solids}

\begin{document}
\begin{frontmatter}

\title{Topological Foundations of Multi-Field Instabilities in Continua: Part 1 — Foundations\\[4pt]
\normalsize\itshape Foundational Part~1 of a series: Topological Foundations of a Parameter-Free Instability Classification}

\author[curtin]{K. Regenauer-Lieb\corref{cor1}}
\ead{Klaus@curtin.edu.au}
\author[savoie]{F. Nicot}

\cortext[cor1]{Corresponding author}

\address[curtin]{ARC Centre of Excellence for Carbon Science \& Innovation,
  WASM: Minerals, Energy and Chemical Engineering,
  Curtin University, Perth WA 6845, Australia}
\address[savoie]{ISTerre, Universit\'e Savoie Mont Blanc, Chamb\'ery, France}

\begin{quote}\small
\textit{This is the foundational paper of a three-part JMPS manuscript series that develops a parameter-free, topological classification of multi-field instability and its realisation in granular matter. The general, representation-free formulation of the underlying principle, the Principle of Ordered Action, is developed in a companion paper submitted to the Proceedings of the Royal Society A; the present JMPS series is its minimal thermodynamic realisation. This paper lays the topological and thermodynamic foundations; its companion JMPS manuscripts carry the concrete realisation through which the architecture becomes falsifiable and benchmarkable against classical elasto-plasticity, for a 1-D granular spin chain in Part 2 (Analytical Formulation) and Part 3 (Numerical Application). The Editor is respectfully asked to arrange a coordinated review of Papers 1 to 3.}
\end{quote}

\begin{abstract}
This series establishes a topologically rigorous foundation for classifying instabilities in granular continua, extending Maxwell’s rigidity constraints from static isostaticity to dynamic, nonequilibrium processes. Rather than relying on empirical energy landscapes, we employ a discrete topological grain contact formulation, based on a Volumetric–Mechanical–Configurational (VMC) triad, to map contact-scale topology directly to macroscopic multiphysics coupling. We demonstrate that configurational failure is an inherent structural necessity in odd-parity systems, identified via the parity of the thermodynamic channel count $N$. By applying the discrete Poincar\'e lemma, we derive the structural necessity of the constraint $L_{VC} = 0$, which acts as a topological "cocycle filter" that isolates configurational failure modes. This framework reveals that the Parity Theorem, $\det(\mathsf{L}) = (-1)^N \det(\mathsf{L})$, guarantees a structural null-mode for all odd-$N$ systems, creating a "Gateway" layer where time-reversal symmetry is broken. When the basis-invariant Gateway number $\mathcal{G}_{\rm inv} \geq 1$, gyroscopic pumping drives deterministic non-modal transient amplification along this null direction. This provides a predictive architecture that bridges the gap between contact-scale dynamics and macroscopic localisation, identifying configurational drift as an emergent consequence of contact-scale topological entanglement rather than a byproduct of phenomenological noise.
\end{abstract}

\end{frontmatter}

\section{Introduction}
\label{sec:intro}
Classical constitutive modelling in multi-field solid mechanics frequently relies on empirical or experimentally driven parameterisations to describe the complex energy landscapes of coupled material systems. While highly successful in stable regimes, such empirical descriptions often encounter fundamental limitations when extended to multi-scale localisation phenomena and material bifurcations. In particular, traditional quasistatic frameworks identify the onset of localised failure with a loss of strong ellipticity. This condition typically renders numerical implementations mathematically ill-posed, non-unique, and pathologically mesh-dependent unless regularisation terms are added ad hoc.

To overcome these limitations, this series of manuscripts departs from empirical curve-fitting by adopting a foundational thermodynamic approach that enlarges the classical solution space into the dynamic evolutionary realm. This shift in approach offers four central advantages for multi-field mechanics:
\begin{enumerate}[label=(\roman*)]
    \item \emph{Dynamic Regularisation of Quasistatic Pathologies:} By embedding a coupled multi-field continuum within a dynamic transport framework, a traditionally ill-posed quasistatic bifurcation problem is recast as a well-posed evolutionary system. Within this expanded space, a static localisation profile naturally resolves as an emergent wave-attractor state or a standing wave mode.
    \item \emph{Compact Mathematical Architecture:} The framework condenses intricate, highly coupled multi-field interactions into an exceptionally compact operator form, where the fundamental symmetries of transport govern the global material evolution.
    \item \emph{Access to System Dynamics Theory:} Elevating the constitutive equations to an evolutionary format allows the direct deployment of the rich analytical framework of Nonlinear Dynamics. Tools such as spectral bifurcation theory, Lyapunov exponents, and non-normal transient growth analysis to assess both linear and nonlinear material bifurcation thresholds become directly available.
  \item \emph{Microphysics-Based Theory Across Scales:} Rather than mapping empirical energy landscapes macroscopically, this approach permits direct multiphysics coupling across scales. The resulting constitutive relationships are built directly from the underpinning contact-scale physics and topological invariants, providing a parameter-free \emph{topological} foundation. This bottom-up approach, initially in 1-D, is the focus of this paper series. Our special aim is to establish an explicit link from the traditional kinematic view, which requires constitutive assumptions, to a topological gatekeeper for dynamic instabilities whose existence is free of them. The series aim is to extend the parameter-free, topological character of Maxwell’s rigidity count, which historically dictates the isostatic limits of static frameworks~\cite{Maxwell1864}, from static isostaticity to dynamic, non-equilibrium instability. The general representation-free statement of this principle is developed in the companion Royal Society paper~\cite{NicotPoA}; here, we establish its minimal thermodynamic realisation via the VMC triad first introduced in Ref. \cite{Nicot2024}.
\end{enumerate}

The mathematical architecture that achieves this synthesis is the top-down Onsager-Casimir decomposition of any constitutive coupling operator into a symmetric dissipative part $\mathsf{D}$ and a skew-symmetric conservative part $\mathsf{L}$, which serves as the thermodynamic foundation underlying gradient flow theories, from phase field to gradient plasticity~\cite{Mielke2013}. 

This paper shows that the decomposition acquires a structural consequence that has not previously been identified: when the number of coupled thermodynamic channels is odd, the skew-symmetric block $\mathsf{L}$ necessarily has a zero eigenvalue, regardless of the values of the coupling coefficients. The resulting null eigenvector defines a direction in thermodynamic force-flux space along which no conservative restoring force acts. This \emph{Parity Theorem}, proved by the single algebraic identity $\det(\mathsf{L}) = (-1)^N \det(\mathsf{L})$, is the central result of the paper.

This thermodynamic synthesis provides a direct, elegant reconciliation with classical landmarks in solid mechanics. In the foundational framework of Hill~\cite{Hill1962} and Rudnicki and Rice~\cite{Rudnicki1975}, the onset of localised deformation is identified by the loss of ellipticity of the acoustic tensor ($\det(\mathbf{Q}(\mathbf{n})) = 0$). At this critical threshold, the material can no longer sustain a homogeneous increment of strain, collapsing into a macroscopic shear band or acceleration wave \cite{Hill1962}. 

While the augmented $\mathsf{A} = \mathsf{D} + \mathsf{L}$ decomposition provides a powerful framework for evaluating stability in entropic space, its thermodynamic thresholds must be strictly distinguished from mechanical bifurcation. When a standard linear stability analysis is conducted about a homogeneous steady state, the skew-symmetric circulation terms encapsulated in the Casimir block $\mathsf{L}$ drop out of the eigenvalue formulation for the onset of stationary modes. The thermodynamic stability of the system is thus left entirely dependent on the symmetric dissipative block $\mathsf{D}$. An entropic instability emerges when $\det(\mathsf{D}) = 0$, representing the point where the inherent self-dissipation of the system (such as when Darcy drainage outpaces internal chemical fluid generation) loses positive definiteness and can no longer damp scalar thermodynamic fluctuations.

A central advantage of the present framework is its capacity to probe the system \emph{before} this classical threshold is crossed. While classical analysis implies unconditional stability as long as $\det(\mathsf{D}) > 0$, our framework reveals that the topology of the skew-symmetric circulation matrix $\mathsf{L}$ governs a highly dynamic sub-threshold regime. For odd-dimensional coupled channels ($N$), the non-normal character of the $\mathsf{D}+\mathsf{L}$ operator triggers a topological null direction that completely evades dissipative regularisation once a basis-invariant Gateway number exceeds unity ($\mathcal{G}_{\rm inv} \geq 1$). This framework, therefore, uncovers a mathematically rigorous, microphysics-based precursor mechanism for localisation, operating well within the classically stable domain.

This parity-driven mechanism provides a formal algebraic extension of the classical bifurcation frameworks of Rice~\cite{Rice1976} and Rudnicki and Rice~\cite{Rudnicki1975}: where those analyses identify shear-band onset with loss of positive semi-definiteness of the \emph{symmetric} part of the acoustic tensor under the classical Legendre--Hadamard condition, the Gateway mechanism operates through the \emph{non-normal transient amplification} of the full non-symmetric operator $\mathsf{A}=\mathsf{D}+\mathsf{L}$, activating before the classical criterion is reached.

It is crucial to properly distinguish the stability bounds of the thermodynamic matrices ($\det(\mathsf{D})$ and $\det(\mathsf{A})$) from the mechanical continuum thresholds. Hill's sufficient condition for stability~\cite{Hill1958}, based on the positivity of second-order work, maps to the positive definiteness of the symmetric part of the continuum elasto-plastic tangent modulus. While conceptually analogous in its role as a symmetric stability limit, this mechanical threshold is distinct from $\det(\mathsf{D}) = 0$, which defines the loss of positive definiteness for the symmetric part of the thermodynamic channel-space operator.

Furthermore, the Rudnicki–Rice criterion evaluates the onset of localisation via the spatial acoustic tensor $\mathbf{Q}(\mathbf{n})$, with components $Q_{ik}(\mathbf{n})$~\cite{Rudnicki1975,Rice1976}. For genuinely non-associated flow, $\mathbf{Q}(\mathbf{n})$ is inherently non-symmetric, and its skew part fundamentally shifts the critical hardening modulus and band orientation. This spatial non-symmetry is distinct from the channel-space Casimir block $\mathsf{L}$ analysed here. 

Consequently, the Gateway mechanism is not a restatement of the acoustic-tensor result, but an independent channel-space precursor. Even when classical mechanical ellipticity is maintained (i.e., $\det(\mathbf{Q}(\mathbf{n})) > 0$ so the Rudnicki–Rice criterion is never triggered) and the symmetric dissipative block remains stable ($\sigma_{\min}(\mathsf{D})>0$), the gyroscopic pumping by $\mathsf{L}$ can bypass classical constraints to initiate a dispersive localisation pathway through the Gateway null direction. Because the quadratic form $\mathbf{X}^T\mathsf{L}\mathbf{X}=0$ removes $\mathsf{L}$ from stationary mode energy considerations, $\det(\mathsf{D})=0$ dictates the absolute boundary for stationary modes. However, the transient, non-normal routing driven by the full augmented operator $\mathsf{A} = \mathsf{D}+\mathsf{L}$ acts dynamically, routing energy and destabilising the system well before the symmetric limit is reached.

The two dynamical classes termed "Stable" and "Gateway" layers derived in this manuscript, are well known from the language of conservative and dissipative mechanics. A system with only conjugate imaginary eigenvalues $\pm i\omega$ in its coupling operator behaves like a conservative gyroscope: the forward and reverse trajectories traverse the same closed orbit and $t\!\to\!-t$ is a symmetry; any sustained forcing is confined to a bounded orbit about a shifted centre, so no secular accumulation occurs.

\begin{figure}[!htbp]
\centering
\resizebox{\linewidth}{!}{%
\begin{tikzpicture}[
  ax/.style={-{Stealth[scale=0.9]}, line width=0.8pt, gray!60},
  eig_stable/.style={circle, fill=stableblue!65, draw=stableblue!80, inner sep=3.2pt,
                     line width=0.6pt},
  eig_zero/.style={circle, fill=red!65, draw=red!80, inner sep=4.0pt,
                   line width=0.9pt},
  eig_imag/.style={circle, fill=stableblue!55, draw=stableblue!75, inner sep=3.0pt,
                   line width=0.6pt},
  orb/.style={stableblue!30, line width=1.0pt, dashed},
  gw_arrow/.style={-{Stealth[scale=1.0]}, red!60, line width=1.2pt},
  lbl/.style={font=\sffamily\footnotesize},
  hdr/.style={font=\sffamily\bfseries\small},
  annbox/.style={rectangle, rounded corners=3pt, inner sep=5pt,
                 font=\sffamily\footnotesize, align=left, line width=0.6pt,
                 text width=4.5cm}
]

\begin{scope}[xshift=0cm]
  \fill[stableblue!4, rounded corners=6pt] (-3.4,-4.0) rectangle (3.4,4.2);
  \draw[stableblue!25, rounded corners=6pt, line width=0.6pt]
        (-3.4,-4.0) rectangle (3.4,4.2);
  \node[hdr, text=stableblue] at (0, 3.9) {Stable Layer \quad ($N$ even)};
  \draw[ax] (-2.8,0) -- (2.8,0) node[right, lbl, text=gray!60] {Re};
  \draw[ax] (0,-3.2) -- (0,3.2) node[above, lbl, text=gray!60] {Im};
  \node[lbl, text=gray!50] at (0.12,-0.22) {$0$};
  \draw[gray!35, line width=0.4pt] (-0.1, 2.0) -- (0.1, 2.0);
  \draw[gray!35, line width=0.4pt] (-0.1,-2.0) -- (0.1,-2.0);
  \node[lbl, left, text=gray!55] at (-0.15, 2.0) {$i\omega$};
  \node[lbl, left, text=gray!55] at (-0.15,-2.0) {$-i\omega$};
  \node[eig_stable] at (0, 2.0) {};
  \node[eig_stable] at (0,-2.0) {};
  \node[lbl, right=2pt, text=stableblue] at (0, 2.0) {$+i\omega_1$};
  \node[lbl, right=2pt, text=stableblue] at (0,-2.0) {$-i\omega_1$};
  \draw[orb] (0,0) circle (1.2cm and 0.75cm);
  \draw[orb, -{Stealth[scale=0.8]}] (1.2,0) arc (0:50:1.2cm and 0.75cm);
  \draw[orb, -{Stealth[scale=0.8]}] (-1.2,0) arc (180:230:1.2cm and 0.75cm);
  \node[lbl, text=stableblue] at ( 1.95, 0.55) {\footnotesize $t\!\to\!+\infty$};
  \node[lbl, text=stableblue] at (-1.95,-0.55) {\footnotesize $t\!\to\!-\infty$};
  \node[annbox, fill=stableblue!9, draw=stableblue!35, text=black] at (0,-6.2) {
    $\mathrm{spec}(\mathsf{L}) = \{\pm i\omega_1, \ldots\}$\\
    No zero eigenvalue\\
    $\det(\mathsf{L}) \neq 0$ generically\\
    \textit{$t\!\to\!-t$ is a symmetry}\\
    \textbf{Forced response confined}
  };
\end{scope}

\begin{scope}[xshift=8.5cm]
  \fill[orange!5, rounded corners=6pt] (-3.4,-4.0) rectangle (3.4,4.2);
  \draw[orange!40, rounded corners=6pt, line width=0.6pt]
        (-3.4,-4.0) rectangle (3.4,4.2);
  \node[hdr, text=orange!70!black] at (0, 3.9) {Gateway Layer \quad ($N$ odd)};
  \draw[ax] (-2.8,0) -- (2.8,0) node[right, lbl, text=gray!60] {Re};
  \draw[ax] (0,-3.2) -- (0,3.2) node[above, lbl, text=gray!60] {Im};
  \node[lbl, text=gray!50] at (0.22,-0.32) {$0$};
  \draw[gray!35, line width=0.4pt] (-0.1, 2.0) -- (0.1, 2.0);
  \draw[gray!35, line width=0.4pt] (-0.1,-2.0) -- (0.1,-2.0);
  \node[lbl, left, text=gray!55] at (-0.15, 2.0) {$i\omega$};
  \node[lbl, left, text=gray!55] at (-0.15,-2.0) {$-i\omega$};
  \node[eig_imag] at (0, 2.0) {};
  \node[eig_imag] at (0,-2.0) {};
  \node[lbl, right=2pt, text=stableblue] at (0, 2.0) {$+i\omega_1$};
  \node[lbl, right=2pt, text=stableblue] at (0,-2.0) {$-i\omega_1$};
  \node[eig_zero] (z) at (0,0) {};
  \node[annbox, fill=red!8, draw=red!40, text=black,
        text width=3.5cm] at (-2.8, 1.4) {
    \textcolor{red!65}{\textbf{Null eigenvalue}}\\
    unconditional\\
    for any $L_{\alpha\beta}$\\
    \emph{Gateway direction}
  };
  \draw[gw_arrow, line width=1.6pt] (0.0, 0.0) -- (2.2, 0.0);
  \node[lbl, text=red!55, above] at (1.5, 0.05) {\footnotesize irreversible pathway};
  \node[annbox, fill=orange!9, draw=orange!45, text=black] at (0,-6.2) {
    $\mathrm{spec}(\mathsf{L}) = \{0, \pm i\omega_1, \ldots\}$\\
    \textcolor{red!65}{Zero eigenvalue unconditional}\\
    $\det(\mathsf{L}) = 0$ always\\
    \textit{null mode unconfined under forcing}\\
    \textbf{Arrow of time (driven)}
  };
\end{scope}

\end{tikzpicture}
}
\caption{Spectral topology of the circulation matrix $\mathsf{L}$ for even ($N=2$) and odd ($N=3$) channel counts. \textit{Left (Stable Layer, even $N$):} All eigenvalues form purely imaginary conjugate pairs $\pm i\omega_j$, yielding closed orbits in force–flux space that preserve time-reversal symmetry ($t \mapsto -t$). This constitutes the dynamical generalisation of Maxwell’s static rigidity count~\cite{Maxwell1864}, where constrained degrees of freedom produce conservative confinement. \textit{Right (Gateway Layer, odd $N$):} The identity $\det(\mathsf{L})=0$ unconditionally forces a structural zero eigenvalue. We identify this as a structural topological property of odd-dimensional conservative coupling. The resulting null direction (red) is the direction along which the \emph{driven} response is unconfined: under sustained forcing with a component along $\mathbf{v}_0$, the state accumulates irreversibly, whereas an even-$N$ system confines any forced response to a bounded orbit. The free operators $e^{\mathsf{L}t}$ are time-reversal symmetric for both parities; the constitutive arrow of time is a property of the forced, open system, not of $\mathsf{L}$ in isolation.}
\label{fig:arrow_of_time}
\end{figure}
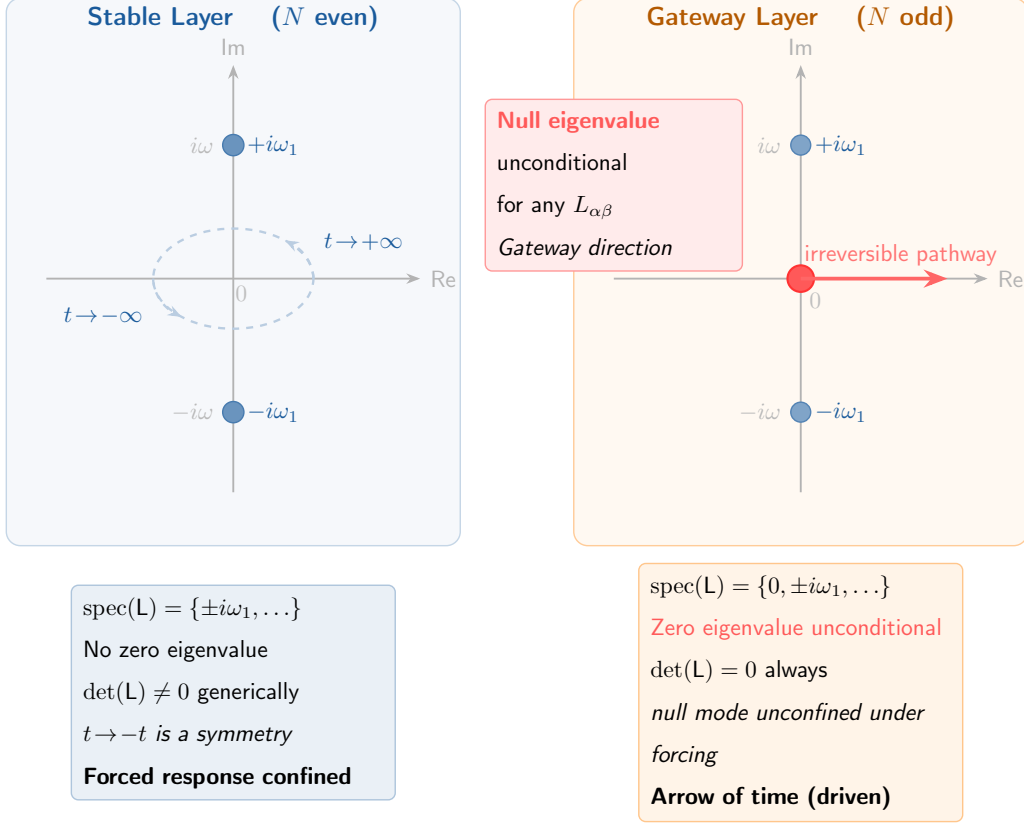

A system that possesses an additional zero eigenvalue behaves differently: one direction in force-flux space carries no conservative restoring force. Under the \emph{free} operator this null component is merely conserved (frozen, $\mathbf{v}_0^T\mathbf{q}=\mathrm{const}$); it is under \emph{sustained forcing} with a component along that direction that the state accumulates without confinement and the forward path cannot be retraced. The constitutive arrow of time is therefore a property of the driven, open system: the free skew operators $e^{\mathsf{L}t}$ are time-reversal symmetric for both parities, and it is the \emph{confinement} of the forced response---bounded for even $N$, unbounded along the null direction for odd $N$---that distinguishes the two classes. The Parity Theorem identifies exactly which class a multi-field system belongs to: \emph{even} channel count $N$ always produces the first (Stable Layer); \emph{odd} $N$ always produces the second (Gateway Layer).
Figure~\ref{fig:arrow_of_time} encodes this translation directly in the eigenvalue spectra of $\mathsf{L}$.

The null direction is a topological feature of odd-dimensional conservative coupling. By itself, it is dynamically inert: it neither produces entropy nor drives instability. These effects emerge only when the activation condition $\mathcal{G}_{\rm inv}\geq 1$ is met (Section~\ref{sec:activation}). Accordingly, the claim in this paper is twofold: odd-dimensional conservative coupling guarantees the existence of a structural null direction, whereas activation of that direction and the associated non-reciprocal energy routing requires the coupling strength to overcome dissipative capacity, as measured by the Gateway number $\mathcal{G}_{\rm inv}$.

The Hasse lattice of thermodynamic channel subsets provides an organising principle for the resulting coupling classes: at every odd-level node the null direction creates a geometric predisposition toward non-reciprocal energy routing when $\mathcal{G}_{\rm inv}\geq 1$.
Sections~\ref{sec:hasse} and~\ref{sec:cascade} develop this hierarchy in full, illustrated in Figs.~\ref{fig:hasse_vmc} and~\ref{fig:satake_vs_hasse}.

When $\mathcal{G}_{\rm inv}\geq 1$ the null direction is no longer suppressed by local dissipation and the boundary value problem requires gradient regularisation to restore well-posedness.
The minimal $N=3$ Gateway realisation used throughout as the motivating example is the three-channel configurational contact framework of Nicot et al.~\cite{Nicot2024}, the VMC triad, introduced in Section~\ref{sec:satake}.

\paragraph*{Relation to odd elasticity and non-reciprocal media}
The skew block $\mathsf{L}$ places this work adjacent to the physics of odd-elastic and non-reciprocal active solids~\cite{Scheibner2020,Fruchart2023}, where antisymmetric moduli likewise generate work-free circulation, zero modes, and non-normal (exceptional-point) amplification. Our contribution is not the elementary observation that an odd-dimensional antisymmetric operator has a nontrivial null space, but three specific results built on it: (i) the identification of the \emph{origin} of that null space in granular contact topology, through the Satake $p$--$c$--$v$ to $V$--$M$--$C$ correspondence; (ii) the resulting parity classification of an entire Hasse lattice of channel subsets into alternating Stable and Gateway layers; and (iii) the reduction of the activation crossover to a single basis-invariant ratio $\mathcal{G}_{\rm inv}$. Crucially, the zero mode here is \emph{mandated} by the parity of the thermodynamic channel count rather than \emph{engineered} through material design, which is the essential distinction from the tuned odd moduli of active-matter systems and from the designed exceptional points of non-Hermitian metamaterials. The transient-growth machinery itself, the Gateway number and the peak amplitude $M(\mathcal{G}_{\rm inv})$, is the mechanics counterpart of non-modal (pseudospectral) stability theory in hydrodynamics, where non-normal operators produce large finite-time amplification despite a stable spectrum; here too the novelty is not that machinery but the parameter-free \emph{topological origin} of the non-normality in a passive, conservative granular contact network.

The paper proceeds through the Onsager framework (Section~\ref{sec:framework}), Hasse lattice and Satake connection (Section~\ref{sec:hasse}), Parity Theorem (Section~\ref{sec:parity}), Gateway number and activation (Section~\ref{sec:activation}), multiscale hierarchy (Section~\ref{sec:cascade}), and finish with the Discussion and Conclusions (Section~\ref{sec:discussion}, \ref{sec:conclusion}).
\section{The Augmented Onsager Framework: Scope and Applicability}
\label{sec:framework}

\begin{figure}[!htbp]
\centering
\resizebox{\linewidth}{!}{%
\begin{tikzpicture}[
  mbox/.style={rectangle, minimum width=0.85cm, minimum height=0.85cm,
               line width=0.55pt, inner sep=0pt},
  label_s/.style={font=\sffamily\small, align=center},
  label_f/.style={font=\sffamily\footnotesize, align=center, text width=4.8cm},
  propbox/.style={rectangle, rounded corners=4pt, line width=0.8pt,
                  minimum width=5.2cm, minimum height=3.6cm,
                  inner sep=8pt, align=left, font=\sffamily\footnotesize},
  eqsym/.style={font=\Large\bfseries, text=gray!55}
]

\begin{scope}[xshift=0cm, yshift=0cm]
  \node[label_s, text=gray!60] at (1.25, 1) {$\mathsf{A}$};
  \node[label_f, text=gray!55] at (1.25, -3.0) {General constitutive\\operator};
  \foreach \r in {0,1,2} {
    \foreach \c in {0,1,2} {
      \node[mbox, fill=gray!18, draw=gray!40] at (\c*0.95, -\r*0.95) {};
    }
  }
\end{scope}

\node[eqsym] at (4.5, -0.95) {$=$};

\begin{scope}[xshift=5.5cm, yshift=0cm]
  \node[label_s, text=stableblue] at (1.25, 1.0) {$\mathsf{D} = \mathsf{D}^T \geq 0$};
  \node[label_f, text=stableblue] at (1.25, -3.0) {Symmetric\\dissipative block};
  \foreach \d in {0,1,2} {
    \node[mbox, fill=stableblue!35, draw=stableblue!60] at (\d*0.95, -\d*0.95) {};
  }
  \foreach \p/\q in {0/1, 0/2, 1/2} {
    \node[mbox, fill=stableblue!14, draw=stableblue!35] at (\q*0.95, -\p*0.95) {};
    \node[mbox, fill=stableblue!14, draw=stableblue!35] at (\p*0.95, -\q*0.95) {};
  }
\end{scope}

\node[eqsym] at (10.0, -0.95) {$+$};

\begin{scope}[xshift=11.0cm, yshift=0cm]
  \node[label_s, text=orange!80!black] at (1.25, 1.) {$\mathsf{L} = -\mathsf{L}^T$};
  \node[label_f, text=orange!70!black] at (1.25, -3.0) {Skew-symmetric\\conservative block};
  \foreach \d in {0,1,2} {
    \node[mbox, fill=gray!5, draw=gray!30] at (\d*0.95, -\d*0.95)
      {\scriptsize $0$};
  }
  \foreach \p/\q in {0/1, 0/2, 1/2} {
    \node[mbox, fill=orange!22, draw=orange!60] at (\q*0.95, -\p*0.95)
      {\tiny $+L$};
    \node[mbox, fill=orange!10, draw=orange!40] at (\p*0.95, -\q*0.95)
      {\tiny $-L$};
  }
\end{scope}

\begin{scope}[yshift=-5.8cm]
  \node[propbox, fill=stableblue!7, draw=stableblue!50, text=black,
        text width=4.8cm] at (6.5, -1.) {
    \textcolor{stableblue}{\textbf{Dissipative D-block}}\\[3pt]
    $\bullet$ Entropy production:\\
    \quad $\sigma_{\rm irr} = \mathbf{X}^T\!\mathsf{D}\,\mathbf{X} \geq 0$\\[2pt]
    $\bullet$ Analogs in solid mechanics:\\
    \quad viscoplastic tangent modulus,\\
    \quad Thermal diffusivity,\\
    \quad Darcy / Fick diffusivity
  };
  \node[propbox, fill=orange!7, draw=orange!60, text=black,
        text width=4.8cm] at (12.5, -1.) {
    \textcolor{orange!80!black}{\textbf{Conservative L-block}}\\[3pt]
    $\bullet$ Work-free coupling:\\
    \quad $\mathbf{X}^T\!\mathsf{L}\,\mathbf{X} = 0$ identically\\[2pt]
    $\bullet$ Analogs in solid mechanics:\\
    \quad Cosserat cross coupling,\\
    \quad rotational forcing,\\
    \quad THMCE cross-coupling
  };
\end{scope}

\end{tikzpicture}
}
\caption{The Onsager-Casimir decomposition of the constitutive operator
  $\mathsf{A} = \mathsf{D} + \mathsf{L}$.
  \textit{Left:} Schematic of the general $N \times N$ operator $\mathsf{A}$,
  shown for $N=3$.
  \textit{Centre:} The symmetric dissipative block $\mathsf{D}$ (blue), with
  strong diagonal entries and symmetric off-diagonal entries; it governs local
  entropy production $\sigma_{\rm irr} = \mathbf{X}^T\!\mathsf{D}\mathbf{X} \geq 0$.
  \textit{Right:} The skew-symmetric conservative block $\mathsf{L}$ (orange),
  with zero diagonal and antisymmetric off-diagonal entries ($+L$ above, $-L$
  below); it performs no work on the system ($\mathbf{X}^T\!\mathsf{L}\mathbf{X}
  = 0$ identically).
  Familiar solid-mechanics operators occupy each slot: the
  viscoplastic tangent modulus and Fourier diffusivity belong to $\mathsf{D}$;
  Cosserat couple-stress coupling, as well as generalised cross-coupled 
  thermo-hydro-mechanical-chemical-electrical (THMCE) processes \cite{RegenHu2024} belong to $\mathsf{L}$.}
\label{fig:DL_decomposition}
\end{figure}
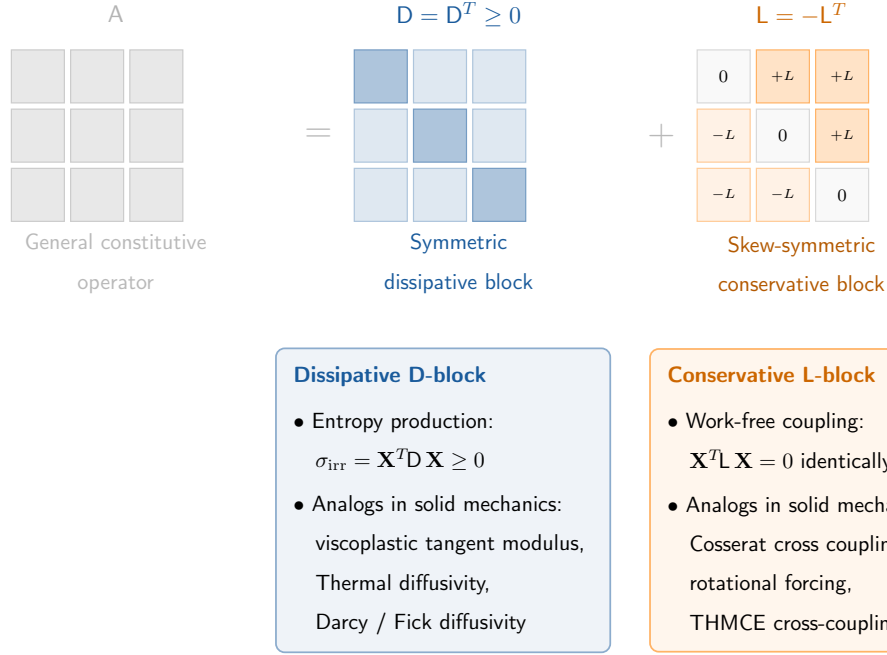

The augmented $\mathsf D+\mathsf L$ framework may be viewed as a natural extension of the gradient-flow structures that underpin modern multi-field solid mechanics. In the gradient-system formulation of Mielke~\cite{Mielke2013}, dissipative processes such as Allen--Cahn, Cahn--Hilliard, and reaction--diffusion evolution are written as

\begin{equation}
\dot U=-\mathcal K(U)\,\mathrm D\Psi(U),
\end{equation}

where $\Psi$ is the Helmholtz free-energy functional, $\mathcal K$ is a symmetric positive-semidefinite Onsager operator, and $\dot U$ represents the material (substantial) time derivative of the internal energy state. Explicitly defining $\dot U$ in the material frame isolates the purely irreversible thermodynamic trajectory driven by $-\mathrm{D}\Psi(U)$ from non-dissipative, reversible spatial transport mechanisms. The present formulation retains this underlying thermodynamic structure but enlarges the admissible constitutive operators to include reversible coupling between transport channels (see Fig. \ref{fig:DL_decomposition}):

\begin{equation}
\mathsf A=\mathsf D+\mathsf L,
\end{equation}

with $\mathsf  D=(\mathsf A+\mathsf A^T)/2\ge0$ the dissipative block and $\mathsf L=(\mathsf A-\mathsf A^T)/2$ the skew-symmetric conservative block.

Throughout, $N$ denotes the number of intensive thermodynamic channels coupled at a material point (a homogenised representative volume element), not a spatial degree of freedom. Mesh refinement or basis enrichment increases the spatial resolution $d$ but leaves $N$ unchanged: the parity classification is a property of the local constitutive operator $\mathsf{A}_{\alpha\beta}$ and is invariant under spatial discretisation.

Because $\mathbf X^T\mathsf L\mathbf X=0$ identically, $\mathsf{L}$ neither produces entropy nor performs net work; it redistributes energy between thermodynamic channels (e.g. thermo-hydro-mechanical-chemical-electrical, THMCE) through conservative interactions, analogous to gyroscopic coupling, convective transport, or Cosserat couple-stress effects~\cite{RegenHu2024}. The Onsager--Casimir reciprocal relations~\cite{Casimir1945} guarantee that every constitutive operator can be decomposed uniquely in this way. The decomposition is illustrated in Fig.~\ref{fig:DL_decomposition}.

Mielke's gradient systems correspond to the special case $\mathsf{L}=0$, whereas the present theory considers the general case $\mathsf{L}\neq0$. In this context, any non-dissipative kinematics arising from material motion or advection map directly into the skew-symmetric block $\mathsf{L}$, leaving $\mathsf{D}$ cleanly mapped to the material-frame dissipation governed by $\mathcal{K}$. The Parity Theorem developed below concerns this conservative component and shows that the topology of $\mathsf{L}$ depends fundamentally on the number of active coupled thermodynamic channels $N$.

Importantly, the Parity Theorem requires far less structure than classical Onsager theory. It is an algebraic property of real skew-symmetric matrices and therefore holds for any $N$, independent of proximity to equilibrium, driving-force magnitude, or the physical interpretation of the channels. The augmented $\mathsf{D}+\mathsf{L}$ framework~\cite{RegenHu2023b,RegenHu2024,RegenEtAl2026a} circumvents the near-equilibrium limitation by operating at the level of thermodynamic transport channels, applying the Onsager--Casimir decomposition locally at the contact scale where $\mathsf{D}$ governs irreversible entropy production and $\mathsf{L}$ governs reversible inter-channel energy exchange.

For rapidly driven systems where local quasi-static equilibrium cannot be assumed, the GENERIC framework of Grmela and \"Ottinger~\cite{Grmela1997,Ottinger1997} provides the fully nonlinear extension; in the linear regime, $L$ identifies with $\mathsf{L}$ and $M$ with $\mathsf{D}$. The Casimir condition $\mathsf{L}\nabla_{\mathbf{X}}S=\mathbf{0}$ identifies $\mathbf{v}_0\in\ker(\mathsf{L})$ as a top-level symmetry of the conservative transport (see Supplementary Material, Section~S1).

The central claim of this paper is that the parity of the channel count $N$ determines whether conservative coupling can be regularised by local dissipation. The Gateway number $\mathcal{G}_{\rm inv}$ provides the corresponding activation criterion and quantifies the proximity to topological activation.

Crucially, while the augmented $\mathsf{D}+\mathsf{L}$ operator acts linearly at a given local thermodynamic tangent state, the addition of the circulation matrix $\mathsf{L}$ constitutes the minimal analytically tractable framework from which macroscopic nonlinearity emerges. In an odd-$N$ system, this framework forces a topological null direction that completely evades dissipative regularisation once $\mathcal{G}_{\rm inv} \geq 1$. By providing an open, non-reciprocal path for energy routing, this localised structure serves as the direct precursor mechanism for the full multiscale, multiphysics coupling and nonlinear bifurcation theory realised in the companion manuscripts for an idealised 1-D case in ~\cite{RegenNicot2026dilatancy,RegenNicot2026numerical}.

In our approach the gradient-flow architecture $\mathbf{J} = \mathsf{A}\mathbf{X}$ provides the natural setting for identifying the topological origin of instability within the Satake void network and not in the traditional geomechanical and elastoplastic frameworks. These are overwhelmingly formulated in terms of stress and Lagrangian strain increments. To bridge these paradigms without obscuring the parameter-free \emph{topological} structure of the main text, a formal Legendre–Fenchel transformation of the local dissipation potential $\Phi(\mathbf{J})$ is detailed in Supplementary Material, Section~S2. There, we establish the exact compliance mapping $\boldsymbol{\varepsilon}_d = V^{-1}\mathbb{S}_{\rm dev} : \mathbf{A}_{\rm conf}$ linking the macroscopic deviatoric strain $\boldsymbol{\varepsilon}_d$ to the configurational force-moment $\mathbf{A}_{\rm conf}$, where the thermodynamic driving forces $\mathbf{X} = -\nabla_{\mathbf{q}}\Psi$ are derived from the Helmholtz free energy $\Psi$. Crucially, this dual representation demonstrates why the internal fabric coordinate $\boldsymbol{\xi}_{\rm conf}$ cannot be algebraically slaved to classical macroscopic strain: at the onset of Gateway activation ($\mathcal{G}_{\rm inv} \geq 1$), the deviatoric tangent stiffness degenerates at the loss of controllability and the effective compliance diverges ($\det(\mathbb{S}_{\rm dev}) \to \infty$). This divergence marks a loss of controllability where standard finite-strain representations fail, whereas the independent topological void network remains mathematically well-posed and non-singular.

\paragraph*{Index notation convention}
Latin subscripts $i,j,k,l$ run over spatial directions ($1\ldots d$) and Greek subscripts $\alpha,\beta$ run over the $N$ thermodynamic channels.
A fourth-order spatial constitutive tensor is denoted $\mathbb{C}_{ijkl}$; an $N\times N$ channel-space operator is denoted $\mathsf{A}_{\alpha\beta}$ or simply $\mathsf{A}$.
The acoustic tensor $Q_{ik}(\mathbf{n})=\mathbb{C}_{ijkl}n_j n_l$ lives in \emph{spatial} index space; the augmented Onsager operator $\mathsf{A}_{\alpha\beta}$ lives in \emph{channel} index space.

\paragraph*{Notation glossary}
The symbols used throughout Part~1 are collected in Table~\ref{tab:notation}. The font distinction is load-bearing: sans-serif $\mathsf{A},\mathsf{D},\mathsf{L}$ always denote the channel-space Onsager operators, while bold upright letters denote tensors and vectors. The same letter in different fonts, therefore, denotes different objects, the clearest case being the channel-space operator $\mathsf{A}$ versus the configurational force-moment $\mathbf{A}_{\rm conf}$. Following the Hill--Rice convention, the spatial acoustic tensor is written $\mathbf{Q}(\mathbf{n})$ (components $Q_{ik}$), which removes any further collision with the operator $\mathsf{A}$.

\begin{table}[htbp]
\centering
\caption{Notation used throughout Part~1.}
\label{tab:notation}
\begin{tabular}{@{}ll@{}}
\hline
\multicolumn{2}{@{}l}{\emph{Channel-space operators ($N\times N$)}}\\
\hline
$\mathsf{A}=\mathsf{D}+\mathsf{L}$ & augmented Onsager operator at a material point\\
$\mathsf{D}=(\mathsf{A}+\mathsf{A}^T)/2\succeq0$ & symmetric dissipative block (entropy production)\\
$\mathsf{L}=(\mathsf{A}-\mathsf{A}^T)/2$ & skew-symmetric reversible block (conservative coupling)\\
$\mathsf{A}_{\alpha\beta}$ & operator in channel indices $\alpha,\beta=1\ldots N$\\
$\mathsf{A}_\sigma,\,\mathsf{D}_\sigma,\,\mathsf{L}_\sigma$ & operators of the channel subset (Hasse node) $\sigma$\\
$\mathsf{A}_{\rm eff}$ & effective reduced operator at the Gateway\\
\hline
\multicolumn{2}{@{}l}{\emph{Tensors and stresses (distinct from the operator)}}\\
\hline
$\mathbf{A}_{\rm conf}$ & configurational force-moment, work-conjugate to $\boldsymbol{\xi}_{\rm conf}$\\
$\mathbf{Q}(\mathbf{n}),\ Q_{ik}(\mathbf{n})$ & acoustic tensor (spatial), Hill--Rice localisation\\
$\mathbb{C}_{ijkl}$ & fourth-order spatial constitutive tensor\\
$\boldsymbol{\xi}_{\rm conf}=\operatorname{dev}\langle\mathbf{n}_{c'}\!\otimes\mathbf{n}_{c'}\rangle$ & configurational variable (deviatoric fabric)\\
\hline
\multicolumn{2}{@{}l}{\emph{Vectors}}\\
\hline
$\mathbf{X}$ & thermodynamic force vector\\
$\mathbf{q}$ & conjugate flux / state vector\\
$\mathbf{v}_0\in\mathcal{N}(\mathsf{L})$ & parity-mandated null mode (Gateway direction)\\
$\mathbf{n}_{c'}$ & contact-normal (fabric) direction\\
\hline
\multicolumn{2}{@{}l}{\emph{Scalars and indices}}\\
\hline
$N$ & number of coupled thermodynamic channels\\
$d$ & spatial dimension / resolution\\
$i,j,k,l\,;\ \alpha,\beta$ & spatial indices ($1\ldots d$); channel indices ($1\ldots N$)\\
$\mathcal{G}_{\rm inv}$ & Gateway number (non-normality activation criterion)\\
$L_{VM},\,L_{MC}$ & off-diagonal coupling entries of $\mathsf{L}$\\
$V,M,C$ & volumetric, mechanical, configurational channels\\
$\mathcal{I},\,\sigma$ & channel index set; a channel subset (Hasse node)\\
\hline
\end{tabular}
\end{table}

\section{The Hasse Lattice as Organising Principle}
\label{sec:hasse}

Drawing upon Schnakenberg's graph-theoretic micro-macro model \cite{Schnakenberg1976}, the parity classification acquires full structural clarity when organised within the Hasse lattice of thermodynamic channel subsets \cite{Stanley2012}, the dynamic algebraic counterpart to Satake's static spatial contact-network topology \cite{Satake1987}.
At every odd-level node the restricted operator $\mathsf{L}_\sigma$ has a one-dimensional null space; at every even-level node the null space is trivial.
The Parity Theorem therefore imposes an alternating pattern of trivial and non-trivial null-space structures across the entire lattice. The odd-level non-trivial null space represents a structural null direction that is intrinsically entropy-neutral and dynamically dormant under standard loading. It defines a path of "least thermodynamic resistance" because it is a subspace entirely uncoupled from the gyroscopic constraint forces that normally recirculate or trap energy within specific thermodynamic channels.

\begin{definition}[Hasse lattice of channel subsets]
\label{def:hasse}
Let $\mathcal{I} = \{1, 2, \ldots, N\}$ be the index set of thermodynamic channels. The Hasse structure $\mathcal{H}_N$ is the partially ordered set of all non-empty subsets $\sigma \subseteq \mathcal{I}$ ordered by inclusion. The level of a node $\sigma$ is its cardinality $|\sigma|$. An edge connects $\sigma$ to $\tau$ if and only if $\sigma \subset \tau$ and $|\tau| = |\sigma| + 1$. 
\end{definition}

The resulting structure is the Boolean lattice on $\mathcal{I}$ with the empty set removed, consisting of $2^N - 1$ nodes arranged in $N$ levels. Each node $\sigma$ at level $m$ represents an $m$-channel subsystem with associated Onsager operator $\mathsf{A}_\sigma = \mathsf{D}_\sigma + \mathsf{L}_\sigma$.

A node of this lattice is not merely a principal submatrix carved from a larger coupled operator. It represents a physical regime in which only the channels in $\sigma$ are dynamically active, the remaining channels being either absent or frozen on the timescale of interest, so that $\mathsf{L}_\sigma$ is the genuine reduced conservative operator of that regime and its parity carries the same physical content as the parity of the full operator. The distinction matters for how the classification is read: if the other channels are present and merely weakly coupled rather than switched off, they are not removed from the count, and the Parity Theorem must be applied to the full channel number $N$ and not to $\mathsf{L}_\sigma$. A weakly coupled even-$N$ system therefore does not inherit the Gateway of an odd-$N$ sub-node; the sub-node describes a different, more restricted regime. Read this way, the Hasse structure classifies admissible \emph{active-channel} regimes, and the alternation of Stable and Gateway layers across levels is a statement about which regimes are dynamically realisable, not about arbitrary algebraic restrictions of a fixed matrix.

\subsection{Connection to Satake's graph-theoretical framework}
\label{sec:satake}

\paragraph*{Nomenclature note}
Satake~\cite{Satake1993,Satake1987,Satake1978} uses an incidence matrix $D_{cp}$ and loop matrix $L_{vc}$ that are discrete integer-valued topological arrays; they share notation with the Onsager tensors $\mathsf{D}$ and $\mathsf{L}$ purely by historical coincidence.

\paragraph*{Satake's framework}
In Satake's formulation, three types of entity are defined: particle centres $p$, contact points $c$, and void centres $v$.
The fundamental topological identity, proved as equation (A5) of Ref.~\cite{Satake1993}, is
\begin{equation}
L_{vc}\,D_{cp} = 0,
\label{eq:satake_identity}
\end{equation}
the discrete Poincaré lemma: the composition of the loop operator (discrete curl) and the incidence operator (discrete gradient) is identically zero for any oriented planar graph, independently of grain shape or mechanical properties.
Satake's framework answers the question \emph{where do forces balance?} It does not address energy routing, thermodynamic channel classification, or the activation conditions for non-reciprocal transfer.

\paragraph*{The VMC configurational contact framework}
Nicot et al.~\cite{Nicot2024} reach the same three-part structure from contact mechanics, resolving the state of a granular assembly into three independent thermodynamic channels, the VMC triad. The \emph{volumetric} channel $V$ carries the normal, compaction part of the contact deformation, with the overlap rate as its flux and the normal contact force as its conjugate. The \emph{mechanical} channel $M$ carries the tangential, shear transmission between grains, with the sliding rate as its flux and the tangential contact force as its conjugate; it is the bridge through which the other two communicate. The \emph{configurational} channel $C$ carries the reorganisation of the contact-network geometry, the fabric, measured by the deviatoric fabric tensor $\boldsymbol{\xi}_{\rm conf}=\operatorname{dev}\langle\mathbf{n}_{c'}\!\otimes\mathbf{n}_{c'}\rangle$, where the average is taken over all active contacts within the representative volume, and the configurational force-moment is its conjugate. We map these channels to a state vector $\dot{\mathbf{q}} = (\dot{\varepsilon}_V, \dot{\varepsilon}_M, \dot{\xi}_C)^T$, where $\dot{\xi}_C$ represents the rate of change of the dominant scalar component of the fabric evolution. The property that Nicot et al.\ establish, and on which the present classification turns, is that this configurational channel is a genuinely independent kinematic degree of freedom. This recognition formally elevates Satake's classical static contact topology into a fully dynamic thermodynamic regime.

Through Satake's duality $L_{vc}\,D_{cp}=0$ (equation~\eqref{eq:satake_identity}), the contact-orientation fabric $\langle\mathbf{n}_{c'}\!\otimes\mathbf{n}_{c'}\rangle$ is the dual descriptor of the void-cell geometry, so $\boldsymbol{\xi}_{\rm conf}$ is independent of the contact \emph{forces}, not of the contacts themselves: it is a geometric degree of freedom carried by the void network rather than slaved to the force-carrying contact network. That independence is what keeps the channel count odd ($N=3$) rather than collapsing $C$ into $M$, and it is the microscopic origin of the Gateway developed in the sections that follow.

\paragraph*{The $p$--$c$--$v$ to $V$--$M$--$C$ identification}
The deep connection between the two frameworks is established by identifying Satake's three entity types with the three thermodynamic channels of the VMC system: particle centres $p \leftrightarrow$ volumetric channel $V$; contact points $c \leftrightarrow$ mechanical channel $M$; void centres $v \leftrightarrow$ configurational channel $C$.

\begin{figure}[!htbp]
\centering
\begin{tikzpicture}[
  scale=0.95,
  grain/.style={circle, draw=gray!55, fill=gray!12, line width=0.7pt,
                minimum size=1.55cm, inner sep=0pt},
  pcent/.style={circle, fill=black, inner sep=1.3pt},
  cpt/.style={circle, draw=black, fill=white, inner sep=1pt, line width=0.7pt},
  vpt/.style={circle, draw=stableblue!70, fill=stableblue!22, line width=0.9pt, inner sep=1.8pt},
  branch/.style={black, line width=1.0pt},
  loop/.style={stableblue!55!black, dashed, line width=0.9pt},
  hnode/.style={rounded corners=2.5pt, draw, line width=0.8pt,
                minimum width=1.1cm, minimum height=0.7cm,
                font=\small, align=center, inner sep=2pt},
  single/.style={hnode, draw=gray!65, fill=gray!10},
  stable/.style={hnode, draw=stableblue!70, fill=stableblue!12},
  gateway/.style={hnode, draw=orange!80, fill=orange!18, minimum width=1.6cm},
  arr/.style={->, gray!65, line width=0.55pt, >=Stealth},
  lbl/.style={font=\footnotesize},
  ttl/.style={font=\sffamily\bfseries\small},
  sttl/.style={font=\sffamily\footnotesize, gray!55!black}
]
\begin{scope}[xshift=0cm]
\node[ttl]  at (1.5, 5.0)  {(a) Satake spatial graph};
\node[sttl] at (1.5, 4.55) {static, Euclidean};
\node[grain] at (0,0)     {};
\node[grain] at (3,0)     {};
\node[grain] at (1.5,2.6) {};
\draw[branch] (0,0) -- (3,0);
\draw[branch] (0,0) -- (1.5,2.6);
\draw[branch] (3,0) -- (1.5,2.6);
\node[cpt,label={[lbl]below:$c_{12}$}]              (c12) at (1.5, 0)    {};
\node[cpt,label={[lbl,xshift=-3pt]above left:$c_{13}$}] (c13) at (0.75,1.3) {};
\node[cpt,label={[lbl,xshift= 3pt]above right:$c_{23}$}](c23) at (2.25,1.3) {};
\node[vpt,label={[lbl]right:$v$}] (v) at (1.5,0.87) {};
\draw[loop] (c12) to[bend left=18] (c13);
\draw[loop] (c13) to[bend left=18] (c23);
\draw[loop] (c23) to[bend left=18] (c12);
\node[pcent,label={[lbl]below left: $p_1$}] at (0,0)     {};
\node[pcent,label={[lbl]below right:$p_2$}] at (3,0)     {};
\node[pcent,label={[lbl]above:       $p_3$}] at (1.5,2.6) {};
\node[align=center, font=\footnotesize] at (1.5,-2.5)
  {Nodes embedded in $\mathbb{R}^2$.\\
   Cycle around void: $L_{vc}\,D_{cp}=0$\\
   Cycle space $\mathcal{Z}\neq\{0\}$\\
   answers: \emph{where do forces balance?}};
\end{scope}
\begin{scope}[xshift=8cm]
\node[ttl]  at (1.8, 5.0)  {(b) Hasse hierarchy};
\node[sttl] at (1.8, 4.55) {dynamic, algebraic};
\node[gateway] (VMC) at (1.8, 3.6) {VMC\\($k{=}3$)};
\node[stable]  (VM)  at (-0.4,1.8) {VM\\($k{=}2$)};
\node[stable]  (MC)  at ( 1.8,1.8) {MC\\($k{=}2$)};
\node[stable]  (VC)  at ( 4.0,1.8) {VC\\($k{=}2$)};
\node[single]  (V)   at (-0.4,0)   {V};
\node[single]  (M)   at ( 1.8,0)   {M};
\node[single]  (C)   at ( 4.0,0)   {C};
\draw[arr] (V) -- (VM); \draw[arr] (M) -- (VM);
\draw[arr] (M) -- (MC); \draw[arr] (C) -- (MC);
\draw[arr] (V) -- (VC); \draw[arr] (C) -- (VC);
\draw[arr] (VM) -- (VMC); \draw[arr] (MC) -- (VMC); \draw[arr] (VC) -- (VMC);
\node[align=center, font=\footnotesize] at (1.8,-2.5)
  {Nodes in poset of subsets.\\
   No independent cycles\\
   Cocycle space $\mathcal{B}^*$ exhausts edge space\\
   answers: \emph{how does energy cascade?}};
\end{scope}
\end{tikzpicture}
\caption{Geometric and algebraic duality in granular multi-field coupling: (a) Satake's spatial graph embedded in $\mathbb{R}^2$ for three contacting grains with particle centres $p_i$, contact points $c_{ij}$, and void centre $v$. The solid edges and dual dashed loop satisfy the compatibility constraint $L_{vc}D_{cp}=0$~\cite{Satake1993}, mapping to thermodynamic fields via $p \leftrightarrow V$ (volumetric), $c \leftrightarrow M$ (mechanical/shear), and $v \leftrightarrow C$ (configurational). The topological absence of a particle--void contact enforces the invariant structural null coupling $L_{VC}=0$. (b) The corresponding algebraic Hasse hierarchy of field subsets. Single channels (singletons) at Level~1 cascade into even-dimensional, stable pairwise layers at Level~2 ($VM$, $MC$, and the inactive $VC$), culminating in the full, odd-dimensional $VMC$ gateway system at Level~3. While the spatial graph (a) resolves the static localisation of local force balances, the acyclic, dynamic Hasse lattice (b) dictates the non-reciprocal cascading of irreversible energy transport through time and space by its channel composition.}
\label{fig:satake_vs_hasse}
\end{figure}
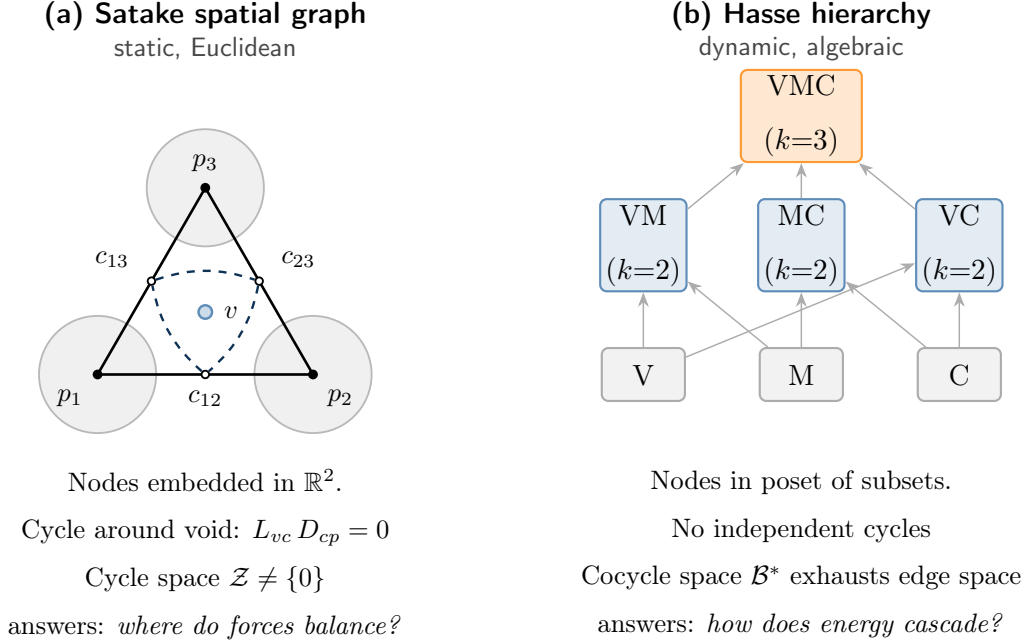

That Satake's spatial topology and Nicot's contact-mechanics channels yield the same triad by independent routes reflects a deep \emph{structural correspondence} that the present classification exploits. Under this identification, Satake's incidence matrix $D_{cp}$ governs the $VM$ coupling and his loop matrix $L_{vc}$ governs the $MC$ coupling in the Hasse lattice of Figure~\ref{fig:satake_vs_hasse}.

\begin{proposition}[Structural vanishing of $L_{VC}$ under the VMC--Satake correspondence]
\label{prop:LVC_zero}
Adopt the topological entity--channel identification $p\leftrightarrow V$, $c\leftrightarrow M$, $v\leftrightarrow C$ together with the constitutive premise that reversible $V$--$C$ exchange is contact-mediated (i.e.\ communicates strictly through the mechanical channel $M$). Under this correspondence, Satake's static--kinematic duality (expressed via the discrete Poincar\'e lemma~Eq.~\eqref{eq:satake_identity}) mandates $L_{VC}=0$ in the dynamic Onsager channel matrix, independently of all grain-stiffness or contact-law parameters. This vanishing fixes the \emph{orientation} of the null direction to the $V$--$C$ axis; it is not required for the \emph{existence} of the Gateway mode, which follows from thermodynamic channel parity alone (Theorem~\ref{thm:parity}) and persists even when $L_{VC}\neq 0$.
\end{proposition}

The proof follows from the fact that $L_{vc}D_{cp}=0$ is the algebraic statement that there is no direct topological path from particle-level entities $p$ to void-level entities $v$ bypassing the contact-level entities $c$. Under the channel identification, and given that the only admitted route for reversible $V$--$C$ exchange is the force-carrying channel $M$, this forbids a direct $V\to C$ coupling: all reversible exchange between $V$ and $C$ is mediated through $M$. The null eigenvector $\mathbf{v}_0 \propto (L_{MC},\,0,\,L_{VM})^T$ is therefore oriented along the $V$--$C$ axis (its existence being guaranteed by parity irrespective of $L_{VC}$). The full argument, and the sense in which the discrete Poincar\'e lemma supplies the cocycle vocabulary rather than forcing the vanishing unaided, is given in Supplementary Material, Section~S3.

To systematically track how cross-diffusion and conservative coupling emerge as independent physical phenomena are linked together, we organise the VMC triad into a Hasse diagram of channel subsets. The VMC triad forms the most elementary structure of a generalised Hasse lattice, which allows the coupling of additional thermodynamic channels higher up in the hierarchy when including them for multi-field continua. In this combinatorial lattice, the vertical hierarchy indexes the number of actively coupled channels ($k=1, 2, 3$), mapping how isolated singletons merge into interactive subsystems. Figure~\ref{fig:satake_vs_hasse} contrasts this algebraic architecture with classical physical space: Satake's graph answers \emph{where do forces balance?} (static, Euclidean, cycle-space $\mathcal{Z}$); the Hasse hierarchy answers \emph{how does energy cascade?} (dynamic, algebraic, cocycle-space $\mathcal{B}^*$). The parity rule is the dynamic analogue of Maxwell's rigidity count: both are parameter-free topological criteria, but Maxwell's governs static rigidity via $\mathcal{Z}$ while parity governs dynamic irreversibility via $\mathcal{B}^*$.

\subsection{Parity classification via the Hasse lattice}
\label{sec:hasse_parity}

The Parity Theorem assigns a qualitative dynamical class to every node of the Hasse lattice based solely on cardinality.
A node $\sigma$ of \emph{even} cardinality hosts a \textbf{Stable Layer}: $\mathsf{L}_\sigma$ generically has full rank, its spectrum consists entirely of conjugate pairs $\pm i\omega_j$, and the dynamics consists of closed conservative orbits in force-flux space.
A node of \emph{odd} cardinality hosts a \textbf{Gateway Layer}: $\mathsf{L}_\sigma$ necessarily has $\det(\mathsf{L}_\sigma) = 0$, its spectrum contains one zero eigenvalue, and the null eigenvector defines a direction along which no conservative restoring force acts.

\begin{figure}[!htbp]
\centering
\begin{tikzpicture}[
  >={Stealth[length=2.4mm]},
  hnode/.style={rounded corners=3pt, minimum width=1.5cm, minimum height=0.9cm,
                align=center, font=\small, line width=0.9pt},
  single/.style={hnode, draw=stableblue!50, fill=stableblue!12, text=stableblue},
  pair/.style={hnode, draw=stableblue!85, fill=stableblue!30, text=stableblue,
               minimum width=1.7cm},
  greyed/.style={hnode, draw=inactivegrey!80, fill=inactivegrey!18,
                 text=inactivegrey!60!black, dashed},
  gw/.style={hnode, draw=gatewayorange!85, fill=gatewayorange!20,
             text=gatewayorange!70!black, minimum width=2.0cm,
             minimum height=1.05cm, line width=1.1pt},
  edge/.style={->, inactivegrey!75, line width=0.7pt},
  edgeg/.style={->, inactivegrey!70, line width=0.7pt, dashed},
  lvl/.style={font=\footnotesize\itshape, text=inactivegrey!70!black, align=left},
]
\node[single] (V) at (0,0)   {$\{V\}$};
\node[single] (M) at (3.3,0) {$\{M\}$};
\node[single] (C) at (6.6,0) {$\{C\}$};
\node[pair]   (VM) at (0,2.5)   {$\{V,M\}$};
\node[pair]   (MC) at (3.3,2.5) {$\{M,C\}$};
\node[greyed] (VC) at (6.6,2.5) {$\{V,C\}$};
\node[gw] (VMC) at (3.3,5.0) {$\{V,M,C\}$};
\draw[edge] (V) -- (VM);  \draw[edge] (M) -- (VM);
\draw[edge] (M) -- (MC);  \draw[edge] (C) -- (MC);
\draw[edgeg] (V) -- (VC); \draw[edgeg] (C) -- (VC);
\draw[edge] (VM) -- (VMC); \draw[edge] (MC) -- (VMC);
\draw[edgeg] (VC) -- (VMC);
\node[font=\scriptsize, text=inactivegrey!70!black] at (6.6,3.5) {$L_{VC}=0$};
\node[lvl, anchor=east] at (-1.1,0)   {$m{=}1$};
\node[lvl, anchor=east] at (-1.1,2.5) {$m{=}2$\\$N$ even};
\node[lvl, anchor=east] at (-1.1,5.0) {$m{=}3$\\$N$ odd};
\end{tikzpicture}
\caption{Hasse diagram of the VMC coupling hierarchy. Nodes are subsets of the three
thermodynamic channels: the volumetric channel~$V$, the mechanical channel~$M$, and the configurational channel~$C$,
ordered by inclusion. \textit{Light-blue nodes} ($m=1$): single uncoupled fields.
\textit{Blue nodes} ($m=2$, $N=2$ even): two-channel pairs form Stable Layers;
$\det\mathsf{L}\neq 0$ and the reversible dynamics traces closed oscillatory orbits.
The $VC$ pair is greyed because the structural zero $L_{VC}=0$ means there is no
direct reversible coupling between $V$ and $C$; all conservative exchange between
them is mediated through $M$, the Mechanical Bridge.
\textit{Orange node} ($m=3$, $N=3$ odd): the full VMC triple is the Gateway Layer;
$\det\mathsf{L}=0$ unconditionally, creating a zero mode that provides the topological
route for non-reciprocal inter-level energy transfer.}
\label{fig:hasse_vmc}
\end{figure}
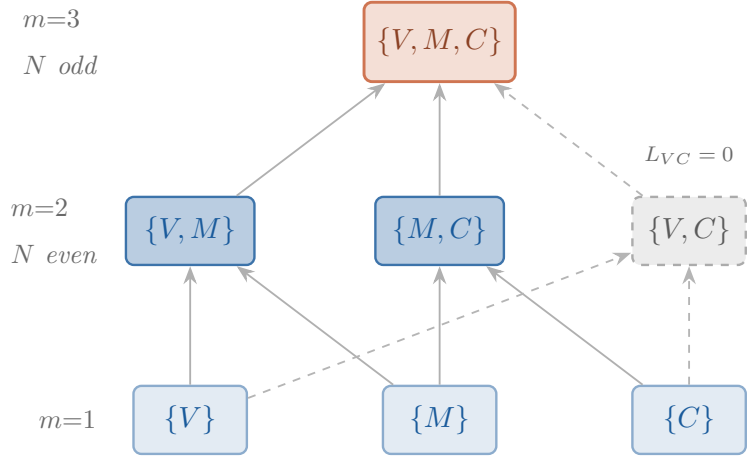

The alternating pattern of Stable and Gateway Layers across the levels is a direct consequence of the alternation of even and odd integers:
\begin{equation}
\underbrace{m=1}_{\text{uncoupled}},\quad
\underbrace{m=2}_{\text{Stable}},\quad
\underbrace{m=3}_{\text{Gateway}},\quad
\underbrace{m=4}_{\text{Stable}},\quad
\underbrace{m=5}_{\text{Gateway}},\quad\ldots
\label{eq:alternating}
\end{equation}
Figure~\ref{fig:hasse_vmc} displays this structure for the VMC case with parity colour-coding at every level.
The $VC$ pair carries a structural zero $L_{VC}=0$ imposed by the contact topology (Proposition~\ref{prop:LVC_zero}): all reversible exchange between $V$ and $C$ is mediated through $M$.

The lattice connectivity mandates that any upward energy migration must transit through a Gateway node; the Gateway number $\mathcal{G}_{\mathrm{inv}}$ is the state-dependent switch determining whether that transit is open or closed.

\section{The Parity Theorem}
\label{sec:parity}

\begin{theorem}[Parity Theorem]
\label{thm:parity}
Let $\mathsf{L} \in \mathbb{R}^{N \times N}$ with $\mathsf{L}^T = -\mathsf{L}$.
If $N$ is odd, then $\det(\mathsf{L}) = 0$ unconditionally and
$\ker(\mathsf{L})$ is non-trivial. If $N$ is even, then
$\det(\mathsf{L}) \neq 0$ generically.
\end{theorem}

The proof follows from the single identity $\det(\mathsf{L}) = \det(\mathsf{L}^T) = \det(-\mathsf{L}) = (-1)^N \det(\mathsf{L})$. For odd $N$, this forces $\det(\mathsf{L})=0$; since the eigenvalues of a real skew-symmetric matrix occur in complex conjugate pairs $\pm i\omega$, an odd-dimensional space cannot be fully partitioned into such pairs, leaving at least one zero eigenvalue. For even $N$ the determinant is unconstrained and generically non-zero. The complete proof is given in Supplementary Material, Section~S4.

For the nearest-neighbour (tridiagonal) coupling chains used throughout this series, the qualifier \emph{generic} can be made exact. Recall that for any skew-symmetric matrix $\mathsf{L}$, the determinant is the square of its Pfaffian, $\det(\mathsf{L}) = \mathrm{Pf}(\mathsf{L})^2$. When $\mathsf{L}$ is tridiagonal with off-diagonal couplings $\gamma_1,\dots,\gamma_{N-1}$, this evaluates explicitly to:
\begin{equation}
\mathrm{Pf}(\mathsf{L}) \;=\; \prod_{k\geq 1}\gamma_{2k-1}\quad (N\text{ even}),
\qquad
\mathrm{Pf}(\mathsf{L}) \;=\; 0\quad (N\text{ odd}),
\label{eq:chain_pfaffian}
\end{equation}
where for even $N$, $\det(\mathsf{L}) = \bigl(\prod_{k}\gamma_{2k-1}\bigr)^2$. The even-$N$ Stable Layer is therefore full-rank precisely when every \emph{odd-indexed} coupling is non-zero (the even-indexed couplings never enter), and is singular otherwise; the odd-$N$ Gateway is singular \emph{unconditionally}. The parity dichotomy is thus unconditional on the Gateway side and conditional (on non-vanishing odd-indexed couplings) on the Stable side, a distinction we preserve throughout.

The spectral consequence is immediate:
\begin{subequations}
\begin{align}
N \text{ odd:}  \quad &
\mathrm{spec}(\mathsf{L}) =
\bigl\{0,\; \pm i\omega_1,\; \ldots,\; \pm i\omega_{(N-1)/2}\bigr\},
\label{eq:spec_odd}\\
N \text{ even:} \quad &
\mathrm{spec}(\mathsf{L}) =
\bigl\{\pm i\omega_1,\; \ldots,\; \pm i\omega_{N/2}\bigr\},
\label{eq:spec_even}
\end{align}
\end{subequations}
holding for all values of the coupling coefficients.

This qualitative difference, and its interpretation as the emergence of an arrow of time, is illustrated in Fig.~\ref{fig:arrow_of_time}. The odd-$N$ topology presupposes that the configurational flux $d\boldsymbol{\xi}_{\rm conf}$ provides an independent degree of freedom, preventing the collapse of the $N=3$ Gateway into an even-$N$ Stable Layer. 

Physical realisation of this reversible backbone is seen in the poromechanical example by the undrained limit, which isolates the reversible block $\mathsf{L}$ by suppressing the dissipative channel $\mathsf{D}$. Within this reversible manifold, system dynamics are governed by symplectic geometry: the closed orbits of the $N=2$ Stable Layer correspond to the reversible exchange of energy between the solid skeleton and the pore fluid, characterised by the skew-symmetric operator $L_{VM} = b\sqrt{M_f}$. This conservative oscillation is a fundamental consequence of the skew-symmetry, with the eigenvalues $\pm iL_{VM}$ representing the underlying symplectic structure. Unlike consolidation, which is the monotone, Darcy-driven relaxation of excess pore pressure, and thus strictly dissipative, this reversible coupling represents the undrained backbone of the system; further poromechanical elaboration is provided in Supplementary Material, Section~S5.

\section{Null-Space Activation and the Gateway Number}
\label{sec:activation}

The Parity Theorem guarantees the existence of the null direction.
Whether it activates depends on the competition between reversible injection and dissipative removal, quantified by the Gateway number.

\subsection{The indirect activation mechanism}
\label{sec:indirect}

In the linear regime, activation of the Gateway null direction is a two-step
process. In the first step, $\mathsf{L}$ mixes energy within the complementary
subspace $\mathcal{S}_\perp = \ker(\mathsf{L})^\perp$, generating oscillatory
components. In the second step, $\mathsf{D}$ projects this oscillatory mixing
onto the null direction $\mathbf{v}_0$ through the cross-correlation
$\hat{\mathbf{v}}_0^T \mathsf{D}\,\mathbf{X}_\perp$. This indirect route is forced
by the null structure itself: because $\hat{\mathbf{v}}_0^T\mathsf{L}=\mathbf{0}$,
so that $\hat{\mathbf{v}}_0^T\mathsf{L}\,\mathbf{X}_\perp=0$ identically, the skew
operator cannot inject into its own null space, and linear energy transfer into
$\mathbf{v}_0$ can proceed only through $\mathsf{D}$.

Crucially, this linear formulation must be understood as the artefact of a simplifying assumption that inherently breaks down as the system approaches the ideal skew-symmetric limit. In this linear setting, without $\mathsf{D}$ there is no activation, while without $\mathsf{L}$ there is no injection to project. However, as the dissipative regularisation $\mathsf{D} \to \mathbf{0}$, the linear approximation yields a singularity in the energy transfer timescale, signalling its own invalidation. The breakdown of this linear case under near-ideal skew symmetry necessitates the transition to a Nonlinear Schr\"odinger (NLS) solution, where weak nonlinearities step in to resolve the asymptotic mismatch. Once nonlinear coupling is present, the conservative dynamics can transfer energy into the null direction unaided, opening a purely $\mathsf{L}$-borne activation route. The degenerate nonlinear case and the ensuing rogue wave instability is discussed elsewhere \cite{RegenHu2024}.

In the linear setting, the structural gateway is therefore activated through a
cooperative cross-diffusion mechanism: without the skew-symmetric operator
($\mathsf{L} = \mathbf{0}$), the system is stripped of its non-normal phase
mixing, and the local dissipation capacity of $\mathsf{D}$ monotonically
suppresses perturbations. Conversely, without $\mathsf{D}$, no macroscopic
instability can grow \emph{linearly} along the null direction.

\subsection{The activation criterion and Gateway number}
\label{sec:gateway_criterion}

Projecting onto the extremal two-dimensional subspace spanned by $\mathbf{v}_0$ and the most-coupled complement direction $\mathbf{v}_1$ (defined such that it obeys the orthogonality constraints $\mathbf{v}_1^T \mathbf{v}_0 = 0$ and $\mathbf{v}_1^T \mathsf{D} \mathbf{v}_0 = 0$) the effective operator reduces to
\begin{equation}
\mathsf{A}_{\rm eff} =
\begin{pmatrix}
d_{\min} & \omega \\
-\omega  & d_{\max}
\end{pmatrix},
\qquad
\det(\mathsf{A}_{\rm eff}) = d_{\min}\,d_{\max} + \omega^2 > 0,    
\end{equation}
where $d_{\min} = \mathbf{v}_0^T\mathsf{D}\mathbf{v}_0$, $d_{\max} = \mathbf{v}_1^T\mathsf{D}\mathbf{v}_1$, and $\omega = |\mathbf{v}_0^T\mathsf{L}\mathbf{v}_1|$. The injection rate into the null direction is $\omega^2/d_{\max}$ and the removal rate is $d_{\min}$.

\begin{definition}[Basis-invariant Gateway number]
\label{def:Ginv}
For $\mathsf{D}$ strictly positive definite, the \emph{basis-invariant Gateway number} is
\begin{equation}
\mathcal{G}_{\rm inv}
\;\equiv\; \bigl\|\mathsf{D}^{-1/2}\,\mathsf{L}\,\mathsf{D}^{-1/2}\bigr\|_2^2
\;=\; \rho\!\bigl(\mathsf{D}^{-1/2}\mathsf{L}\,\mathsf{D}^{-1}\mathsf{L}^T\mathsf{D}^{-1/2}\bigr),
\label{eq:G_inv_def}
\end{equation}
where $\|\cdot\|_2$ is the matrix 2-norm and $\rho(\cdot)$ is the spectral radius.
$\mathcal{G}_{\rm inv}$ is invariant under any invertible change of basis for the thermodynamic channel variables.
In the isotropic projected case ($d_{\min}=d_{\max}=d$) it reduces to $\mathcal{G}_{\rm inv}=\omega^2/d^2$, yielding the activation criterion
\begin{equation}
\omega/d = \sqrt{\mathcal{G}_{\rm inv}} \;\geq\; 1,
\label{eq:gateway_yield}
\end{equation}
the thermodynamic crossover criterion. We caution that this is a \emph{soft} routing crossover rather than a sharp yield surface: at $\mathcal{G}_{\rm inv}=1$ the null mode captures only $M^\ast\approx0.32$ of the complement-space energy (Section~\ref{sec:activation}). The formal resemblance to the von Mises ratio $\sqrt{J_2}/\bar{\sigma}\geq1$ is dimensional-analytic. Both compare a driving magnitude to a resistance and should not be read as identifying $\mathcal{G}_{\rm inv}=1$ with a yield point.
\end{definition}

The Gateway criterion identifies not a spectral instability ($\mathsf{A}_{\rm eff}$ is always spectrally stable: $\mathrm{tr}>0$, $\det>0$) but a regime of strong \emph{non-normal transient growth} driven by the skew-symmetry of $\mathsf{L}$.

\begin{theorem}[Null-mode transient amplification]
\label{thm:transient_amplification}
Consider the two-dimensional reduced linear system
\begin{equation}
\dot{x}_1 = -d_{\min}\,x_1 - \omega\,x_2, \qquad
\dot{x}_2 = +\omega\,x_1 - d_{\max}\,x_2,
\label{eq:2D_reduced}
\end{equation}
with $d_{\min}, d_{\max}, \omega > 0$ and $\omega > \tfrac{1}{2}|d_{\max}-d_{\min}|$.
Starting from unit energy in $\mathcal{S}_\perp$ (zero null-mode component):
all eigenvalues have a negative real part (spectral stability); the null-mode amplitude undergoes a transient excursion with isotropic peak
\begin{equation}
M(\mathcal{G}_{\rm inv}) \;=\;
\sqrt{\dfrac{\mathcal{G}_{\rm inv}}{1+\mathcal{G}_{\rm inv}}}
\;\exp\!\!\left(-\dfrac{\arctan\!\sqrt{\mathcal{G}_{\rm inv}}}{\sqrt{\mathcal{G}_{\rm inv}}}\right),
\label{eq:M_isotropic}
\end{equation}
which is strictly increasing from $M(0)=0$ to $M(\infty)=1$.
\end{theorem}

The full derivation, the general anisotropic form of $M$, and the non-normality analysis via the commutator $[\mathsf{A}_{\rm eff},\mathsf{A}_{\rm eff}^T]$ are given in Supplementary Material, Sections~S5 and~S6.

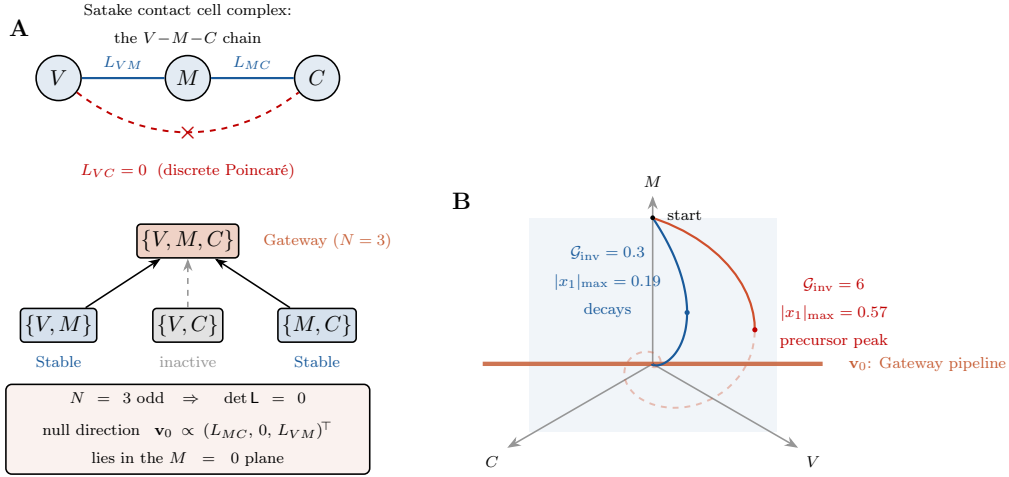
\begin{figure}[!htbp]
\centering
\resizebox{\linewidth}{!}{%
\begin{tabular}{c@{\hspace{8mm}}c}
\begin{tikzpicture}[
    every node/.style={font=\small},
    chan/.style={circle,draw=black,thick,minimum size=8mm,inner sep=0pt,fill=stableblue!12},
    lat/.style={rectangle,rounded corners=2pt,draw,thick,minimum width=12mm,minimum height=6mm,inner sep=2pt},
    >={Stealth[length=2.2mm]},
  ]
  \node[font=\scriptsize,align=center] at (2.3,3.65)
        {Satake contact cell complex:\\ the $V\!-\!M\!-\!C$ chain};
  \node[chan] (V) at (0,2.7)   {$V$};
  \node[chan] (M) at (2.3,2.7) {$M$};
  \node[chan] (C) at (4.6,2.7) {$C$};
  \draw[stableblue,line width=1.1pt] (V) -- node[above,font=\scriptsize]{$L_{VM}$} (M);
  \draw[stableblue,line width=1.1pt] (M) -- node[above,font=\scriptsize]{$L_{MC}$} (C);
  \draw[red!75!black,dashed,line width=1pt]
        (V) to[bend right=38] coordinate[pos=0.5](mid) (C);
  \node[red!75!black] at (mid) {$\boldsymbol{\times}$};
  \node[red!75!black,font=\scriptsize] at (2.3,1.05)
        {$L_{VC}=0$ \;(discrete Poincar\'e)};

  \node[lat,fill=gatewayorange!28] (VMC) at (2.3,-0.2) {$\{V,M,C\}$};
  \node[gatewayorange,font=\scriptsize,anchor=west] at (3.5,-0.2) {Gateway ($N=3$)};
  \node[lat,fill=stableblue!18]   (VM) at (0.0,-1.7) {$\{V,M\}$};
  \node[lat,fill=inactivegrey!28] (VC) at (2.3,-1.7) {$\{V,C\}$};
  \node[lat,fill=stableblue!18]   (MC) at (4.6,-1.7) {$\{M,C\}$};
  \draw[->,thick] (VM) -- (VMC);
  \draw[->,thick] (MC) -- (VMC);
  \draw[->,thick,inactivegrey,dashed] (VC) -- (VMC);
  \node[font=\scriptsize,stableblue]   at (0.0,-2.35) {Stable};
  \node[font=\scriptsize,inactivegrey] at (2.3,-2.35) {inactive};
  \node[font=\scriptsize,stableblue]   at (4.6,-2.35) {Stable};

  \node[draw,thick,rounded corners=3pt,fill=gatewayorange!8,
        text width=62mm,align=center,font=\scriptsize] at (2.3,-3.55)
       {$N=3$ odd $\;\Rightarrow\; \det\mathsf{L}=0$\\[2pt]
        null direction $\;\mathbf{v}_0\propto(L_{MC},\,0,\,L_{VM})^{\!\top}$
        lies in the $M=0$ plane};

  \node[font=\bfseries] at (-0.7,3.6) {A};
\end{tikzpicture}
&
\begin{tikzpicture}[
    every node/.style={font=\small},
    >={Stealth[length=2.4mm]},
  ]
  \def\orbitsub{(0.000,2.600) -- (0.063,2.503) -- (0.122,2.409) -- (0.176,2.317) -- (0.226,2.228) -- (0.272,2.141) -- (0.314,2.057) -- (0.353,1.976) -- (0.388,1.896) -- (0.419,1.819) -- (0.448,1.745) -- (0.474,1.673) -- (0.497,1.603) -- (0.518,1.535) -- (0.536,1.469) -- (0.552,1.406) -- (0.565,1.345) -- (0.577,1.285) -- (0.587,1.228) -- (0.595,1.173) -- (0.601,1.119) -- (0.606,1.068) -- (0.610,1.018) -- (0.612,0.971) -- (0.613,0.924) -- (0.612,0.880) -- (0.611,0.837) -- (0.609,0.796) -- (0.605,0.757) -- (0.601,0.719) -- (0.596,0.682) -- (0.591,0.647) -- (0.585,0.613) -- (0.578,0.581) -- (0.570,0.550) -- (0.563,0.520) -- (0.554,0.492) -- (0.546,0.464) -- (0.537,0.438) -- (0.527,0.413) -- (0.518,0.389) -- (0.508,0.366) -- (0.498,0.344) -- (0.488,0.323) -- (0.478,0.304) -- (0.467,0.284) -- (0.457,0.266) -- (0.446,0.249) -- (0.436,0.232) -- (0.425,0.217) -- (0.415,0.202) -- (0.404,0.187) -- (0.394,0.174) -- (0.383,0.161) -- (0.373,0.149) -- (0.363,0.137) -- (0.352,0.126) -- (0.342,0.116) -- (0.332,0.106) -- (0.323,0.097) -- (0.313,0.088) -- (0.303,0.079) -- (0.294,0.071) -- (0.285,0.064) -- (0.276,0.057) -- (0.267,0.050) -- (0.258,0.044) -- (0.250,0.038) -- (0.241,0.033) -- (0.233,0.028) -- (0.225,0.023) -- (0.217,0.018) -- (0.210,0.014) -- (0.202,0.010) -- (0.195,0.007) -- (0.188,0.003) -- (0.181,-0.000) -- (0.174,-0.003) -- (0.168,-0.006) -- (0.161,-0.008) -- (0.155,-0.010) -- (0.149,-0.013) -- (0.143,-0.015) -- (0.137,-0.016) -- (0.132,-0.018) -- (0.127,-0.019) -- (0.121,-0.021) -- (0.116,-0.022) -- (0.112,-0.023) -- (0.107,-0.024) -- (0.102,-0.025) -- (0.098,-0.026) -- (0.094,-0.026) -- (0.089,-0.027) -- (0.085,-0.027) -- (0.082,-0.028) -- (0.078,-0.028) -- (0.074,-0.028) -- (0.071,-0.028) -- (0.068,-0.028) -- (0.064,-0.028) -- (0.061,-0.028) -- (0.058,-0.028) -- (0.055,-0.028) -- (0.053,-0.028) -- (0.050,-0.028) -- (0.047,-0.028) -- (0.045,-0.027) -- (0.043,-0.027) -- (0.040,-0.027) -- (0.038,-0.026) -- (0.036,-0.026) -- (0.034,-0.026) -- (0.032,-0.025) -- (0.030,-0.025) -- (0.029,-0.024) -- (0.027,-0.024) -- (0.025,-0.024) -- (0.024,-0.023) -- (0.023,-0.023) -- (0.021,-0.022) -- (0.020,-0.022) -- (0.019,-0.021) -- (0.017,-0.021) -- (0.016,-0.020) -- (0.015,-0.020) -- (0.014,-0.019) -- (0.013,-0.019) -- (0.012,-0.018) -- (0.011,-0.018) -- (0.010,-0.017) -- (0.010,-0.017) -- (0.009,-0.016) -- (0.008,-0.016) -- (0.007,-0.015) -- (0.007,-0.015) -- (0.006,-0.015) -- (0.006,-0.014) -- (0.005,-0.014) -- (0.004,-0.013) -- (0.004,-0.013) -- (0.004,-0.012) -- (0.003,-0.012) -- (0.003,-0.012) -- (0.002,-0.011) -- (0.002,-0.011) -- (0.002,-0.010) -- (0.001,-0.010) -- (0.001,-0.010) -- (0.001,-0.009) -- (0.000,-0.009) -- (0.000,-0.009) -- (-0.000,-0.008) -- (-0.000,-0.008) -- (-0.000,-0.008) -- (-0.001,-0.007) -- (-0.001,-0.007) -- (-0.001,-0.007) -- (-0.001,-0.007) -- (-0.001,-0.006)}
  \coordinate (pksub) at (0.613,0.913);
  \def\orbitsupSolid{(0.000,2.600) -- (0.283,2.493) -- (0.543,2.370) -- (0.778,2.233) -- (0.989,2.085) -- (1.176,1.927) -- (1.337,1.762) -- (1.474,1.593) -- (1.587,1.420) -- (1.676,1.247) -- (1.743,1.074) -- (1.788,0.903) -- (1.813,0.736) -- (1.818,0.574)}
  \def\orbitsupFade{(1.818,0.574) -- (1.806,0.419) -- (1.777,0.271) -- (1.734,0.130) -- (1.677,-0.001) -- (1.607,-0.122) -- (1.528,-0.234) -- (1.440,-0.335) -- (1.344,-0.426) -- (1.242,-0.506) -- (1.136,-0.575) -- (1.027,-0.634) -- (0.915,-0.683) -- (0.803,-0.721) -- (0.692,-0.750) -- (0.582,-0.769) -- (0.474,-0.780) -- (0.370,-0.782) -- (0.270,-0.777) -- (0.174,-0.764) -- (0.084,-0.745) -- (-0.001,-0.721) -- (-0.079,-0.691) -- (-0.151,-0.657) -- (-0.217,-0.619) -- (-0.275,-0.578) -- (-0.327,-0.534) -- (-0.371,-0.488) -- (-0.409,-0.441) -- (-0.440,-0.393) -- (-0.465,-0.345) -- (-0.484,-0.297) -- (-0.496,-0.250) -- (-0.503,-0.204) -- (-0.504,-0.159) -- (-0.501,-0.116) -- (-0.493,-0.075) -- (-0.481,-0.036) -- (-0.465,0.001) -- (-0.446,0.034) -- (-0.423,0.065) -- (-0.399,0.093) -- (-0.372,0.118) -- (-0.344,0.141) -- (-0.315,0.160) -- (-0.284,0.176) -- (-0.253,0.189) -- (-0.222,0.200) -- (-0.192,0.208) -- (-0.161,0.213) -- (-0.131,0.216) -- (-0.102,0.217) -- (-0.074,0.215) -- (-0.048,0.212) -- (-0.023,0.207) -- (0.001,0.200) -- (0.022,0.192) -- (0.042,0.182) -- (0.060,0.171) -- (0.076,0.160) -- (0.091,0.148) -- (0.103,0.135) -- (0.114,0.122) -- (0.122,0.109) -- (0.129,0.096) -- (0.134,0.082) -- (0.138,0.069) -- (0.139,0.056) -- (0.140,0.044) -- (0.139,0.032) -- (0.137,0.021) -- (0.133,0.010) -- (0.129,-0.000) -- (0.123,-0.010) -- (0.117,-0.018) -- (0.111,-0.026) -- (0.103,-0.033) -- (0.095,-0.039) -- (0.087,-0.044) -- (0.079,-0.049) -- (0.070,-0.053) -- (0.062,-0.056) -- (0.053,-0.058) -- (0.045,-0.059) -- (0.036,-0.060) -- (0.028,-0.060) -- (0.021,-0.060) -- (0.013,-0.059) -- (0.006,-0.057) -- (-0.000,-0.055) -- (-0.006,-0.053) -- (-0.012,-0.050) -- (-0.017,-0.048) -- (-0.021,-0.044) -- (-0.025,-0.041) -- (-0.029,-0.037) -- (-0.032,-0.034) -- (-0.034,-0.030) -- (-0.036,-0.026) -- (-0.037,-0.023) -- (-0.038,-0.019) -- (-0.039,-0.016) -- (-0.039,-0.012) -- (-0.039,-0.009) -- (-0.038,-0.006) -- (-0.037,-0.003) -- (-0.036,0.000) -- (-0.034,0.003) -- (-0.033,0.005) -- (-0.031,0.007) -- (-0.029,0.009) -- (-0.026,0.011) -- (-0.024,0.012) -- (-0.022,0.014) -- (-0.019,0.015) -- (-0.017,0.015) -- (-0.015,0.016) -- (-0.012,0.016) -- (-0.010,0.017) -- (-0.008,0.017) -- (-0.006,0.017) -- (-0.004,0.016) -- (-0.002,0.016) -- (0.000,0.015) -- (0.002,0.015) -- (0.003,0.014) -- (0.005,0.013) -- (0.006,0.012) -- (0.007,0.011) -- (0.008,0.010) -- (0.009,0.009) -- (0.009,0.008) -- (0.010,0.007) -- (0.010,0.006) -- (0.011,0.005) -- (0.011,0.004) -- (0.011,0.003) -- (0.011,0.002) -- (0.011,0.002) -- (0.010,0.001) -- (0.010,-0.000) -- (0.009,-0.001) -- (0.009,-0.001) -- (0.008,-0.002) -- (0.008,-0.003) -- (0.007,-0.003) -- (0.007,-0.003)}
  \coordinate (pksup) at (1.819,0.606);
  \coordinate (axV) at (2.589,-1.495);
  \coordinate (axM) at (0.000,2.990);
  \coordinate (axC) at (-2.589,-1.495);
  \coordinate (orig) at (0.000,0.000);
  \coordinate (pipeplus) at (3.025,0.000);
  \coordinate (pipeminus) at (-3.025,-0.000);
  \coordinate (pipelabel) at (3.344,0.000);
  \coordinate (start) at (0.000,2.600);

  \fill[stableblue!6]
        ($(orig)+(2.2,-1.2)$) -- ($(orig)+(-2.2,-1.2)$)
        -- ($(orig)+(-2.2,2.6)$) -- ($(orig)+(2.2,2.6)$) -- cycle;

  \draw[->,inactivegrey,line width=0.8pt] (orig) -- (axV) node[below right,font=\scriptsize,black]{$V$};
  \draw[->,inactivegrey,line width=0.8pt] (orig) -- (axM) node[above,font=\scriptsize,black]{$M$};
  \draw[->,inactivegrey,line width=0.8pt] (orig) -- (axC) node[below left,font=\scriptsize,black]{$C$};

  \draw[gatewayorange,line width=2.2pt,opacity=0.85] (pipeminus) -- (pipeplus);
  \node[gatewayorange,anchor=west,font=\scriptsize] at (pipelabel)
       {$\mathbf{v}_0$: Gateway pipeline};

  \draw[stableblue,line width=1.1pt] \orbitsub;
  \fill[stableblue] (pksub) circle (1.3pt);
  \node[stableblue,font=\scriptsize,align=center,anchor=east] at ($(pksub)+(-0.32,0.55)$)
       {$\mathcal{G}_{\rm inv}=0.3$\\$|x_1|_{\max}=0.19$\\decays};

  \draw[gatewayorange!85!red,line width=1.1pt] \orbitsupSolid;
  \draw[gatewayorange!85!red,line width=1.0pt,dashed,opacity=0.35] \orbitsupFade;
  \fill[red!75!black] (pksup) circle (1.3pt);
  \node[red!75!black,font=\scriptsize,align=center,anchor=west] at ($(pksup)+(0.30,0.30)$)
       {$\mathcal{G}_{\rm inv}=6$\\$|x_1|_{\max}=0.57$\\precursor peak};

  \fill[black] (start) circle (1.2pt);
  \node[font=\scriptsize,anchor=west] at ($(start)+(0.12,0.05)$) {start};

  \node[font=\bfseries] at ($(orig)+(-3.4,2.9)$) {B};
\end{tikzpicture}
\end{tabular}}
\caption{The Gateway mechanism: topological cause and dynamic effect in a $(V,M,C)$ continuum. \textbf{(A)} The Satake complex and Hasse lattice reveal that the discrete Poincar\'e lemma forces energy exchange through the mechanical channel $M$. This creates an odd-dimensional $\{V,M,C\}$ Gateway, unconditionally yielding a null direction $\mathbf{v}_0$ in the $M=0$ plane. \textbf{(B)} Linear trajectories from a common perturbation. For sub-critical coupling ($\mathcal{G}_{\rm inv}=0.3$), dissipation dominates and the null amplitude decays. For super-critical coupling ($\mathcal{G}_{\rm inv}=6$), non-normal amplification drives a large transient excursion along the Gateway pipeline $\mathbf{v}_0$. This peak is the proposed precursor to localisation: when $\mathcal{G}_{\rm inv}\geq1$, the state is pushed beyond the linearly regularised neighbourhood, nucleating non-linear bifurcation before the return to equilibrium.}
\label{fig:gateway}
\end{figure}

At $\mathcal{G}_{\rm inv}=1$ (coupling frequency equals the geometric mean of the dissipation rates), the transient peak equals $M^* = \tfrac{1}{\sqrt{2}}\,e^{-\pi/4} \approx 0.322$.
For $\mathcal{G}_{\rm inv}<1$ the null mode receives less than $32\%$ of the complement-space energy and dissipation dominates.
For $\mathcal{G}_{\rm inv}>1$ coupling-dominated transient routing progressively dominates; the threshold $\mathcal{G}_{\rm inv}=1$ marks the transition from dissipation-dominated to coupling-dominated energy routing.
Figure \ref{fig:gateway} shows this criterion as cause and effect. Panel~A traces it to the topology, where the discrete Poincar\'e lemma forces the odd-dimensional $VMC$ Gateway and hence the structural null direction $\mathbf{v}_0$. Panel~B shows the resulting orbits: the sub-critical case ($\mathcal{G}_{\rm inv}=0.3$) decays almost directly to the origin, while the super-critical case ($\mathcal{G}_{\rm inv}=6$) is pumped along $\mathbf{v}_0$ through a non-normal transient excursion whose peak null-mode amplitudes ($0.19$ and $0.57$) reproduce the closed form of equation~\eqref{eq:M_isotropic}.

\begin{heuristic}[Gateway activation in the $N$-dimensional system]
\label{prop:activation}
Under a linear Ansatz, the proposed activation criterion for the full
$N$-dimensional $\mathsf{D}+\mathsf{L}$ system ($N$ odd) is
$\mathcal{G}_{\rm inv}\geq 1$ (Definition~\ref{def:Ginv}).
When $\mathcal{G}_{\rm inv}\geq 1$, transient energy routing into the null
direction dominates dissipative decay; this is a constitutive instability
driven by the interplay of $\mathsf{L}$ and $\mathsf{D}$, not a collapse of
dissipation alone, but a breakdown of the balance between nonlocal energy
pumping and local energy removal.
\end{heuristic}

\emph{Linear stipulation:} The threshold and its associated growth metrics are
stipulated under a linear Ansatz: the operators $\mathsf{D}$ and $\mathsf{L}$
are held as static, state-independent tensors evaluated at the current
thermodynamic tangent, and the complementary subspace $\mathcal{S}_\perp$ is
treated as an effectively unbounded energy reservoir. Under this Ansatz the
null direction is fed only indirectly, through the cross-diffusive projection
$\hat{\mathbf{v}}_0^T\mathsf{D}\,\mathbf{X}_\perp$, since
$\hat{\mathbf{v}}_0^T\mathsf{L}=\mathbf{0}$ forbids any direct skew injection
into the null space (Section~\ref{sec:indirect}). The criterion
$\mathcal{G}_{\rm inv}\geq 1$ therefore locates the \emph{onset} of activation,
not its saturation.

\emph{Enforced gateway in the nonlinear regime.} Once the linear Ansatz is
relaxed the obstruction $\hat{\mathbf{v}}_0^T\mathsf{L}=\mathbf{0}$ no longer
applies: nonlinear coupling transfers energy into the null direction without
the mediation of $\mathsf{D}$, so activation ceases to be conditional on the
linear threshold and becomes enforced by the odd-$N$ null topology itself. As
the ideal skew-symmetric limit is approached,
$\mathcal{G}_{\rm inv}=\|\mathsf{D}^{-1/2}\mathsf{L}\mathsf{D}^{-1/2}\|_2^2$
diverges and ceases to discriminate, the infinite-reservoir assumption fails,
and the growth along $\mathbf{v}_0$ is no longer a passive regularised
projection but a dynamic feedback in which nonlinear terms actively re-route and
saturate the energy flux, mirroring the asymptotic behaviour of the focusing
Nonlinear Schr\"odinger (NLS) envelope. Nonlinear saturation then bounds the
amplitude that the linear metric $M(\mathcal{G}_{\rm inv})$ leaves unbounded.

By Definition~\ref{def:Ginv},
$\mathcal{G}_{\rm inv}=\|\mathsf{D}^{-1/2}\mathsf{L}\mathsf{D}^{-1/2}\|_2^2$ is an
exact basis-invariant property of the full $N$-dimensional operator. The
extremal 2D subspace $(\mathbf{v}_0,\mathbf{v}_1)$ realises this 2-norm in the
field of values of $\mathsf{D}^{-1/2}\mathsf{L}\mathsf{D}^{-1/2}$, so the maximal
instantaneous null-mode growth rate of the full linear operator is captured
exactly. What remains a conjecture is whether this extremal plane also dominates
the routing under generic loading, and whether the closed-form peak
$M(\mathcal{G}_{\rm inv})$ transfers globally to the $N$-dimensional system once
nonlinear energy routing bounds the linear growth.

\begin{figure}[!htbp]
\centering
\resizebox{0.9\linewidth}{!}{%
\begin{tikzpicture}[
  panelbg_s/.style={fill=stableblue!5,   draw=stableblue!30,   rounded corners=6pt, line width=0.7pt},
  panelbg_g/.style={fill=orange!6, draw=orange!50, rounded corners=6pt, line width=0.9pt},
  chan/.style={rectangle, rounded corners=3pt, minimum width=2.2cm,
               minimum height=0.85cm, line width=0.7pt,
               font=\sffamily\footnotesize, align=center},
  nullbox/.style={chan, fill=gray!12, draw=gray!45, text=gray!60},
  stabbox/.style={chan, fill=stableblue!12, draw=stableblue!50, text=stableblue},
  gwbox/.style={chan, fill=orange!14, draw=orange!65, text=orange!75!black},
  arr_in/.style={-{Stealth[scale=1.1]}, line width=2.0pt, orange!70!black},
  arr_out/.style={-{Stealth[scale=1.1]}, line width=2.0pt, stableblue!60},
  arr_in_big/.style={-{Stealth[scale=1.3]}, line width=3.8pt, red!65},
  arr_out_thin/.style={-{Stealth[scale=1.0]}, line width=1.2pt, stableblue!40},
  lbl/.style={font=\sffamily\footnotesize},
  hdr/.style={font=\sffamily\bfseries\small},
  eqlbl/.style={font=\sffamily\footnotesize\itshape, text=gray!60, align=center},
  rate/.style={font=\sffamily\scriptsize, align=center, text width=2.8cm}
]

\begin{scope}[xshift=0cm]
  \fill[panelbg_s] (-3.8,-5.8) rectangle (3.8,5.5);

  \node[hdr, text=stableblue] at (0, 4.9) {$\mathcal{G}_{\rm inv} < 1$\quad Gateway regularised};

  \node[stabbox] (s_perp) at (0,  2.8) {$\mathcal{S}_\perp$ \\ (reversible subspace)};
  \node[nullbox] (null)   at (0,  0.8) {$\ker(\mathsf{L})$\\ null direction $\mathbf{v}_0$};
  \node[stabbox] (xamp)   at (0, -1.2) {null amplitude $x_0$};

  \draw[arr_in] (-1.6, 1.85) -- (-1.6, 0.05);
  \node[rate, text=orange!65, left=0pt] at (-1.2, 0.95) {
    injection rate\\
    $\dfrac{\|\mathsf{L}\|_2^2}{\sigma_{\max}(\mathsf{D})}$
  };

  \draw[arr_out] (1.6,-0.05) -- (1.6,-0.9);
  \node[rate, text=stableblue, right=0pt] at (1.3, -0.5) {
    dissipation\\
    $\sigma_{\min}(\mathsf{D})$
  };

  \node[lbl, text=gray!55] at (0, -2.5) {$\dfrac{\|\mathsf{L}\|_2^2}{\sigma_{\max}} \;<\; \sigma_{\min}$};

  \node[rectangle, rounded corners=4pt, fill=stableblue!14, draw=stableblue!50, line width=0.8pt,
        minimum width=6.4cm, minimum height=1.0cm, font=\sffamily\footnotesize,
        align=center, text=stableblue] at (0,-4.0)
    {$x_0 \to 0$ \quad null mode decays\\ system remains in reversible subspace};

  \node[font=\Large, text=stableblue] at (0,-5.2) {$\checkmark$\;stable};

\end{scope}

\draw[gray!30, line width=0.5pt, dashed] (4.1,-5.8) -- (4.1,5.5);

\begin{scope}[xshift=8.5cm]
  \fill[panelbg_g] (-3.8,-5.8) rectangle (3.8,5.5);

  \node[hdr, text=red!65] at (0, 4.9) {$\mathcal{G}_{\rm inv} \geq 1$\quad Gateway activated};

  \node[stabbox] (s_perp2) at (0,  2.8) {$\mathcal{S}_\perp$ \\ (reversible subspace)};
  \node[gwbox]   (null2)   at (0,  0.8) {$\ker(\mathsf{L})$\\ null direction $\mathbf{v}_0$};
  \node[gwbox]   (xamp2)   at (0, -1.2) {null amplitude $x_0$\;{\color{red!65}$\nearrow$}};

  \draw[arr_in] (-1.6, 1.85) -- (-1.6, 0.05);
  \node[rate, text=red!60, left=0pt] at (-1.2, 0.95) {
    injection rate\\
    $\dfrac{\|\mathsf{L}\|_2^2}{\sigma_{\max}(\mathsf{D})}$
  };

  \draw[arr_out_thin] (1.9,-0.05) -- (1.9,-0.9);
  \node[rate, text=stableblue, right=0pt] at (1.5, -0.5) {
    dissipation\\
    $\sigma_{\min}(\mathsf{D})$
  };

  \node[lbl, text=red!55] at (0, -2.5) {$\dfrac{\|\mathsf{L}\|_2^2}{\sigma_{\max}} \;>\; \sigma_{\min}$};

  \node[rectangle, rounded corners=4pt, fill=orange!14, draw=orange!60,
        line width=0.9pt,
        minimum width=6.4cm, minimum height=1.0cm, font=\sffamily\footnotesize,
        align=center, text=orange!75!black] at (0,-4.0)
    {$x_0(t)$ large transient excursion\\
     non-normal amplification dominates\\
     proposed precursor to localisation};

  \node[font=\Large, text=red!55] at (0,-5.4) {$\times$\;Gateway activates};

\end{scope}

\node[rectangle, rounded corners=4pt, fill=gray!8, draw=gray!40, line width=0.6pt,
     font=\sffamily\footnotesize, align=center, inner sep=6pt,
     text width=5.8cm] at (4.35, 8.0) {
  \textbf{Two-step activation mechanism}\\[3pt]
  \textbf{\circled{1}} $\mathsf{L}$ mixes energy within $\mathcal{S}_\perp$\\[1pt]
  \textbf{\circled{2}} $\mathsf{D}$ projects mixing onto $\ker(\mathsf{L})$\\[3pt]
  $\mathcal{G}_{\rm inv} = \bigl\|\mathsf{D}^{-1/2}\mathsf{L}\mathsf{D}^{-1/2}\bigr\|_2^2$
};

\end{tikzpicture}
}
\caption{Gateway activation mechanism. The Gateway number $\mathcal{G}_{\rm inv} = \Vert{}\mathsf{D}^{-1/2}\mathsf{L}\mathsf{D}^{-1/2}\Vert{}_2^2$ dictates the stability of the null direction $\mathbf{v}_0$. For $\mathcal{G}_{\rm inv}<1$ (\textit{left}), local dissipation dominates and the null amplitude decays. For $\mathcal{G}_{\rm inv}\geq1$ (\textit{right}), conservative coupling triggers a large transient excursion, proposed as the precursor to localisation.}
\label{fig:gateway_activation}
\end{figure}

\section{The Alternating Energy Routing Hierarchy}
\label{sec:cascade}

\begin{figure}[!htbp]
\centering
\resizebox{0.6 \linewidth}{!}{%
\begin{tikzpicture}[
    node distance=1.8cm,
    box/.style={rectangle, rounded corners=5pt, minimum width=7.0cm,
                minimum height=1.2cm, text centered,
                font=\sffamily\small, line width=0.8pt},
    stable/.style={box, fill=stableblue!10, draw=stableblue!60, text width=6.5cm},
    gateway/.style={box, fill=orange!15, draw=orange!70, text width=6.5cm},
    arr_d/.style={-{Stealth[scale=1.2]}, line width=1.2pt, stableblue!70!black},
    arr_u/.style={-{Stealth[scale=1.2]}, line width=1.2pt, dashed, red!70!black},
    lf/.style={font=\sffamily\bfseries\small}
]
    \node[gateway] (L5) {
        \textbf{Level 5: Five-field} ($N=5$, odd)\\
        \textit{Gateway: $\mathcal{G}_5\geq1$ \\ triggers global irreversibility}
    };
    \node[stable]  (L4) [below=of L5] {
        \textbf{Level 4: Four-field} ($N=4$, even)\\
        \textit{Stable Layer: mesoscale energy reservoir}
    };
    \node[gateway] (L3) [below=of L4] {
        \textbf{Level 3: Three-field VMC} ($N=3$, odd)\\
        \textit{Gateway: $\mathcal{G}_3\geq1$ triggers contact-scale irreversibility}
    };
    \node[stable]  (L2) [below=of L3] {
        \textbf{Level 2: Two-field pairwise} ($N=2$, even)\\
        \textit{Stable Layer: invariant-contact conservative oscillation}
    };
    \node[lf, stableblue!70!black]
        at ($(L5.east)!0.5!(L2.east)+(1.5,0)$)
        {\rotatebox{-90}{DISSIPATIVE CASCADE $(\mathsf{D})$}};
    \foreach \top/\bot in {L5/L4, L4/L3, L3/L2}
        \draw[arr_d] ([xshift=2.8cm]\top.south) -- ([xshift=2.8cm]\bot.north);
    \node[lf, red!70!black]
        at ($(L5.west)!0.5!(L2.west)-(3.5,0)$)
        {\rotatebox{90}{CONSERVATIVE PUMPING $(\mathsf{L})$}};
    \draw[arr_u] ([xshift=-2.8cm]L2.north) --
        node[left, lf, xshift=-4pt, pos=0.5]{\circled{1} Eigenstate excitation}
        ([xshift=-2.8cm]L3.south);
    \draw[arr_u] ([xshift=-2.8cm]L3.north) --
        node[left, lf, xshift=-4pt, pos=0.5]{\circled{2} Energy pumping}
        ([xshift=-2.8cm]L4.south);
    \draw[arr_u] ([xshift=-2.8cm]L4.north) --
        node[left, lf, xshift=-4pt, pos=0.5]{\circled{3} Gateway transfer}
        ([xshift=-2.8cm]L5.south);
    \draw[arr_u] ([xshift=-2.8cm]L5.north) -- ++(0,1.2)
        node[left, lf, xshift=-4pt, pos=0.6]{\circled{4} Zero-mode activation};
\end{tikzpicture}
}
\caption{Proposed energy routing hierarchy of the parity classification for a granular
  system, following the alternating Stable/Gateway pattern of
  equation~\eqref{eq:alternating}. Blue solid arrows: dissipative cascade
  driven by $\mathsf{D}$, producing positive entropy at every level. Dashed
  red arrows: conservative pumping driven by $\mathsf{L}$, carrying energy
  upward without local entropy production. Gateway Layers (orange, odd $N$)
  act as topological routing junctions: when the Gateway number $\mathcal{G}_N \geq 1$ at that
  level the null direction is predicted to activate and energy may be non-reciprocally routed to the
  next hierarchical scale. Stable Layers (blue, even $N$) are energy reservoirs
  sustaining reversible oscillation.}
\label{fig:cascade}
\end{figure}
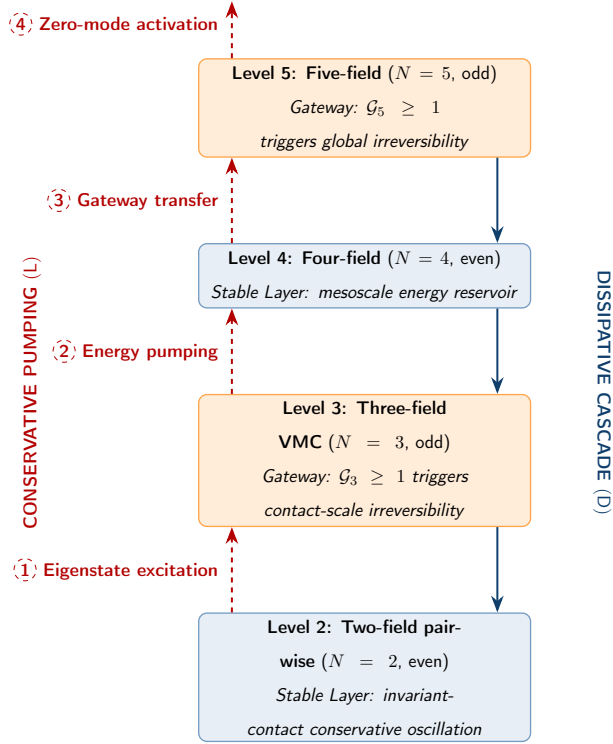

The Hasse lattice provides the natural language for the proposed multiscale energy routing hierarchy.
Figure~\ref{fig:cascade} illustrates the bidirectional structure for the granular hierarchy; the same structure holds for any multi-field system governed by the alternating pattern of equation~\eqref{eq:alternating}.

\begin{definition}[Micro, Meso, and Macro scales via the Hasse lattice]
\label{def:scales}
The Hasse lattice level $|\sigma|$ identifies physical scales: \emph{Micro} ($|\sigma|=1,2$) are individual channels and pairwise Stable Layer oscillations; \emph{Meso} ($|\sigma|=3,4$) are Level-3 Gateways (minimal topological irreversibility unit, $\mathcal{G}_{\rm inv}\geq 1$ required for activation) and Level-4 Stable Layer energy reservoirs; \emph{Macro} ($|\sigma|\geq 5$) are collective Gateway activations corresponding to localisation onset.
\end{definition}

Energy in a Stable Layer at level $2k$ can only reach level $2k+1$ through a Gateway node.
Conservative pumping by $\mathsf{L}$ builds coherent oscillations upward without local entropy production; dissipative cascade by $\mathsf{D}$ drains coherent structure downward.

\section{Discussion}
\label{sec:discussion}

\subsection{Scope of the parity criterion}
\label{sec:scope_disc}

The Parity Theorem applies to any physical system described by $N$ coupled thermodynamic forces and fluxes with a skew-symmetric reversible Onsager block.
The Gateway criterion $\mathcal{G}_{\rm inv} \geq 1$ is related to but distinct from both Hill's second-order work criterion~\cite{Hill1958} and the classical Rudnicki--Rice criterion~\cite{Rudnicki1975,Rice1976}.
Hill's condition $W_2 > 0$ involves only the symmetric part of the tangent operator ($\sigma_{\min}(\mathsf{D}) > 0$); the Rudnicki--Rice acoustic-tensor criterion is likewise governed by $\sigma_{\min}(\mathsf{D})\to 0$.
The Gateway identifies a precursor instability arising from the antisymmetric coupling $\mathsf{L}$ overwhelming local dissipation \emph{before} any single dissipative channel closes, operating in the parameter regime where neither Hill's nor Rudnicki--Rice's criterion is yet triggered.
The physical scope extends from granular mechanics to gyroscopic coupling in entangled multiphysics coupling~\cite{RegenHu2023b}, to odd-viscosity fluids and active matter~\cite{Scheibner2020,Fruchart2023}, and to biological signal-transduction networks requiring odd-cardinality coupling subsets~\cite{Regenauer2025}.

\subsection{Relationship to Schnakenberg's network thermodynamics}
\label{sec:schnakenberg_disc}

The Schnakenberg cycle-current decomposition~\cite{Schnakenberg1976} decomposes entropy production into independent cycle-current contributions; a cycle of odd length can sustain a non-zero steady-state current without violating detailed balance at any edge.
Within the cocycle-borne routing mechanism addressed here, odd-cardinality Hasse nodes correspond to channel configurations with non-trivial $\ker(\mathsf{L})$, and $\mathcal{G}_{\rm inv}\geq 1$ is the continuum counterpart of the threshold above which a Schnakenberg cycle current becomes self-sustaining.
Formalising this correspondence is ongoing work.

\subsection{Distinction from non-associated plasticity and Cosserat theories}
\label{sec:plasticity_distinction}

The Gateway mechanism is structurally distinct from non-associated plasticity~\cite{Rudnicki1975,deBorst2022} on two counts.
First, the asymmetry of $\mathsf{A}$ is \emph{structural}: $\mathsf{L}=(\mathsf{A}-\mathsf{A}^T)/2$ is fully determined by the channel specification with no fitting freedom.
Second, the Gateway null-mode excursion is \emph{conservative and work-free}: $\mathbf{q}^T\mathsf{L}\mathbf{q}\equiv 0$; the instability is a topological consequence of $\det\mathsf{L}=0$ for odd $N$, not a tuned failure surface.
The descriptor work-free applies strictly to the conservative routing by $\mathsf{L}$ ($\mathbf{q}^T\mathsf{L}\mathbf{q}\equiv 0$). It does not imply an energy source: the amplified null-mode excursion draws on perturbation energy resident in $\mathcal{S}_\perp$, supplied through boundary loading, which the non-normal interplay of $\mathsf{D}$ and $\mathsf{L}$ transiently redirects along $\mathbf{v}_0$ rather than dissipates. Bond rupture and surface creation are downstream nonlinear consequences of this precursor and are not claimed to be work-free.
The Gateway criterion is accordingly parameter-free in its existence (parity of $N$) and depends on $\mathcal{G}_{\rm inv}$ only for activation.

Cosserat rotational coupling contributes a $2\times 2$ non-singular block to $\mathsf{L}$; the parity-mandated null eigenvalue appears only when a third channel is added, constituting a Level-3 Gateway.

\subsection{Limitations and perspectives}
\label{sec:limitations}

The parity argument requires $N$ to be a finite integer; for continuum field theories a channel-counting argument must be applied to a finite-dimensional projection. The activation criterion $\mathcal{G}_{\rm inv}\geq 1$ is a sufficient condition; its necessity must be established by separate analysis. The identification of the correct channels and the evaluation of $\sigma_{\min}(\mathsf{D})$ and $\|\mathsf{L}\|_2$ in terms of measurable material properties is the main challenge in applying the framework to specific physical systems.

We close with a conjecture (in the sense of Section~\ref{sec:claim_status}) rather than a result, to indicate how the parity idea might extend beyond granular matter. If the coupled multi-field engine of a planetary interior is modelled as an augmented Onsager system, the parity of its reversible coupling count separates bounded, cyclic evolution (even $N$, a Stable Layer confined to an invariant torus) from secularly drifting dynamics (odd $N$, a Gateway). On this reading the thermo-mechanical convection that drives Earth's mantle is a multiscale even system, sustaining motion through bounded convective and supercontinent cycles without secular runaway, a hallmark of long-term geodynamic stability. This stability is in stark contrast to the evolution of our sister planet. Venus \cite{Solomon1999,Turcotte1999} has experienced catastrophic resurfacing during its tectonic history. We posit that the endurance of a tectonically active world may favour an \emph{even} dominant coupling count. A change of one channel in that count, e.g. through the gain or loss of a strong volatile, chemical, or electromagnetic coupling, opens a Gateway to the odd-counted localised secular instabilities analysed here at the granular scale. Testing the hypothesis would require identifying the genuinely independent reversible channels of a planetary energy budget; that task lies beyond the present scope, and we record the conjecture only to indicate the broader trajectory of the parity classification. 

\subsection{Theoretical framework validation and assumptions}
\label{sec:claim_status}

The claims of the framework fall into three distinct levels.
\emph{Theorems}: exact algebraic or analytic results proved in this paper or cited standard references.
\emph{Model interpretations}: physical readings of the algebraic structure, which are internally consistent with the assumed constitutive class but not derived from first principles of any specific material system.
\emph{Conjectures}: candidate criteria compatible with the framework but requiring independent verification by continuum PDE analysis, DEM simulation, or laboratory experiment, with explicit falsification protocols developed in the companion Parts~2 and~3~\cite{RegenNicot2026dilatancy,RegenNicot2026numerical}.

\section{Conclusion}
\label{sec:conclusion}

The identity $\det(\mathsf{L}) = (-1)^N \det(\mathsf{L})$ forces the zero determinant of any odd-dimensional real skew-symmetric coupling matrix and defines a null direction, the Gateway direction, in thermodynamic force-flux space.
The Parity Theorem converts this algebraic identity into a classification of multi-field systems into Stable Layers and Gateway Layers, and the Hasse lattice of channel subsets organises this classification into a multiscale hierarchy that connects to Satake's graph-theoretical tradition in granular mechanics through the cycle/cocycle duality illustrated in Fig.~\ref{fig:satake_vs_hasse}.

A structural side result is Proposition~\ref{prop:LVC_zero}: under the VMC identification and the premise that $V$--$C$ exchange is force-mediated, the discrete Poincar\'e lemma fixes $L_{VC}=0$. The Gateway itself does not depend on this vanishing---only on the parity of $N$---so $L_{VC}=0$ fixes the \emph{orientation} of the irreversible pathway rather than creating it.
The discrete Poincar\'e lemma applied to Satake's contact cell complex forbids a direct reversible coupling between the volumetric channel $V$ and the configurational channel $C$ that bypasses the mechanical channel $M$.
This fixes the chain form of the VMC coupling matrix, the orientation of the null eigenvector $\mathbf{v}_0\propto(L_{MC},0,L_{VM})^T$, and hence the Gateway topology of the three-channel system, from the contact-graph topology together with the force-mediation premise.

The basis-invariant Gateway number $\mathcal{G}_{\rm inv}$ is the single dimensionless parameter governing whether the system is in the coupling-dominated transient-routing regime.
When $\mathcal{G}_{\rm inv} \geq 1$, the null-mode transient excursion is large relative to dissipative decay and the governing equations lose strong dissipative regularisation of the null direction: this is the candidate precursor mechanism for localisation, operating before the Rudnicki--Rice threshold is reached.

\bibliographystyle{elsarticle-num}
\bibliography{bib_merged}

@article{KumarLuding2016,
  author  = {Kumar, N. and Luding, S.},
  title   = {Memory of jamming{--}multiscale models for soft and granular materials},
  journal = {Granular Matter},
  volume  = {18},
  pages   = {58--80},
  year    = {2016}
}

@article{Goldhirsch2010,
   author = {Goldhirsch, Isaac},
   title = {Stress, stress asymmetry and couple stress: from discrete particles to continuous fields},
   journal = {Granular Matter},
   volume = {12},
   number = {3},
   pages = {239-252},
   ISSN = {1434-5021
1434-7636},
   DOI = {10.1007/s10035-010-0181-z},
   year = {2010},
   type = {Journal Article}
}

@article{Weinhart2012,
   author = {Weinhart, Thomas and Thornton, Anthony R. and Luding, Stefan and Bokhove, Onno},
   title = {From discrete particles to continuum fields near a boundary},
   journal = {Granular Matter},
   volume = {14},
   number = {2},
   pages = {289-294},
   ISSN = {1434-5021
1434-7636},
   DOI = {10.1007/s10035-012-0317-4},
   year = {2012},
   type = {Journal Article}
}

@article{Peyneau2008,
  author  = {Peyneau, P.-E. and Roux, J.-N.},
  title   = {Frictionless bead packs have macroscopic friction, but no dilatancy},
  journal = {Physical Review E},
  volume  = {78},
  number  = {1},
  pages   = {011307},
  year    = {2008},
  doi     = {10.1103/PhysRevE.78.011307}
}

@article{Azema2015,
  author  = {Az{\'e}ma, {\'E}. and Radjai, F. and Roux, J.-N.},
  title   = {Internal friction and absence of dilatancy of packings of frictionless polygons},
  journal = {Physical Review E},
  volume  = {91},
  number  = {1},
  pages   = {010202},
  year    = {2015},
  doi     = {10.1103/PhysRevE.91.010202}
}

@article{Henkes2009,
  author  = {Henkes, S. and Chakraborty, B.},
  title   = {Statistical mechanics framework for static granular matter},
  journal = {Physical Review E},
  volume  = {79},
  number  = {6},
  pages   = {061301},
  year    = {2009},
  doi     = {10.1103/PhysRevE.79.061301}
}

@article{Sarkar2013,
  author  = {Sarkar, S. and Bi, D. and Zhang, J. and Behringer, R. P. and Chakraborty, B.},
  title   = {Origin of rigidity in dry granular solids},
  journal = {Physical Review Letters},
  volume  = {111},
  number  = {6},
  pages   = {068301},
  year    = {2013},
  doi     = {10.1103/PhysRevLett.111.068301}
}

@article{Eshelby1975,
   author = {Eshelby, J. D.},
   title = {The elastic energy-momentum tensor},
   journal = {Journal of Elasticity},
   volume = {5},
   number = {3},
   pages = {321-335},
   ISSN = {1573-2681},
   DOI = {10.1007/BF00126994},
   year = {1975},
   type = {Journal Article}
}

@book{Kroner1958,
  author    = {Kr{\"o}ner, E.},
  title     = {Kontinuumstheorie der Versetzungen und Eigenspannungen},
  publisher = {Springer},
  address   = {Berlin},
  year      = {1958},
  ISBN =  {978-3-540-02261-9}
}

@article{Acharya2001,
   author = {Acharya, Amit},
   title = {A model of crystal plasticity based on the theory of continuously distributed dislocations},
   journal = {Journal of the Mechanics and Physics of Solids},
   volume = {49},
   number = {4},
   pages = {761-784},
   ISSN = {0022-5096},
   DOI = {https://doi.org/10.1016/S0022-5096(00)00060-0},
   year = {2001},
   type = {Journal Article}
}

@article{Solomon1999,
   author = {Solomon, S. C. and Bullock, M. A. and Grinspoon, D. H.},
   title = {Climate change as a regulator of tectonics on Venus},
   journal = {Science},
   volume = {286},
   number = {5437},
   pages = {87-90},
   year = {1999},
   type = {Journal Article}
}

@article{Regen2010,
   author = {Regenauer-Lieb, K. and Karrech, A. and Chua, H. T. and Horowitz, F. G. and Yuen, D.},
   title = {Time-dependent, irreversible entropy production and geodynamics},
   journal = {Philosophical Transactions of the Royal Society A: Mathematical, Physical and Engineering Sciences},
   volume = {368},
   number = {1910},
   pages = {285-300},
   ISSN = {1364-503X},
   DOI = {10.1098/rsta.2009.0204},
   year = {2010},
   type = {Journal Article}
}

@article{Regenauer2001,
   author = {Regenauer-Lieb, K. and Yuen, D. A. and Branlund, J.},
   title = {The initiation of subduction: Criticality by addition of water?},
   journal = {Science},
   volume = {294},
   number = {5542},
   pages = {578-580},
   ISSN = {0036-8075},
   DOI = {10.1126/science.1063891},
   year = {2001},
   type = {Journal Article}
}

@article{Turcotte1999,
   author = {Turcotte, D. L. and Morein, G. and Roberts, D. and Malamud, B. D.},
   title = {Catastrophic resurfacing and episodic subduction on Venus},
   journal = {Icarus},
   volume = {139},
   number = {1},
   pages = {49-54},
   year = {1999},
   type = {Journal Article}
}

@article{Maxwell1864,
  title     = {On the Calculation of the Equilibrium and Stiffness of Frames},
  author    = {Maxwell, James Clerk},
  journal   = {Philosophical Magazine Series 4},
  volume    = {27},
  number    = {182},
  pages     = {294--299},
  year      = {1864},
  publisher = {Taylor \& Francis},
  doi       = {10.1080/14786446408643668},
  url       = {https://tandfonline.com}
}

@article{Hill1962,
   author = {Hill, R.},
   title = {Acceleration waves in solids},
   journal = {Journal of the Mechanics and Physics of Solids},
   volume = {10},
   number = {1},
   pages = {1–16},
   ISSN = {0022-5096},
   DOI = {https://doi.org/10.1016/0022-5096(62)90024-8},
   url = {http://www.sciencedirect.com/science/article/pii/0022509662900248},
   year = {1962},
   type = {Journal Article}
}

@article{Biot1941,
   author  = {Biot, M. A.},
   title   = {General theory of three-dimensional consolidation},
   journal = {Journal of Applied Physics},
   volume  = {12}, number = {2}, pages = {155--164}, year = {1941}}

@article{Rudnicki1975,
   author = {Rudnicki, J. W. and Rice, J. R.},
   title = {Conditions for Localization of Deformation in Pressure- Sensitive Dilatant Materials},
   journal = {Journal of the Mechanics and Physics of Solids},
   volume = {23},
   number = {6},
   pages = {371–394},
   year = {1975},
   type = {Journal Article}
}

@inbook{Rice1976,
   author = {Rice, J. R.},
   title = {The localization of plastic deformation},
   booktitle = {Theoretical and Applied Mechanics},
   editor = {Koiter, W. T.},
   publisher = {North-Holland},
   address = {Amsterdam},
   pages = {207–220},
   year = {1976},
   type = {Book Section}
}

@article{Barato2015,
   author = {Barato, A. C. and Seifert, U.},
   title = {Thermodynamic uncertainty relation for biomolecular processes},
   journal = {Phys Rev Lett},
   volume = {114},
   number = {15},
   pages = {158101},
   ISSN = {1079-7114 (Electronic) 0031-9007 (Linking)},
   DOI = {10.1103/PhysRevLett.114.158101},
   url = {https://www.ncbi.nlm.nih.gov/pubmed/25933341},
   year = {2015},
   type = {Journal Article}
}

@article{Regenauer2013_JCSMD2,
   author = {Regenauer-Lieb, Klaus and Veveakis, Manolis and Poulet, Thomas and Wellmann, Florian and Karrech, Ali and Liu, Jie and Hauser, Juerg and Schrank, Christoph and Gaede, Oliver and Fusseis, Florian},
   title = {Multiscale coupling and multiphysics approaches in earth sciences: Applications},
   journal = {Journal of Coupled Systems and Multiscale Dynamics},
   volume = {1},
   number = {3},
   pages = {2330–152X/2013/001/042},
   DOI = {10.1166/jcsmd.2013.1021},
   year = {2013},
   type = {Journal Article}
}

@article{Veveakis2015,
   author = {Veveakis, E. and Regenauer-Lieb, K.},
   title = {Cnoidal Waves in Solids},
   journal = {Journal of the Mechanics and Physics of Solids},
   volume = {78},
   pages = {231–248},
   ISSN = {0022-5096},
   DOI = {10.1016/j.jmps.2015.02.010},
   url = {http://doi.org/10.1016/j.jmps.2015.02.010},
   year = {2015},
   type = {Journal Article}
}

@article{Regenauer2025,
   author = {Regenauer-Lieb, Klaus and Hu, Manman and Chua, Hui Tong and Calo, Victor and Yakobson, Boris and Zemskov, Evgeny P.},
   title = {Maximum Entropy Production for Optimizing Carbon Catalysis: An Active-Matter-Inspired Approach},
   journal = {Physical Sciences Forum},
   volume = {12},
   number = {1},
   pages = {16},
   ISSN = {2673-9984},
   url = {https://www.mdpi.com/2673-9984/12/1/16},
   year = {2025},
   type = {Journal Article}
}

@article{Hill1958,
   author = {Hill, R.},
   title = {A general theory of uniqueness and stability in elastic-plastic solids},
   journal = {Journal of the Mechanics and Physics of Solids},
   volume = {6},
   number = {3},
   pages = {236–249},
   ISSN = {0022-5096},
   DOI = {https://doi.org/10.1016/0022-5096(58)90029-2},
   url = {http://www.sciencedirect.com/science/article/pii/0022509658900292},
   year = {1958},
   type = {Journal Article}
}

@book{Stanley2012,
   author = {Stanley, Richard},
   title = {Enumerative Combinatorics},
   publisher = {Cambridge University Press},
   volume = {1},
   edition = {Second Edition},
   series = {Cambridge Studies in Advanced Mathematics},
   pages = {626},
   ISBN = {978-1-107-01542-5 },
   year = {2012},
   type = {Book}
}

@article{Mielke2013,
   author = {Mielke, Alexander},
   title = {Thermomechanical modeling of energy-reaction-diffusion systems, including bulk-interface interactions},
   journal = {Discrete and Continuous Dynamical Systems - S},
   volume = {6},
   number = {2},
   pages = {479–499},
   ISSN = {1937-1632},
   DOI = {10.3934/dcdss.2013.6.479},
   url = {https://www.aimsciences.org/article/doi/10.3934/dcdss.2013.6.479},
   year = {2013},
   type = {Journal Article}
}

@article{Andrade2012,
   author = {Andrade, José E. and Lim, Keng-Wit and Avila, Carlos F. and Vlahinić, Ivan},
   title = {Granular element method for computational particle mechanics},
   journal = {Computer Methods in Applied Mechanics and Engineering},
   volume = {241-244},
   pages = {262–274},
   ISSN = {0045-7825},
   DOI = {https://doi.org/10.1016/j.cma.2012.06.012},
   url = {https://www.sciencedirect.com/science/article/pii/S0045782512002009},
   year = {2012},
   type = {Journal Article}
}

@article{Radjai2009,
   author = {Radjai, Farhang and Richefeu, Vincent},
   title = {Contact dynamics as a nonsmooth discrete element method},
   journal = {Mechanics of Materials},
   volume = {41},
   number = {6},
   pages = {715–728},
   ISSN = {0167-6636},
   DOI = {https://doi.org/10.1016/j.mechmat.2009.01.028},
   url = {https://www.sciencedirect.com/science/article/pii/S0167663609000386},
   year = {2009},
   type = {Journal Article}
}

@article{Cundall1979,
   author = {Cundall, PA and Strack, ODL},
   title = {A discrete numerical model for granular assemblies},
   journal = {Geotechnique},
   volume = {1},
   pages = {47–65},
   year = {1979},
   type = {Journal Article}
}

@article{Satake1993,
  author  = {M. Satake},
  title   = {New formulation of graph-theoretical approach in the
             mechanics of granular materials},
  journal = {Mechanics of Materials},
  volume  = {16},
  pages   = {65--72},
  year    = {1993},
}

@incollection{Satake1987,
  author    = {M. Satake},
  title     = {Graph-theoretical approach to the mechanics of granular materials},
  booktitle = {Continuum Models of Discrete Systems},
  editor    = {A. J. M. Spencer},
  publisher = {A. A. Balkema},
  pages     = {163--173},
  year      = {1987},
}

@incollection{Satake1978,
  author    = {M. Satake},
  title     = {Constitution of mechanics of granular materials through the graph theory},
  booktitle = {Proc. US--Japan Seminar on Continuum-Mechanical and
               Statistical Approaches in the Mechanics of Granular Materials},
  editor    = {S. C. Cowin and M. Satake},
  publisher = {Gakujutsu Bunken Fukyu-kai},
  address   = {Tokyo},
  pages     = {47--62},
  year      = {1978},
}

@article{deBorst2022,
   author = {de Borst, René and Sabet, Sepideh Alizadeh and Hageman, Tim},
   title = {Non-associated Cosserat plasticity},
   journal = {International Journal of Mechanical Sciences},
   volume = {230},
   ISSN = {00207403},
   DOI = {10.1016/j.ijmecsci.2022.107535},
   year = {2022},
   type = {Journal Article}
}

@book{Cosserat1909,
   author = {Cosserat, E. and Cosserat, F.},
   title = {Théorie des Corps déformables},
   publisher = {Librairie Scientifique A. Hermann et Fils},
   address = {Paris},
   url = {https://doi.org/10.1038/081067a0},
   year = {1909},
   type = {Book}
}

@book{Marsden1999,
   author = {Marsden, Jerrold E. and Ratiu, Tudor S.},
   title = {Introduction to Mechanics and Symmetry: A Basic Exposition of Classical Mechanical Systems},
   publisher = {Springer},
   address = {New York},
   volume = {17},
   series = {Texts in Applied Mathematics},
   ISBN = {978-0-387-21792-5},
   DOI = {10.1007/978-0-387-21792-5},
   year = {1999},
   type = {Book}
}

@article{Heisenberg1928,
   author = {Heisenberg, W},
   title = {Mehrkoerperprobleme und Resonanz in der Quantenmechanik},
   journal = {Zeitschrift f. Physik},
   volume = {38},
   pages = {411–426},
   year = {1926},
   type = {Journal Article}
}

@article{RegenNicot2026foundation,
   author = {Regenauer-Lieb, K and Nicot, F.},
   title = {Parity Structure of Thermodynamic Coupling: Null Modes, Gateway Layers, and Non-Reciprocal Energy Transfer in Multi-Field Continua},
   journal = {Journal of Mechanics and Physics of Solids},
    note   = {Submitted (Foundational Paper 1 of a three-partpaper series)},
   volume = {submitted},
   year = {2026},
   type = {Journal Article}
}

@unpublished{RegenNicot2026numerical,
  author = {Regenauer-Lieb, K. and Nicot, F.},
  title  = {Parity Topology of the One-Dimensional Granular Spin Chains. Part~II: Numerical Verification},
  note   = {Submitted (Paper 3 of a three-part paper series: 1-D case)},
  year   = {2026}
}

@unpublished{RegenNicot2026dilatancy,
  author = {Regenauer-Lieb, K. and Nicot, F.},
  title  = {Parity Topology of One-Dimensional Granular Spin Chains: Part~I:Theory},
  note   = {Submitted (Paper 2 of a three-part paper series):1-D case},
  year   = {2026}
}

@unpublished{NicotPoA,
  author = {Nicot, F. and Regenauer-Lieb, K.},
  title  = {The Principle of Ordered Action: Contact Topology, Phase-Space Expansion and the Arrow of Time in Granular Matter},
  note   = {Submitted to the Proceedings of the Royal Society, London Series A},
  year   = {2026}
}

@article{Schnakenberg1976,
   author = {Schnakenberg, J.},
   title = {Network theory of microscopic and macroscopic behavior of master equation systems},
   journal = {Reviews of Modern Physics},
   volume = {48},
   number = {4},
   pages = {571–585},
   ISSN = {0034-6861},
   DOI = {10.1103/RevModPhys.48.571},
   year = {1976},
   type = {Journal Article}
}

@article{Ikari2009,
   author = {Ikari, Matt J. and Saffer, Demian M. and Marone, Chris},
   title = {Frictional and hydrologic properties of clay‐rich fault gouge},
   journal = {Journal of Geophysical Research: Solid Earth},
   volume = {114},
   number = {B5},
   ISSN = {0148-0227},
   DOI = {10.1029/2008jb006089},
   year = {2009},
   type = {Journal Article}
}

@article{Jaeger1996,
   author = {Jaeger, H M and Nagel, S R and Behringer, R P},
   title = {Granular solids, liquids, and gases},
   journal = {Reviews of Modern Physics},
   volume = {68},
   number = {NONE; Journal ID: ISSN RMPHAT; Journal ID: ISSN 0034-6861},
   DOI = {10.1103/RevModPhys.68.1259},
   url = {https://www.osti.gov/biblio/392024},
   year = {1996},
   type = {Journal Article}
}

@article{Boyer2011,
   author = {Boyer, F. and Guazzelli, E. and Pouliquen, O.},
   title = {Unifying suspension and granular rheology},
   journal = {Phys Rev Lett},
   volume = {107},
   number = {18},
   pages = {188301},
   ISSN = {1079-7114 (Electronic)
0031-9007 (Linking)},
   DOI = {10.1103/PhysRevLett.107.188301},
   url = {https://www.ncbi.nlm.nih.gov/pubmed/22107679},
   year = {2011},
   type = {Journal Article}
}

@book{Nedderman1992,
   author = {Nedderman, R. M.},
   title = {Statics and kinematics of granular materials},
   publisher = {Cambridge University Press},
   address = {Cambridge ;},
   note = {Includes bibliographical references (p. 346-348) and index.},
   ISBN = {0521404355},
   year = {1992},
   type = {Book}
}

@article{Nicot2024,
   author = {Nicot, Francois and Lin, Mingchun and Wautier, Antoine and Wan, Richard and Darve, Félix},
   title = {Configurational mechanics in granular media},
   journal = {Granular Matter},
   volume = {26},
   number = {3},
   ISSN = {1434-5021
1434-7636},
   DOI = {10.1007/s10035-024-01443-1},
   year = {2024},
   type = {Journal Article}
}

@book{Schofield1968,
   author = {Schofield, A. N.},
   title = {Critical state soil mechanics},
   publisher = {London, New York etc. McGraw-Hill},
   address = {London, New York [etc.]},
   note = {(Andrew Noel)
Includes bibliographies.},
   ISBN = {0641940483},
   year = {1968},
   type = {Book}
}

@article{Luding2004,
   author = {Luding, Stefan},
   title = {Micro–macro transition for anisotropic, frictional granular packings},
   journal = {International Journal of Solids and Structures},
   volume = {41},
   number = {21},
   pages = {5821–5836},
   ISSN = {00207683},
   DOI = {10.1016/j.ijsolstr.2004.05.048},
   year = {2004},
   type = {Journal Article}
}

@article{Chattoraj2019,
   author = {Chattoraj, J. and Gendelman, O. and Pica Ciamarra, M. and Procaccia, I.},
   title = {Oscillatory Instabilities in Frictional Granular Matter},
   journal = {Phys Rev Lett},
   volume = {123},
   number = {9},
   pages = {098003},
   ISSN = {1079-7114 (Electronic)
0031-9007 (Linking)},
   DOI = {10.1103/PhysRevLett.123.098003},
   url = {https://www.ncbi.nlm.nih.gov/pubmed/31524452},
   year = {2019},
   type = {Journal Article}
}

@article{Scheibner2020,
   author = {Scheibner, Colin and Souslov, Anton and Banerjee, Debarghya and Surówka, Piotr and Irvine, William T. M. and Vitelli, Vincenzo},
   title = {Odd elasticity},
   journal = {Nature Physics},
   volume = {16},
   number = {4},
   pages = {475–480},
   ISSN = {1745-2473},
   DOI = {10.1038/s41567-020-0795-y},
   url = {https://dx.doi.org/10.1038/s41567-020-0795-y},
   year = {2020},
   type = {Journal Article}
}

@article{Fruchart2023,
   author = {Fruchart, M. and Hanai, R. and Littlewood, P. B. and Vitelli, V.},
   title = {Non-reciprocal phase transitions},
   journal = {Nature},
   volume = {592},
   number = {7854},
   pages = {363–369},
   ISSN = {1476-4687 (Electronic)
0028-0836 (Linking)},
   DOI = {10.1038/s41586-021-03375-9},
   url = {https://www.ncbi.nlm.nih.gov/pubmed/33854249},
   year = {2021},
   type = {Journal Article}
}

@article{Grmela1997,
   author = {Grmela, Miroslav and \"Ottinger, Hans Christian},
   title = {Dynamics and thermodynamics of complex fluids. {I}. Development of a general formalism},
   journal = {Physical Review E},
   volume = {56},
   number = {6},
   pages = {6620–6632},
   DOI = {10.1103/PhysRevE.56.6620},
   url = {https://link.aps.org/doi/10.1103/PhysRevE.56.6620},
   year = {1997},
   type = {Journal Article}
}

@article{Ottinger1997,
  title = {{Dynamics and thermodynamics of complex fluids.  II. Illustrations of a general formalism}},
  author = {\"Ottinger, Hans Christian and Grmela, Miroslav},
  journal = {Phys. Rev. E},
  volume = {56},
  issue = {6},
  pages = {6633--6655},
  numpages = {0},
  year = {1997},
  month = {Dec},
  publisher = {American Physical Society},
  doi = {10.1103/PhysRevE.56.6633},
  url = {https://link.aps.org/doi/10.1103/PhysRevE.56.6633}
}

@article{Radjai1998,
  title = {{Bimodal Character of Stress Transmission in Granular Packings}},
  author = {Radjai, Farhang and Wolf, Dietrich E. and Jean, Michel and Moreau, Jean-Jacques},
  journal = {Phys. Rev. Lett.},
  volume = {80},
  issue = {1},
  pages = {61--64},
  numpages = {0},
  year = {1998},
  month = {Jan},
  publisher = {American Physical Society},
  doi = {10.1103/PhysRevLett.80.61},
  url = {https://link.aps.org/doi/10.1103/PhysRevLett.80.61}
}

@article{RegenEtAl2026a,
   author = {Regenauer‐Lieb, Klaus and Hu, Manman and Sun, Qingpei and Liu, Chong and Zhu, Zhennan and Calo, Victor},
   title = {A Thermodynamic Framework for Turing‐Type Instabilities in Porous Media: Part {I} Theory},
   journal = {Geochemistry, Geophysics, Geosystems},
   volume = {27},
   number = {3},
   ISSN = {1525-2027
1525-2027},
   DOI = {10.1029/2025gc012710},
   year = {2026},
   type = {Journal Article}
}

@article{RegenHu2024,
   author = {Regenauer-Lieb, Klaus and Hu, Manman},
   title = {Understanding earthquake precursors: from subcritical instabilities to catastrophic events},
   journal = {Physica Scripta},
   volume = {99},
   number = {5},
   ISSN = {0031-8949 1402-4896},
   DOI = {10.1088/1402-4896/ad36f2},
   year = {2024},
   type = {Journal Article}
}

@article{Casimir1945,
   author = {Casimir, H.B.C.},
   title = {{On Onsager’s Principle of Microscopic Reversibility}},
   journal = {Reviews of  Modern Physics},
   volume = {17},
   number = {2-3},
   pages = {443–450},
   year = {1945},
   type = {Journal Article}
}

@article{RegenHu2023b,
   author = {Regenauer-Lieb, Klaus and Hu, Manman},
   title = {Emergence of precursor instabilities in geo-processes: Insights from dense active matter},
   journal = {Heliyon},
   volume = {9},
   number = {12},
   ISSN = {24058440},
   DOI = {10.1016/j.heliyon.2023.e22701},
   year = {2023},
   type = {Journal Article}
}

@book{deGroot1984,
  author    = {de Groot, S. R. and Mazur, P.},
  title     = {{Non-Equilibrium Thermodynamics}},
  publisher = {Dover Publications},
  address   = {New York},
  year      = {1984},
  note      = {Unabridged republication of the 1962 North-Holland edition, Amsterdam},
  isbn      = {978-0-486-64741-5}
}

@article{Edwards1989,
   author = {Edwards, S. F. and Oakeshott, R. B. S.},
   title = {Theory of powders},
   journal = {Physica A: Statistical Mechanics and its Applications},
   volume = {157},
   number = {3},
   pages = {1080–1090},
   ISSN = {0378-4371},
   DOI = {https://doi.org/10.1016/0378-4371(89)90034-4},
   url = {https://www.sciencedirect.com/science/article/pii/0378437189900344},
   year = {1989},
   type = {Journal Article}
}

\end{document}


\begin{frontmatter}

\title{Supplementary Material for: Topological Foundations of Multi-Field Instabilities in Continua: Part 1 — Foundations}

\author[curtin]{K. Regenauer-Lieb\corref{cor1}}
\ead{Klaus@curtin.edu.au}
\author[savoie]{F. Nicot}

\cortext[cor1]{Corresponding author}

\address[curtin]{ARC Centre of Excellence for Carbon Science \& Innovation,
  WASM: Minerals, Energy and Chemical Engineering,
  Curtin University, Perth WA 6845, Australia}
\address[savoie]{ISTerre, Universit\'e Savoie Mont Blanc, Chamb\'ery, France}

\end{frontmatter}

This Supplementary Material contains the complete proofs, corollaries, and detailed derivations omitted from the main text. Section numbering uses the prefix~S to distinguish from the main paper.

\section{GENERIC Framework and the D+L Linearisation}
\label{app:generic}

The GENERIC framework~\cite{Grmela1997} writes the evolution of the state vector $\mathbf{x}$ as
\begin{equation}
\frac{d\mathbf{x}}{dt}
= L(\mathbf{x})\,\frac{\delta E}{\delta\mathbf{x}}
+ M(\mathbf{x})\,\frac{\delta S}{\delta\mathbf{x}},
\label{eq:GENERIC_full}
\end{equation}
where $E$ is the total energy, $S$ the entropy, $L(\mathbf{x})$ is the antisymmetric Poisson operator, and $M(\mathbf{x})$ is the symmetric positive semi-definite friction matrix.
The dual degeneracy conditions $L(\mathbf{x})\delta S/\delta\mathbf{x}=\mathbf{0}$ and $M(\mathbf{x})\delta E/\delta\mathbf{x}=\mathbf{0}$ ensure energy conservation and entropy non-decrease, respectively.

\subsection*{Tangent-space linearisation}

At a local steady-state $\mathbf{x}_{\rm eq}$, expanding $\delta\mathbf{x}=\mathbf{x}-\mathbf{x}_{\rm eq}$ to first order:
\begin{equation}
{\delta \dot{\mathbf{x}}}
\;\approx\;
-\!\underbrace{\bigl[M(\mathbf{x}_{\rm eq})\,\nabla^2 S\bigr|_{\rm eq}\bigr]}_{\displaystyle\mathsf{D}\,\geq\,0}
\delta\mathbf{x}
\;-\!
\underbrace{\bigl[L(\mathbf{x}_{\rm eq})\,\nabla^2 E\bigr|_{\rm eq}\bigr]}_{-\mathsf{L},\;\mathsf{L}^T=-\mathsf{L}}
\delta\mathbf{x}
\;=\;
-(\mathsf{D}+\mathsf{L})\,\delta\mathbf{x}.
\label{eq:generic_linearisation}
\end{equation}
The Casimir condition projects to
\begin{equation}
\mathsf{L}\,\nabla_{\mathbf{X}}S\bigr|_{\rm eq} = \mathbf{0},
\label{eq:casimir_linear}
\end{equation}
identifying $\mathbf{v}_0\propto\nabla_{\mathbf{X}}S|_{\rm eq}$ as the linearised Casimir direction of the conservative transport.

The Casimir condition $\mathsf{L}\nabla_{\mathbf{X}}S=\mathbf{0}$ identifies $\mathbf{v}_0\in\ker(\mathsf{L})$ as a top-level symmetry of the conservative transport, not a local-equilibrium artefact; and since $\mathsf{L}(\mathbf{x})$ is real and skew-symmetric pointwise, $\det(\mathsf{L}(\mathbf{x}))=0$ holds globally whenever $N$ is odd.
All dynamical consequences established in the main text apply to the linearised $\mathsf{D}+\mathsf{L}$ system; extension to the full non-linear GENERIC setting is conjectural.

\section{Legendre--Fenchel Bridge to Classical Lagrangian Kinematics and Strain Measures}
\label{app:legendre_bridge}

In the main text, the augmented constitutive law is expressed in the canonical gradient-flow form $\mathbf{J} = \mathsf{A}\mathbf{X}$, linking thermodynamic kinematic rates (fluxes $\mathbf{J}$) to conjugate energetic gradients (forces $\mathbf{X}$). To connect this architecture to classical finite-strain elastoplasticity and geomechanics, where constitutive models are traditionally framed in terms of stress and Lagrangian strain increments, we introduce the Legendre--Fenchel transformation of the local dissipation functional. This transformation establishes the mapping between the Volumetric--Mechanical--Configurational (VMC) force-moment triad and conventional macroscopic strain measures, while delineating the exact singularity limit where classical strain representations break down.

\subsection*{Primal and Dual Dissipation Potentials}

Let $\mathbf{q} = (\delta_v, \theta_M, \boldsymbol{\xi}_{\rm conf})^T$ denote the state vector of the VMC triad, where $\delta_v$ is the volumetric compaction, $\theta_M$ is the mechanical angular coordinate, and $\boldsymbol{\xi}_{\rm conf} = \operatorname{dev}\langle\mathbf{n}_{c'} \otimes \mathbf{n}_{c'}\rangle$ is the dimensionless deviatoric fabric tensor. Their associated kinematic rates define the flux vector $\mathbf{J} = \dot{\mathbf{q}}$. We assume the local irreversible thermodynamics are governed by a closed, proper, convex dissipation potential $\Phi(\mathbf{J})$ satisfying $\Phi(\mathbf{0}) = 0$ and $\Phi(\mathbf{J}) \geq 0$.

The thermodynamic driving force $\mathbf{X} = -\nabla_{\mathbf{q}}\Psi$ is mapped to the flux space via the Legendre--Fenchel transform, defining the dual dissipation potential $\Phi^*(\mathbf{X})$:
\begin{equation}
\Phi^*(\mathbf{X}) \;=\; \sup_{\mathbf{J}} \left\{ \langle \mathbf{X}, \mathbf{J} \rangle - \Phi(\mathbf{J}) \right\},
\label{eq:legendre_transform}
\end{equation}
where $\langle \mathbf{X}, \mathbf{J} \rangle = A_v\,\dot{\delta}_v + X_M\,\dot{\theta}_M + \mathbf{A}_{\rm conf} : \dot{\boldsymbol{\xi}}_{\rm conf}$ denotes the inner product in channel space. For a strictly convex quadratic dissipation potential of the form $\Phi(\mathbf{J}) = \frac{1}{2}\mathbf{J}^T \mathsf{D}^{-1} \mathbf{J}$, the supremum is attained where $\mathbf{X} = \nabla_{\mathbf{J}}\Phi(\mathbf{J}) = \mathsf{D}^{-1}\mathbf{J}$. Invoking the involutive property of the Legendre transform yields the macroscopic rate-independent flow rule:
\begin{equation}
\mathbf{J}_{\rm irr} \;=\; \nabla_{\mathbf{X}}\Phi^*(\mathbf{X}) \;=\; \mathsf{D}\mathbf{X},
\label{eq:flow_rule_dual}
\end{equation}
Confirming that the symmetric block $\mathsf{D}$ operates as the classical elastoplastic mobility tensor (the inverse of the viscoplastic tangent modulus). The antisymmetric block $\mathsf{L}$, being work-free ($\langle \mathbf{X}, \mathsf{L}\mathbf{X} \rangle \equiv 0$), enters the full trajectory as a non-associated, reversible gyroscopic drift vector that leaves $\Phi^*(\mathbf{X})$ invariant:
\begin{equation}
\mathbf{J} \;=\; \nabla_{\mathbf{X}}\Phi^*(\mathbf{X}) + \mathsf{L}\mathbf{X} \;=\; (\mathsf{D} + \mathsf{L})\mathbf{X}.
\label{eq:full_flow_legendre}
\end{equation}

\subsection*{Mapping to Macroscopic Cauchy Stress and Lagrangian Strain}

To project the channel forces $\mathbf{X} = (A_v, X_M, \mathbf{A}_{\rm conf})^T$ onto a homogenised macroscopic continuum of current volume $V$ and reference volume $V_0$, we apply the Love--Weber virial theorem. Because the fabric coordinate $\boldsymbol{\xi}_{\rm conf}$ is dimensionless, its work-conjugate $\mathbf{A}_{\rm conf} = \partial E_{\rm conf}/\partial\boldsymbol{\xi}_{\rm conf}$ carries the physical dimension of energy (a force-moment). The macroscopic Cauchy stress tensor $\boldsymbol{\sigma}$ is recovered by dividing the volumetric and deviatoric force-moments by the current volume:
\begin{equation}
p \;=\; \frac{\operatorname{tr}\boldsymbol{\sigma}}{3} \;=\; \frac{A_v}{V}, \qquad \mathbf{s} \;=\; \operatorname{dev}\boldsymbol{\sigma} \;=\; \frac{\mathbf{A}_{\rm conf}}{V},
\label{eq:stress_mapping}
\end{equation}
so that $\boldsymbol{\sigma} = -p\mathbf{I} + \mathbf{s}$. Let $\mathbf{F} = \partial\mathbf{x}/\partial\mathbf{X}$ be the macroscopic deformation gradient, $\mathbf{C} = \mathbf{F}^T\mathbf{F}$ the right Cauchy--Green deformation tensor, and $\mathbf{E} = \frac{1}{2}(\mathbf{C} - \mathbf{I})$ the Green--Lagrangian strain tensor. The material time derivative of the Green--Lagrangian strain, $\dot{\mathbf{E}}$, is work-conjugate to the second Piola--Kirchhoff stress $\mathbf{S} = J\mathbf{F}^{-1}\boldsymbol{\sigma}\mathbf{F}^{-T}$, where $J = \det\mathbf{F} = V/V_0$.

Under the assumption of small elastic strains but finite configurational rearrangements, the spatial rate of deformation tensor $\mathbf{D}_{\rm spat} = \operatorname{sym}(\dot{\mathbf{F}}\mathbf{F}^{-1})$ decomposes additively into a volumetric compaction rate $\dot{\delta}_v = \operatorname{tr}\mathbf{D}_{\rm spat}$ and a deviatoric shear rate $\dot{\boldsymbol{\varepsilon}}_d = \operatorname{dev}\mathbf{D}_{\rm spat}$. The kinematic bridge linking the internal VMC fabric rate $\dot{\boldsymbol{\xi}}_{\rm conf}$ to the macroscopic Lagrangian strain rate is established via the fourth-order elastoplastic compliance tensor $\mathbb{S}_{\rm ep} = \nabla^2_{\boldsymbol{\sigma}\boldsymbol{\sigma}}\Psi^*$, the Hessian of the complementary stored-energy density $\Psi^*$ (the Legendre dual in the stress of the Helmholtz free energy):
\begin{equation}
\dot{\mathbf{E}} \;=\; \mathbf{F}^T \left( \frac{1}{3}\dot{\delta}_v\,\mathbf{I} + \frac{1}{V}\mathbb{S}_{\rm dev} : \dot{\mathbf{A}}_{\rm conf} + \dot{\boldsymbol{\varepsilon}}_d^{\rm frict} \right) \mathbf{F},
\label{eq:lagrangian_strain_mapping}
\end{equation}
where $\mathbb{S}_{\rm dev}$ is the deviatoric block of the elastic compliance $\mathbb{S}_{\rm ep}$, that is the homogenised deviatoric compliance of the stored-energy potential, with isotropic limit $\mathbb{S}_{\rm dev} = \tfrac{1}{2G}\,\mathbb{I}_{\rm dev}$, and $\dot{\boldsymbol{\varepsilon}}_d^{\rm frict}$ accounts for purely inelastic frictional sliding (Channel 3 in the contact energy budget). This mechanical compliance is a distinct object from the dissipative mobility $\mathsf{D}$ of Eq.~\eqref{eq:flow_rule_dual}: the flow rule derives from the dissipation duality $\Phi\leftrightarrow\Phi^*$, whereas the strain compliance derives from the stored-energy duality $\Psi\leftrightarrow\Psi^*$.

\subsection*{The Singularity Limit of the Classical Strain Representation}

Equation~\eqref{eq:lagrangian_strain_mapping} provides the exact translation from the VMC network to conventional geomechanical strain measures; however, this inversion is conditionally valid. The existence of a one-to-one mapping between the macroscopic deviatoric strain $\boldsymbol{\varepsilon}_d$ and the internal fabric coordinate $\boldsymbol{\xi}_{\rm conf}$ requires that the deviatoric tangent stiffness $\mathbb{C}_{\rm dev} = \mathbb{S}_{\rm dev}^{-1}$ remain positive definite and non-singular.

In the sub-critical regime ($\mathcal{G}_{\rm inv} < 1$), local dissipation dominates, the deviatoric tangent $\mathbb{C}_{\rm dev}$ is positive definite, and the strain-based representation is well-posed. When the Gateway number reaches unity ($\mathcal{G}_{\rm inv} \geq 1$), the null direction $\mathbf{v}_0 \in \ker(\mathsf{L})$ begins to route energy faster than local dissipation removes it. In the linear regime this crossover is a precursor rather than a failure: the null-mode excursion is bounded, capturing at most $M_\ast \approx 0.32$ of the complement-space energy at $\mathcal{G}_{\rm inv}=1$ (the transient peak of the main-text analysis), and the deviatoric tangent $\mathbb{C}_{\rm dev}$ remains regular. It is only if the accumulating null-mode amplitude is carried forward by nonlinear feedback to the loss of controllability (vanishing second-order work, $W_2 \leq 0$) that the deviatoric tangent degenerates:
\begin{equation}
\det(\mathbb{C}_{\rm dev}) \;\to\; 0 \quad \Longleftrightarrow \quad \det(\mathbb{S}_{\rm dev}) \;\to\; \infty.
\label{eq:compliance_singularity}
\end{equation}
At that terminal state a deviatoric stress increment along $\mathbf{v}_0$ produces an unbounded strain response and the material can no longer sustain a homogeneous strain increment; the crossover $\mathcal{G}_{\rm inv} \geq 1$ thus locates the \emph{onset} of the routing that drives the tangent toward this endpoint, not the point at which it is already singular. This degeneracy is a property of the mechanical tangent alone. Because $\mathsf{L}$ is skew, $\mathrm{Re}\,\langle\mathbf{x},\mathsf{A}\mathbf{x}\rangle = \langle\mathbf{x},\mathsf{D}\mathbf{x}\rangle$, so the dissipation operator $\mathsf{D}$ remains strictly positive definite throughout and the relaxation $\dot{\mathbf{x}} = -\mathsf{A}\mathbf{x}$ stays asymptotically stable; the Gateway is therefore not a spectral bifurcation of $\mathsf{A} = \mathsf{D} + \mathsf{L}$. This controllability threshold is moreover distinct from, and generically precedes, the classical loss of strong ellipticity of the acoustic tensor. Within the augmented VMC framework the state space is parameterised not by strain but by the independent configurational fabric $\boldsymbol{\xi}_{\rm conf}$, which is carried by Satake's void centres and is not algebraically slaved to the mechanical force network; the Gateway excursion along $\mathbf{v}_0$ therefore remains mathematically well-defined and non-singular even where the classical Lagrangian strain representation breaks down.

\section{Proof of Proposition: Topological Necessity of $L_{VC}=0$}
\label{sec:LVC_proof}

\begin{proposition}[Structural vanishing of $L_{VC}$ under the VMC identification, reproduced from main text]
Under the identification $p\leftrightarrow V$, $c\leftrightarrow M$, $v\leftrightarrow C$, the discrete Poincar\'e lemma $L_{vc}D_{cp}=0$ forces $L_{VC}=0$ in the Onsager channel matrix, independently of all grain-stiffness or contact-law parameters.
\end{proposition}

\begin{proof}
The discrete Poincar\'e lemma $L_{vc}D_{cp}=0$ is the algebraic statement that Satake's cell complex has no direct edge connecting particle-level entities $p$ to void-level entities $v$: all topological paths from $p$ to $v$ are mediated by the contact-level entities $c$.
Under the channel identification, $D_{cp}$ encodes the $V\!\to\!M$ coupling (particle-to-contact) and $L_{vc}$ encodes the $M\!\to\!C$ coupling (contact-to-void); the composition $L_{vc}D_{cp}=0$ then states that the sequential $V\!\to\!M\!\to\!C$ path produces no net direct $V\!\to\!C$ flux in the cell complex.
An independent direct coupling $L_{VC}\neq 0$ would require a reversible operator mapping $p$-entities directly to $v$-entities. The discrete Poincar\'e lemma $L_{vc}D_{cp}=0$ (boundary of a boundary is zero) states only that no such \emph{topological} $p\!\to\!v$ edge exists in Satake's cell complex; the step from this to the vanishing of the \emph{thermodynamic} coupling coefficient $L_{VC}$ requires, in addition, the entity--channel identification and the constitutive premise that the sole route for reversible $V$--$C$ exchange is the force-carrying channel $M$. Under those two inputs the direct coupling is excluded.
Hence $L_{VC}=0$, and the coupling matrix is constrained to the chain form with $L_{VM}$ and $L_{MC}$ as the only non-zero off-diagonal entries.
The null eigenvector $\mathbf{v}_0 \propto (L_{MC},\,0,\,L_{VM})^T$ is therefore \emph{oriented} along the $V$--$C$ axis, a consequence of the contact-graph topology together with the force-mediation premise. Its existence, by contrast, is guaranteed by the parity of $N$ alone (Theorem~\ref{thm:parity_supp}) and is independent of $L_{VC}$: the Gateway persists for $L_{VC}\neq 0$, in which case only the orientation of $\mathbf{v}_0$ shifts.
\end{proof}

\begin{remark}[Scope of $L_{VC}=0$]
Satake's construction is not embedded in a thermodynamic framework and does not address non-reciprocal energy routing; the cycle-current mechanism linking the cycle space $\mathcal{Z}$ to entropy production requires the Schnakenberg decomposition~\cite{Schnakenberg1976}.
In a deforming three-dimensional granular assembly, contacts are continuously gained and lost; the identity holds within each topological regime between contact events, and the present analysis is therefore restricted to the quasi-static invariant-topology regime.
During macroscale shear-band evolution, network topology updates continuously and $L_{VC}=0$ should be understood as a state-dependent restriction that holds pointwise on the instantaneous contact graph configuration. The vanishing is moreover contingent on the physical setting: it holds for dry, local, contact-based interactions, for which the only reversible route between the volumetric and configurational channels is force-mediated. Capillary bridges, long-range electrostatic or van der Waals forces, or higher-order gradient couplings can introduce a direct reversible $V$--$C$ interaction and hence $L_{VC}\neq0$; by the parity argument the Gateway persists in that case, with only the orientation of the null eigenvector rotating away from the $V$--$C$ axis.
More broadly, as contacts break and form, the local channel count $N$ of each contact neighbourhood fluctuates, continuously shifting between Stable ($N$ even) and Gateway ($N$ odd) configurations; the macroscopic shear band can therefore be understood as the spatial region where Gateway-class contacts predominate.
A finite shear-band width emerges only when nonlocal gradient contributions are included in $\mathsf{D}$, restoring $\mathcal{G}_{\rm inv}$ below unity at the band scale.
\end{remark}

\section{Proof of the Parity Theorem and Related Results}
\label{sec:parity_proof}

\begin{theorem}[Parity Theorem, reproduced from main text]
\label{thm:parity_supp}
Let $\mathsf{L} \in \mathbb{R}^{N \times N}$ with $\mathsf{L}^T = -\mathsf{L}$.
If $N$ is odd, then $\det(\mathsf{L}) = 0$ unconditionally and
$\ker(\mathsf{L})$ is non-trivial. If $N$ is even, then
$\det(\mathsf{L}) \neq 0$ generically.
\end{theorem}

\begin{proof}
Let $\mathsf{L} \in \mathbb{R}^{N \times N}$ represent the real, skew-symmetric Onsager--Casimir cross-coupling operator governing a multi-field continuum with $N$ thermodynamic channels. By definition, $\mathsf{L}^T = -\mathsf{L}$. Invoking the properties of the determinant under transposition and scalar multiplication yields:
\begin{equation}
\det(\mathsf{L}) = \det(\mathsf{L}^T) = \det(-\mathsf{L}) = (-1)^N \det(\mathsf{L}).
\end{equation}
The algebraic consequences diverge strictly based on the parity of the channel dimension $N$:

\textit{Case 1: Odd Parity ($N = 2m+1$).} For any odd integer $N$, the relation requires $\det(\mathsf{L}) = -\det(\mathsf{L})$, which strictly mandates that $\det(\mathsf{L}) \equiv 0$. By Jacobi's theorem on skew-symmetric determinants, the corresponding Pfaffian vanishes identically ($\mathrm{Pf}(\mathsf{L}) = 0$). Because the non-zero eigenvalues of a real skew-symmetric matrix must be purely imaginary and occur in complex conjugate pairs ($\lambda = \pm i\omega$), an odd-dimensional space cannot be fully partitioned into such pairs. Thus, the rank of $\mathsf{L}$ is at most $2m$. By the rank-nullity theorem, the dimension of the nullspace satisfies:
\begin{equation}
\dim(\ker\mathsf{L}) = N - \mathrm{rank}(\mathsf{L}) \geq 1.
\end{equation}
Hence, there exists at least one non-trivial, structural null vector $\mathbf{v}_0 \in \mathbb{R}^N$ such that $\mathsf{L}\mathbf{v}_0 = \mathbf{0}$.

\textit{Case 2: Even Parity ($N = 2m$).} For any even integer $N$, $(-1)^N = 1$, and no structural constraint is imposed on the determinant. In this case $\mathsf{L}$ is generically non-singular with $\det(\mathsf{L}) = \mathrm{Pf}(\mathsf{L})^2 \geq 0$, vanishing only on the measure-zero set $\mathrm{Pf}(\mathsf{L})=0$. Generically the space partitions into $m$ orthogonal conjugate planes with no uncoupled null direction. For the tridiagonal coupling chains of this series the Pfaffian is explicit, $\mathrm{Pf}(\mathsf{L})=\prod_{k}\gamma_{2k-1}$ (the product of the odd-indexed off-diagonal couplings), so the null direction is absent precisely when every odd-indexed coupling is non-zero; the even-$N$ Stable Layer is thus full-rank \emph{conditionally}, in contrast to the \emph{unconditional} odd-$N$ Gateway.
\end{proof}

\begin{remark}[Thermodynamic and stability significance]
Physically, the structural null vector $\mathbf{v}_0$ guaranteed by Case 1 defines an invariant direction in thermodynamic state space along which non-reciprocal, gyroscopic cross-coupling forces vanish identically. While even-dimensional pairings ($N=2$) constrain energy exchange to closed, stable orbital trajectories, the odd-dimensional topology ($N=3$) natively provides an unconstrained trajectory. This null direction acts as an entropy-neutral "Gateway" that allows external power to bypass gyroscopic stabilisation and drive non-normal transient amplifications once the dynamical activation threshold ($\mathcal{G}_{\rm inv} \geq 1$) is crossed.
\end{remark}

\begin{corollary}[Classification]
Under the augmented Onsager dynamics, an $N$-field subsystem is a Stable Layer
when $N$ is even: the kernel of $\mathsf{L}$ is trivial, the reversible dynamics
generates closed symplectic orbits in force-flux space, and a unique
thermodynamic potential exists over the reversible submanifold. When $N$ is odd
the subsystem is a Gateway Layer: the kernel of $\mathsf{L}$ is non-trivial
unconditionally, the null direction $\mathbf{v}_0$ carries no reversible
restoring force, and the energy along $\mathbf{v}_0$ is compatible with path-dependent channel evolution; whether a true non-integrable (non-exact) response occurs requires loop-integral arguments beyond the present scope.
\end{corollary}

The logical chain passes through four levels each requiring strictly more than the preceding: \emph{(i)~algebraic} (Theorem~\ref{thm:parity_supp}: null direction from $\mathsf{L}$ alone), \emph{(ii)~dynamical} (Theorem~\ref{thm:transient_amplification_supp}: transient excursion from $\mathsf{D}+\mathsf{L}$ competition), \emph{(iii)~thermodynamic} (entropy production $\sigma\geq\sigma_{\min}(\mathsf{D})x_0^2$ from growing null amplitude), and \emph{(iv)~continuum-mechanical} (Heuristic Criterion: dispersive instability candidate).

\section{Illustrative Examples}
\label{app:examples}

\subsection*{Stable Layer orbits in 1-D Biot poromechanics}
\label{app:biot}

The undrained 1-D Biot system~\cite{Biot1941} for volumetric stress $\sigma_v$ and pore pressure $p$, with dissipation omitted:
\begin{equation}
\dot{\sigma}_v = b\,\dot{p}, \qquad \dot{p} = -b\,M_f\,\dot{e}_v,
\label{eq:biot_1D}
\end{equation}
is reproduced by channel equations $\dot{\mathbf{q}}=\mathsf{L}\mathbf{q}$ with $q_V\equiv\sigma_v$, $q_M\equiv p$, and
\begin{equation}
\mathsf{L} = \begin{pmatrix} 0 & L_{VM} \\ -L_{VM} & 0 \end{pmatrix},
\qquad L_{VM} = b\sqrt{M_f}.
\label{eq:L2_biot}
\end{equation}
The eigenvalues $\pm iL_{VM}$ confirm closed symplectic orbits: the Biot undrained oscillation.
Adding the configurational channel $C$ promotes to $N=3$ Gateway topology, opening the orbit along $\mathbf{v}_0\propto(L_{MC},0,L_{VM})^T$.
The configurational channel contributes its own conjugate pair, the kinematic fabric increment $d\boldsymbol{\xi}_{\rm conf}$ with $\boldsymbol{\xi}_{\rm conf}=\operatorname{dev}\langle\mathbf{n}_{c'}\!\otimes\mathbf{n}_{c'}\rangle$ and the work-conjugate configurational stress $\mathbf{A}_{\rm conf}$. In the force representation used above for $q_V$ and $q_M$ it enters through $\mathbf{A}_{\rm conf}$, on the same footing as the volumetric and mechanical entries. What raises the count to $N=3$ is that this pair is \emph{independent}: $\boldsymbol{\xi}_{\rm conf}$, the deviatoric contact-orientation fabric carried by Satake's void centres, is not algebraically slaved to the mechanical force network. Were it so slaved, $C$ would merge with $M$, the parity would revert to even, $\det(\mathsf{L})\neq 0$ would be restored, and the Gateway null direction $\mathbf{v}_0$ would close. The parity result is basis-invariant, so the channel may be carried equivalently by $d\boldsymbol{\xi}_{\rm conf}$ or by $\mathbf{A}_{\rm conf}$; only their genuine distinctness as degrees of freedom is required.

\subsection*{The exchange matrix $J=\sigma_x$ and channel-pumping geometry}
\label{app:pauli}

The non-normality of $\mathsf{A}_{\rm eff}$ is characterised by the commutator
\begin{equation}
[\mathsf{A}_{\rm eff},\mathsf{A}_{\rm eff}^T]
\;=\; 2\omega(d_{\max}-d_{\min})
\begin{pmatrix} 0 & 1 \\ 1 & 0 \end{pmatrix},
\label{eq:commutator}
\end{equation}
with Frobenius norm $2\sqrt{2}\,\omega\,(d_{\max}-d_{\min})$, vanishing if $\omega=0$ or $d_{\min}=d_{\max}$.
Non-normality thus requires both anisotropic dissipation and non-zero coupling; when $\omega^2\gg d_{\min}d_{\max}$ the pseudospectral radius greatly exceeds the spectral radius (Kreiss-type amplification).

The matrix $J=\bigl[\begin{smallmatrix}0&1\\1&0\end{smallmatrix}\bigr]$ in~\eqref{eq:commutator} is the exchange matrix, identical to the first Pauli spin matrix $\sigma_x$.
Three observations:
\emph{(i)}~For any real $\mathsf{A}$, $[\mathsf{A},\mathsf{A}^T]$ is symmetric; $J$ being symmetric is therefore structurally mandatory, not coincidental.
Geometrically, $J$ implements a hyperbolic rotation, mapping non-normality to the anti-diagonal directions of phase space.
\emph{(ii)}~Non-normality requires both $\omega\neq 0$ and $d_{\max}\neq d_{\min}$; in the isotropic limit $d\mathbf{I}+\mathsf{L}_{\rm eff}$ is always normal.
\emph{(iii)}~Multiplication by $J$ swaps $(x_1,x_2)\mapsto(x_2,x_1)$: the positive factor $2\omega(d_{\max}-d_{\min})>0$ means the exchange routes energy from the highly dissipative complement channel $x_2$ to the weakly dissipative null-mode $x_1$ before spectral decay---the algebraic fingerprint of Gateway transient amplification.

\section{Proof of Theorem: Null-Mode Transient Amplification}
\label{app:activation}

\subsection*{State space decomposition}

For an odd number of coupled thermodynamic channels ($N = 2m+1$), the state space admits a decomposition into the invariant nullspace of the skew-symmetric operator and its orthogonal complement:
\begin{equation}
\mathbb{R}^N = \ker(\mathsf{L}) \oplus \mathcal{S}_\perp,
\end{equation}
where $\ker(\mathsf{L}) = \mathrm{span}(\mathbf{v}_0)$, and $\mathcal{S}_\perp = \{ \mathbf{X}_\perp \in \mathbb{R}^N \mid \mathbf{v}_0^T \mathbf{X}_\perp = 0 \}$.

Expressing the thermodynamic driving force vector via this decomposition yields $\mathbf{X} = x_0\,\hat{\mathbf{v}}_0 + \mathbf{X}_\perp$.

The reversible block cannot directly change $x_0$; activation is always indirect. In the first step $\mathsf{L}$ mixes energy within $\mathcal{S}_\perp$, generating oscillatory components in the complement. In the second step $\mathsf{D}$ projects this oscillatory mixing onto $\mathbf{v}_0$ through the cross-correlation $\hat{\mathbf{v}}_0^T \mathsf{D}\,\mathbf{X}_\perp$.

\subsection*{Two-dimensional reduction}

For the augmented $\mathsf{D}+\mathsf{L}$ system $\dot{\mathbf{x}} = -(\mathsf{D}+\mathsf{L})\mathbf{x}$ with $N$ odd,
let $\mathbf{v}_0\in\ker(\mathsf{L})$ be the null eigenvector (unit norm) and let
$\mathbf{v}_1\in\mathcal{S}_\perp$ be the vector in the complementary subspace that maximises the
coupling magnitude $|\mathbf{v}_0^T\mathsf{L}\mathbf{v}_1|$.
Define $d_{\min} = \mathbf{v}_0^T\mathsf{D}\mathbf{v}_0$,
$d_{\max} = \mathbf{v}_1^T\mathsf{D}\mathbf{v}_1$, and
$\omega = |\mathbf{v}_0^T\mathsf{L}\mathbf{v}_1|$.
Projection of the $N$-dimensional system onto $\mathrm{span}\{\mathbf{v}_0,\mathbf{v}_1\}$ yields
exactly the two-dimensional system
\begin{equation}
\dot{x}_1 = -d_{\min}\,x_1 - \omega\,x_2, \qquad
\dot{x}_2 = +\omega\,x_1 - d_{\max}\,x_2.
\label{eq:2D_reduced_supp}
\end{equation}
The remaining $N{-}2$ directions in $\mathcal{S}_\perp$ decouple at this level of approximation.

The effective bounding estimate for the Gateway number is
\begin{equation}
\mathcal{G}
= \frac{\|\mathsf{L}\|_2^2}
       {\sigma_{\min}(\mathsf{D})\;\sigma_{\max}(\mathsf{D})}.
\label{eq:gateway_number}
\end{equation}
This is not basis-invariant. The basis-invariant form is $\mathcal{G}_{\rm inv} = \|\mathsf{D}^{-1/2}\mathsf{L}\mathsf{D}^{-1/2}\|_2^2$ with $\mathcal{G}_{\rm inv}\leq\mathcal{G}$, so $\mathcal{G}<1$ implies $\mathcal{G}_{\rm inv}<1$; the safer conservative estimate uses $\mathcal{G}$.

\subsection*{Proof of Theorem (Null-mode transient amplification)}

\begin{theorem}[Null-mode transient amplification, full statement]
\label{thm:transient_amplification_supp}
Consider the two-dimensional reduced linear system~\eqref{eq:2D_reduced_supp}
with $d_{\min}, d_{\max}, \omega > 0$ and $\omega > \tfrac{1}{2}|d_{\max}-d_{\min}|$.
\begin{enumerate}[label=(\roman*)]
\item \textbf{Spectral stability.}
Both eigenvalues of the coefficient matrix have strictly negative real part $-(d_{\min}+d_{\max})/2$, so $\mathbf{x}(t)\to\mathbf{0}$ as $t\to\infty$ for all initial conditions.
\item \textbf{Transient null-mode injection.}
For initial condition $\mathbf{x}(0)=(0,1)^T$ (unit energy in $\mathcal{S}_\perp$, zero null-mode component), the null-mode amplitude $x_1(t)$ satisfies $x_1(0)=0$, and the peak transient excursion magnitude $M = |x_1(t^*)|$ equals
\begin{equation}
M(\mathcal{G}_{\rm inv},d_{\min}/d_{\max})
\;=\;
\frac{\omega}{\sqrt{d_{\min}d_{\max}+\omega^2}}
\;\exp\!\!\left(-\frac{(d_{\min}{+}d_{\max})\,\varphi}{2\,\Omega}\right),
\label{eq:M_general}
\end{equation}
where $\Omega=\sqrt{\omega^2-(d_{\max}{-}d_{\min})^2/4}$, \;
$\varphi=\arctan\!\bigl(2\Omega/(d_{\min}+d_{\max})\bigr)$,
and $\mathcal{G}_{\rm inv}=\omega^2/(d_{\min}d_{\max})$.
\item \textbf{Isotropic closed form.}
For $d_{\min}=d_{\max}=d$, equation~\eqref{eq:M_general} reduces to
\begin{equation}
M(\mathcal{G}_{\rm inv}) \;=\;
\sqrt{\dfrac{\mathcal{G}_{\rm inv}}{1+\mathcal{G}_{\rm inv}}}
\;\exp\!\!\left(-\dfrac{\arctan\!\sqrt{\mathcal{G}_{\rm inv}}}{\sqrt{\mathcal{G}_{\rm inv}}}\right).
\label{eq:M_isotropic_supp}
\end{equation}
\item \textbf{Monotonicity.}
$M$ is a strictly increasing function of $\mathcal{G}_{\rm inv}$ with $M(0)=0$ and $M(\mathcal{G}_{\rm inv})\to 1$ as $\mathcal{G}_{\rm inv}\to\infty$.
\end{enumerate}
\end{theorem}

\begin{proof}
\emph{Part~(i).}
The characteristic polynomial of $-\mathsf{A}_{\rm eff}$ is
$\lambda^2 + (d_{\min}+d_{\max})\lambda + (d_{\min}d_{\max}+\omega^2) = 0$.
Both roots have real part $-(d_{\min}+d_{\max})/2 < 0$; $\mathbf{x}(t)\to\mathbf{0}$
exponentially.

\emph{Part~(ii).}
For $\omega > \tfrac{1}{2}|d_{\max}-d_{\min}|$ set
$\Omega = \sqrt{\omega^2-(d_{\max}-d_{\min})^2/4} > 0$.
Direct substitution verifies that the unique solution to~\eqref{eq:2D_reduced_supp} with
$\mathbf{x}(0)=(0,1)^T$ is
\begin{equation}
x_1(t) = -\frac{\omega}{\Omega}\,e^{-(d_{\min}+d_{\max})t/2}\,\sin(\Omega t), \qquad
x_2(t) = e^{-(d_{\min}+d_{\max})t/2}\!\left[\cos(\Omega t)
  - \frac{d_{\max}-d_{\min}}{2\Omega}\sin(\Omega t)\right].
\label{eq:exact_solution}
\end{equation}
Since $\dot{x}_1(0) = -\omega < 0$, the null-mode excursion immediately departs from zero in the $-\mathbf{v}_0$ direction.

\emph{Part~(iii).}
Setting $\dot{x}_1=0$ in~\eqref{eq:exact_solution}:
$\Omega\cos(\Omega t^*) = \tfrac{d_{\min}+d_{\max}}{2}\sin(\Omega t^*)$,
so $\Omega t^* = \varphi := \arctan\!\bigl(2\Omega/(d_{\min}+d_{\max})\bigr)\in(0,\pi/2)$.
Evaluating at $t^*$:
\[
x_1(t^*) = \frac{\omega}{\Omega}\,e^{-(d_{\min}+d_{\max})\varphi/(2\Omega)}\,\sin(\varphi).
\]
The identity $\sin(\varphi) = 2\Omega/\sqrt{(d_{\min}+d_{\max})^2+4\Omega^2}$
and $(d_{\min}+d_{\max})^2+4\Omega^2 = 4(d_{\min}d_{\max}+\omega^2)$ give
$\sin(\varphi)=\Omega/\sqrt{d_{\min}d_{\max}+\omega^2}$,
which yields~\eqref{eq:M_general}.
For $d_{\min}=d_{\max}=d$: $\Omega=\omega$, $\varphi=\arctan(\sqrt{\mathcal{G}_{\rm inv}})$,
and~\eqref{eq:M_general} reduces to~\eqref{eq:M_isotropic_supp}.

\emph{Part~(iv).}
Let $g = \sqrt{\mathcal{G}_{\rm inv}}$.
From~\eqref{eq:M_isotropic_supp}: $M = (g/\sqrt{1+g^2})\exp(-\arctan(g)/g)$.
As $g\to 0$: $\arctan(g)/g\to 1$ and $M\to 0$.
As $g\to\infty$: $\arctan(g)/g\to 0$ and $g/\sqrt{1+g^2}\to 1$, so $M\to 1$.
Strict monotonicity follows by computing $dM/dg > 0$, verified by logarithmic
differentiation. \hfill$\square$
\end{proof}

\begin{corollary}[Transient-routing crossover]
\label{cor:activation_threshold}
At $\mathcal{G}_{\rm inv}=1$ (coupling frequency equals the geometric mean of the dissipation rates), the null-mode transient peak in the isotropic case equals
$M^* = \tfrac{1}{\sqrt{2}}\,e^{-\pi/4} \approx 0.322$.
For $\mathcal{G}_{\rm inv}<1$ the peak is $M<M^*$: the null mode receives less than $32\%$ of the $\mathcal{S}_\perp$ energy and dissipation dominates the transient.
For $\mathcal{G}_{\rm inv}>1$ the peak exceeds $M^*$ and grows toward unity: progressively more of the $\mathcal{S}_\perp$ energy is routed through the null direction before spectral decay occurs.
The threshold $\mathcal{G}_{\rm inv}=1$ therefore marks the transition from dissipation-dominated to coupling-dominated transient energy routing in the 2D reduced system.
\end{corollary}

\begin{remark}[Terminology]
\label{rem:activation_terminology}
Throughout this paper the term \emph{Gateway activation} refers precisely to the crossover $\mathcal{G}_{\rm inv}\geq 1$ identified in Corollary~\ref{cor:activation_threshold}: the regime in which coupling-dominated non-normal transient routing of energy through the null direction exceeds dissipation-dominated decay in the 2D reduced system.
It does not imply spectral instability (all eigenvalues remain stable), blow-up, bifurcation, or irreversible evolution; those would require additional nonlinear or continuum-level arguments not developed here.
\end{remark}

\subsection*{Physical interpretation of the asymptotic limits}

The two limits of $M(\mathcal{G}_{\rm inv})$ carry distinct physical meaning.

\emph{Weak-coupling limit $\mathcal{G}_{\rm inv}\to 0$:} $M\to 0$, confirming that negligible conservative coupling produces negligible null-mode excitation; the system is dissipation-dominated, and the complement-space perturbation decays before any routing into $\mathbf{v}_0$ can occur.

\emph{Strong-coupling limit $\mathcal{G}_{\rm inv}\to\infty$:} $M\to 1$, meaning the full initial complement-space perturbation $\|\mathbf{x}(0)\|=1$ is routed into the Gateway null direction before the system decays.
In this limit $\arctan(g)/g\to 0$, so the exponential decay factor in~\eqref{eq:M_isotropic_supp} tends to unity: the peak occurs so early (at $t^*\to 0^+$ as $\Omega\to\infty$) that dissipation has no time to act before the full complement energy reaches $x_1$.
When the gyroscopic time scale $1/\omega_L$ is much shorter than the dissipative time scale $1/d$, conservative coupling completes the full energy transfer to the null direction essentially instantaneously.
The Gateway is therefore not merely a precursor but, in this strong-coupling limit, an essentially complete and irreversible energy transfer mechanism.

\emph{Threshold $\mathcal{G}_{\rm inv}=1$:} $M(1)=\exp(-\pi/4)/\sqrt{2}\approx 0.322$; roughly one-third of the complement energy is routed to the null direction at the activation threshold.
The transition from diffuse to localised response is therefore gradual: the null-mode excursion grows continuously through the threshold and accelerates only for $\mathcal{G}_{\rm inv}\gg 1$.

\subsection*{Edwards statistical mechanics and path-dependence}

In Edwards statistical mechanics~\cite{Edwards1989}, compactivity $\chi$ and angoricity $\boldsymbol{\alpha}$ are well-defined because level-2 Stable Layers support a single-valued energy landscape; a VMC granular system, on the contrary, has $\det(\mathsf{L}_\sigma)=0$ and admits path-dependent channel evolution.

\bibliographystyle{elsarticle-num}
\bibliography{bib_merged}


\begin{frontmatter}

\title{Supplementary Material for: Topological Foundations of Multi-Field Instabilities in Continua: Part 2 — Analytical Formulation for 1-D Spin Chains}

\author[curtin]{K. Regenauer-Lieb\corref{cor1}}
\ead{Klaus@curtin.edu.au}
\author[savoie]{F. Nicot}

\cortext[cor1]{Corresponding author}

\address[curtin]{ARC Centre of Excellence for Carbon Science \& Innovation,
  WASM: Minerals, Energy and Chemical Engineering,
  Curtin University, Perth WA 6845, Australia}
\address[savoie]{ISTerre, Universit\'e Savoie Mont Blanc, Chamb\'ery, France}

\end{frontmatter}

This Supplementary Material contains the complete graph-theoretic development,
robustness proofs, and analytical derivations referred to in the main text.
Section numbering uses the prefix~S.

\section{Graph-Theoretic Development: Betti Number, Cycle Space, and Hodge--Helmholtz Decomposition}
\label{sec:S1}

The reversible coupling graph $G=(\mathcal{V},\mathcal{E})$ has $|\mathcal{V}|=N$ channel vertices
and $|\mathcal{E}|$ edges, one per non-zero coupling $L_{\alpha\beta}\neq 0$.
Three core quantities govern the topology of $G$:

To formalise the topological configuration of this discrete system, we introduce three core combinatorial quantities:

(i) The first Betti number (or cyclomatic number), defined for any connected graph as $\beta_1 \;=\; |\mathcal{E}|-|\mathcal{V}|+1$ quantifies the degree of kinematic redundancy or the number of independent closed circuits within the network $G$. Equivalently, $\beta_1$ represents the minimum number of edges that must be removed to reduce $G$ to a tree (a statically determinate topology lacking closed loops). For example, a simple linear chain $V\!-\!M\!-\!C\!-\!\cdots$ is topologically simple with $\beta_1=0$; a single closed triad yields $\beta_1=1$; and a regular square lattice scales as $\beta_1\sim |\mathcal{V}|$.

(ii) The cycle space $\mathcal{Z}$, with dimension equal to $\beta_1$, is the linear subspace of edge vectors representing closed circulation loops. Kinematically, vectors within $\mathcal{Z}$ carry solenoidal (curl) information, representing the algebraic description of localised rotational and shear-like micro-dynamics that do not result in net nodal accumulation.

(iii) The coboundary (or cocycle) space $\mathcal{B}^*$, of dimension $|\mathcal{V}|-1$, is the orthogonal complement of $\mathcal{Z}$ within the edge space. Vectors in $\mathcal{B}^*$ are defined strictly as the discrete gradients of vertex-indexed scalar potentials. This subspace captures the irrotational (divergence) mechanics, governing the translational kinematics and volumetric/dilatational deformations of the system. 
Every edge-indexed vector $\mathbf{u}\in\mathbb{R}^{|\mathcal{E}|}$ decomposes uniquely via the orthogonal direct sum ($\oplus$) of its constituent subspaces as: $\mathbf{u} \;=\; \mathbf{u}_{\mathcal{B}^*}\oplus \mathbf{u}_{\mathcal{Z}}$
establishing the \emph{discrete Hodge--Helmholtz decomposition}. This is the graph-theoretic analogue of the classical Stokes–Helmholtz theorem in continuum mechanics, which states that any sufficiently smooth vector field decomposes into an irrotational gradient and a solenoidal curl. Crucially, because the skew-symmetric coupling operator $\mathsf{L}$ preserves this decomposition, its structural action remains uncoupled between the fields: potential-driven gradient inputs remain strictly within the cocycle space $\mathcal{B}^*$, while cyclic, shear-driven inputs remain confined to the cycle space $\mathcal{Z}$. This exact, basis-independent, and natural decomposition ensures that volumetric and rotational modes do not cross-contaminate, a physical property we exploit throughout this paper and its companions.

\section{Robustness and Limits of the Chord-Removal}
\label{sec:S2}

The following remark (originally Remark~5.6 of the main text) analyses four robustness
questions: finite aspect ratio, spontaneous arrest, state-dependence of couplings,
and the conjugate manifestation under kinematic constraint.

\begin{remark}[Robustness and limits of the chord-removal]
\label{rem:chord_robustness}
The cocycle-filter argument that underlies the present analysis relies on $L_{VC}=0$ holding strictly, which the constitutive derivation of Ref.~\cite{RegenNicot2026foundation} (Supplementary Material, Section~S1, Proposition~S1.5) establishes as a structural theorem in the ideal 1-D limit, with quantitative bounds on perturbations away from this limit given in Ref.~\cite{RegenNicot2026foundation} (Supplementary Material, Section~S1.6). Four robustness questions then arise naturally and are worth stating explicitly.
\end{remark}

\begin{enumerate}[label=(\roman*),leftmargin=*,itemsep=2pt]
\item \emph{Finite aspect ratio.} In a physical column of finite lateral extent, an effective $L_{VC}^{\rm eff}$ may emerge from residual transverse coupling, producing a continuous crossover between cocycle-dominated dilatancy (high aspect ratio, $L_{VC}^{\rm eff}\to 0$) and cycle-mediated shear localisation (low aspect ratio, $L_{VC}^{\rm eff}$ finite). The parity-mandated Gateway is robust because $\det(\mathsf{L})=0$ for any odd-dimensional skew-symmetric matrix; it is the \emph{isolation} of the cocycle pathway, not the existence of the null mode, that depends on strict chord-removal.
\item \emph{Spontaneous arrest under fabric realignment.} The boundary-work injection rate $\mathcal{W}_{\rm inject}=L_{MC}f_0/\sqrt{L_{VM}^2+L_{MC}^2}$ vanishes in the limit $L_{MC}\to 0$, corresponding to a fabric configuration in which contact-force chains align perfectly with grain-contact normals and no torque is exchanged between the mechanical and configurational channels. In that limit the rotation axis $\boldsymbol\omega$ rotates from the $V$--$s$ plane onto the pure spin axis, the null-mode vector $\mathbf{v}_0$ becomes orthogonal to the axial boundary load, and the secular drift collapses to zero. The Gateway is \emph{parity-mandated and topologically robust}; the \emph{coupling of boundary work to the Gateway} is fabric-dependent and can spontaneously vanish under specific microstructural reorganisations. This is a non-trivial falsifiable prediction: a granular column whose contact statistics evolve toward $L_{MC}\to 0$ should cease drifting, even though it still possesses a topological Gateway, and the framework therefore predicts both the dominant failure mode under generic conditions and the conditions under which that failure mode is silenced.
\item \emph{State-dependence of the couplings.} The linear analysis treats $L_{VM}, L_{MC}$ as constants of a frozen reference fabric. As the column dilates and the contact network reconfigures, these coefficients evolve. The secular drift therefore identifies the \emph{initial bifurcation direction}, not an infinite post-failure trajectory; geometric jamming, contact loss, and fabric realignment of type~(ii) all act to terminate the drift at finite amplitude. The framework is a theory of \emph{onset} of internal dilatancy, with the nonlinear post-bifurcation regime deferred to a separate analysis.
\item \emph{Conjugate manifestation under kinematic constraint.} The null direction $\mathbf{v}_0$ is specified in channel-flux space, and its physical realisability depends on whether the boundary conditions permit motion along $\mathbf{v}_0$ in position space. Under stress control (free piston, sustained boundary load), the system drifts along $\mathbf{v}_0$ as a kinematic secular strain, which is the dilatancy picture used throughout the present paper. Under strain control (fully clamped configuration), the kinematic strain is geometrically forbidden, and the topologically guaranteed energy injection $\mathcal{W}_{\rm inject}$ accumulates instead in the conjugate variable: the boundary-reaction stress grows linearly in time. The two signatures are related by Legendre transform, and the Gateway null direction $\mathbf{v}_0$ is invariant under the transform because the skew-symmetry $\mathsf{L}^T=-\mathsf{L}$ implies $\ker(\mathsf{L})=\ker(\mathsf{L}^T)$. The framework therefore predicts a precursor signature in any boundary geometry: dilatancy creep for free-piston configurations, stress concentration for clamped configurations, and partitioned signatures for mixed boundaries such as the laboratory oedometer (axial stress-controlled, transverse strain-controlled). This conjugate-variable duality enlarges the family of accessible DEM falsification tests beyond the strain creep documented in the companion paper~\cite{RegenNicot2026numerical}. The linear stress-concentration picture under full clamping is itself only the precursor to a genuinely nonlinear post-bifurcation regime, in which the unbounded growth of the constraint Lagrange multiplier resolves into propagating Cnoidal-wave structures -- the topological-framework signature of compaction bands -- analysed analytically by Veveakis and Regenauer-Lieb~\cite{Veveakis2015,Regenauer2013_JCSMD2}.
\end{enumerate}

\section{Robustness of the Gateway under $L_{VC}\neq 0$ Perturbations}
\label{sec:S3}

The following remark (originally Remark~4.1 of the main text) establishes that the Gateway
null mode is insensitive to perturbations of the structural zero.

\begin{remark}[Robustness of the Gateway under $L_{VC}\neq 0$ perturbations]
\label{rem:LVC_robustness}
It rests to show that the analysis does not hinge on the strict structural zero $L_{VC}=0$ and that any minute direct volumetric--configurational coupling does not close the Gateway and invalidates the dilatancy mechanism. This is not the case, and the distinction is worth making explicit. The parity-mandated zero eigenvalue $\det\mathsf{L}=0$ is forced by the algebraic identity $\det\mathsf{L}=(-1)^N\det\mathsf{L}$ for any odd-dimensional real skew-symmetric matrix; it depends only on $N$ being odd, not on which entries of $\mathsf{L}$ vanish. The null mode therefore exists for every choice of $L_{VC}$, including the perturbed case $L_{VC}\neq 0$. What the structural zero controls is the \emph{orientation} of the null eigenvector, not its existence: for $L_{VC}=0$, $\mathbf{v}_0$ lies in the $V$--$C$ plane and identifies the irreversible pathway with configurational dissipation; for $L_{VC}\neq 0$, $\mathbf{v}_0$ rotates within the full $V$--$M$--$C$ space and the irreversible pathway acquires a mechanical component, but the Gateway remains topologically open. The constitutive derivation of Ref.~\cite{RegenNicot2026foundation} (Supplementary Material, Section~S1), which establishes $L_{VC}=0$ as the structural theorem of Proposition~S1.5 of Ref.~\cite{RegenNicot2026foundation} from the additive separability of the contact-scale free energy and the canonical Poisson structure of the rotational sector, is therefore an orientation theorem, not an existence theorem: the Parity Theorem is structurally robust to perturbations of $L_{VC}$ at first order, and the shear-localisation modes that emerge when an effective $L_{VC}^{\rm eff}$ becomes finite at low aspect ratio (e.g.\ in genuinely two-dimensional geometries) are treated separately as cycle-borne irreversibility in future work on shear instabilities. Quantitative bounds on $L_{VC}^{\rm eff}$ as a function of aspect ratio are given in Ref.~\cite{RegenNicot2026foundation} (Supplementary Material, Section~S1.6). The Gateway is parameter-free in $\mathsf{L}$: only the parity of $N$ matters for its existence.
\end{remark}

\section{The Cocycle Filter and the Absence of Curl-Driven Modes}
\label{sec:S4}

The following remark (originally Remark~4.3 of the main text) provides the formal
cycle-space argument that excludes Schnakenberg cycle currents in the 1-D chain.

\begin{remark}[The cocycle filter and the absence of curl-driven modes]
\label{rem:cocycle_filter}
In the 1-D chain, the cocycle space $\mathcal{B}^* \equiv \operatorname{range}(\mathsf{B}^T)$ of the reversible-graph incidence matrix $\mathsf{B}$ exhausts the entire edge 1-form space, and the cycle space $\mathcal{Z} \equiv \ker(\mathsf{B})$ is trivial ($\beta_1 = 0$). The parity-mandated null mode $\mathbf{v}_0$ therefore lies entirely in $\mathcal{B}^*$ and supports cocycle-borne (gradient) dynamics only. Schnakenberg cycle currents~\cite{Schnakenberg1976}, which require $\beta_1 \geq 1$ to be non-zero, are mathematically excluded by the chain topology. The 1-D Cartesian restriction is the exact algebraic mechanism that filters out curl-driven shear-localisation modes, isolating the cocycle-borne dilatancy instability. The lifting of this filter and the curl-driven shear theorem it enables are discussed in the foundational work \cite{RegenNicot2026foundation}.
\end{remark}

\section{Phase-Space Setup, Governing Equations, and Boundary Forcing}
\label{sec:S5}

The underlying mathematical model describes the non-dissipative core (the $\mathsf{L}$-channel) of a multi-scale framework mapping the granular contact-scale couplings, with localised atomic lattice friction ($\mathsf{D}$-channel dissipation) treated separately following the Onsager reciprocal relations and the Legendre transforms of the dissipation potential.

Let $\mathbf{q}(t) = [q_0, q_1, \dots, q_{N-1}]^T \in \mathbb{R}^N$ define the generalised coordinate vector of the microstructural state space, where $q_0 = V$ represents the macroscopic volumetric change channel, $q_1 = M$ represents the macro-scale structural shear or moment, and $q_j$ (for $j \geq 2$) represents local internal fabric rearrangements, contact-normal rotations, and multi-scale localised topology.

The dynamic trajectory through phase space is governed by the first-order non-homogeneous linear system:
\begin{equation}\label{eq:core_governing}
\frac{d\mathbf{q}}{dt} = \mathsf{L}\mathbf{q} + \mathbf{f}_0
\end{equation}
where $\mathsf{L} \in \mathbb{R}^{N \times N}$ is a tridiagonal skew-symmetric structural coupling operator ($\mathsf{L}^T = -\mathsf{L}$), and $\mathbf{f}_0$ is the macroscopic boundary forcing vector.

\subsection{The boundary forcing condition ($\mathbf{f}_0$)}
Across both configurations, the external boundary work enters the system through a highly localised, non-homogeneous loading constraint. The constant macroscopic volumetric force $f_0 = 1$ is applied \textit{exclusively to the first structural coordinate ($q_0 = V$)} adjacent to the driving wall.

The boundary forcing vector is defined uniformly across all parities as:
\begin{equation}
\mathbf{f}_0 = \begin{bmatrix} 1 \\ 0 \\ 0 \\ \vdots \\ 0 \end{bmatrix}_{N \times 1}
\label{eq:initial_cond}
\end{equation}
The remaining internal structural coordinates ($q_1, q_2, \dots, q_{N-1}$) receive no direct energy from the external environment. All downstream microstructural perturbations rely completely on internal transmission across neighbouring dimensions mediated by the skew-symmetric coupling operator $\mathsf{L}$. The system initialises from a state of total rest:
\begin{equation}
\mathbf{q}(0) = \mathbf{0}.
\end{equation}

\section{The $N=3$ Odd Operator: Permanent Geometric Leakage}
\label{sec:S6}

In the three-dimensional configuration, the microstructural state space is tracked by three coupled coordinates: Macroscopic Volume ($V$), Macro-Shear ($M$), and Internal Fabric Shear ($C$).

\subsection{Internal coupling network}
The skew-symmetric interaction layout forms a directional chain:
\begin{itemize}
    \item Boundary work directly injects energy into the Volume channel ($V$).
    \item $V$ couples to Macro-Shear ($M$) via the coefficient $+L_{VM}$, while $M$ exerts an equal and opposite internal reaction on $V$ via $-L_{VM}$.
    \item Simultaneously, $M$ couples downstream to the Fabric Shear channel ($C$) via $+L_{MC}$, while $C$ pushes back on $M$ via $-L_{MC}$.
\end{itemize}

\subsection{Structural defect and secular motion}
Because the system dimension $N$ is odd, the structural coupling matrix is singular ($\det(\mathsf{L}_3) = 0$). This singularity guarantees the conservation of a non-trivial \textit{topological zero-mode}. The network contains an unconstrained directional pathway spanned by the null vector $\mathbf{v}_0 = (L_{MC},\,0,\,L_{VM})^T/\sqrt{L_{VM}^2+L_{MC}^2}$.

When the continuous boundary load is applied to $q_0$, the coordinate combination $(L_{MC} \cdot V + L_{VM} \cdot C)$ moves completely unresisted by the network's internal constraints. While the intermediate coordinate $M(t)$ safely oscillates in a closed loop, it acts as a passive kinematic pump. It continuously transfers net energy from the boundary wall directly into $V(t)$ and $C(t)$, triggering a state of \textit{secular escape}. As time progresses ($t \to \infty$), the assembly undergoes unbounded linear growth, meaning the topological gateway remains permanently open. This is the frictionless $\mathsf{D}=0$ idealisation of the present section: any $\mathsf{D}\succ0$ bounds the growth at the finite offset $\mathbf{q}^\ast=(\mathsf{D}-\mathsf{L})^{-1}\mathbf{f}_0$, and the unbounded drift is recovered dynamically only once nonlinear routing supersedes the linear damped transient (see the three-regime discussion in the main text).

\section{The $N=4$ Even Operator: Perfect Harmonic Containment}
\label{sec:S7}
These constructions are thermodynamic channel-graph operators rather than clusters of grains: the index $N$ counts coupled channels, and the relevant geometry is that of the coupling graph on those channels, not of any particle arrangement. The minimal VMC triad of the main text \cite{RegenNicot2026foundation} is one admissible choice of three channels, $(V,M,C)$, and by the Parity Theorem it is odd and hence a Gateway. An even Stable Layer is obtained simply by extending the channel set with one further channel, and which channel is added is fixed by the scale of description rather than by the framework. The instance carried through the analytical treatment below is the electric channel $E$ of the thermal-hydraulic-mechanical-chemical-electric (THMCE) pentade, coupled to the fabric channel $C$ through $L_{CE}$; at another scale the added channel might equally be a Cosserat contact rotation or a thermal channel, as in the planetary reductions of Part~3~\cite{RegenNicot2026numerical}. What produces the harmonic containment established below is the change of parity from odd to even, which by the Parity Theorem is generic to any even channel set; the identity of the added channel does not enter. This parameter-free character, in which the dynamics follows from the parity of the coupling graph alone, is the sense in which the augmented operator $\mathsf{A}=\mathsf{D}+\mathsf{L}$ extends the static Maxwell relations to the coupled dynamic system. Appending $E$ as a further link, so that the acyclic ($\beta_1=0$) coupling path becomes $V \longleftrightarrow M \longleftrightarrow C \longleftrightarrow E$, the operator $\mathsf{L}_4$ is the corresponding tridiagonal matrix.

\subsection{Structural confinement mechanism}
Because the system dimension $N$ is even, the operator $\mathsf{L}_4$ is non-singular ($\det(\mathsf{L}_4) \neq 0$). The system possesses \textit{no zero-mode}. Every axis of motion is bound to a restoring skew-symmetric partner element.

The continuous boundary force shifts the equilibrium centre away from the origin to a stable focal point offset in the mechanical and $E$ channels:
\begin{equation}
\mathbf{q}_{\text{center}} = -\mathsf{L}_4^{-1}\mathbf{f}_0 = \begin{bmatrix} 0 \\ -1/L_{VM} \\ 0 \\ -L_{MC}/(L_{VM}L_{CE}) \end{bmatrix}
\end{equation}
The trajectory orbits this centre.

\subsection{Generic torus and the role of the 2:1 resonance}

For any even $N=4$ with generic (incommensurate) coupling coefficients, the spectrum $\{\pm i\omega_1, \pm i\omega_2\}$ generates two independent oscillations at irrational frequency ratio. The trajectory winds on a two-torus but never exactly closes in finite time.

The coupling values $L_{VM}=2.0$, $L_{MC}=1.5$, $L_{CE}=2.5$ are chosen specifically to enforce the commensurate ratio $\omega_{\text{fast}}/\omega_{\text{slow}} = 2:1$ ($\omega_{\text{slow}} \approx 1.5811$, $\omega_{\text{fast}} \approx 3.1623$). This is a \emph{visualisation device}: by making the two modes return to zero phase simultaneously, the simulation can complete the trajectory in the phase-space portrait within a tractable window. At any integer multiple of the slow period $T = 2\pi/\omega_{\text{slow}}$, the matrix exponential satisfies $e^{\mathsf{L}_4 T} = \mathsf{I}$, returning $\mathbf{q}$ to zero.

The topological boundedness, marked by the absence of secular drift, is \emph{not a consequence of the resonance}. It holds for all even-$N$ systems regardless of whether the frequencies are commensurate. The resonance only determines whether the orbit closes visibly in a given time window; the torus itself exists and confines the trajectory for all coupling choices.

\section{Analytical Solutions for $N=3$ and $N=4$}
\label{sec:S8}

This section provides the mathematical proof of the closed-form solutions for the baseline $N=3$ and $N=4$ cases, whose validation against direct numerical integration is shown in the main-text parity-contrast figure.

\subsection{The odd parity architecture ($N=3$)}
For $N=3$, the tridiagonal operator $\mathsf{L}_3$ tracks the three channels of the minimal triad, Volume ($V$), Macro-Shear ($M$), and Fabric Shear ($C$):
\begin{equation}
\mathsf{L}_3 = \begin{bmatrix} 0 & L_{VM} & 0 \\ -L_{VM} & 0 & L_{MC} \\ 0 & -L_{MC} & 0 \end{bmatrix}
\end{equation}
Because $N$ is odd, $\det(\mathsf{L}_3) = 0$. The non-trivial null space is spanned by the topological zero-mode vector $\mathbf{v}_0 = (L_{MC},\,0,\,L_{VM})^T/\sqrt{L_{VM}^2+L_{MC}^2}$. Projecting the dynamic system Eq. \eqref{eq:core_governing} directly along this zero-mode vector isolates the secular, non-oscillatory trajectory component:
\begin{equation}
\mathbf{v}_0 \cdot \frac{d\mathbf{q}}{dt} = \mathbf{v}_0 \cdot \mathsf{L}_3\mathbf{q} + \mathbf{v}_0 \cdot \mathbf{f}_0
\end{equation}
By skew-symmetry, $\mathbf{v}_0 \cdot \mathsf{L}_3\mathbf{q} = 0$. Thus, the projection simplifies to:
\begin{equation}
\frac{d}{dt}(\mathbf{v}_0 \cdot \mathbf{q}) = \mathbf{v}_0 \cdot \mathbf{f}_0 = \frac{L_{MC}}{\sqrt{L_{VM}^2+L_{MC}^2}}.
\end{equation}
Integrating with respect to time under initial condition Eq. \eqref{eq:initial_cond} yields the explicit linear invariant drift equation:
\begin{equation}\label{eq:linear_drift_inv}
\mathbf{v}_0 \cdot \mathbf{q}(t) = \frac{L_{MC} \cdot V(t) + L_{VM} \cdot C(t)}{\sqrt{L_{VM}^2+L_{MC}^2}} = \frac{L_{MC}}{\sqrt{L_{VM}^2+L_{MC}^2}}\,t.
\end{equation}
Decoupling the remaining orthogonal subspace variables yields the complete time-dependent analytical solutions:
\begin{align}
\label{eq:sol_v} V(t) &= \frac{L_{MC}^2}{\omega^2}t + \frac{L_{VM}^2}{\omega^3}\sin(\omega t) \\
\label{eq:sol_m} M(t) &= \frac{L_{VM}}{\omega^2}\left(\cos(\omega t) - 1\right) \\
\label{eq:sol_c} C(t) &= \frac{L_{VM}L_{MC}}{\omega^2}t - \frac{L_{VM}L_{MC}}{\omega^3}\sin(\omega t)
\end{align}
where $\omega = \sqrt{L_{VM}^2 + L_{MC}^2}$. As $t \to \infty$, $V(t)$ and $C(t)$ exhibit unbounded secular growth, meaning the topological gateway is permanently characterised by continuous leakage.

\subsection{The even parity architecture ($N=4$)}
For $N=4$, the channel set is extended by one further channel $E$, so that $\mathbf{q} = [V, M, C, E]^T$ and the tridiagonal operator becomes:
\begin{equation}
\mathsf{L}_4 = \begin{bmatrix} 0 & L_{VM} & 0 & 0 \\ -L_{VM} & 0 & L_{MC} & 0 \\ 0 & -L_{MC} & 0 & L_{CE} \\ 0 & 0 & -L_{CE} & 0 \end{bmatrix}
\end{equation}
The matrix is non-singular with determinant $\det(\mathsf{L}_4) = (L_{VM}L_{CE})^2 \neq 0$. This invertibility yields a fixed, stable stationary orbit focus in phase space:
\begin{equation}
\mathbf{q}_{\text{center}} = -\mathsf{L}_4^{-1}\mathbf{f}_0 = \begin{bmatrix} 0 \\ -\frac{1}{L_{VM}} \\ 0 \\ -\frac{L_{MC}}{L_{VM}L_{CE}} \end{bmatrix}
\end{equation}
The spectrum contains exclusively imaginary conjugate roots ($\pm i\omega_{\text{slow}}, \pm i\omega_{\text{fast}}$). The analytical trajectory is strictly bounded and oscillates around the offset focal centre:
\begin{equation}
\mathbf{q}(t) = \mathbf{q}_{\text{center}} + \sum_{j=1}^2 \left( \mathbf{A}_j \cos(\omega_j t) + \mathbf{B}_j \sin(\omega_j t) \right)
\end{equation}
The analytical trajectory is bounded for any coupling choice. For the specific commensurate values $L_{VM}=2.0$, $L_{MC}=1.5$, $L_{CE}=2.5$ (which enforce $\omega_{\text{fast}}/\omega_{\text{slow}}=2$), the matrix exponential satisfies $e^{\mathsf{L}_4 T_{\text{slow}}} = \mathsf{I}$ at every integer multiple of the slow period, providing exact finite-time closure. For incommensurate couplings, no finite $T$ achieves exact closure, but the trajectory remains confined to the compact torus $\|\mathbf{q}(t)-\mathbf{q}_{\text{center}}\|=\text{const}$ for all $t$.

\bibliographystyle{elsarticle-num}
\bibliography{bib_merged}


\begin{frontmatter}

\title{Supplementary Material for: Topological Foundations of Multi-Field Instabilities in Continua: Part 3 — Numerical Upscaling}

\author[inrae]{F. Nicot\corref{cor1}}
\ead{francois.nicot@univ-smb.fr}
\author[inrae]{Amir Saker}
\author[curtin]{K. Regenauer-Lieb}

\cortext[cor1]{Corresponding author.}

\address[inrae]{Universit\'e Savoie Mont Blanc, INRAE, ETNA, France.}
\address[curtin]{WA School of Mines: Minerals, Energy and Chemical Engineering,
  Curtin University, Perth WA 6845, Australia.}

\end{frontmatter}

This Supplementary Material contains the Thermodynamic Uncertainty Relation
analysis, quad-precision verification protocol, Newton-to-Onsager derivation,
and the cnoidal-wave mathematical outlook referred to in the main text.
Section numbering uses the prefix~S.

\section{Thermodynamic Uncertainty Relation and Spectral Scaling Consistency}
\label{sec:S1}

The following analysis (originally the closing paragraph of Section~5 of the main text) connects the Toeplitz spectral scaling to the macroscopic fluctuation--dissipation framework via the Thermodynamic Uncertainty Relation.

This $T_{\rm slow}\propto N$ behaviour is the classical signature of critical slowing down at a marginal stability threshold, and it has a direct fluctuation-theoretic counterpart that connects the discrete tridiagonal spectrum to the macroscopic fluctuation--dissipation framework expected of any continuum thermodynamic limit. The Thermodynamic Uncertainty Relation (TUR) bounds the relative variance of any steady-state current from below by the inverse of the cumulative entropy production~\cite{Barato2015}: $\mathrm{Var}(J)/\langle J\rangle^2\geq 2/\Sigma$, where $\Sigma$ is the total entropy produced over the observation window. In the augmented $\mathsf{D}+\mathsf{L}$ system the conservative drift along the parity-mandated null direction produces no entropy through $\mathsf{L}$ ($\mathbf{q}^T\mathsf{L}\mathbf{q}\equiv 0$), so the only contribution to $\Sigma$ comes from $\mathsf{D}$, and the TUR floor on relative current fluctuations is set entirely by the dissipative spectrum. The even-parity spectral scaling $|\lambda_{\min}^{\rm even}|\sim\gamma\pi/N$ then implies that the harmonic energy-storage capacity per slow mode decays as $1/N$ in the continuum limit, so the relative fluctuations of the boundary-driven current must widen as $N$ grows in order to remain compatible with the TUR floor at fixed $\Sigma$. The discrete matrix-spectral statement and the macroscopic fluctuation--dissipation statement are therefore quantitatively consistent, and the parity contrast survives the continuum limit not as a singular feature but as a smooth thermodynamic boundary between Stable Layers, whose slow modes broaden with $N$ but never close, and Gateway Layers, whose null mode is structurally identical at every $N$.


\section{Quad-Precision Verification for $N=5$ and $N=6$}
\label{app:quadprec}

\subsection{Description and network mechanics}
This model resolves the intermediate evolutionary step ($N=5$ versus $N=6$) to isolate the challenge of multi-frequency incommensurability. For any even chain larger than $N=4$, the frequencies given by the tridiagonal Toeplitz roots are irrationally related. This script addresses this out-of-sync phase accumulation by implementing a 128-bit arbitrary-precision tracking routine to locate an optimal Poincar\'e recurrence window. 

\subsection{Boundary conditions and temporal resolution}
The external boundary load matches the standard form: $\mathbf{f}_0 = [1.0, 0.0, \dots, 0.0]^T$ acting on $\mathbf{q}(0) = \mathbf{0}$ for a uniform coupling matrix element $L_{i, i+1} = \gamma = 2.0$. The eigenvalues follow the analytical equation of Part I. Evaluating it for $N=6$ yields three distinct positive frequency modes: $\omega_3 = 4.0\cos(3\pi/7)$, $\omega_2 = 4.0\cos(2\pi/7)$, and $\omega_1 = 4.0\cos(\pi/7)$. To achieve absolute zero-return alignment, the integration window is bounded to an ultra-high-resolution synchronisation time ($T_{\text{perfect}} \approx 735.1432$) calculated using the Python \texttt{decimal} module to 35 decimal places. This represents the exact moment where the system executes exactly $k=104$ slow cycles, while the faster modes complete near-perfect integer cycle counts, minimising the cumulative phase drift residual $\epsilon$:
\begin{equation}
\epsilon = \sum_{m=1}^2 \left| \left( k \frac{\omega_m}{\omega_3} \right) - \text{round}\left( k \frac{\omega_m}{\omega_3} \right) \right|.
\end{equation}

\subsection{Interpretation of runtime results}
\begin{itemize}
    \item \textbf{Even system ($N=6$):} The three natural frequency pairs are irrationally related (roots of the Chebyshev polynomial of the second kind of degree~7), so the orbit densely fills a three-torus in the six-dimensional phase space. This \emph{is} the generic even-parity behaviour: structural containment on a compact invariant torus. At the precision-synchronised window $T_{\text{perfect}}$, all three modes are simultaneously near-integer multiples of their periods, driving the terminal residual to $\sim 10^{-9}$. This near-return is the closest rational approximation within a finite time horizon, not exact closure; the orbit remains bounded regardless.
    \item \textbf{Odd system ($N=5$):} The system completely bypasses the multi-frequency torus structure. The zero-mode overrides the harmonic components, causing the trajectory to exit the localised domain and drift continuously into infinite space along a steady secular path. Under the closed-boundary bottleneck of the present setup, this drift cannot be absorbed by any upscale routing channel and accumulates without bound.
\end{itemize}

\subsection{Commentary on the finite-time residual in the N=6 simulation}
The $N=6$ orbit never closes exactly in finite time because the three natural frequencies are irrationally related. This is not a flaw but the expected behaviour of quasi-periodic motion on a dense torus. The precision-synchronised window $T_{\text{perfect}}$ is chosen to minimise the near-return residual, not to achieve exact closure (which would require $T\to\infty$). The small remaining mismatch has two contributing causes:
\begin{enumerate}
    \item \textbf{Irrational phase asymptotics:} The natural frequencies are irrationally related because they are determined by the roots of the 7th Chebyshev polynomial of the second kind. True mathematical closure requires an infinite time horizon ($T \to \infty$) to achieve a zero-error alignment. The quadruple-precision algorithm samples a finite scan window ($k \leq 300$), picking the absolute best rational approximation available. The remaining analytical phase drift ($\epsilon \approx 10^{-5}$) represents the fundamental limitation of fitting a finite number of integer cycles to irrational frequency ratios.
    \item \textbf{Integrator precision bottlenecks:} While the target recurrence time ($T_{\text{perfect}}$) is evaluated to 35 decimal places, standard high-order Runge--Kutta differential equation solvers (such as \texttt{DOP853}) operate internally using standard IEEE 754 \texttt{float64} double precision. When the solver steps through time, the 128-bit boundary condition is truncated to roughly 16 digits of precision. Local truncation errors and floating-point round-off accumulate across the 15{,}000 discrete integration steps, leaving a tiny residual offset at the final state. This demonstrates that while the topological gate is theoretically closed, capturing that closure within a numerical simulation remains bound by the limits of double-precision floating-point arithmetic.
\end{enumerate}

\section{Newton-to-Onsager Mapping for DEM Implementations}
\label{sec:S3}

This appendix derives the analytical mapping by which the entries of the skew-symmetric Onsager block $\mathsf{L}$ are read off directly from the gyroscopic coefficients of a DEM implementation that satisfies the channel-coupling structure of Part I. The derivation supplies the practical bridge between a Newton-equation DEM simulation, in which the contact-scale degrees of freedom are positions and velocities, and the first-order Onsager equation $\dot{\mathbf{q}}=\mathsf{L}\mathbf{q}+\mathbf{f}_0$ of Paper~I~\cite{RegenNicot2026dilatancy} and the numerical simulations of Section~4 in the main text, in which the state vector consists of thermodynamic fluxes.

\subsection{Setup: Newton equations with antisymmetric gyroscopic coupling}

Consider a 1-D chain of grains forming $N_c$ contacts, each grain carrying three contact-scale degrees of freedom: axial position $x_i$, tangential shear displacement $s_i$, and a contact-normal rotation $\theta_i$. The Newton equations of motion in the linear regime, with tridiagonal stiffness matrices $\mathsf{K}_{xx}, \mathsf{K}_{ss}, \mathsf{K}_{\theta\theta}$ and antisymmetric gyroscopic couplings $\alpha_{VM}$ (overlap--shear) and $\alpha_{MC}$ (shear--rotation), read
\begin{align}
  m\,\ddot{x}_i    &= -\,[\mathsf{K}_{xx}\,\mathbf{x}]_i    + \alpha_{VM}\,\dot{s}_i + f_{x,i}\,, \label{eq:nm_x}\\
  m\,\ddot{s}_i    &= -\,[\mathsf{K}_{ss}\,\mathbf{s}]_i    - \alpha_{VM}\,\dot{x}_i + \alpha_{MC}\,\dot{\theta}_i\,, \label{eq:nm_th}\\
  I\,\ddot{\theta}_i &= -\,[\mathsf{K}_{\theta\theta}\,\boldsymbol{\theta}]_i - \alpha_{MC}\,\dot{s}_i\,. \label{eq:nm_y}
\end{align}
The antisymmetric placement of the $\alpha$ couplings ($+\alpha_{VM}\dot s$ in the $\ddot x$ equation versus $-\alpha_{VM}\dot x$ in the $\ddot s$ equation, and analogously for $\alpha_{MC}$) is the kinematic signature of a conservative cross-coupling: the work performed by the gyroscopic force is identically zero, which is the time-domain statement of the Onsager--Casimir relation $\mathsf{L}=-\mathsf{L}^T$. The structural choice $\alpha_{VC}=0$ in the above system enforces the cocycle filter of Paper~I~\cite{RegenNicot2026dilatancy}: the chain topology of the reversible coupling network has first Betti number $\beta_1=0$, and the cycle space $\mathcal{Z}$ is trivial.

\subsection{Contact-level flux variables}

The Onsager state vector is built from \emph{relative velocities at each contact}, not from individual-grain positions. Let $c=0,\ldots, N_c-1$ index the contacts (with $c=0$ the wall--grain contact). Define
\begin{equation}\label{eq:flux_definition}
  V_c \;=\; \dot{x}_{c+1}-\dot{x}_c\,,\qquad
  M_c \;=\; \dot{s}_{c+1}-\dot{s}_c\,,\qquad
  C_c \;=\; \dot{\theta}_{c+1}-\dot{\theta}_c\,,
\end{equation}
with $\dot{x}_0=\dot s_0=\dot\theta_0=0$ at the wall. Differentiating once in time and substituting the Newton equations~\eqref{eq:nm_x}--\eqref{eq:nm_y}, one obtains the contact-level evolution
\begin{align}
  m\,\dot{V}_c &= +\alpha_{VM}\,M_c     + (\text{stiffness terms})    + \Delta_c f_x, \label{eq:Vdot}\\
  m\,\dot{M}_c &= -\alpha_{VM}\,V_c + \alpha_{MC}\,C_c + (\text{stiffness terms}), \label{eq:Mdot}\\
  I\,\dot{C}_c &= -\alpha_{MC}\,M_c                                 + (\text{stiffness terms}), \label{eq:Cdot}
\end{align}
where "(stiffness terms)" denotes the contributions from $\mathsf{K}_{xx},\mathsf{K}_{ss},\mathsf{K}_{\theta\theta}$ acting on time-integrated fluxes. The crucial observation is that the gyroscopic $\alpha$ terms couple fluxes \emph{at the same contact} to fluxes \emph{at the same contact}, with no involvement of positions or stiffness. After the Onsager decomposition $\mathsf{A}=\mathsf{D}+\mathsf{L}$ (Section~2 in the main text), the $\alpha$ contributions feed the antisymmetric block $\mathsf{L}$, and the stiffness contributions feed the symmetric block $\mathsf{D}$.

\subsection{Energy-normalised channels and the skew-symmetric L}

Direct comparison of Eqs.~\eqref{eq:Vdot}--\eqref{eq:Cdot} shows that the gyroscopic coupling matrix in $(V_c, M_c, C_c)$ coordinates is \emph{not} exactly antisymmetric: the prefactors $m$, $m$, $I$ in the time-derivative terms break the symmetry. Energy normalisation restores it. Define
\begin{equation}\label{eq:energy_normalised}
  \widetilde{V}_c \;=\; \sqrt{m}\,V_c\,,\qquad
  \widetilde{M}_c \;=\; \sqrt{m}\,M_c\,,\qquad
  \widetilde{C}_c \;=\; \sqrt{I}\,C_c\,,
\end{equation}
which makes the kinetic energy of each channel equal to $\tfrac{1}{2}\widetilde\bullet_c^2$. Substituting into Eqs.~\eqref{eq:Vdot}--\eqref{eq:Cdot} and isolating the conservative (gyroscopic) contribution gives the contact-level Onsager block in normalised channels:
\begin{equation}\label{eq:L_block}
  \mathsf{L}_c \;=\; \begin{pmatrix}
     0    & L_{VM} & 0      \\
    -L_{VM} & 0    & L_{MC} \\
     0    & -L_{MC} & 0
  \end{pmatrix}\!,
  \qquad
  L_{VM}=\frac{\alpha_{VM}}{m}\,,\qquad
  L_{MC}=\frac{\alpha_{MC}}{\sqrt{m\,I}}\,.
\end{equation}
This block is exactly skew-symmetric, has structural zero $L_{VC}=0$, and possesses the parity-mandated null vector $\mathbf{v}_0=(L_{MC},0,L_{VM})^T/\sqrt{L_{VM}^2+L_{MC}^2}$ in the $V$--$C$ plane, as predicted by Paper~I~\cite{RegenNicot2026dilatancy}. For a chain of $N_c$ contacts with no direct gyroscopic coupling between distinct contacts (the simplest DEM implementation), the full Onsager block is block-diagonal: $\mathsf{L}_{\rm full}=\mathrm{diag}(\mathsf{L}_1,\ldots,\mathsf{L}_{N_c})$, of dimension $3N_c\times 3N_c$, with $N_c$ independent zero modes -- one per contact. Augmenting the DEM contact law with inter-contact gyroscopic terms (couplings between $V_c$ and $M_{c\pm 1}$, or analogous) generates the genuine spin-chain extension of Section~5 in the main text with tridiagonal Toeplitz structure on the full $N$-channel state and a single parity-mandated null eigenvalue. This block-diagonal versus tridiagonal alternative is the same distinction drawn, at the coarse-grained continuum level, in the remark closing Section~6.2  of the main Part 3 text: the block-diagonal column is a stack of contact-scale VMC Gateways, whereas the tridiagonal chain is the operator whose parity the high-$N$ scaling certifies. Under axial-end boundary forcing only the boundary-coupled contact is loaded, so only one of the $N_c$ block-diagonal null directions is activated; uniform body forcing activates the wall-side contact instead, and a linearly graded body force activates all $N_c$ null directions simultaneously. The forcing-pattern dependence distinguishes the block-diagonal from the tridiagonal Toeplitz realisation and serves as a discriminating DEM test.

\subsection{Practical recipe}

The mapping above gives a practical recipe for any DEM implementation that satisfies the conditions of Section~\ref{sec:S3}:
\begin{enumerate}
  \item Identify the gyroscopic coefficients $\alpha_{VM},\alpha_{MC}$ in
    the contact force law (typically a rolling-resistance constant times a
    geometric factor).
  \item Compute the energy-normalised entries
    $L_{VM}=\alpha_{VM}/m$ and $L_{MC}=\alpha_{MC}/\sqrt{m\,I}$
    directly; no covariance estimation is required.
  \item Verify $\det(\mathsf{L}_c)=0$ to machine precision and that
    the null vector $\mathbf{v}_0$ has the predicted $V$--$C$
    orientation. This is a one-line sanity check at the start of
    the analysis.
  \item Extract the dissipative block $\mathsf{D}$ from the
    fluctuation--dissipation analysis of the equilibrated reference
    state, and compute the Gateway number
    $\mathcal{G}=\|\mathsf{L}\|_2^2/[\sigma_{\min}(\mathsf{D})\sigma_{\max}(\mathsf{D})]$.
    Whether the secular drift of Section~4 in the main text is
    visible in a given DEM simulation depends on whether
    $\mathcal{G}\geq 1$ in that simulation.
\end{enumerate}
Cross-correlation--based statistical estimators for $\mathsf{L}$ are not required when the DEM contact law is constructed with explicit gyroscopic terms; the analytical mapping above replaces them. The same cross-correlation concept is, however, the correct route to the symmetric dissipative block $\mathsf{D}$, which is not read off the contact law and must be extracted from the equilibrated fluctuation spectrum.

\subsection{Numerical verification of the analytical mapping}
\label{sec:S3_verification}

\begin{figure}[htbp]
  \centering
  \includegraphics[width=0.9\linewidth]{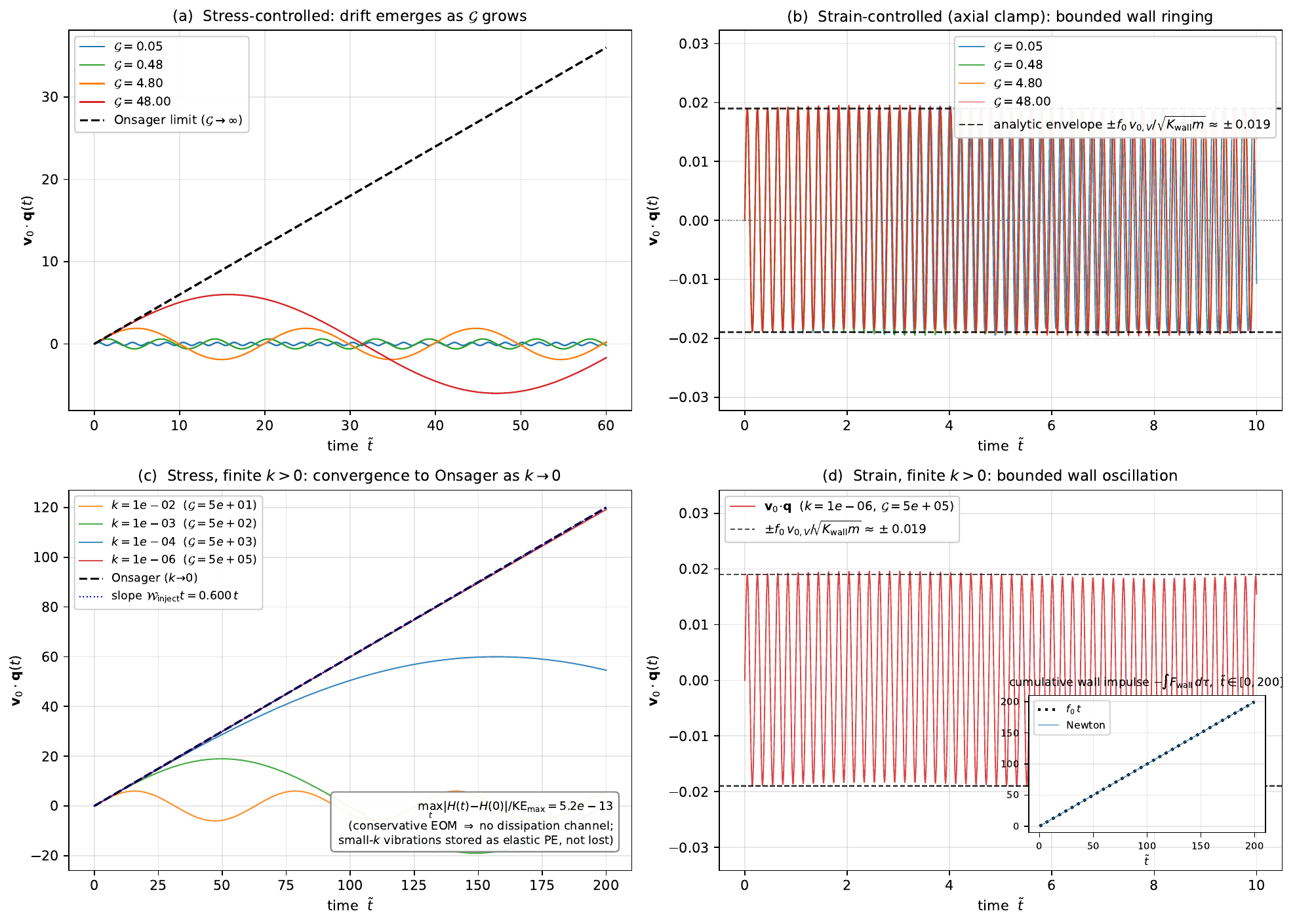}
  \caption{Numerical verification of the Newton-to-Onsager mapping (Eqs.~\eqref{eq:nm_x}--\eqref{eq:nm_y}). \textbf{Top row, Gateway-number sweep:} \textbf{(a)} Stress-controlled ($\tilde t\in[0,60]$): finite stiffness $k$ converts unbounded Onsager drift into slow oscillations of period $T_{\rm slow} \approx 2\pi\sqrt{m/k}$; the Onsager limit ($\mathcal{G} \to \infty$) is the envelope-tangent slope. \textbf{(b)} Strain-controlled ($K_{\rm wall}=10^3, \tilde t\in[0,10]$): axial clamping caps drift at the analytic envelope $\pm f_0\,v_{0,V}/\sqrt{K_{\rm wall}\,m}$. \textbf{Bottom row, convergence:} \textbf{(c)} Stress-controlled, $k \to 0$ ($\tilde t\in[0,200]$): Newton trajectories converge to the Onsager drift, conserving mechanical energy to machine precision. \textbf{(d)} Strain-controlled, $k=10^{-6}$ ($\tilde t\in[0,10]$): drift is suppressed within the analytic envelope; inset confirms impulse-momentum consistency ($\int F_{\rm wall}\,d\tau = f_0\,t$).}
  \label{fig:newton_verification}
\end{figure}

The analytical derivation above is verified directly by integrating the linear Newton equations~\eqref{eq:nm_x}--\eqref{eq:nm_y} for the VMC triad (single-contact limit $N_c=1$, $\mathsf{K}_{\bullet\bullet}=k\,\mathbb{I}$) and comparing the resulting velocity trajectory to the first-order Onsager integration of Section~4 in the main text. 
Figure~\ref{fig:newton_verification} summarises the result. Two boundary conditions are integrated side-by-side: a stress-controlled BC in which the axial channel is free and the constant force $f_0$ is applied to it, and a strain-controlled BC in which the axial channel is pinned by a stiff wall spring $K_{\rm wall}\gg k,\alpha_{\bullet\bullet}$, so that the $V$-component of $\mathbf{v}_0$ is geometrically blocked.

Three structural features of the linear Newton chain are read off Figure~\ref{fig:newton_verification}. First, in the conservative $k\!\to\!0$ limit (panel~c, red curve) the Newton velocity vector $(\sqrt{m}\dot x,\sqrt{m}\dot s,\sqrt{I}\dot\theta)$ coincides with the energy-normalised flux $\mathbf{q}$ of the Onsager dynamics by construction. The finite-$k$ trajectories at $k=10^{-2},10^{-3},10^{-4}$ converge to this limit smoothly as $k$ decreases, with no residual deviation that survives the limit. This is the practical realisation of the structural identities $L_{VM}=\alpha_{VM}/m$, $L_{MC}=\alpha_{MC}/\!\sqrt{mI}$ established in the derivation above, and recovers the analytical Onsager drift slope $\mathcal{W}_{\rm inject}=\mathbf{v}_0\!\cdot\!\tilde{\mathbf{f}}_0$ to all observable orders.

Second, for any $k>0$ the elastic spring perturbs the parity-mandated zero eigenvalue of $\mathsf{L}$ into a small imaginary eigenvalue, and the unbounded Onsager drift along $\mathbf{v}_0$ is converted into a slow oscillation of period $T_{\rm slow}\!\approx\!2\pi\sqrt{m/k}$ (panels~a and~c). The Onsager drift slope $\mathcal{W}_{\rm inject}$ is recovered as the initial tangent slope of the slow oscillation, but the spring restoring force returns the trajectory toward the origin within one half-period. As $\mathcal{G}$ grows the oscillation slows; the Onsager limit is recovered only as $\mathcal{G}\!\to\!\infty$. This is the in-model demonstration of the $\mathcal{G}\geq 1$ visibility criterion of Equation 3 in the main text and the source of the elastic-clamping effect referenced in the subsection that follows.

The convergence in panel~(c) also disposes of an obvious objection. One might worry that the small vibrations resolved at finite $k$ in panels~(a) and~(c) are transferred to phonon dissipation, and that the Gateway drift is therefore an artifact of the conservative $k\!\to\!0$ idealisation that would be eliminated in any physically realisable contact mechanics. Two observations preempt this objection. The Newton equations~\eqref{eq:nm_x}--\eqref{eq:nm_y} are purely conservative by construction: the symmetric stiffness terms are elastic (not dissipative), the antisymmetric gyroscopic couplings conserve kinetic energy, and the total mechanical energy $H=\tfrac{1}{2}m v^2+\tfrac{1}{2}k|\mathbf{q}|^2-f_0\,x$ satisfies $dH/dt\equiv 0$ as an exact identity. Numerical integration verifies this conservation to integrator precision ($\sim\!5\!\times\!10^{-13}$ relative to peak kinetic energy) over the full $k$-sweep of panel~(c), including the largest oscillation amplitudes. The finite-$k$ vibrations are therefore stored reversibly in the elastic spring potential energy and recovered every oscillation period; they are not lost to phonons because there is no dissipation channel in the law. A physically realised contact mechanics with a genuinely dissipative $\mathsf{D}$ block (friction, viscoelastic damping, Hertzian radiation) is treated separately in the fluctuation--dissipation extraction of the procedure above; the present conservative test is the diagnostic baseline against which any such dissipative contribution is to be measured.

Third, the conjugate-variable duality of Section~6.5 in the main text is recovered without modification of the integrator: the algebraic null direction $\mathbf{v}_0$ is invariant under the choice of boundary condition, but its manifestation flips between strain creep (stress-controlled, panels~a,c) and stress concentration (strain-controlled, panels~b,d). Under the axial clamp the $V$-component of $\mathbf{v}_0$ is geometrically blocked, and the drift projection is capped at the analytic envelope $\pm f_0\,v_{0,V}/\!\sqrt{K_{\rm wall}\,m}$ for all $\mathcal{G}$ (panel~b). The applied force is absorbed by the wall as a linearly accumulating impulse $-\!\int\!F_{\rm wall}\,d\tau=f_0\,t$ (panel~d inset), which is the linear-Newton realisation of the stress-concentration signature. We note that the fully-clamped post-bifurcation prediction of linearly-growing wall stress in the augmented-operator analysis of Section~6.5 in the main text is itself a feature of genuinely dissipative $\mathsf{D}$ and of the nonlinear Cnoidal-wave regime~\cite{Veveakis2015,Regenauer2013_JCSMD2}; it is not captured by the linear contact mechanics used here, which saturate at the bounded envelope above. The Cnoidal regime is the appropriate testing ground for the nonlinear strain-controlled falsification protocol of Section~6.3.

\subsection{Why direct Onsager integration is used in Section~4 in the main text}

The simulations reported in Section~4 in the main text integrate the conservative Onsager dynamics $\dot{\mathbf{q}}=\mathsf{L}\mathbf{q}+\mathbf{f}_0$ directly, rather than the full Newton equations~\eqref{eq:nm_x}--\eqref{eq:nm_y}. The reason is presentation rather than principle, as Figure~\ref{fig:newton_verification} shows directly. At any finite spring stiffness $k>0$ the secular drift signature of the activated Gateway competes against the symmetric stiffness terms that feed $\mathsf{D}$, and unless the Gateway number $\mathcal{G}\geq 1$ the dissipative side suppresses the parity-mandated drift into a slow elastic oscillation of period $T_{\rm slow}\!\approx\!2\pi\sqrt{m/k}$ (panel~a). In the conservative limit $k\to 0$ (equivalently $\mathcal{G}\to\infty$, panel~c) the drift is direct, clean, and matches the closed-form analytical solutions of Paper~I~\cite{RegenNicot2026dilatancy} and the Onsager integration to integrator precision.

Crucially, the finite-$k$ to $k\to 0$ convergence is conservative and dissipation-free: the Newton equations~\eqref{eq:nm_x}--\eqref{eq:nm_y} contain no friction or damping operator, the total mechanical energy $H=\tfrac{1}{2}m v^2+\tfrac{1}{2}k|\mathbf{q}|^2-f_0\,x$ is conserved as an exact identity, and the small elastic vibrations resolved at finite $k$ in panels~(a)~and~(c) are stored reversibly in spring potential energy and recovered each oscillation period -- they are not transferred to a phonon dissipation channel that would obstruct the $\mathcal{G}\to\infty$ limit. The numerical integration verifies this conservation to $\sim\!5\!\times\!10^{-13}$ relative to peak kinetic energy across the full $k$-sweep (annotation in panel~c). A genuinely dissipative $\mathsf{D}$ block -- friction, viscoelastic damping, Hertzian radiation, or the phonon channel of a full lattice model -- is treated separately via the fluctuation--dissipation extraction described above, and would shift the visibility threshold $\mathcal{G}\geq 1$ but not the algebraic invariance of $\mathbf{v}_0$ under boundary-condition choice. The Newton-equation simulation is therefore the physically realised limit in which $\mathcal{G}<\infty$, and is the appropriate testing ground for the DEM falsification protocol of Section~6.3. The conjugate-variable duality discussed in Section~6.5 in the main text further determines whether the manifested signature is strain creep (stress-controlled boundary, Figure~\ref{fig:newton_verification}a,c) or stress concentration (strain-controlled boundary, Figure~\ref{fig:newton_verification}b,d), with the null direction $\mathbf{v}_0$ invariant under the choice.

This Reynolds-number-like structure of the Gateway criterion has a direct geophysical implication that pre-empts a related objection. A reader might object that in any engineering or geological application the local stiffness $k$ is large and the friction-borne $\mathsf{D}$ block is substantial, so $\mathcal{G}$ ought to be far below unity in any steady state, and the topological Gateway should therefore be choked out by ordinary dissipation in all physically realised settings. The Gateway number plays the role here that the Reynolds and P\'eclet numbers play in fluid and transport mechanics: it does not need to exceed unity in steady state for the conservative routing to dominate. It need only exceed unity \emph{transiently}, during episodes in which the effective stiffness drops sharply -- exactly what happens at the onset of a local unjamming transition, at the rupture of an asperity, or in the slip phase of a stick--slip cycle. During such episodes the denominator of $\mathcal{G}$ collapses faster than the numerator, the system crosses into the $\mathcal{G}\!\geq\!1$ regime for a finite time, and the parity-mandated null direction is briefly available as an energy-routing channel. This transient activation is the topological-framework interpretation of the slip-rate acceleration observed at the onset of stick--slip failure and of the macroscopic precursors that precede laboratory-scale faulting events. The framework therefore predicts not steady-state Gateway dominance, which would be inconsistent with the macroscopic stability of granular packings, but episodic activation tied to local stiffness collapse -- a prediction that is directly accessible to DEM and laboratory experiments designed to track $k$ and $\mathcal{G}$ through stick--slip cycles.


\section{Mathematical Outlook: Cnoidal Wave Extension}
\label{sec:S4}

The following section (formerly Section~10, Concluding Remarks and Future Outlook, of the main text) develops the mathematical bridge between the linear topological invariants and the nonlinear cnoidal-wave regime.


The topological parity framework verified herein assumes a strictly linear state-space cross-coupling regime. A compelling and logical extension of this work is to relax the assumption of linear contact laws and investigate the role of high-order power-law nonlinearities ($n \ge 3$), which naturally arise from non-smooth asperity deformation and Hertzian grain-contact physics. Preliminary structural considerations suggest that a transition to non-linear contact metrics will fundamentally alter the nature of the space; it maps the global, invariant null-space rays of our linear operator onto localised, state-dependent tangent manifolds. 

In such a higher-order integer non-linear lattice, the balance between non-linear sharpening and the inherent geometric dispersion of the network is well known to support periodic cnoidal wave solutions \cite{Veveakis2015}, of the Jacobi-elliptic form $\mathrm{cn}(u\,|\,m)$ with modulus parameter $m\in[0,1]$ that interpolates between the linear-dispersion limit $m\to 0$ (sinusoidal small-amplitude waves recovered as the Toeplitz limit of the present paper) and the strongly-nonlinear limit $m\to 1$ (localised solitary-wave compaction structures of the Veveakis--Regenauer-Lieb regime~\cite{Veveakis2015,Regenauer2013_JCSMD2}). The linear model developed here is therefore the structurally necessary $m\to 0$ baseline against which the cnoidal nonlinear extension is to be measured, providing the topological invariants ($\det\mathsf{L}=0$ for $N$ odd, the null direction $\mathbf{v}_0$, and the Gateway number $\mathcal{G}$) that remain meaningful at every $m$ because skew-symmetry of the state-dependent $\mathsf{L}(\mathbf{q})$ is preserved at every state. Exploring these closed-form elliptic solutions offers a rich mathematical pathway to translate our discrete, boundary-confined state-space drift into travelling non-linear solitary wave trains. This approach would effectively generalise the static macro-poroplastic creep signature identified in this paper into a highly dynamic regime of localised acoustic soliton transport and periodic stick-slip fabric instability.

\bibliographystyle{elsarticle-num}
\bibliography{bib_merged}


\begin{frontmatter}

\title{ Topological Foundations of Multi-Field Instabilities in Continua: \\ Part 2 — Analytical Formulation for 1-D Spin Chains}

\author[curtin]{K. Regenauer-Lieb\corref{cor1}}
\ead{Klaus@curtin.edu.au}
\author[savoie]{F. Nicot}

\cortext[cor1]{Corresponding author}

\address[curtin]{ARC Centre of Excellence for Carbon Science \& Innovation,
  WASM: Minerals, Energy and Chemical Engineering,
  Curtin University, Perth WA 6845, Australia}
\address[savoie]{ISTerre, Universit\'e Savoie Mont Blanc, Chamb\'ery, France}

\begin{quote}\small
\textit{This is the second paper of a three-part JMPS manuscript series that develops a parameter-free, topological classification of multi-field instability and its realisation in granular matter. Building upon the foundational framework established in Part 1, this manuscript provides an explicit analytical formulation of the theory for the specific case of a 1-D granular spin chain. We derive the governing equations and theoretical predictions, framing the behaviour of the dynamic system against classical elasto-plasticity. The numerical application and falsification protocols for these analytical results are detailed in the final companion paper, Part 3. The Editor is respectfully asked to arrange coordinated review of Papers 1 to 3.}
\end{quote}

\begin{abstract}
Granular materials exhibit irreversible macroscopic deformation despite the time-reversibility of individual contact interactions. This paper demonstrates that this phenomenon is rooted in a parity-dependent bifurcation of coupling structures: even-dimensional systems form Stable Layers with preserved time-reversal symmetry, while odd-dimensional systems form "Gateway Layers" whose parity-mandated null direction leaves the driven response unconfined, resulting in a constitutive arrow of time under sustained forcing. Building upon the Parity Theorem [Part 1], we identify the three-channel Volumetric–Mechanical–Configurational (VMC) granular contact as the minimal Gateway. Specifically, for $N=3$, the skew-symmetric coupling block $\mathsf{L} \in \mathfrak{so}(3)$ forces a persistent zero eigenvalue, providing an unresisted configurational drift channel that operates even in the absence of friction. We show that in a one-dimensional chain, the acyclic graph topology, characterised by the absence of force-bearing loops (vanishing first Betti number $\beta_1=0$), algebraically constrains this irreversibility to volumetric gradient modes, isolating dilatancy as the sole failure pathway. We derive closed-form analytical solutions for $N=3$ (secular drift) and $N=4$ (harmonic confinement), demonstrating that boundary work projected onto the null axis of $\mathsf{L}$ accumulates as secular configurational drift in odd-dimensional systems. These analytical formulations provide the theoretical basis for the numerical application and DEM correspondence presented in Part 3. \\
\end{abstract}

\begin{keyword}
Granular mechanics \sep Configurational mechanics \sep
Classical spin chain \sep Skew-symmetric operators \sep
Lie algebra $\mathfrak{so}(3)$ \sep
Parity theorem \sep Gateway irreversibility \sep
Cocycle filter \sep One-dimensional chains
\end{keyword}

\end{frontmatter}

\section{Introduction}
\label{sec:introduction}

Granular materials are among the most fundamental objects in mechanics. They underlie nearly every model of soil failure~\cite{Schofield1968}, powder compaction~\cite{Nedderman1992}, fault gouge~\cite{Ikari2009}, and dense suspension rheology~\cite{Boyer2011}, yet the microscopic origin of their irreversible behaviour remains incompletely understood~\cite{Jaeger1996}. Conventional approaches attribute irreversibility primarily to frictional dissipation at grain contacts. This paper argues that this attribution is incomplete: geometric coupling mechanisms at the grain-contact scale play an equally fundamental role. They operate even in the complete absence of friction, following from the algebraic necessity of the three-channel topology of grain interactions.

\begin{figure}[!htbp]
\centering
\includegraphics[width=1.0\linewidth]{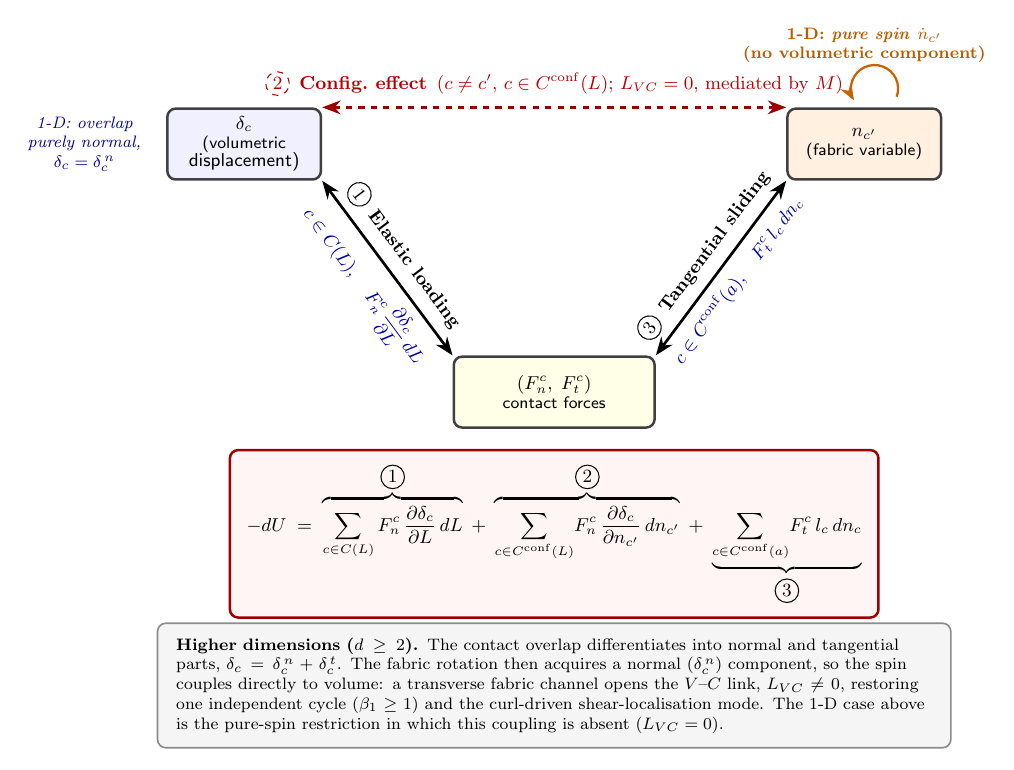}
\caption{Structural decomposition of the internal energy increment $dU$ and the dissipation potential $d\phi$ into three contact-mechanics channels. State nodes: volumetric displacement $\delta_c$ and fabric variable $n_{c'}$; double-headed arrows indicate the work-conjugate coupling structure identified in this paper. \circled{1}~\textit{Elastic loading channel} ($c\in C(L)$): reversible normal-force work $F_n^c\,\partial\delta_c/\partial L\,dL$, whose structure is compatible with hyperelastic formulations. \circled{2}~\textit{Configurational channel} ($c\neq c'$, dashed, $c\in C^{\mathrm{conf}}(L)$): normal-force work mediated by the fabric variable. \circled{3}~\textit{Tangential sliding channel} ($c\in C^{\mathrm{conf}}(a)$): work term $F_t^c\,l_c\,dn_c$, whose structure is compatible with frictional constitutive laws not included here. Under $dL\to -dL$, channel~\circled{1} cancels identically, leaving $d\phi = -dU$ as the sum of the configurational dissipation $dE^{\mathrm{conf}}$ and the frictional dissipation. $-dU$ is positive because boundary work done on the forward stroke exceeds the elastic energy recovered on the reverse stroke.}
\label{fig:contact_decomp}
\end{figure}

Recent developments in configurational contact mechanics~\cite{Nicot2024} have demonstrated that grain interactions cannot be reduced to force transmission alone (see Fig.~\ref{fig:contact_decomp}). Each contact transmits three independent mechanical actions: a force, a torque, and a configurational moment associated with the evolving geometry of the contact network. The present paper identifies this coupling structure as the irreducible source of one-dimensional configurational irreversibility, and embeds it within the general augmented Onsager framework developed in~\cite{RegenNicot2026foundation}.

\paragraph{A graph-theoretic representation of reversible coupling networks}
The algebraic structure of the reversible cross-coupling $\mathsf{L}$ used throughout this paper admits a transparent representation as a graph, the combinatorial vocabulary of which is briefly recalled here for the convenience of readers. The construction is elementary. We treat each thermodynamic channel ($V$, $M$, $C$, and additional channels in higher-dimensional extensions) as a \emph{vertex}, and each non-zero reversible coupling coefficient $L_{\alpha\beta}\neq 0$ as an undirected \emph{edge} joining the two coupled channels. The resulting object is the \emph{reversible coupling graph} $G=(\mathcal{V},\mathcal{E})$ of the system, with $|\mathcal{V}|=N$ vertices and $|\mathcal{E}|$ edges. Each edge carries a sign that records the orientation of the antisymmetric coupling, but for the topological counting that follows only its absolute value, i.e., its presence or absence matters.

To characterise the coupling topology, we use two quantities. The \emph{first Betti number}
\begin{equation}
\beta_1 = |\mathcal{E}|-|\mathcal{V}|+1  
\label{eq:betti_def}
\end{equation}
counts the independent closed circuits of the graph; when $\beta_1=0$ the network is a tree and every edge vector is a pure gradient (cocycle), whereas $\beta_1\geq 1$ admits solenoidal (curl) cycle modes. The resulting Hodge--Helmholtz decomposition 
\begin{equation}
\mathbf{u}=\mathbf{u}_{\mathcal{B}^*}\oplus\mathbf{u}_{\mathcal{Z}}   
\label{eq:hodge_helmholtz_intro}
\end{equation}
 into cocycle and cycle subspaces is the graph-theoretic analogue of the classical Stokes--Helmholtz theorem: because $\mathsf{L}$ preserves this decomposition, gradient and rotational modes decouple exactly.

 Here, $\mathbf{u}_{\mathcal{B}^*}$ denotes the cocycle subspace, spanned by the image of the coboundary operator $\delta$, representing the gradient-like (irrotational) flux of the granular state variable. Conversely, $\mathbf{u}_{\mathcal{Z}}$ represents the cycle subspace, the kernel of the boundary operator $\partial$ ($\partial \mathbf{u}_{\mathcal{Z}} = 0$), corresponding to the solenoidal (divergence-free) or rotational modes. Because the augmented Onsager operator $\mathsf{L}$ is constructed as a topological operator on the graph, it commutes with the projection operators onto these subspaces. Consequently, the gradient-like (dilatant) and solenoidal (circulatory) coupling channels carried by $\mathsf{L}$ decouple exactly, providing a rigorous topological basis for the observed independence of shear and volumetric strain rates in the linear regime.  Formal definitions of the orthogonal direct sum ($\oplus$), proofs, and the full Hodge--Helmholtz development are given in Supplementary Material, Section~S1.

\begin{figure}[!htbp]
\centering
\begin{tikzpicture}[
  vertex/.style={circle, draw=black, fill=blue!10, line width=0.6pt, minimum size=9mm, inner sep=0pt, font=\small\bfseries},
  edgeL/.style={line width=1.0pt, draw=blue!70!black},
  chord/.style={line width=1.0pt, dashed, draw=orange!85!black},
  cycle_arrow/.style={->, >=Stealth, line width=0.8pt, draw=red!75!black},
  scale=1.0
]
\begin{scope}[xshift=0cm]
\node[vertex] (V1) at (0, 0) {V};
\node[vertex] (M1) at (2.0, 0) {M};
\node[vertex] (C1) at (4.0, 0) {C};
\draw[edgeL] (V1) -- (M1) node[midway, above=1pt, font=\footnotesize] {$L_{VM}$};
\draw[edgeL] (M1) -- (C1) node[midway, above=1pt, font=\footnotesize] {$L_{MC}$};
\draw[dashed, draw=gray!50, line width=0.5pt] (V1) to[bend right=35] node[midway, below, font=\footnotesize, text=gray] {$L_{VC}=0$} (C1);
\node[align=center, font=\small] at (2.0, -2.2)
  {\textbf{(a)} 1-D chain $P_3$ (VMC triad)\\[1pt]
   $|\mathcal{V}|=3$, $|\mathcal{E}|=2$, $\beta_1=0$\\[1pt]
   $\mathcal{Z}=\{0\}$, $\mathcal{B}^*=\mathbb{R}^2$};
\end{scope}
\begin{scope}[xshift=8.5cm]
\node[vertex] (V2) at (0, 0) {V};
\node[vertex] (M2) at (2.0, 1.2) {M};
\node[vertex] (C2) at (4.0, 0) {C};
\draw[edgeL] (V2) -- (M2) node[midway, above left=-1pt, font=\footnotesize] {$L_{VM}$};
\draw[edgeL] (M2) -- (C2) node[midway, above right=-1pt, font=\footnotesize] {$L_{MC}$};
\draw[chord] (V2) to[bend right=25] node[midway, below=2pt, font=\footnotesize, text=orange!75!black] {$L_{VC}^{(2D)}\neq 0$} (C2);
\draw[cycle_arrow] (1.0, 0.2) to[bend left=20] (2.0, 0.8);
\node[font=\footnotesize, text=red!70!black] at (2.0, 0.4) {$\gamma_1$};
\node[align=center, font=\small] at (2.0, -2.2)
  {\textbf{(b)} 2-D extension with chord\\[1pt]
   $|\mathcal{V}|=3$, $|\mathcal{E}|=3$, $\beta_1=1$\\[1pt]
   $\dim\mathcal{Z}=1$, one cycle $\gamma_1$};
\end{scope}
\end{tikzpicture}
\caption{Graph-theoretic illustration of the cocycle filter at the Level-3 Hasse apex (the VMC triad). \textbf{(a)} The 1-D reversible coupling graph $G$ has vertices = channels $\{V,M,C\}$ and edges = non-zero reversible couplings $\{L_{VM}, L_{MC}\}$. The structural zero $L_{VC}=0$ (faded gray) is the absence of a chord. By Eq.~\eqref{eq:betti_def}, $\beta_1 = |\mathcal{E}|-|\mathcal{V}|+1 = 2-3+1 = 0$: no independent closed loop exists in the graph, so the cycle space $\mathcal{Z}$ is trivial and every edge-indexed vector is a pure gradient (cocycle). At every Hasse level above the apex, the 1-D induced subgraph is itself a path, so this acyclicity propagates through the entire Hasse hierarchy: the 1-D chain is the maximally degenerate Hasse-lattice configuration. \textbf{(b)} The simplest non-trivial extension, available only in dimension $d\geq 2$~, restores the $V$--$C$ closure through a transverse fabric channel (orange dashed chord, $L_{VC}^{(2D)}\neq 0$), producing one independent cycle $\gamma_1$ ($V\!\to\!M\!\to\!C\!\to\!V$) and raising $\beta_1$ to 1. The cycle space is now one-dimensional, supporting Schnakenberg cycle currents and the curl-driven shear-localisation to be discussed in details in future work. The 1-D Cartesian restriction of this paper is therefore identifiable as the precise algebraic act of \emph{removing the chord} from the Level-3 Hasse-apex graph.}
\label{fig:graph_primer}
\end{figure}

\paragraph{The 1-D Cartesian restriction as a topological filter}
The 1-D granular chain analysed in this paper is the path graph $P_N$ on $N$ vertices with $N-1$ edges, the simplest non-trivial connected graph. Eq.~\eqref{eq:betti_def} immediately gives
\begin{equation}\label{eq:beta1_chain}
  \beta_1(P_N) \;=\; (N-1)-N+1 \;=\; 0\qquad\text{for every }N,
\end{equation}
so the cycle space is trivial, $\mathcal{Z}=\{0\}$, and the orthogonal Hodge--Helmholtz decomposition~Eq. \eqref{eq:hodge_helmholtz_intro} collapses onto the cocycle subspace alone, $\mathcal{B}^*=\mathbb{R}^{N-1}$ (Fig.~\ref{fig:graph_primer}\,a). The 1-D Cartesian coordinate restriction imposed in the present paper must therefore not be viewed as a descriptive oversimplification but rather as an exact \emph{topological filter}: all curl-driven, cyclic shear-localisation mechanisms are mathematically filtered out by the absence of any chord in $G$. 

In the Hasse-lattice classification of channel subsets developed in Ref.~\cite{RegenNicot2026foundation}, the path graph $P_N$ is the most degenerate possible sub-pattern: at every Hasse level above the apex the induced subgraph is itself a path, so $\beta_1=0$ propagates through every level of the multi-channel hierarchy without remediation. The complete Volumetric--Mechanical--Configurational (VMC) triad is present at the contact scale, yet its structural evolution is forced entirely into the cocycle (gradient) space. Consequently, when the Gateway activation criterion of Ref.~\cite{RegenNicot2026foundation} is met, the system cannot dissipate energy via macro-shear; it is forced to trigger a purely volumetric instability. The macroscopic signature is a secular dilatancy drift along the parity-mandated null direction, with no shear-localisation component available as an alternative pathway. The lifting of this cocycle filter, achieved by introducing transverse fabric channels in dimension $d\geq 2$ that restore the $V$--$C$ closure and raise $\beta_1$ to $\geq 1$ (Fig.~\ref{fig:graph_primer}\,b), opens a second, algebraically independent irreversibility mode: the curl-driven shear-localisation instability is discussed in Ref.\cite{RegenNicot2026foundation} and its demonstration on a 2-D ladder granular chain is subject of ongoing work.

\paragraph{Inheritance from the general framework}
The augmented Onsager operator $\mathsf{A} = \mathsf{D} + \mathsf{L}$ with $\mathsf{D} = \mathsf{D}^T \succeq 0$ symmetric dissipative and $\mathsf{L} = -\mathsf{L}^T$ skew-symmetric reversible, the Parity Theorem stating that $\det(\mathsf{L}) = 0$ for any real skew-symmetric $\mathsf{L}$ of odd dimension, the Gateway activation number $\mathcal{G} = \|\mathsf{L}\|_2^2/[\sigma_{\min}(\mathsf{D})\,\sigma_{\max}(\mathsf{D})]$ governing null-mode activation, the transient amplification theorem for the 2-D reduced system, the Hasse-lattice organisation of channel subsets, and the Satake graph-theoretical identification of the $p$--$c$--$v$ entity types with the $V$--$M$--$C$ thermodynamic channels are all developed in Ref.~\cite{RegenNicot2026foundation}. The present paper takes these as given and develops their granular-mechanical content at the contact scale.

\paragraph{Central object and minimal model}
The central object of this manuscript is the \emph{VMC triad}, a minimal thermodynamic representation comprising Volumetric ($V$), Mechanical ($M$), and Configurational ($C$) channels. Its evolution is governed by an augmented Onsager coupling $\mathsf{A} = \mathsf{D} + \mathsf{L}$, where the skew-symmetric operator $\mathsf{L} \in \mathfrak{so}(3)$ is topologically constrained by its odd dimensionality to possess a structural zero eigenvalue. This minimal model is self-contained: by deriving its dynamics from contact-scale topological invariants rather than empirical interaction coefficients, we establish a parameter-free \emph{topological} foundation for one-dimensional granular irreversibility. We demonstrate that the VMC triad provides the microscopic origin of one-dimensional irreversibility, identifying the topological Gateway as the fundamental mechanism driving granular dilatancy. Mirroring the way ideal plasticity emerges from dislocation geometry, our framework reveals that granular irreversibility is an emergent property of the contact-space topology. While classical models rely on internal friction to dictate yielding, this framework builds on the topological groundwork of Satake \cite{Satake1987} and its dynamic evolution principles \cite{Nicot2024}. We demonstrate here that irreversibility arises as a geometric requirement of the contact-space connectivity, appearing even in the absence of internal friction.

The present paper establishes the analytical correspondence between the discrete VMC contact triad and the macroscopic augmented thermodynamic framework introduced in Part 1. The manuscript covers:
\begin{itemize}\item \textbf{Topological Irreducibility:} Proof that the $N=3$ configuration of the VMC triad represents the irreducible generator of one-dimensional irreversibility.
\item \textbf{Lie-Algebraic Mapping:} A spin-reinterpretation of the contact where the configurational channel $C$ is treated as a rotational degree of freedom. Here, $\mathsf{L}$ acts as the generator of rotations in $\mathfrak{so}(3)$, where the parity-mandated null mode corresponds exactly to the preserved rotation axis.
\item \textbf{Analytical Bifurcation:} Closed-form solutions for $N=3$ (secular drift/Gateway) and $N=4$ (harmonic confinement/Stable Layer), providing an explicit algebraic demonstration of the parity-driven transition.\end{itemize}\noindent 

The scope of the numerical verification and physical validation (Part 3~\cite{RegenNicot2026numerical}) is summarised as follows:
\begin{itemize}
\item \textbf{Spectral Scaling:} High-precision numerical validation of the parity contrast for chains ranging from $N=3$ to $50$, confirming the asymptotic frequency scaling $|\lambda_{\min}^{\rm even}|\sim\gamma\pi/N$.
\item \textbf{DEM Correspondence:} A direct mapping of VMC channels to discrete-element method (DEM) observables, providing falsifiable protocols for oedometer and triaxial simulations.
\item \textbf{Thermodynamic Bottlenecking:} Interpretation of the global tridiagonal boundary conditions as a dynamic thermodynamic bottleneck, constraining the system's evolution.\end{itemize}

\begin{remark}
 This paper focuses exclusively on the linear operator evolution; nonlinear saturation and mappings to the Nonlinear Schr\"odinger equation are reserved for forthcoming work. 
\end{remark}
\section{Configurational Contact Mechanics}
\label{sec:contact_mechanics}

\subsection{Microstructure and contact decomposition}

Consider a granular assembly of deformable spheres under external loading $\mathbf{L}$. The microstructure is encoded in the contact ensemble~\cite{Nicot2024}
\begin{equation}
  \mu = \bigl\{\mathbf{n}_c,\,\delta_c\bigr\}_{c\in\mathcal{C}(\mathbf{L})},
  \label{eq:microstructure}
\end{equation}
where $\mathbf{n}_c$ is the unit normal to contact $c$ and $\delta_c$ is the overlap index. Under incremental loading $d\mathbf{L}$, the contact set splits into two disjoint subsets,
\begin{equation}
  \mathcal{C}(\mathbf{L}) =
    \mathcal{C}^{\rm inv}(\mathbf{L})\cup\mathcal{C}^{\rm conf}(\mathbf{L}),
  \label{eq:contact_split}
\end{equation}
where at \emph{invariant} contacts $c\in\mathcal{C}^{\rm inv}$ the contact normal does not rotate and the response is reversible, whereas at \emph{configurational} contacts $c\in\mathcal{C}^{\rm conf}$ the normal rotates by $d\mathbf{n}_c$ constrained by neighbouring fabric grains $c'$. This kinematic differentiation is related to the bimodal organisation of force chains in dense granular assemblies, where strong-force networks carry most of the shear resistance and weak contacts provide lateral structural support~\cite{Radjai1998,Luding2004}.

\subsection{Configurational irreversibility}

The key result of Ref.~\cite{Nicot2024} is that during frictional sliding the reverse rotation vanishes, $d\mathbf{n}_c^-=\mathbf{0}$, so that the microstructure cannot return to its original state under load reversal $(d\mathbf{L},-d\mathbf{L})$:
\begin{equation}
  d\mu^{(d\mathbf{L},-d\mathbf{L})}
    \neq \{\mathbf{0},0\}_{c\in\mathcal{C}(\mathbf{L})}.
  \label{eq:irreversibility}
\end{equation}
The dissipation increment $d\phi = -dU$ after a closed loading cycle equals
\begin{align}
  d\phi &=
    \underbrace{\sum_{c\in\mathcal{C}^{\rm conf}}
      F_n^c\,\frac{\partial\delta_c}{\partial\mathbf{n}_{c'}}\,
      d\mathbf{n}_{c'}^+}_{\text{configurational dissipation}\;(dE^{\rm conf})}
    \notag\\&\quad+
    \underbrace{\sum_{c\in\mathcal{C}^{\rm conf}}
      \mathbf{F}_t^c\,l_c\,d\mathbf{n}_c^+}_{\text{frictional dissipation}},
  \label{eq:dEconf}
\end{align}
where $F_n^c$ and $\mathbf{F}_t^c$ are the normal and tangential contact forces, $l_c$ is the branch-vector length, and $\partial\delta_c/\partial \mathbf{n}_{c'}$ is the sensitivity of the overlap to the fabric grain orientation.

\subsection{The three-channel contact structure}
\label{sec:three_channels}

The energy increment~Eq. \eqref{eq:dEconf} couples three channels at each configurational contact:
\begin{enumerate}[label=(\roman*)]
  \item the \emph{volumetric channel} $V$: the overlap $\delta_c$, work-conjugate to the normal contact force $F_n^c$;
  \item the \emph{mechanical channel} $M$: the contact-level tangential (shear) transmission, work-conjugate to the tangential force;
  \item the \emph{configurational channel} $C$: the fabric orientation $\mathbf{n}_{c'}$ from neighbouring grains, i.e.\ the rotation of the contact normal, work-conjugate to the configurational moment.
\end{enumerate}
These three channels form a minimal interaction unit. The V--M and M--C pairs each admit a direct reversible coupling through overlap--force elastic exchange and force--fabric torque exchange, while the V--C pair does not: a change in volumetric overlap cannot rotate the contact fabric without force mediation, so $L_{VC}=0$. The reversible coupling network is therefore a chain $V$--$M$--$C$ rather than a fully connected triangle. The constitutive derivation of $\mathsf{A}=\mathsf{D}+\mathsf{L}$ from a contact-scale free-energy and dissipation-potential pair, including the proof that $L_{VC}=0$ is a structural theorem rather than a kinematic idealisation, is given in Ref.~\cite{RegenNicot2026foundation} (Supplementary Material, Section~S1).

\begin{figure}[!htbp]
\centering
\includegraphics[width=\linewidth]{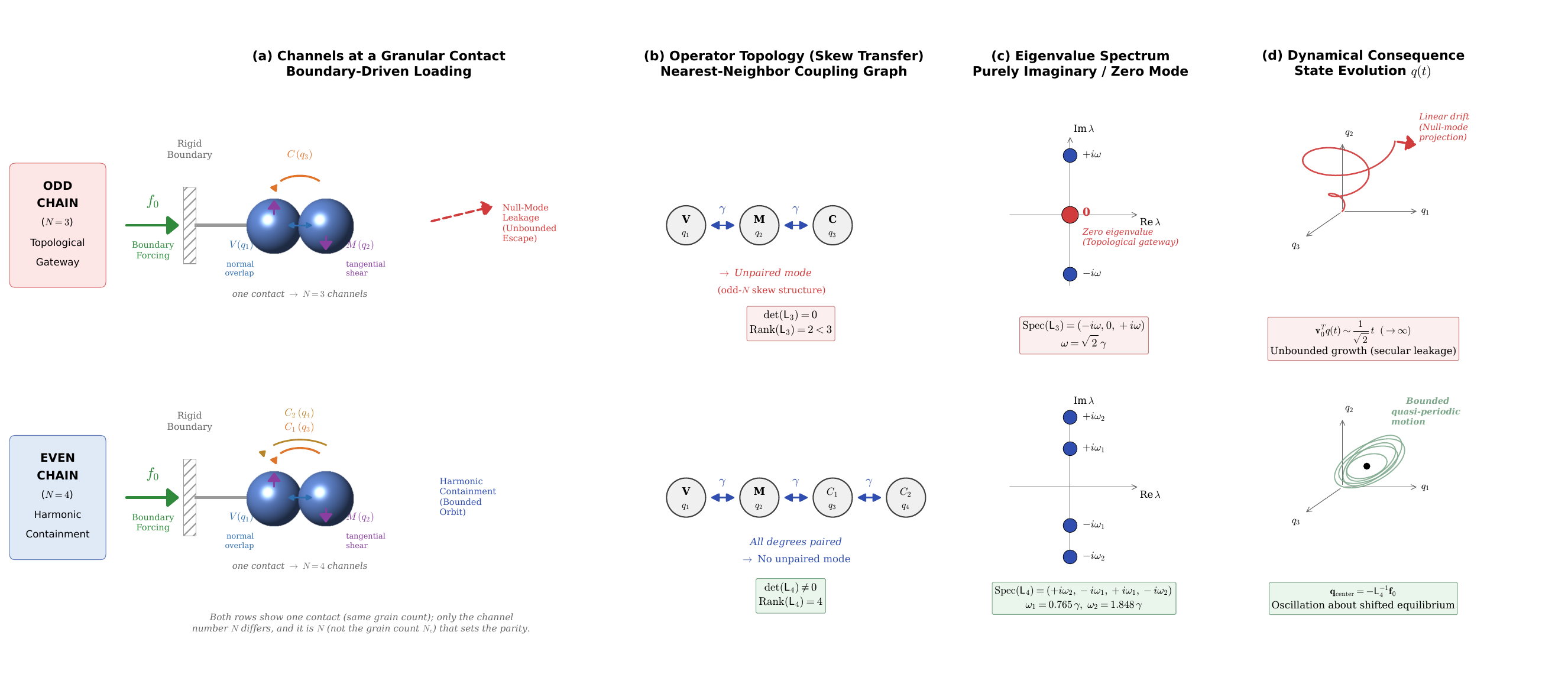}
\caption{Parity-governed dynamics of the granular contact coupling operator: the odd $N=3$ VMC triad and its even $N=4$ extension. \textbf{(a)} A single granular contact (two grains) and the thermodynamic channels it carries: the volumetric ($V$), mechanical ($M$) and configurational ($C$) channels, with the even case adding a higher-order configurational channel $C_2$ at the \emph{same} contact. Both rows show the same contact and the same number of grains; only the channel count $N$ differs, and it is $N$, not the grain count $N_c$, that sets the parity, so odd $N$ admits the null-mode expulsion (dashed arrow) and even $N$ forces containment. \textbf{(b)} Operator topology of $\mathsf{L}_N$: the channels form a chain \emph{in channel space}, with the missing $V$--$C$ chord ($\beta_1=0$) in the odd case. \textbf{(c)} Eigenvalue spectra, with the Gateway null-mode (red dot) at the origin for $N=3$. \textbf{(d)} Dynamical trajectory: $N=3$ drifts secularly along the null-mode (dilatancy instability), whereas $N=4$ stays bounded and quasi-periodic. Adding a single thermodynamic channel, whether electrical, chemical, thermal, or, as here, a higher-order configurational one, turns an odd unstable oscillator into a stable even one. The distinction between the channel count $N$ and the grain count $N_c$ is developed in Part~3~\cite{RegenNicot2026numerical}.}
\label{fig:model_setup}
\end{figure}

\section{The Augmented Onsager Framework as Used Here}
\label{sec:DL_framework}

The framework underlying the present paper, including the derivation of the augmented Onsager operator from the GENERIC formalism~\cite{Grmela1997,Ottinger1997}, the scope and applicability conditions for granular media operating far from global thermodynamic equilibrium, the Onsager--Casimir reciprocal relations that fix $\mathsf{D}=\mathsf{D}^T$ and $\mathsf{L}=-\mathsf{L}^T$ at the channel level, and the physical interpretation of the dissipative ($\mathsf{D}$) and conservative ($\mathsf{L}$) blocks as carriers of entropy-producing transport and reversible cross-coupling respectively, is developed in Ref.~\cite{RegenNicot2026foundation}. We summarise here only the inputs required for the granular content of the present paper.

\paragraph{The decomposition}
At each contact neighbourhood, taken to be in local quasi-static equilibrium~\cite{deGroot1984}, the conductance matrix decomposes uniquely as
\begin{equation}
  \mathsf{A} = \mathsf{D} + \mathsf{L}, \qquad
  \mathsf{D} = \tfrac{1}{2}(\mathsf{A} + \mathsf{A}^T) \succeq 0, \qquad
  \mathsf{L} = \tfrac{1}{2}(\mathsf{A} - \mathsf{A}^T) = -\mathsf{L}^T.
  \label{eq:DL_decomp}
\end{equation}
The evolution of thermodynamic forces $\mathbf{X} = (X_1, \ldots, X_N)^T$ is governed by $\dot{\mathbf{X}} = \mathsf{A}\,\mathbf{X}$. The dissipative block $\mathsf{D}$ generates entropy production $\sigma_{\rm irr} = \mathbf{X}^T \mathsf{D}\,\mathbf{X} \geq 0$; the conservative block $\mathsf{L}$ is work-free, $\mathbf{X}^T \mathsf{L}\,\mathbf{X} \equiv 0$, with spectrum on the imaginary axis (or at zero) generating oscillatory inter-channel exchange. The Onsager--Casimir relations for time-odd transport coefficients require $L_{\alpha\beta} = -L_{\beta\alpha}$~\cite{deGroot1984,Casimir1945}. Such conservative cross-couplings appear from gyroscopic forcing in rotating machinery to the antisymmetric viscous moduli of active matter~\cite{Scheibner2020,Fruchart2023}.

\paragraph{Scope}
The local quasi-static equilibrium assumption holds whenever the macroscopic gradient length scale is large compared with the contact scale, as occurs in slowly loaded granular assemblies. For rapidly driven far-from-equilibrium systems the GENERIC extension applies; the parity argument we use is purely algebraic and survives the nonlinear extension. Both scope arguments are developed in Ref. \cite{RegenNicot2026foundation}; we proceed in the linear, local-quasi-static regime.

\section{The Parity Theorem and the VMC Gateway}
\label{sec:VMC_gateway}

\subsection{Parity Theorem (recalled from \cite{RegenNicot2026foundation})}
\label{sec:parity_recall}

The central algebraic input from the general framework is:

\begin{theorem}[Parity Theorem, \cite{RegenNicot2026foundation}]
\label{thm:parity}
Let $\mathsf{L} \in \mathbb{R}^{N\times N}$ be a real skew-symmetric matrix, $\mathsf{L}^T = -\mathsf{L}$.
If $N$ is odd, then $\det(\mathsf{L}) = 0$ unconditionally and $\ker(\mathsf{L})$ is non-trivial. If $N$ is even, $\det(\mathsf{L}) \neq 0$ generically. The spectral structures are
\begin{subequations}
\begin{align}
N \text{ odd:}  \quad & \mathrm{spec}(\mathsf{L}) = \bigl\{0,\; \pm i\omega_1,\; \ldots,\; \pm i\omega_{(N-1)/2}\bigr\}, \\
N \text{ even:} \quad & \mathrm{spec}(\mathsf{L}) = \bigl\{\pm i\omega_1,\; \ldots,\; \pm i\omega_{N/2}\bigr\},
\end{align}
\end{subequations}
holding for all values of the coupling coefficients $L_{\alpha\beta}$.
\end{theorem}

The proof is the single algebraic identity $\det(\mathsf{L}) = \det(\mathsf{L}^T) = \det(-\mathsf{L}) = (-1)^N \det(\mathsf{L})$, which forces $\det(\mathsf{L}) = 0$ when $N$ is odd. Full development of the proof and its corollaries is given in Ref.~\cite{RegenNicot2026foundation}.

\begin{definition}[Stable and Gateway Layers]
A coupled subsystem with \emph{even}-dimensional $\mathsf{L}$ is a \textbf{Stable Layer}: spectrum $\{\pm i\omega_1,\ldots,\pm i\omega_{N/2}\}$, closed symplectic orbits \footnote{ A closed symplectic orbit is a periodic trajectory within a phase space equipped with a symplectic 2-form $\omega$ that represents a stable, non-equilibrium thermodynamic cycle (a symplectic oscillator). In a multi-field coupling framework (such as 1-D poromechanics), these orbits are driven by the internal momenta generated through the direct feedback, e.g., in the fluid-solid poromechanical problem between solid stress/pressure and fluid displacement. Governed by a bilinear Poisson bracket containing explicit mixed solid-fluid terms, the closed orbit acts as a dynamic attractor where macroscopic ordered action is maintained against microscopic dissipation, with its relative fluctuations bounded by the rate of entropy production.}

A coupled subsystem with \emph{odd}-dimensional $\mathsf{L}$ is a \textbf{Gateway Layer}: the spectrum necessarily contains a zero eigenvalue, defining a null direction $\mathbf{v}_0 \in \ker(\mathsf{L})$ along which no reversible restoring force acts.
\end{definition}

The zero eigenvalue is a structural property of the coupling topology, not of the coupling magnitude. It exists even when all $L_{\alpha\beta}$ are infinitesimally small. This is the sense in which irreversibility is \emph{topological}: the Gateway is written into the dimension of the coupling network, not its strength.

\subsection{The minimal three-channel Gateway}
\label{sec:VMC_minimal}

\paragraph{The sense in which the triad is minimal.}
The claim of minimality is topological, not phenomenological. It does not assert that three channels suffice to \emph{describe} a granular contact, which plainly requires friction, particle shape, size, and restitution; it asserts that three is the smallest channel count that \emph{supports} a parity-mandated Gateway. The hierarchy is fixed by the Parity Theorem: a single channel carries no reversible coupling ($\mathsf{L}\equiv0$), two channels form an even Stable Layer whose spectrum is purely imaginary and confined, and three channels are the first odd case, for which $\det\mathsf{L}=0$ guarantees a null direction. Removing any one of $V$, $M$, $C$ returns the system to $N=2$, pairs the reversible couplings, and destroys the null mode, so the triad cannot be reduced without eliminating the mechanism it exists to describe. Two points make this precise, and are established in the remainder of this section: the null mode is independent of the volumetric--configurational coupling $L_{VC}$, so that a non-zero particle-scale $L_{VC}$ only reorients it and can neither create nor remove the Gateway; and the sole load-bearing assumption is that the configurational channel $C$ is a genuine degree of freedom, not one kinematically slaved to $V$ and $M$.

Ordering the contact state vector as $(V,M,C)^T$, the reversible coupling matrix of the three-body contact structure (Section~\ref{sec:three_channels}) is
\begin{equation}
  \mathsf{L}^{\rm contact}_{3\times3} =
  \begin{pmatrix} 0 & L_{VM} & 0 \\ -L_{VM} & 0 & L_{MC} \\ 0 & -L_{MC} & 0 \end{pmatrix}
  \begin{array}{l}\leftarrow V\\[4pt]\leftarrow M\\[4pt]\leftarrow C\end{array},
  \label{eq:L3contact}
\end{equation}
with $\mathrm{spec}(\mathsf{L}^{\rm contact}_{3\times3})=\{0,\pm i\sqrt{L_{VM}^2+L_{MC}^2}\}$.

The V--M coupling $L_{VM}$ governs overlap--force exchange and is reversible because reversing the contact spin retraces the overlap. The M--C coupling $L_{MC}$ governs force--fabric rotation and is reversible because reversing the mechanical trajectory reverses the fabric rotation. The V--C entry is zero ($L_{VC}=0$) because a change in volumetric overlap cannot rotate the contact fabric without force mediation, detailed in Ref.~\cite{RegenNicot2026foundation} (Supplementary Material, Section~S1).

By the Parity Theorem, $\mathsf{L}^{\rm contact}_{3\times3}$ necessarily has zero determinant. Its spectrum comprises a conjugate pair $\pm i\omega$ describing reversible V--M and M--C oscillations, and a zero eigenvalue with null eigenvector
\begin{equation}
  \mathbf{v}_0 \;=\; \frac{1}{\sqrt{L_{VM}^2+L_{MC}^2}}\,(L_{MC},\;0,\;L_{VM})^T.
  \label{eq:null_vector}
\end{equation}
The null eigenvector points in the $V$--$C$ plane with vanishing $M$ component: irreversibility is localised to the coupling of volumetric compaction to fabric rotation, which is exactly the configurational dissipation $dE^{\rm conf}$ of Eq.~\eqref{eq:dEconf}.

This identification rests on the configurational channel being a genuine, independent degree of freedom. The variable carried by $C$ is the kinematic fabric rotation $d\mathbf{n}_{c'}$, work-conjugate to the configurational moment $X_C=F_n^c\,\partial\delta_c/\partial\mathbf{n}_{c'}$ of Eq.~\eqref{eq:dEconf}: a rotational degree of freedom in the Cosserat sense, not a stress and not algebraically slaved to the mechanical force network. This independence is what keeps the channel count at $N=3$. Were the fabric rotation fixed by the volumetric and mechanical variables, $C$ would merge with $M$, the count would fall to $N=2$, the parity would turn even, and by the Parity Theorem the null eigenvector, and with it the Gateway, would disappear. The classification is itself basis-invariant: the channel may be carried equivalently by the flux $d\mathbf{n}_{c'}$ or by its conjugate moment $X_C$, the two being related by a Legendre transform with $\mathbf{v}_0$ invariant since $\ker(\mathsf{L})=\ker(\mathsf{L}^T)$ (Supplementary Material, Section~S2). What the construction requires is therefore not a particular representation, but that the fabric rotation be a degree of freedom genuinely distinct from $V$ and $M$.

The Gateway null mode exists for every value of $L_{VC}$, including $L_{VC}\neq 0$: what $L_{VC}=0$ controls is the \emph{orientation} of the null eigenvector, not its existence. Quantitative bounds on the effective $L_{VC}^{\rm eff}$ as a function of column aspect ratio are given in Supplementary Material, Section~S3.

The structural zero $L_{VC}=0$ does not cause the null eigenvalue (which is guaranteed by parity alone); rather, it fixes the \emph{orientation} of the null eigenvector to the $V$--$C$ axis. The topological origin of this orientation, via Satake's discrete Poincar\'e lemma~\cite{Satake1993} and the $p$--$c$--$v$ to $V$--$M$--$C$ identification, is developed in~\cite{RegenNicot2026foundation}.

\begin{figure}[!htbp]
\centering
\includegraphics[width=0.7\linewidth]{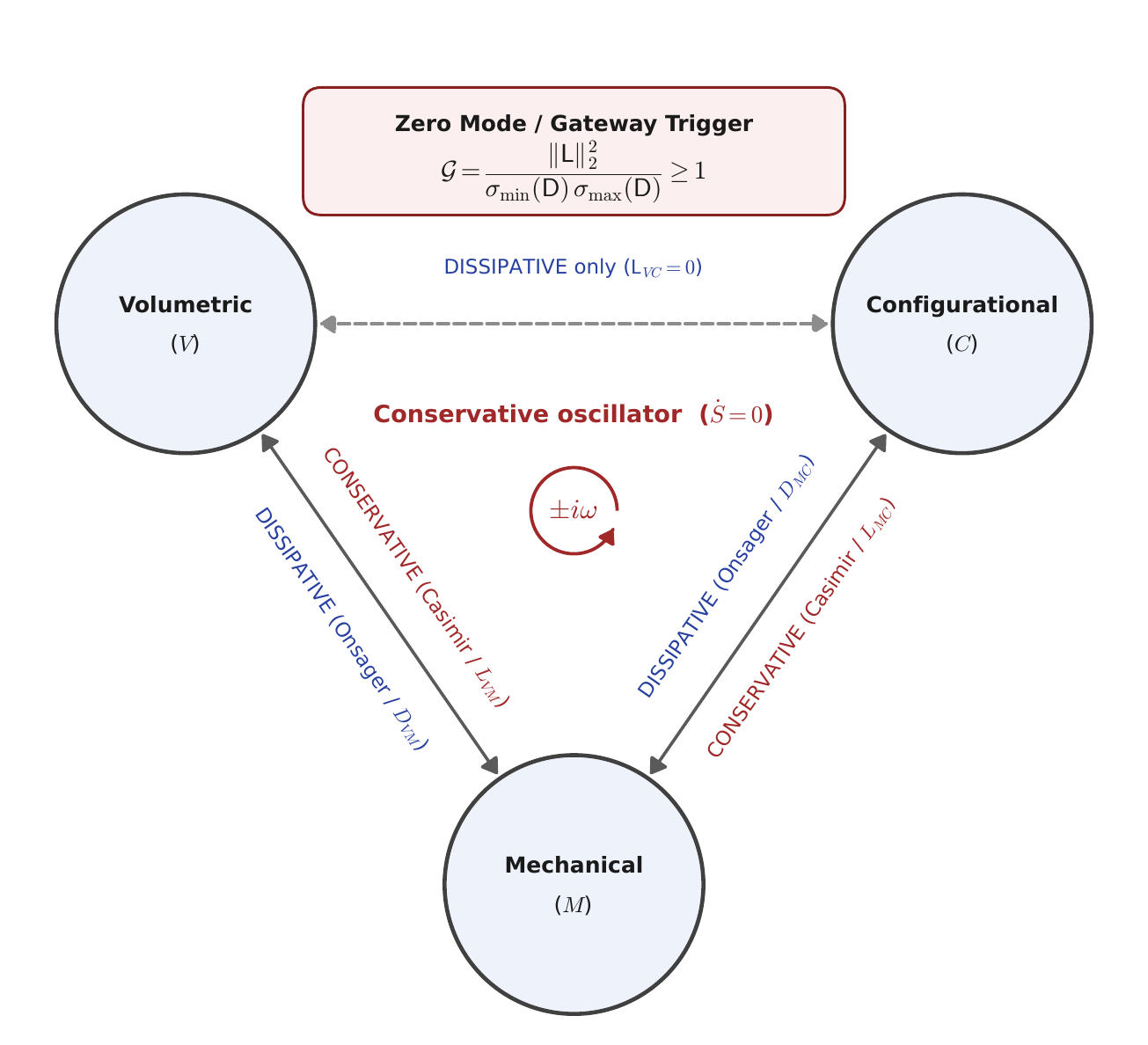}
\caption{Minimal V--M--C energy-channel diagram for the three-body Gateway at a granular contact. Solid arrows (V--M and M--C): direct reversible energy transfer (overlap--force exchange; force--fabric coupling). Dashed arrow (V--C): no direct reversible pathway; V--C coupling is mediated entirely by $M$, giving $L_{VC}=0$. The central $\pm i\omega$ denotes the conservative oscillator with zero entropy production. The Gateway criterion $\mathcal{G}\geq1$ marks zero-mode activation.}
\label{fig:grain_gateway}
\end{figure}

\begin{remark}[The friction pitfall]
Simplifying the mechanical interactions without the Gateway mechanism one may expect irreversibility to vanish as $\mathbf{F}_t^c\to\mathbf{0}$. A striking counter-example from classical plasticity is instructive: a von~Mises or Tresca yield criterion carries zero internal friction ($\phi=0$), yet plastic deformation is fully irreversible. The present framework explains why.
\end{remark}

The dissipative block $\mathsf{D}$ is the rate-sink: it sets how fast energy is removed, and friction enters here. The reversible block $\mathsf{L}$ encodes the connectivity of the interaction network --- topology. The Gateway \emph{exists} whenever $N=3$, regardless of $\|\mathsf{D}\|$; it \emph{activates} whenever $\mathcal{G}\geq1$, regardless of whether the threshold is crossed by reducing $\sigma_{\min}(\mathsf{D})$ or increasing $\|\mathsf{L}\|_2$. Friction can delay activation by raising $\sigma_{\min}(\mathsf{D})$, but removing friction does not eliminate the Gateway because its topology is set by $N$, not by $\|\mathsf{D}\|$. The arrow of time is written in the dimension of coupling, not in the coefficient of friction.

\paragraph{Independence of the configurational channel}
The construction requires the configurational channel $C$ to be a genuine kinematic degree of freedom rather than a quantity algebraically slaved to the volumetric and mechanical channels. This is not a formality. If the deviatoric fabric $\boldsymbol\xi_{\rm conf}$, with work-conjugate moment $A_{\rm conf}$, were a single-valued function of the stress and strain increments, the effective channel count would collapse from $N=3$ to $N=2$, restoring an even-parity Stable Layer and removing the null direction entirely. Three independent observations establish that it is not slaved. First, induced anisotropy is path-dependent: in both experiment and DEM the same stress state is reached with materially different fabrics according to loading history, so the fabric tensor $\boldsymbol\Phi=\langle\mathbf{n}\otimes\mathbf{n}\rangle$ cannot be recovered from the current stress alone and must be carried as an independent state variable, as in critical-state~\cite{Schofield1968} and anisotropic critical-state descriptions of granular media. Second, the fabric-rate is in general non-coaxial with the stress-rate, a relation that no scalar slaving multiplier can reproduce. Third, fabric reorganisation lags volumetric strain by a finite delay in one-dimensional DEM oedometer tests~\cite{RegenNicot2026numerical}, which is the direct dynamical signature of an independent relaxation channel rather than instantaneous kinematic slaving. The three-channel count is therefore a measured property of granular fabric~\cite{Nicot2024}, not a modelling convenience, and the odd parity that follows from it is correspondingly robust to the slaving objection.

\subsection{Gateway activation criterion}
\label{sec:activation_recall}

The activation criterion derived in~\cite{RegenNicot2026foundation} is summarised as follows. Decompose the state space as $\mathbb{R}^N = \mathrm{span}(\mathbf{v}_0) \oplus \mathcal{S}_\perp$. Projecting onto $\mathbf{v}_0$ shows that $\mathsf{L}$ cannot directly change $x_0 = \mathbf{v}_0^T\mathbf{X}$ (since $\mathbf{v}_0^T \mathsf{L} = \mathbf{0}^T$); the null-mode amplitude evolves only through the dissipative block $\mathsf{D}$. Activation is therefore an indirect, two-step process: $\mathsf{L}$ mixes energy within $\mathcal{S}_\perp$, and $\mathsf{D}$ then projects this mixing onto $\mathbf{v}_0$. With $d_{\min} = \sigma_{\min}(\mathsf{D})$, $d_{\max} = \sigma_{\max}(\mathsf{D})$, and $\omega = \|\mathsf{L}\|_2$, the effective injection rate into the null direction is $\omega^2/d_{\max}$ and the removal rate is $d_{\min}$. The transient-routing crossover occurs when these balance, yielding the \emph{Gateway number}
\begin{equation}
  \mathcal{G} \;=\; \frac{\|\mathsf{L}\|_2^2}{\sigma_{\min}(\mathsf{D})\,\sigma_{\max}(\mathsf{D})}
  \label{eq:gateway_number}
\end{equation}
and the activation criterion $\mathcal{G} \geq 1$. The full proof, the basis-invariant form $\mathcal{G}_{\rm inv} = \|\mathsf{D}^{-1/2}\mathsf{L}\mathsf{D}^{-1/2}\|_2^2$, and the transient amplification theorem (an explicit closed-form expression for the null-mode peak amplitude in the 2-D reduced system) are developed in Ref.~\cite{RegenNicot2026foundation}.

The mechanical interpretation is direct: $\mathcal{G} \geq 1$ means that conservative nonlocal coupling injects energy into the weakest dissipative channel faster than local dissipation can remove it, activating the null direction. At this threshold, the system enters the non-normal transient-amplification regime identified in Ref.~\cite{RegenNicot2026foundation} as the dispersive Gateway precursor, which precedes, and is algebraically independent of, any loss of ellipticity in the classical Legendre--Hadamard sense; the latter remains governed exclusively by $\sigma_{\min}(\mathsf{D})\to 0$~\cite{RegenNicot2026foundation}.

In the 1-D chain the cycle space is trivial ($\beta_1=0$), so Schnakenberg cycle currents are mathematically excluded and the only irreversibility pathway is the cocycle-borne null-mode drift. The full argument is in Supplementary Material, Section~S4.

\section{The Spin Interpretation: \texorpdfstring{$\mathsf{L}$}{L} as a Generator of Rotations}
\label{sec:spin}

Section~\ref{sec:VMC_gateway} established the algebraic structure of the contact-scale gateway: a $3\times 3$ skew-symmetric reversible block $\mathsf{L}$ with non-zero entries $L_{VM}$ and $L_{MC}$, structural zero $L_{VC}=0$, and the null mode $\mathbf{v}_0 \propto (L_{MC},0,L_{VM})^T$ that defines the topological escape direction. This section provides a physically transparent reformulation by mapping the system to the foundational structures of Lie theory.

Specifically, the configurational channel $C$ is treated as a classical spin degree of freedom, and the coupling operator $\mathsf{L}$ is recognised as an element of the Lie algebra $\mathfrak{so}(3)$ defined by the three-dimensional vector space of real skew-symmetric matrices representing infinitesimal rotations. Because any element $A \in \mathfrak{so}(3)$ satisfies $A^T = -A$, its operation on a state vector is isomorphic to a standard vector cross product.

Consequently, $\mathsf{L}$ acts as a continuous rotation generator on the joint $(V,M,\mathrm{spin})$ state space. Under this geometric interpretation, the parity-mandated null vector $\mathbf{v}_0$ is revealed to be the literal rotation axis of the algebra. Because a rotation operator exerts zero torque and induces no displacement along its own axis of rotation, the restoring forces vanish identically along $\mathbf{v}_0$, providing the exact mathematical mechanism for the uninhibited, secular configurational drift.

\subsection{Translational and rotational degrees of freedom of a grain contact}

The three channels of the VMC triad are kinematically of two distinct characters. The volumetric channel $V$ describes a scalar overlap or displacement at the contact ($\delta_c$, a translational degree of freedom). The mechanical channel $M$ describes the conjugate force/momentum/torque variable that mediates inter-grain transfer (also translational in nature, with the dimensions of velocity or momentum). The configurational channel $C$, by contrast, describes the rotation of the contact normal $\mathbf{n}_c$ (a rotational degree of freedom). Its time derivative $J_C=\dot{\mathbf{n}}_{c'}$ is an angular velocity, and its conjugate force $X_C$ is a torque-like configurational moment. The channel $C$ is therefore a classical spin variable in the precise sense of granular Cosserat mechanics~\cite{Cosserat1909}: a microrotational degree of freedom attached to each contact, distinct from the translational degrees of freedom of the grains.

We will use $s\equiv C$ interchangeably in this section to emphasise the spin character of the channel. The state vector of the contact is then
\begin{equation}\label{eq:state_spin}
  \mathbf{q} \;=\; (V,\,M,\,s)^T \;\in\; \mathbb{R}^3,
\end{equation}
with $V$ and $M$ translational and $s$ rotational.

\subsection{\texorpdfstring{$\mathsf{L}$}{L} as an element of \texorpdfstring{$\mathfrak{so}(3)$}{so(3)}}

Any $3\times 3$ real skew-symmetric matrix is an element of the Lie algebra $\mathfrak{so}(3)$ of generators of rotations in $\mathbb{R}^3$. The VMC coupling block~Eq. \eqref{eq:L3contact} written in the spin notation $(V,M,s)$ is
\begin{equation}\label{eq:Lso3}
  \mathsf{L}
  \;=\;
  \begin{pmatrix}
    0      & L_{VM}  & 0      \\
    -L_{VM}& 0       & L_{Ms} \\
    0      & -L_{Ms} & 0
  \end{pmatrix}
  \;\in\;\mathfrak{so}(3),
\end{equation}
with $L_{Ms}\equiv L_{MC}$. The structural zero $L_{Vs}=0$ (no direct reversible coupling between volumetric translation and spin) is the graph-theoretic statement that the only path between $V$ and $s$ in the reversible network passes through the mechanical channel $M$. This is the kinematic-shielding content of the VMC chain re-expressed in spin language: spin and volumetric translation are not directly coupled; their interaction is mediated by the mechanical translational channel.

A general element of $\mathfrak{so}(3)$ admits the representation
\begin{equation}\label{eq:omega_so3}
  \mathsf{L}(\boldsymbol\omega)
  \;=\;
  \begin{pmatrix}
    0 & -\omega_3 & \omega_2 \\
    \omega_3 & 0 & -\omega_1 \\
    -\omega_2 & \omega_1 & 0
  \end{pmatrix},
\end{equation}
parametrised by an axial vector $\boldsymbol\omega\in\mathbb{R}^3$ that we will call the \emph{coupling vector}. Matching Eqs.~\eqref{eq:Lso3} and~Eq. \eqref{eq:omega_so3} and adopting the orientation convention aligned with the physically meaningful secular drift direction under positive axial forcing yields
\begin{equation}\label{eq:coupling_vector}
  \boldsymbol\omega
  \;=\;
  \bigl(L_{Ms},\,0,\,L_{VM}\bigr)^T.
\end{equation}
The coupling vector $\boldsymbol\omega$ has zero mechanical component because $L_{Vs}=0$: it lies entirely in the $V$--$s$ plane.

\subsection{The cross-product representation: \texorpdfstring{$\mathsf{L}\mathbf{q}=\boldsymbol\omega\times\mathbf{q}$}{L q = omega x q}}

The Lie-algebra identification of Eq.~\eqref{eq:omega_so3} provides the cross-product representation of the reversible dynamics:
\begin{equation}\label{eq:cross_product}
  \mathsf{L}\mathbf{q}
  \;=\;
  \boldsymbol\omega \times \mathbf{q}.
\end{equation}
The reversible part of the augmented Onsager dynamics is therefore a precession of the state vector $\mathbf{q}$ about the coupling axis $\boldsymbol\omega$, with angular speed $|\boldsymbol\omega|$. This is the same algebraic structure as the precession of a classical gyroscope or the Larmor precession of a magnetic moment in an external field: the skew-symmetric coupling block produces no work, only rotation.

The square magnitude of the coupling vector is
\begin{equation}\label{eq:omega_squared}
  |\boldsymbol\omega|^2
  \;=\;
  L_{VM}^2 + L_{Ms}^2
  \;=\;
  \tfrac{1}{2}\|\mathsf{L}\|_F^2,
\end{equation}
where $\|\cdot\|_F$ is the Frobenius norm. The precession frequency $\omega=|\boldsymbol\omega|$ is therefore the natural oscillation frequency of the reversible dynamics in the absence of forcing.

\subsection{The null vector is the rotation axis}

The kernel of any cross-product operator $\boldsymbol\omega\times(\cdot)$ is, by elementary vector calculus, the one-dimensional subspace spanned by $\boldsymbol\omega$ itself:
\begin{equation}\label{eq:ker_cross}
  \ker\bigl(\boldsymbol\omega\times(\cdot)\bigr)
  \;=\;
  \mathrm{span}\{\boldsymbol\omega\}.
\end{equation}
The null direction of $\mathsf{L}$ is therefore precisely the rotation axis $\boldsymbol\omega$:
\begin{equation}\label{eq:nullvec_as_axis}
  \mathbf{v}_0
  \;=\;
  \frac{\boldsymbol\omega}{|\boldsymbol\omega|}
  \;=\;
  \frac{(L_{Ms},\,0,\,L_{VM})^T}{\sqrt{L_{VM}^2+L_{Ms}^2}}.
\end{equation}
The parity-mandated null vector of Section~\ref{sec:VMC_minimal} and the rotation axis of the cross-product representation are the same object.

This identification is the central physical content of the spin reinterpretation. The Parity Theorem can now be restated in geometric terms: an odd-dimensional skew-symmetric operator is a rotation generator, and a rotation generator in odd dimensions has a preserved axis. The preserved axis is the topological gateway.

\subsection{Boundary forcing and secular drift along the axis}

The augmented Onsager dynamics with forcing, $\dot{\mathbf{q}}=\mathsf{L}\mathbf{q}+\mathbf{f}_0$, has the cross-product form
\begin{equation}\label{eq:forced_cross}
  \dot{\mathbf{q}} \;=\; \boldsymbol\omega\times\mathbf{q}\,+\,\mathbf{f}_0.
\end{equation}
Decomposing the forcing into components parallel and perpendicular to the axis,
\begin{equation*}
  \mathbf{f}_0 \;=\; f_0^\parallel\,\hat{\boldsymbol\omega}
                 + \mathbf{f}_0^\perp,
\end{equation*}
the parallel component cannot be rotated by $\boldsymbol\omega\times$ and accumulates as secular drift along the axis:
\begin{equation}\label{eq:secular_drift_axis}
  \mathbf{v}_0^T\mathbf{q}(t)
  \;=\;
  f_0^\parallel\,t.
\end{equation}
The perpendicular component undergoes precession about the axis at frequency $\omega$, producing bounded oscillation in the plane normal to $\boldsymbol\omega$. The decomposition is exact and analytically identical to the standard solution of the forced gyroscope: a uniform secular drift along the precession axis superposed on a circular orbit transverse to it.

The parity-mandated zero mode is thus recognised, in the spin language, as the universal fact that a precession cannot dissipate or redirect a force aligned with its own axis. The boundary loading of the granular column, projected onto the $\boldsymbol\omega$ axis, has no rotational pathway to relax and must instead accumulate as configurational drift.

\subsection{Energetic coupling of the microstructural null mode and the boundary work}
\label{sec:null_mode_coupling}

The previous subsection established that the parity-mandated null direction $\mathbf{v}_0$ of $\mathsf{L}$ coincides with the rotation axis of the contact-scale generator, and that any state-vector component parallel to $\boldsymbol\omega$ grows secularly under sustained boundary forcing. We now make the activation criterion concrete: under a macroscopically applied axial compression, the external boundary forcing forms a non-vanishing work-conjugate projection onto the microstructural null mode whenever the mechanical-spin coupling coefficient $L_{Ms}$ is non-zero. Consequently, the topological gateway is kinematically activated under any arbitrary boundary traction, rendering the volumetric dilatancy instability an inherent structural feature of the granular column.

Take the macroscopic boundary forcing in the contracted $(V,M,s)$ state space to be
\begin{equation}\label{eq:boundary_forcing}
  \mathbf{f}_0
  \;=\; \bigl(f_0,\; 0,\; 0\bigr)^T,
\end{equation}
i.e.\ a purely axial volumetric stress applied at the piston, with no independent external torque or spin moment acting directly on the internal micro-rotation channel. The dual pairing of the generalised boundary forcing $\mathbf{f}_0$ and the null-mode vector $\mathbf{v}_0$ quantifies the rate of external work injected into this unresisted kinematic channel:
\begin{equation}\label{eq:boundary_coupling_proof}
  \mathcal{W}_{\rm inject}
  \;\equiv\; \mathbf{v}_0^T\,\mathbf{f}_0
  \;=\;
  \frac{L_{Ms}\,f_0}{\sqrt{L_{VM}^2 + L_{Ms}^2}}.
\end{equation}

\begin{proposition}[Non-orthogonality of $\mathbf{f}_0$ and $\mathbf{v}_0$]
\label{prop:nonorth}
For the prescribed axial boundary traction of Eq.~\eqref{eq:boundary_forcing}, the inner product $\mathbf{v}_0^T\mathbf{f}_0$ is non-vanishing if and only if the force--rotation cross-coupling $L_{Ms}$ is non-zero, demonstrating that external boundary work natively injects power into the microstructural null mode.
\end{proposition}

The proof is immediate: the only non-zero component of $\mathbf{f}_0$ is its first ($V$) component, and the first component of $\mathbf{v}_0$ in Eq.~\eqref{eq:nullvec_as_axis} is proportional to $L_{Ms}$. Hence $\mathbf{v}_0^T\mathbf{f}_0 = L_{Ms} f_0/\sqrt{L_{VM}^2+L_{Ms}^2}$, which vanishes if and only if $L_{Ms}=0$. The underlying geometric mechanics are clear: the rotation axis of the operator $\mathsf{L}$ is confined to the Volumetric--Spin ($V$--$s$) plane due to the structural zero $L_{Vs} = 0$. Because the macroscopically applied axial boundary traction natively possesses a non-vanishing volumetric component, the force vector and the microstructural null-mode vector are inherently non-orthogonal whenever the rotation axis tilts away from the spin axis. This geometric alignment ensures a non-vanishing work-conjugate projection, enabling external boundary forcing to directly drive the internal null mode.

This non-orthogonality is the physical content of the activation criterion. Projecting the boundary-forced dynamics $\dot{\mathbf{q}}=\mathsf{L}\mathbf{q}+\mathbf{f}_0$ onto $\mathbf{v}_0$ and using $\mathbf{v}_0^T\mathsf{L}=\mathbf{0}^T$ gives
\begin{equation}\label{eq:secular_drift_along_axis}
  \frac{d}{dt}\bigl(\mathbf{v}_0^T\mathbf{q}(t)\bigr)
  \;=\; \mathcal{W}_{\rm inject},
\end{equation}
so the null-mode component of $\mathbf{q}(t)$ grows linearly in time at the rate $\mathcal{W}_{\rm inject}$ for as long as the boundary forcing is sustained. This is the geometric statement of the parity-mandated drift: a uniform axial load, projected onto the rotation axis, produces uniform translation along that axis with no oscillatory pathway available for the conservative block to absorb it.

\paragraph{Frictional damping and the three activation regimes}
The reversible block $\mathsf{L}$ is normal, but the augmented operator $\mathsf{D}+\mathsf{L}$ is generically non-normal because $[\mathsf{D},\mathsf{L}]\neq\mathbf{0}$. The non-normality is not the cause of the drift identified by Eq.~\eqref{eq:secular_drift_along_axis}, which follows directly from $\mathsf{L}\mathbf{v}_0=\mathbf{0}$ and Proposition~\ref{prop:nonorth} alone. The role of the Gateway number $\mathcal{G}$ is, however, regime-dependent, and it is worth stating the three regimes explicitly to avoid conflating the exact drift of Eq.~\eqref{eq:secular_drift_along_axis} with the damped transient of Part~1.

\begin{figure}[!htbp]
\centering
\includegraphics[width=\linewidth]{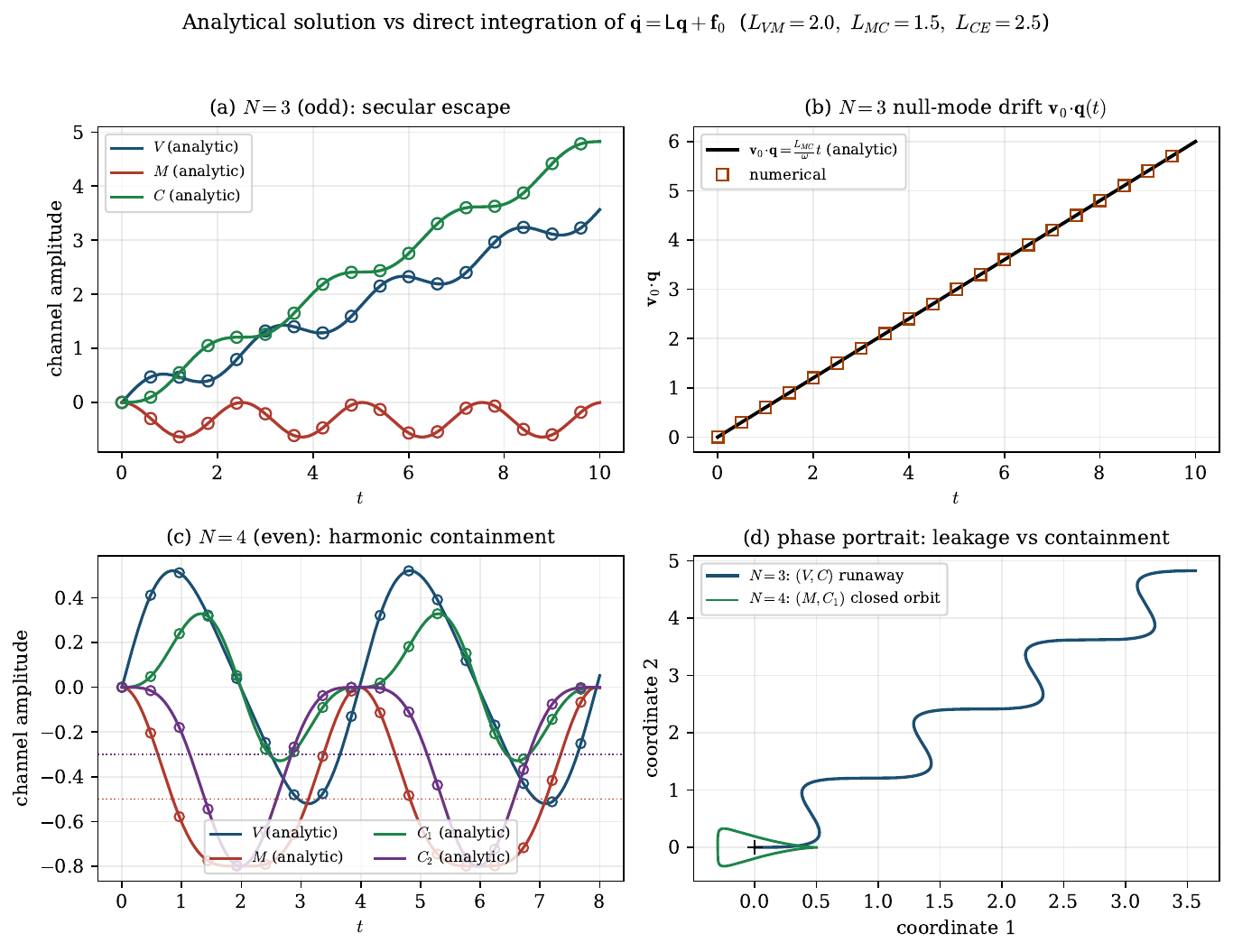}
\caption{Closed-form analytical solutions (lines) compared to direct numerical integration (markers) of $\dot{\mathbf{q}}=\mathsf{L}\mathbf{q}+\mathbf{f}_0$ in the frictionless limit, for the baseline couplings $L_{VM}=2.0$, $L_{MC}=1.5$, $L_{CE}=2.5$; the two agree to machine precision. \textbf{(a)}~Odd chain ($N=3$): the volumetric ($V$) and configurational ($C$) channels grow secularly while the mechanical channel ($M$) remains bounded. \textbf{(b)}~The null-mode projection $\mathbf{v}_0^{T}\mathbf{q}(t)=(L_{MC}/\omega)\,t$ grows strictly linearly, the direct signature of the open Gateway. \textbf{(c)}~Even chain ($N=4$): all channels remain bounded, oscillating about the shifted centre $\mathbf{q}_{\rm center}=-\mathsf{L}_4^{-1}\mathbf{f}_0$. \textbf{(d)}~Phase-space contrast: the odd chain drifts without bound along $(V,C)$, whereas the even chain is confined to a closed orbit. The underlying closed forms are derived in Supplementary Material, Section~S8.}
\label{fig:analytical_proof}
\end{figure}

\emph{(i) Frictionless limit ($\mathsf{D}=0$).} The drift $\mathbf{v}_0^T\mathbf{q}(t)=\mathcal{W}_{\rm inject}\,t$ is exact and unbounded, and is \emph{unconditional}: it does not require $\mathcal{G}\geq1$. Indeed $\mathcal{G}=\|\mathsf{D}^{-1/2}\mathsf{L}\mathsf{D}^{-1/2}\|_2^2$ diverges and ceases to discriminate as $\mathsf{D}\to0$. This is the frictionless configurational irreversibility that is the subject of the present paper, and the phase-space construction and closed-form solutions of Supplementary Material, Sections~S5--S8 are computed in exactly this limit and are validated against direct numerical integration in Fig.~\ref{fig:analytical_proof}.

\emph{(ii) Damped linear regime ($\mathsf{D}\succ0$).} Augmenting the reversible dynamics with dissipative relaxation, $\dot{\mathbf{q}}=(\mathsf{L}-\mathsf{D})\mathbf{q}+\mathbf{f}_0$ (the stabilising sign, consistent with the asymptotic stability of Part~1's amplification theorem), the system is asymptotically stable and relaxes to the \emph{finite} offset $\mathbf{q}^\ast=(\mathsf{D}-\mathsf{L})^{-1}\mathbf{f}_0$; there is no unbounded secular drift. What $\mathcal{G}\geq1$ controls here is not the existence of drift but the size of the transient \emph{overshoot} along $\mathbf{v}_0$ before that relaxation, quantified by the peak $M(\mathcal{G})$ of Ref.~\cite{RegenNicot2026foundation} (at $\mathcal{G}=1$, $M^\ast\approx0.32$, a soft crossover rather than a sharp instability).

\emph{(iii) Nonlinear regime.} Once the overshoot of regime (ii) carries the state out of the linear neighbourhood, the obstruction $\hat{\mathbf{v}}_0^T\mathsf{L}=\mathbf{0}$ no longer applies: nonlinear coupling sustains growth along $\mathbf{v}_0$ without the mediation of $\mathsf{D}$, so activation becomes enforced by the odd-$N$ null topology itself rather than gated by the linear threshold.

The physically observed precursor is the damped transient (ii) triggering the nonlinear localisation (iii); the exact unbounded drift (i) is its frictionless idealisation, in which the gating role of $\mathcal{G}$ is vacuous because no dissipation is present to compete with. The transient-amplification theorem of Ref.~\cite{RegenNicot2026foundation} quantifies the competition in regime (ii) exactly.

\paragraph{The physical origin of $\mathsf{D}$, and dissipation-free irreversibility}
It is worth pre-empting a natural objection: if the reversible block is work-free ($\mathbf{q}^T\mathsf{L}\mathbf{q}\equiv0$) and $\mathsf{D}$ is the only entropy-producing block, what constitutes $\mathsf{D}$ in a \emph{frictionless} chain? The answer clarifies the status of the three regimes above. In the strictly frictionless, elastic case there is \emph{no} dissipative block ($\mathsf{D}=0$), and the irreversibility is purely \emph{configurational}: it is the loading-path dependence of Nicot et al.~\cite{Nicot2024}, in which the reverse fabric-rotation increment fails to retrace the forward one ($d\mathbf{n}_c^-\neq d\mathbf{n}_c^+$), and it produces zero entropy since $\mathbf{q}^T\mathsf{L}\mathbf{q}\equiv0$. This anholonomy is a reorientation of existing contact normals and requires no gain, loss, or exchange of contacts, so it is fully compatible with the fixed neighbour topology of the one-dimensional chain. For odd $N$, therefore, geometric irreversibility and thermodynamic dissipation come apart: the arrow of time along $\mathbf{v}_0$ is carried by the non-retraceability of the \emph{forced} path, not by entropy production. A non-zero $\mathsf{D}$, and with it the Gateway number $\mathcal{G}$, enters only when a physical dissipation mechanism is present, such as rate-dependent contact damping or pore-fluid drag, and, once local quasi-staticity is relaxed, radiation of elastic energy away from the contact neighbourhood. Frictional sliding is one such mechanism but not the only one, and its removal does not restore reversibility, because the configurational drift of regime~(i) survives it. In this respect the mechanism is distinct from Rowe's stress--dilatancy relation, in which dilatancy is inseparable from frictional interlocking: here the configurational free-volume generation is topological and persists at zero friction.

The secular null-mode drift is the precursor signature; boundary conditions (stress- vs strain-control, mixed oedometer) modulate its observable form, and the $L_{MC}\to 0$ limit silences it without closing the Gateway. Quantitative robustness bounds and the full four-item analysis are in Supplementary Material, Section~S2.

\subsection{Extension to the spin chain (\texorpdfstring{$N>3$}{N>3})}

A chain of $N$ coupled channels generalises the $\mathfrak{so}(3)$ structure of the VMC triad to $\mathfrak{so}(N)$, the Lie algebra of skew-symmetric $N\times N$ matrices; here $N$ counts the thermodynamic channels of the coupling graph rather than grains, and in the spin picture of this section each channel carries a rotation generator.

For $N$ odd, the rank of $\mathsf{L}\in\mathfrak{so}(N)$ is at most $N-1$ (even), so the dimension of $\ker\mathsf{L}$ is at least $1$: a topological rotation axis persists. The state vector $\mathbf{q}$ precesses in the $(N-1)$-dimensional orthogonal complement of the null direction, while a single forced axis accumulates secular drift. This is the spin-chain version of the Parity Theorem.

For $N$ even, $\mathsf{L}$ has full rank generically; the spectrum consists of $N/2$ pairs $\pm i\omega_k$, and the dynamics is a $(N/2)$-frequency torus in $\mathbb{R}^N$. There is no preserved axis.

The $N$-channel spin chain therefore exhibits a parity contrast that is geometrically transparent in the rotation-generator language: odd chains have an unpaired axis, even chains are fully paired. The quad-precision numerical verification of this large-$N$ contrast, and the asymptotic spectral scaling $|\lambda_{\min}^{\rm even}|\sim\gamma\pi/N$ for the Toeplitz coupling chain, is presented in Part~3~\cite{RegenNicot2026numerical}.

\subsection{Physical implications of the spin reformulation}

The spin reinterpretation has three concrete physical implications.

First, the configurational energy increment $dE^{\rm conf}$ of Eq.~\eqref{eq:dEconf} acquires the geometric meaning of work done \emph{along the rotation axis} of the contact-scale generator. It is the component of the boundary work that the precession cannot absorb.

Second, the Mechanical Bridge structure of the VMC chain ($M$ as the cut vertex of the reversible graph) becomes physically intuitive: the rotation axis lies in the $V$--$s$ plane, with zero $M$ component, so energy flowing along the axis transfers directly between volumetric translation and spin rotation, bypassing the mechanical channel entirely. The mechanical channel mediates the precession but does not store the secular drift.

Third, the spin reinterpretation provides a direct bridge to the established physics of classical and quantum spin chains~\cite{Heisenberg1928}, to the Cosserat continuum of generalised micropolar elasticity~\cite{Cosserat1909,deBorst2022}, and to gyroscopic-stability arguments in classical mechanics~\cite{Marsden1999}. The Parity Theorem in spin form is the granular counterpart of the standard observation that any classical gyroscope has a preserved spin axis, with the topological gateway identified as the granular realisation of that axis.

\begin{remark}[Scale selection of the channel set]
The channel count $N$ is a property of the scale at which the coupled system is described, not of the material. As the microscopic channel count grows without bound, scale separation collapses the active dynamics onto a small effective subset of the thermal-hydraulic-mechanical-chemical-electric (THMCE) pentade, and the parity of that subset is what the Parity Theorem classifies: an even reduction is a Stable Layer with confined orbits, an odd reduction a Gateway with a parity-mandated null mode. The statement that makes the three-channel VMC contact a Gateway therefore extends without modification across scales.
\end{remark}

At planetary scale, Earth reduces to a thermal-mechanical coupling (TM, $N=2$, a stable convective TM heat engine \cite{Regen2010}) that is even and confined. In contrast, Venus reduces to a thermal-mechanical-electric coupling (TME, $N=3$, the lack of water in Venus' mantle keeps the electronic binding forces of the elastic core of the Venus surface intact and prevents recycling \cite{Regenauer2001}) that is odd and carries the null-mode drift, a contrast developed for global convection in Parts~1 and~3~\cite{RegenNicot2026foundation,RegenNicot2026numerical}. Adding one channel flips the parity of the reduced operator, precisely as extending the granular triad by a single channel does; whether the flip lands on a Stable Layer or a Gateway depends only on the resulting count. Earth's even TM reduction and Venus's odd TME reduction, like the granular triad and its even extension, are therefore instances of a single classification, differing only in which channels scale separation retains.

\section{Frictionless Irreversibility and Cross-Scale Correspondence}
\label{sec:frictionless}

\subsection{The contact-scale and continuum-scale matrices are identical}

The formal correspondence between the contact-scale Gateway matrix~Eq. \eqref{eq:L3contact} and the Level-3 VMC Gateway derived from the augmented Onsager framework~\cite{RegenNicot2026foundation} is exact rather than merely analogical: both operate in the same three-channel space $(V,M,C)$ because Ref.~\cite{Nicot2024} identifies precisely these channels at both the contact and continuum scales. Entry by entry, the two matrices are identical:
\begin{equation}
  \mathsf{L}^{\rm contact}_{3\times3}
  \;\cong\;
  \mathsf{L}^{\rm gateway}_{\rm VMC},
  \label{eq:contact_gateway_iso}
\end{equation}
with $L_{VM}$ the overlap--force coupling and $L_{MC}$ the force--fabric-rotation coupling at both scales. The contact-scale matrix derives its entries bottom-up from contact-scale physics; the continuum-scale matrix derives them top-down from the augmented Onsager framework; Eq.~\eqref{eq:contact_gateway_iso} closes the loop.

The two terms of Eq.~\eqref{eq:dEconf} map directly onto the two cascade directions: the geometric-constraint term $F_n^c(\partial\delta_c/\partial \mathbf{n}_{c'})\,d\mathbf{n}_{c'}^+$ corresponds to conservative upward routing through the null direction driven by $\mathsf{L}$ without local entropy production, while the frictional-dissipation term $\mathbf{F}_t^c l_c\,d\mathbf{n}_c^+$ corresponds to downward dissipative transfer driven by $\mathsf{D}$ with positive entropy production.

\subsection{Topological origin of loading-path irreversibility}

Ref.~\cite{Nicot2024} establishes that reversibility in granular media is a function of the loading path rather than a simple temporal scaling. The parity framework provides the formal thermodynamic basis: a loading path confined to even-parity Stable Layers ($N=2,4$) remains theoretically reversible as $\mathsf{D}\to0$. Any path traversing an odd-parity Gateway Layer ($N=3,5$) activates a topological zero mode, rendering the process irreversible regardless of loading rate.

The parity argument applies specifically to \emph{configurational irreversibility} that persists even when frictional tangential forces vanish, that is, in the case of frictionless elastic grains. For frictionless elastic grains, the fabric-rotation constraint $d\mathbf{n}_c^-=\mathbf{0}$ of Ref.~\cite{Nicot2024} is the sole source of irreversibility, and the Parity Theorem identifies it as topological. Note that granular media with inelastic collisions are not time-reversible at the microscopic level; the parity argument provides a separate and weaker claim that covers the frictionless case.

That irreversibility can be geometric rather than frictional is not new. It has a direct precedent in crystal plasticity, where the irreversibility of ideal, rate-independent plastic flow originates in the configurational forces on defects~\cite{Eshelby1975} and, more fundamentally, in the geometry of the dislocation-density field~\cite{Kroner1958}, independently of any interfacial friction; a modern field-theoretic rendition is given by Acharya~\cite{Acharya2001}. The present construction is the granular counterpart, with the role played there by dislocation geometry played here by the Satake contact-network topology~\cite{Satake1987}. Within granular physics the same topological viewpoint underlies the force-space description of rigidity, in which the force network is represented through height-field gauge potentials and force tiles~\cite{Henkes2009,Sarkar2013}; the reversible-coupling decomposition developed here is the conservative-block complement to that force-balance picture.

When a trajectory evolves along $\mathbf{v}_0$, energy is routed from the $V$ and $M$ basis into the $C$ channel. Because $L_{VC}=0$, no conservative mechanism pumps configurational energy back. The trajectory carries no restoring tendency; the inverse path is topologically blocked. The configurational energy $dE^{\rm conf}$ of Eq.~\eqref{eq:dEconf} is therefore path-dependent rather than state-dependent and defines the precise thermodynamic signature of irreversibility encoded in the coupling topology.

\subsection{Stable Layers as thermodynamic substrates}

In an even-dimensional subsystem, $\det(\mathsf{L})\neq0$ and the reversible coupling operator is invertible, guaranteeing closed orbits in force-flux space. This is the algebraic prerequisite for a unique thermodynamic potential. In the Edwards statistical mechanics of granular media~\cite{Edwards1989}, the intensive variables compactivity $\chi$ and angoricity $\boldsymbol{\alpha}$ are well-defined precisely because the underlying $N=2$ Stable Layers support a single-valued energy landscape. Stable Layers are therefore the thermodynamic substrate on which configurational energy can be evaluated unambiguously.

\section{Discussion}
\label{sec:discussion}

\subsection{The VMC triad as a complete minimal model in one dimension}

The central result of this paper can be stated in one sentence: the three channels $V$, $M$, $C$ of a granular contact, coupled through the skew-symmetric Onsager operator and embedded in a one-dimensional chain under sustained boundary forcing, contain everything necessary to predict the onset of dilatancy-driven matrix instability, with no free parameters and no dependence on the friction coefficient. The spin reinterpretation of Section~\ref{sec:spin} sharpens this statement geometrically: the contact-scale reversible coupling is an element of $\mathfrak{so}(3)$, the rotation-axis coupling vector $\boldsymbol\omega=(L_{MC},0,L_{VM})^T$ lies in the $V$--spin plane, and the secular drift of the boundary-forced dynamics is the work done along this preserved rotation axis.

This completeness is not an accident of the particular granular system. It follows from the algebraic structure of the coupling topology alone: three channels, chain connectivity ($L_{VC}=0$), odd dimension. The same structure would produce the same Gateway irreversibility in any physical system with three mutually coupled channels satisfying these constraints: mechanical, chemical, optical, or biological. In the spin-chain language, the granular contact is the simplest non-trivial classical spin chain with two translational and one rotational degree of freedom, and the Parity Theorem is the granular analogue of the standard observation that a precessing gyroscope has a preserved spin axis.

A complementary dynamical perspective is provided by Ref.~\cite{Chattoraj2019}, which shows that the stability operator of frictional granular contacts acquires eigenvalue pairs $\pm i\omega$ at a discrete critical strain, triggering oscillatory growth that culminates in sudden failure. The present topological Gateway is the frictionless analogue: it identifies the same $\pm i\omega$ oscillatory modes as the reversible $\mathsf{L}$ block, and the zero mode as the topological route for irreversible escape, without requiring frictional tangential forces.

\subsection{The cocycle filter and the algebraic isolation of dilatancy}

\paragraph{Terminological note on "1-D dilatancy"}
Before proceeding, we clarify the sense in which the word "dilatancy" is used here, which is distinct from the classical Reynolds definition. Reynolds dilatancy refers to transverse volume expansion of a dense granular medium in response to applied shear, and is by construction a multi-dimensional kinematic phenomenon requiring transverse degrees of freedom. The 1-D mechanism analysed in the present paper refers instead to \emph{internally generated configurational free volume}: the matrix expansion of the chain required to accommodate the unresisted secular drift along the parity-mandated null axis. The two share a common physical origin in the configurational reorganisation of granular fabric, but they differ in geometry: Reynolds dilatancy is shear-induced transverse expansion, whereas the 1-D instability is axial compaction-driven internal expansion of the matrix into the configurational free-volume channel. Throughout the present paper, "dilatancy" refers exclusively to the 1-D internal-expansion mechanism unless otherwise noted, and is more precisely characterised as a configurational matrix instability whose macroscopic signature in a one-dimensional column is the secular axial drift documented in Sec.~\ref{sec:spin} and Part~3~\cite{RegenNicot2026numerical}. This distinction removes an apparent tension with the numerical finding that frictionless granular assemblies support deviatoric stress yet exhibit no macroscopic dilatancy~\cite{Peyneau2008,Azema2015}. That finding concerns Reynolds dilatancy proper: a shear-driven response of the full deviatoric fabric tensor, defined only in two or more dimensions, recording the absence of any net volumetric bias when a frictionless assembly is sheared. The one-dimensional instability analysed here is not Reynolds dilatancy but a single scalar constituent of it, namely the volumetric--configurational mode that survives once the deviatoric fabric tensor is projected out by the acyclic ($\beta_1=0$) chain topology. It is a directed drift along one forced axis rather than a net fabric-averaged volume change, and it is mandated by the parity of $N=3$ rather than by any frictional bias. There is therefore no conflict: the frictionless shear studies measure the fabric-averaged Reynolds response, whereas the present mechanism isolates its axial constituent; the full deviatoric response, and with it any direct comparison to the frictionless-shear result, re-enters only in the two-dimensional extension where the cycle-space shear-localisation modes are restored~\cite{RegenNicot2026foundation}.

\paragraph{Two algebraically independent instabilities}
The mechanism analysed here is a one-dimensional matrix instability of dilatancy type. The acyclic topology of the chain ($\beta_1=0$, the cocycle filter excludes Schnakenberg cycle currents and forbids any curl-driven failure mode: in a purely axial column under sustained boundary load, the only available irreversible pathway is the cocycle-borne null-mode drift. Macroscopic shear localisation, by contrast, resides within the cycle space of higher-genus Hasse-lattice subgraphs that appear only in dimension $\geq 2$ when the cocycle filter is lifted. The dilatancy instability and the shear-localisation instability are therefore distinct algebraic objects: the first is the cocycle (gradient) projection of the boundary forcing onto the parity-mandated null mode of a chain; the second is the cycle (curl) current driven by non-zero Schnakenberg cycle affinities in a graph with $\beta_1\geq 1$. They can and do superpose at finite dimension but they are independent in the parity classification, and they require separate validation pathways. The present paper establishes the one-dimensional dilatancy mechanism in full; the curl-driven shear-localisation mechanism, the associated non-associated flow law $\tan(\phi-\psi)\sim\sqrt{\mathcal{G}}$, and triaxial-DEM tests are subject of ongoing work.

\subsection{Scope and limitations}

The framework assumes that coupled subsystems can be parametrised by a finite number of well-separated thermodynamic channels and that local thermodynamic equilibrium holds within each channel. The linear activation criterion $\mathcal{G}\geq 1$ identifies the onset of failure; the full nonlinear post-failure trajectory requires a separate analysis. The criterion is a sufficient condition for instability; its necessity in specific granular systems remains an open question for one-dimensional DEM simulations to resolve. The 1-D Cartesian restriction is, as developed throughout this paper, an exact algebraic filter rather than a modelling simplification; the higher-dimensional extension is subject of ongoing work.

\subsection{Measurables and one-dimensional DEM implementation}

The framework translates directly into measurable quantities accessible in one-dimensional discrete-element simulations of a granular column under sustained axial load. The ratio $|\mathcal{C}^{\rm conf}|/|\mathcal{C}|$ is a dimensionless scalar proximity to the contact-scale Gateway: it measures what fraction of contacts are configurational and thus contributing to the null-mode activation. The coupling coefficients $L_{VM}$ and $L_{MC}$ are extractable from the force-moment statistics of jammed columns as the off-diagonal entries of the augmented Onsager matrix fitted to the column's fluctuation spectrum. The Gateway number $\mathcal{G}=\|\mathsf{L}\|_2^2/(\sigma_{\min}(\mathsf{D})\, \sigma_{\max}(\mathsf{D}))$ is then computable without free parameters.

The falsifiable one-dimensional predictions are developed in Part~3~\cite{RegenNicot2026numerical} in terms of these DEM-accessible quantities. A single well-instrumented one-dimensional DEM oedometer simulation (varying coordination number and grain angularity along the column to change $|\mathcal{C}^{\rm conf}|/|\mathcal{C}|$, then sustaining axial $\sigma_n$ past primary compaction) can test all four simultaneously, providing a comprehensive validation of the one-dimensional VMC triad model and isolating the dilatancy mechanism from any superposed shear contribution.

\begin{remark}
The analytical formulation presented here is restricted to the linear regime of the augmented Onsager operator $(\mathsf{D}+\mathsf{L})$. While the Gateway activation ($\mathcal{G} \geq 1$) represents a threshold for instability, the evolution is treated as a linear operator problem. This restriction ensures mathematical clarity and rigour for the fundamental coupling topology. Extensions into the nonlinear realm, specifically the saturation of the null-mode growth and potential mappings to nonlinear wave equations, are naturally suggested by the Lie-algebraic structure, but are reserved for a subsequent contribution to ensure the validity of the linear foundation is established first.   
\end{remark}
\section{Conclusions}
\label{sec:conclusions}

We have established the V--M--C triad as the minimal complete model of one-dimensional granular irreversibility: three channels, one skew-symmetric $3\times3$ coupling operator, one algebraically guaranteed zero eigenvalue. Nothing more is needed to describe configurational irreversibility and the onset of dilatancy-driven matrix instability in a granular chain under sustained boundary forcing. The cocycle-filter interpretation of the 1-D Cartesian restriction shows that this isolation is not a modelling approximation but an exact algebraic consequence of the chain topology ($\beta_1=0$). The foundational paper \cite{RegenNicot2026foundation} discusses the construction to dimension $\geq 2$, where the cocycle filter is lifted and Schnakenberg cycle currents in the Hasse-lattice subgraphs generate a distinct, algebraically independent shear-localisation instability with the dilatancy-deficit scaling $\tan(\phi-\psi)\sim\sqrt{\mathcal{G}}$.

The theory establishes seven results:
\begin{enumerate}
  \item \textbf{Minimality.} The VMC triad is the smallest granular
    model that contains a topological irreversibility: $N=2$ pairwise
    contacts are reversible; $N=3$ VMC contacts are not. This boundary
    is exact, algebraic, and parameter-free.
  \item \textbf{Zero mode as Gateway.} The null eigenvector
    $\mathbf{v}_0=(L_{MC},0,L_{VM})^T /\\ \sqrt{L_{VM}^2+L_{MC}^2}$ points in the $V$--$C$ plane,
    which is the precise direction of configurational dissipation $dE^{\rm conf}$.
    This is not a modelling choice; it is determined by the $L_{VC}=0$
    constraint and the Parity Theorem together.
  \item \textbf{Activation criterion.} The Gateway number
    $\mathcal{G}=\|\mathsf{L}\|_2^2/(\sigma_{\min}(\mathsf{D})\,
    \sigma_{\max}(\mathsf{D}))$ determines whether the zero mode is
    activated. It is computable from DEM simulations without free
    parameters.
  \item \textbf{Spin-chain reformulation.} The configurational channel
    $C$ is a classical spin degree of freedom, the skew-symmetric
    block $\mathsf{L}$ is an element of $\mathfrak{so}(3)$ acting as
    a rotation generator with the cross-product representation
    $\mathsf{L}\mathbf{q}=\boldsymbol\omega\times\mathbf{q}$, and the
    parity-mandated null mode is precisely the rotation axis
    $\boldsymbol\omega/|\boldsymbol\omega|$. A chain of $N$ coupled channels
    extends the triad to a classical spin chain with $\mathfrak{so}(N)$
    coupling, in which the parity contrast is the universal observation
    that odd-$N$ rotation generators preserve an axis.
  \item \textbf{Closed-form analytical solutions at $N=3$ and $N=4$.}
    For the two smallest cases the dynamics admit explicit
    closed-form expressions (Supplementary Material, Section~S8) that exhibit the
    parity contrast in algebraic form: the $N=3$ orbit drifts
    linearly along $\mathbf{v}_0$ under axial forcing while the
    $N=4$ orbit lies on a compact invariant torus. The contrast is
    visible in the analytical solution itself, without any
    numerical step. Numerical verification at higher $N$ is reported in
    Part~3~\cite{RegenNicot2026numerical}.
  \item \textbf{Cocycle filter.} The 1-D Cartesian restriction
    is an exact topological filter that forces $\beta_1=0$, excludes
    curl-driven shear modes by mathematical necessity, and isolates
    the cocycle-borne dilatancy mechanism. The lifting of this
    filter is the subject of ongoing work.
  \item \textbf{Frictionless irreversibility.} Neither the parity
    constraint nor the activation criterion contains the friction
    coefficient. One-dimensional dilatancy instability is a consequence
    of contact topology, not contact friction.
\end{enumerate}

The VMC triad has all the hallmarks of more complex multi-physics systems: a Hasse diagram, a parity-controlled Gateway, a single basis-invariant activation criterion. It serves as the stepping-stone for both the higher-dimensional extension to curl-driven shear localisation and the linear superposition synthesis with fabric-anisotropy corrections, subjects of future work.

\section*{Acknowledgements}

K.R.L.\ acknowledges research support from ARC grants CE230100032, DP200102517, and LP170100233, and discussions with Victor Calo, Manman Hu, and his PhD students on the augmented Onsager D+L formalism. F.N.\ acknowledges support from the GeoMech group. Both authors contributed equally to the synthesis presented here.


\bibliographystyle{elsarticle-num}
\bibliography{bib_merged}


\begin{frontmatter}

\title{Topological Foundations of Multi-Field Instabilities in Continua: Part 3 — Numerical Upscaling}

\author[inrae]{F. Nicot \corref{cor1}}
\ead{francois.nicot@univ-smb.fr}
\author[inrae]{Amir Saker}
\author[curtin]{K. Regenauer-Lieb}

\cortext[cor1]{Corresponding author.}

\address[inrae]{Universit\'e Savoie Mont Blanc, INRAE, ETNA, France.}
\address[curtin]{WA School of Mines: Minerals, Energy and Chemical Engineering,
  Curtin University, Perth WA 6845, Australia, and Australian Research Council
  Centre of Excellence for Carbon Science and Innovation.}

\begin{quote}\small
\textit{This is the final paper of a three-part JMPS manuscript series that develops a parameter-free, topological classification of multi-field instability and its realisation in granular matter. This manuscript provides the numerical application of the analytical formulation derived in Part 2, demonstrating the computational implementation, benchmarking, and falsification of the theoretical framework established in Part 1. By comparing numerical simulations against our analytical derivations, we quantify the dynamic response of the system and validate the predictive capacity of the topological approach. The Editor is respectfully asked to arrange coordinated review of Papers 1 to 3.}
\end{quote}

\begin{abstract}
This paper provides the numerical upscaling and discrete-element method (DEM) correspondence for the parity-mandated 1-D configurational matrix-dilatancy mechanism derived in Part 2, establishing its physical application to granular dilatancy and compaction bands. Using quad-precision integration of tridiagonal skew-symmetric Onsager chains ($N=3$ to $50$), we confirm an absolute topological contrast: odd-$N$ chains exhibit permanent secular drift along a structural null-mode axis, driving localised matrix expansion under rigid boundaries, whereas even-$N$ chains maintain bounded harmonic confinement on compact invariant tori. In the continuum limit ($N\to\infty$), even-chain frequencies scale asymptotically as $|\lambda_{\min}^{\rm even}|\sim\gamma\pi/N$, demonstrating scale-dependent critical slowing down. We map these Volumetric–Mechanical–Configurational (VMC) channels directly to measurable DEM observables, enabling the parameter-free evaluation of the Gateway activation criterion $\mathcal{G}$ from contact statistics. Finally, four falsifiable protocols, including the spontaneous arrest of dilatancy drift at $L_{MC}\to 0$, are presented for oedometer DEM validation.
\end{abstract}

\begin{keyword}
Granular mechanics \sep One-dimensional chains \sep
Parity theorem \sep Quad-precision integration \sep Toeplitz spectrum \sep
Classical discrete-element method \sep Oedometer validation \sep
Thermodynamic bottleneck
\end{keyword}

\end{frontmatter}

\section{Introduction}
\label{sec:intro}

Part~2 ~\cite{RegenNicot2026dilatancy} of this paper series establishes the upscaling framework for parity-mandated configurational irreversibility of one-dimensional granular chains. Its central result is that the top-down augmented Onsager $\mathsf{D}+\mathsf{L}$ framework possesses a Gateway operator $\mathsf{L^{gateway}}$ that is entry by entry identical to the bottom up contact matrix $\mathsf{L^{contact}}$. This operator necessarily possesses a zero eigenvalue whenever the channel count is odd, which provides a topological Gateway for the configurational energy increment $dE^{\rm conf}$ to accumulate as secular drift under sustained boundary forcing. The mechanism is purely topological with no friction content, and the smallest non-trivial realisation is the three-channel Volumetric--Mechanical--Configurational (VMC) triad with the spin-chain reinterpretation as a rotation generator in $\mathfrak{so}(3)$. The general augmented-Onsager framework, including the proof of the Parity Theorem, the derivation of the Gateway number, the transient amplification theorem, and the Hasse-lattice classification of channel subsets, is developed in~\cite{RegenNicot2026foundation}. We emphasise at the outset that the channel-space Gateway is a \emph{precursor} in the sense of preceding, and being algebraically distinct from, the loss of strong ellipticity of the spatial acoustic tensor ($\det\mathbf{A}=0$) that defines classical localisation~\cite{Rudnicki1975}. The present series does not claim that $\mathcal{G}\geq 1$ implies $\det\mathbf{A}=0$; the quantitative projection from channel-space transient amplification onto the spatial acoustic tensor in a multi-dimensional boundary-value problem is deferred to the two-dimensional extension, where the cocycle filter is lifted ($\beta_1\geq 1$) and shear-band modes appear.

The present paper carries this framework forward through three complementary lines of numerical and structural evidence.

\paragraph{Quad-precision numerical upscaling}
The Parity Theorem is an algebraic statement about real skew-symmetric matrices of arbitrary odd dimension. Section~\ref{sec:numerical} examines its dynamical consequences numerically by direct integration of $\dot{\mathbf{q}}=\mathsf{L}\mathbf{q}+\mathbf{f}_0$ on tridiagonal skew-symmetric chains, from $N=3$ to $N=50$, using quad-precision arithmetic and an implicit Radau~IIA integrator. The simulations confirm exactly the parity contrast: odd $N$ produces unbounded null-mode drift, even $N$ produces bounded harmonic containment, and the even-parity minimum eigenvalue obeys the asymptotic Toeplitz scaling $|\lambda_{\min}^{\rm even}|\sim\gamma\pi/N$. Since the vanishing determinant for odd $N$ is an exact algebraic identity, the value of these integrations is not the confirmation of that identity but the demonstration of its \emph{dynamical} consequence under sustained forcing, the bounded-torus versus secular-drift dichotomy, together with the spectral scaling, which is not evident from the Parity Theorem alone.

\paragraph{Thermodynamic-bottleneck interpretation of the closed boundary}
A clear distinction must be made between the unconstrained algebraic possibilities of the Hasse tensor hierarchy of~\cite{RegenNicot2026foundation} and the explicit numerical simulations presented herein. While the Hasse lattice demonstrates that an active Gateway layer natively ejects energy upscale into higher-order topological spaces, the boundary conditions chosen for this 1-D tridiagonal study intentionally prohibit this upscale migration. Operating under this strict kinematic and thermodynamic confinement, the system possesses no open multiscale safety valves. The energy driven through the Gateway cannot escape upscale, forcing its continuous, destructive accumulation within the lowest-level null vector direction. This boundary-induced bottleneck explains the unbounded null-mode drift observed in our quad-precision integrations, and it provides an ideal baseline validation: it shows precisely how a material is forced into a localised macroscopic breakdown when its natural, multiscale energy-routing channels are kinematically choked. The lifting of this bottleneck is achieved (i) by extending the network to dimension $\geq 2$ so that cycle-space currents become available, or (ii) by embedding the chain within the unconstrained multi-field continuum hierarchy of Ref.~\cite{RegenNicot2026foundation}.

\paragraph{Classical one-dimensional DEM correspondence}
The theoretical framework predicts that the topological dilatancy instability is observable in standard one-dimensional discrete-element simulations of a granular column under sustained axial load. Section~\ref{sec:DEM_classical} develops this correspondence: each VMC channel is mapped onto a measurable DEM observable, the Gateway activation number $\mathcal{G}$ is shown to be computable from the contact statistics of a jammed column without free parameters, and four explicit falsification protocols are formulated.

\paragraph{Organisation}
Section~\ref{sec:recap} provides a brief recap of the upscaling framework of Part~ 2 and Paper~1, with no proofs (they are in the cited papers). Section~\ref{sec:numerical} reports the numerical upscaling of the parity contrast through three simulation studies. Section~\ref{sec:spectral_scaling} derives and numerically confirms the asymptotic spectral scaling for $N\to\infty$. Section~\ref{sec:DEM_classical} develops the classical one-dimensional DEM correspondence. Section~\ref{sec:predictions} states the four falsifiable predictions accessible to DEM simulation. Section~\ref{sec:future} sketches the extensions to higher dimension. Section~\ref{sec:conclusions} concludes. Supplementary Material, Section~S2 provides the quad-precision verification protocol and benchmark results at $N=5$ and $N=6$.

\section{Brief Recap}
\label{sec:recap}

For self-containedness we summarise here the algebraic objects on which the numerical and DEM work below depends. All derivations and proofs are in~\cite{RegenNicot2026foundation,RegenNicot2026dilatancy}.

\paragraph{The augmented Onsager operator}
A granular contact is described by the matrix equation
\begin{equation}\label{eq:DL}
  \mathbf{J} \;=\; \bigl[\mathsf{D}+\mathsf{L}\bigr]\,\mathbf{X},
\end{equation}
with $\mathsf{D}=\mathsf{D}^T\succeq 0$ symmetric dissipative and $\mathsf{L}=-\mathsf{L}^T$ skew-symmetric reversible. The conjugate flux--force pairs at a configurational contact are: volumetric ($V$) overlap rate vs.\ normal force; mechanical ($M$) tangential sliding rate vs.\ tangential force; configurational ($C$) fabric rotation rate vs.\ configurational moment. The off-diagonal entries $L_{VM}$ and $L_{MC}$ are non-zero in general; $L_{VC}=0$ by the kinematic-shielding argument detailed in Ref.~\cite{RegenNicot2026dilatancy}.

\paragraph{The Parity Theorem}
For any real skew-symmetric $\mathsf{L}\in\mathbb{R}^{N\times N}$ with $N$ odd, $\det\mathsf{L}=0$. The corresponding null vector $\mathbf{v}_0\in\ker(\mathsf{L})$ is the parity-mandated Gateway null mode, and for the VMC triad it is
\begin{equation}\label{eq:null_vector_recap}
  \mathbf{v}_0
  \;=\;
  \frac{1}{\sqrt{L_{VM}^2+L_{MC}^2}}\,\bigl(L_{MC},\,0,\,L_{VM}\bigr)^T,
\end{equation}
aligned in the $V$--$C$ plane. Boundary forcing $\mathbf{f}_0$ with $\mathbf{v}_0\cdot\mathbf{f}_0\neq 0$ produces secular drift $\mathbf{v}_0^T\mathbf{q}(t)\propto t$ along this axis. Even-$N$ chains have generically full-rank $\mathsf{L}$ and no Gateway null mode; for the tridiagonal chain this is exact, with $\det\mathsf{L}=\bigl(\prod_k\gamma_{2k-1}\bigr)^2$ (Part~1~\cite{RegenNicot2026foundation}), so full rank holds precisely when every odd-indexed coupling is non-zero.

\paragraph{The Gateway activation criterion}
Because the symmetric dissipative operator $\mathsf{D}$ and the skew-symmetric reversible operator $\mathsf{L}$ do not commute ($\mathsf{L}\mathsf{D} \neq \mathsf{D}\mathsf{L}$), the complete augmented Onsager operator $\mathsf{D}+\mathsf{L}$ is inherently non-normal. This structural non-normality has significant dynamical consequences for mechanics: it decouples the short-term energetic growth of the system from its long-term eigenvalue spectrum, permitting significant non-modal transient amplification even in nominally stable configurations. To govern the competition between these non-orthogonal modes, the reversible-to-dissipative spectral ratio

\begin{equation}\label{eq:gateway_number}
  \mathcal{G}
  \;=\;
  \frac{\|\mathsf{L}\|_2^2}{\sigma_{\min}(\mathsf{D})\,\sigma_{\max}(\mathsf{D})}
\end{equation}

distinguishes stable ($\mathcal{G}<1$) from active ($\mathcal{G}\geq 1$) Gateway dynamics. The criterion is parameter-free and computable directly from the contact statistics of a jammed granular column without phenomenological fitting. Here \emph{parameter-free} refers to the topological classification: the channel count, and hence the parity-mandated existence of the Gateway, follows from graph connectivity alone and is invariant to the grain-scale contact parameters ($k_n$, $k_s$, $\mu$, and damping), which set only the numerical values of the couplings entering $\mathcal{G}$, that is the quantitative activation threshold, and not the qualitative parity classification. The full proof, including the transient amplification theorem that leverages this operator non-normality to yield the explicit closed-form expression for the null-mode peak amplitude in the 2-D reduced system, is detailed in~\cite{RegenNicot2026foundation}.

\paragraph{The spin reformulation}
The skew-symmetric block $\mathsf{L}\in\mathfrak{so}(3)$ is a generator of rotations on the $(V,M,C)$ state space, with cross-product representation $\mathsf{L}\mathbf{q}=\boldsymbol\omega\times\mathbf{q}$ and coupling vector $\boldsymbol\omega=(L_{MC},0,L_{VM})^T$. The null vector of $\mathsf{L}$ coincides with the rotation axis $\boldsymbol\omega/|\boldsymbol\omega|$. A chain of $N$ coupled channels extends the triad to a classical spin chain with $\mathfrak{so}(N)$ coupling, where $N$ is the channel count; odd-$N$ rotation generators preserve an axis, even-$N$ rotation generators are fully paired.

\paragraph{Cocycle filter and bottleneck framing}
The 1-D Cartesian restriction acts as a topological cocycle filter~\cite{RegenNicot2026dilatancy}: the path-graph topology $P_N$ has first Betti number $\beta_1=0$, so the cycle space is trivial and Schnakenberg cycle currents are mathematically excluded. The parity-mandated null mode therefore resides in the cocycle (gradient) subspace $\mathcal{B}^*$. The mechanistic contrast between Satake's static spatial loop graph~\cite{Satake1993} and the dynamic Hasse hierarchy used in the present formulation, including the explicit $p$--$c$--$v$ to $V$--$M$--$C$ identification under Satake's discrete Poincar\'e lemma $L_{vc}D_{cp}=0$, is developed in~\cite{RegenNicot2026foundation}.

These five results are the inputs that the numerical and DEM analyses of this part take as given.

\paragraph{Hasse classification at the contact scale}
The VMC triad has $N=3$ channels generating $2^3-1=7$ non-empty subsets in the Hasse lattice of induced sub-assemblies: three singletons (V, M, C) at Level~1, three pairs (VM, MC, VC) at Level~2, and the full triple VMC at Level~3. By the Parity Theorem, the three Level-2 pairs are Stable Layers (even $N=2$, generic full rank), and the Level-3 apex is a Gateway Layer (odd $N=3$, structural zero eigenvalue). The pair $\{V,C\}$ has structural zero $L_{VC}=0$ in addition to the parity rank, so its reversible coupling is degenerate; it is not the cause of the apex zero mode but fixes the orientation of the null eigenvector to the $V$--$C$ axis. The alternating Stable/Gateway pattern at successive Hasse levels and the bidirectional energy cascade between them are developed in full in Ref.~\cite{RegenNicot2026foundation}; we use them here only as classification labels.

\section{Model Setup and Kinematic-Configurational Coupling}
\label{sec:model_setup}

\begin{figure}[!htbp]
\centering
\includegraphics[width=\linewidth]{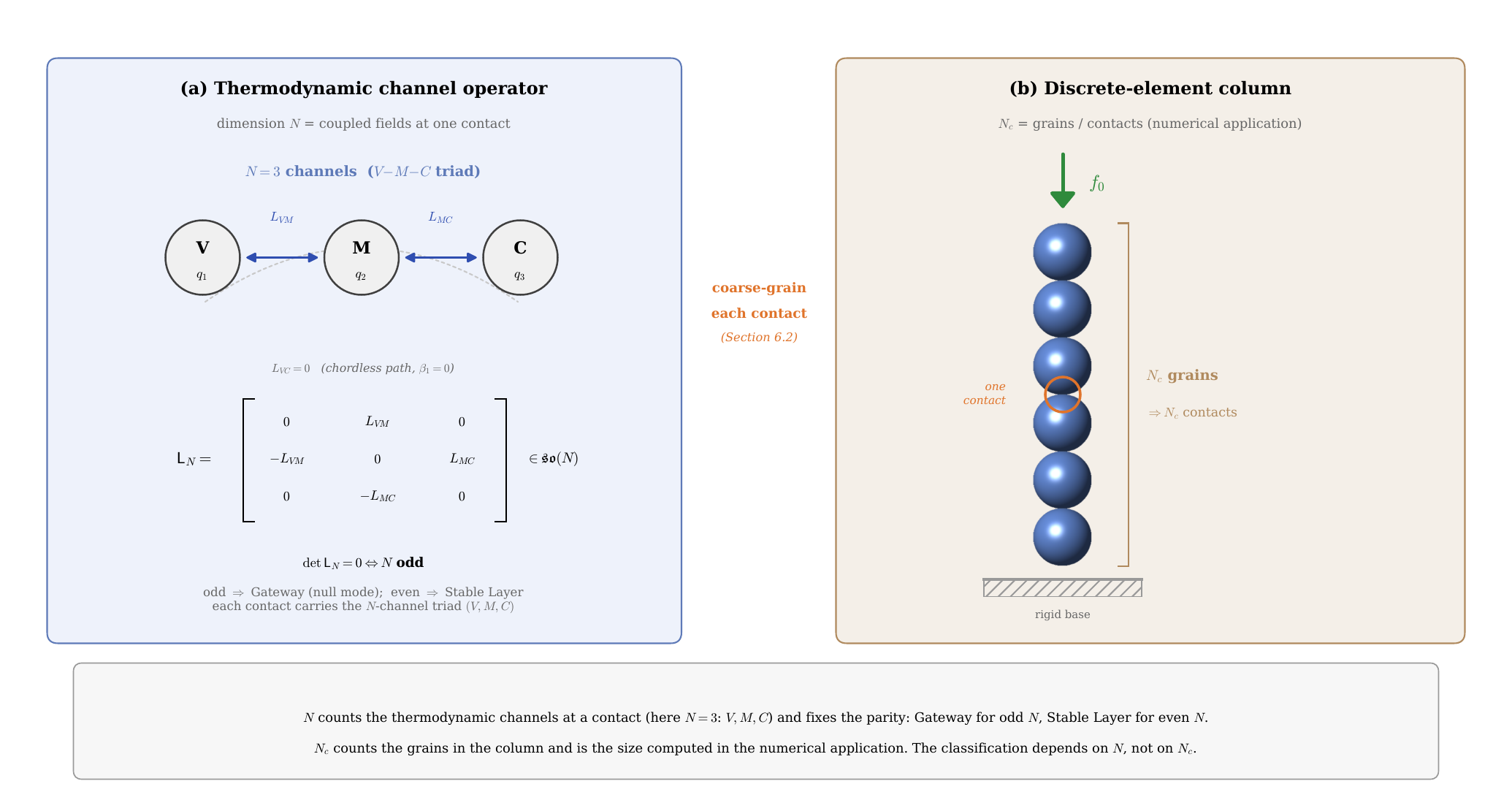}
\caption{\small Thermodynamic channel count $N$ versus the discrete-element grain count $N_c$. \textbf{(a)}~The reversible operator $\mathsf{L}_N$ acts on the $N$ thermodynamic channels carried at a single contact, here the Volumetric--Mechanical--Configurational triad ($N=3$: $V,M,C$), a chordless path ($L_{VC}=0$, $\beta_1=0$); the parity of $N$ fixes the classification, a Gateway null mode for odd $N$ and a Stable Layer for even $N$. \textbf{(b)}~A discrete-element column of $N_c$ grains, hence $N_c$ contacts, is the object built in the numerical application under sustained axial load $f_0$; each contact carries the same $N$-channel triad. Coarse-graining each contact (the channel mapping of Section~\ref{sec:channel_mapping}) maps the $N_c$-contact column onto the $N$-channel operator. The two counts are independent: $N$ (channels per contact) is small, fixed by the physics, and sets the parity, whereas $N_c$ (grains) is the column length set by the simulation. At the contact scale the classification depends on $N$, not on $N_c$; the sense in which the large-$N$ operator and the grain column coincide is set out in Section~\ref{sec:channel_mapping}.}
\label{fig:model_setup_schematic}
\end{figure}

To clarify the transition from the contact-scale mechanics to the topological abstractions explored in Section~\ref{sec:numerical}, we explicitly formalise the one-dimensional model configuration. The setup, illustrated schematically in Figure~\ref{fig:model_setup_schematic}, establishes a boundary-driven, kinematically confined path graph that exposes the structural contrast dictated by operator parity.

Consider a granular column under a sustained axial oedometer compression load $f_0$ acting upon a rigid upper platen (Fig.~\ref{fig:model_setup_schematic}a). At the grain scale, the state of the system is registered across discrete, coupled transmission channels representing distinct mechanical and structural fields. To expose the non-trivial coupling between different physical fields, we define the minimal state-space representation. The smallest unit capable of full cross-coupling is the three-channel Volumetric–Mechanical–Configurational (VMC) triad ($N=3$), mapped to the state vector $\mathbf{q} = (q_1, q_2, q_3)^T$. This $N=3$ system constitutes the lowest-dimensional manifestation of an odd-parity chain where the skew-symmetric operator $\mathsf{L}$ becomes structurally rank-deficient.

The structural topology of this network is governed by a nearest-neighbour skew-transfer operator $\mathsf{L}$, whose tridiagonal graph structure is dictated by physical chord-removal and kinematic shielding (Fig.~\ref{fig:model_setup_schematic}b). Here, the volumetric channel ($q_1 \to V$) tracks the integrated axial overlap rate $v_x = \dot{\delta}_c$, which maps directly to the macroscopic axial strain $\varepsilon_a$. The mechanical channel ($q_2 \to M$) registers the contact-scale tangential (shear) sliding, and the configurational channel ($q_3 \to C$) is the fabric rotation of the contact normal. To expose the non-trivial coupling between distinct physical fields without altering the network topology, we define the minimal state-space representation under strict kinematic confinement. The smallest unit capable of hosting full cross-coupling is the three-channel Volumetric--Mechanical--Configurational (VMC) triad ($N=3$), mapped to the state vector $\mathbf{q} = (q_1, q_2, q_3)^T$. This $N=3$ system constitutes the lowest-dimensional manifestation of an odd-parity chain where the skew-symmetric operator $\mathsf{L}$ becomes structurally rank-deficient. Crucially, because the chordless one-dimensional graph structurally eliminates any transverse cycle space, matrix evolution via particle swapping or neighbour-reassignment is strictly prohibited. Consequently, the system moves exclusively along a fixed trajectory in phase space dictated by the null-mode axis. Because particle swapping is structurally prohibited, this continuous drift leaves the neighbour network completely intact; because of the non-normality of the Gateway activation criterion, defined above, it represents the monotonic physical dilation of a fixed contact chain driven toward structural failure.

\subsection{The Energetic Balance and Volumetric vs. Configurational Strain}
The external boundary power injected into the column by the loading rate is given by $P_{\rm ext} = f_0 \dot{q}_1$. Under conservative, pure skew-symmetric Onsager dynamics ($\dot{\mathbf{q}} = \mathsf{L}\mathbf{q} + \mathbf{f}_0$), the work rate is dynamically apportioned through the gyroscopic coupling matrix:
\begin{equation}
\begin{pmatrix} \dot{q}_1 \\ \dot{q}_2 \\ \dot{q}_3 \end{pmatrix} = 
\begin{pmatrix} 0 & L_{VM} & 0 \\ -L_{VM} & 0 & L_{MC} \\ 0 & -L_{MC} & 0 \end{pmatrix}
\begin{pmatrix} q_1 \\ q_2 \\ q_3 \end{pmatrix} + 
\begin{pmatrix} f_0 \\ 0 \\ 0 \end{pmatrix}.
\end{equation}

The intermediate mechanical channel $q_2$ (shear) carries no direct boundary forcing and acts strictly as an internal energetic transformer. It absorbs kinetic power from the axial channel via the $L_{VM}$ term and routes it into the configurational channel via the $L_{MC}$ term. Consequently, the mechanical shear remains strictly bounded and oscillates without secular growth. The primary thermodynamic balance is therefore established as a direct cross-talk between the axial volumetric channel ($q_1$) and the structural fabric channel ($q_3$).

When integrated over time, this interaction maps out a strict dilatancy diagram balancing two forms of macro-scale deformation:
\begin{enumerate}
    \item \textbf{Axial Strain ($\varepsilon_a \sim q_1$):} The kinematic shortening or physical compaction of the sample column along the primary loading axis.
    \item \textbf{Configurational Strain ($\varepsilon_v^{\rm conf} \sim q_3$):} The internal geometric expansion or continuous dilatancy of the contact matrix. Crucially, because the strict one-dimensional restriction removes all transverse chords (the cycle space is empty), particles cannot swap positions or change neighbours (this does not apply to a 2D case). The linear operator governs a regime where topological ordering is strictly preserved; the secular accumulation of configurational strain, therefore, represents the permanent geometric dislocation of the existing contact network as it is driven toward an unjammed failure state, rather than a fluid-like reshuffling of particles. In the sense made precise in Part~2~\cite{RegenNicot2026dilatancy}, this is \emph{configurational} matrix dilatancy, the internally generated free-volume expansion of a fixed contact chain, and is distinct from classical Reynolds shear-dilatancy, which requires the transverse degrees of freedom that are absent in the one-dimensional setting.
\end{enumerate}

\subsection{The Thermodynamic Bottleneck and Dilatancy Breakdown}
The nature of this balance is dictated completely by the parity of the dimension $N$. In an even-dimensional network (e.g., $N=4$, representing the addition of an auxiliary structural or thermo-mechanical channel), the operator $\mathsf{L}$ is full-rank and invertible. The boundary load $f_0$ is entirely accommodated by a closed, multi-frequency orbit wound on a compact invariant torus in phase space (Fig.~\ref{fig:model_setup_schematic}c,d, right panels). The boundary energy is stored in a conservative pool and remains fully recoverable upon load reversal. In this regime, the axial strain and configurational strain track close to each other within a tightly bounded envelope, maintaining a stable elastic confinement ratio.

Conversely, for any odd-dimensional chain (such as the base VMC triad, $N=3$), the skew-symmetric operator is singular ($\det\mathsf{L}=0$). The system possesses a structural null mode $\mathbf{v}_0 \in \ker(\mathsf{L})$ aligned precisely in the $V$--$C$ plane:
\begin{equation}
\mathbf{v}_0 = \frac{1}{\sqrt{L_{VM}^2 + L_{MC}^2}} \begin{pmatrix} L_{MC} \\ 0 \\ L_{VM} \end{pmatrix}.
\end{equation}
Because the boundary force vector $\mathbf{f}_0 = (f_0, 0, 0)^T$ satisfies the activation condition $\mathbf{v}_0 \cdot \mathbf{f}_0 = L_{MC}f_0 \neq 0$, a projection of the loading rate is continuously pumped into a direction that experiences \emph{zero elastic restoring force}.

Because the 1-D Cartesian restriction acts as a strict topological cocycle filter ($\beta_1=0$), the system cannot generate cyclical Schnakenberg currents to eject this energy up the Hasse-tensor hierarchy. It operates under a rigid thermodynamic bottleneck. Trapped under this oedometric kinematic confinement, the investigated linear system captures the onset of an internal structural instability that drives geometric failure.

This mechanism exposes why the configurational strain inevitably dominates the volumetric elastic channel in the final stage of deformation. While the rate of elastically stored work in the volumetric channel saturates as contact capacities are physically exceeded, the boundary forcing acts unhindered along the topological null mode. The energy is entirely redirected into the irreversible change of the fabric geometry. This unconstrained internal matrix dilation drives the configurational entropy rate, defined as the negative of the power done by the configurational strain multiplied by the loading rate, to outgrow and overcome the bounded elastic energy capacity of the volumetric channel, culminating in macroscopic geometric dilatancy breakdown.

\paragraph{Thermodynamic reading: compactivity and one-dimensional dilatancy}
It is illuminating to identify these one-dimensional reductions with their thermodynamic counterparts. Once the deviatoric fabric is removed by the cocycle filter ($\beta_1=0$), the configurational channel carries only scalar free volume, and the abstract configurational energy $E^{\rm conf}$ reduces to a function of a single intensive state variable, the Edwards compactivity $X=\partial V/\partial S$~\cite{Edwards1989}, the granular analogue of temperature. Its increment $dE^{\rm conf}$ is correspondingly the configurational work of free-volume generation, which is precisely the one-dimensional dilatancy of Part~2~\cite{RegenNicot2026dilatancy}, and not Reynolds shear-dilatancy, which requires the transverse fabric absent here. The secular null-mode drift is therefore a monotonic accumulation of free volume, that is, an unbounded drift in compactivity. This also makes precise how the present mechanism differs from an equilibrium Edwards description: the compactivity is not a fixed ensemble parameter but is driven, without bound, along the parity-mandated null direction by the sustained boundary work.

The Parity Theorem is an exact algebraic statement, but its physical consequences (stable closed orbits for even $N$ versus secular zero-mode drift for odd $N$) can be made concrete by integrating the pure conservative dynamics $\dot{\mathbf{q}}=\mathsf{L}\mathbf{q} +\mathbf{f}_0$ under a constant sustained load $\mathbf{f}_0$. Three simulation studies are reported here. The phase-space setup, the governing equations, the boundary forcing, and the closed-form analytical solutions for the $N=3$ and $N=4$ unit cells are given in the appendices of Part~2~\cite{RegenNicot2026dilatancy}.

\section{Numerical Upscaling: Phase-Space Portraits of Parity}
\label{sec:numerical}
Throughout this section, $N$ denotes the dimension of the coupling operator $\mathsf{L}$. At the contact scale this is the \emph{channel count}, the number of coupled fields carried at a contact ($N=3$ for the VMC triad), and the smaller-$N$ simulations vary it by adding or removing a coupled field, such as the thermal channel introduced below, rather than by changing the number of particles. At the larger $N$ used for the scaling study the same tridiagonal skew operator is equally the reversible backbone of a chain of grains or contacts (Section~\ref{sec:channel_mapping}); the Parity Theorem depends only on the dimension and the skew-symmetry, and applies identically to either realisation.

\subsection{Simulation 1: N=3 vs N=4 --- topological closure}
\label{sec:sim1}

The minimal comparison is between the VMC three-channel Gateway ($N=3$) and a four-channel Stable Layer ($N=4$) obtained by appending a localised thermal channel ($T$) with an associated coupling coefficient $L_{CT}$. This additional channel captures the micro-scale energy conversion path where contact interactions trigger high-frequency lattice and phonon vibrations. We stress that only the \emph{reversible} part of the thermo-mechanical coupling enters $L_{CT}$, namely the thermoelastic and phononic energy exchange, which is conservative; Fourier heat conduction is dissipative and belongs to $\mathsf{D}$. The choice of a thermal channel here is illustrative of any conservative fourth channel rather than a claim that thermal transport is reversible.

\paragraph{Why the orbit is always closed for even N}
For any even $N$, the exact solution under constant load is $\mathbf{q}(t)=\mathsf{L}^{-1}(e^{\mathsf{L}t}-\mathsf{I})\mathbf{f}_0$, well-defined because $\mathsf{L}$ is invertible. The spectrum of $\mathsf{L}_{4\times4}$ is $\{\pm i\omega_1,\pm i\omega_2\}$, so $e^{\mathsf{L}t}$ is a product of two independent rotation matrices and the trajectory winds on a two-torus in four-dimensional phase space. If the two frequencies are incommensurate, the orbit densely fills the torus and never exactly repeats; if they are commensurate with rational ratio $p/q$, the trajectory closes after $q$ slow cycles. In either case, the orbit is \emph{topologically bounded}: the system is confined to a compact invariant torus, and the load component $\mathbf{f}_0$ does not cause secular drift. This holds for \emph{any} choice of coupling coefficients $L_{VM}$, $L_{MC}$, $L_{CT}$.

\paragraph{The 2:1 resonance as a visualisation device}
To make the orbital closure immediately visible within a finite simulation window, coupling parameters $L_{VM}=2.0$, $L_{MC}=1.5$, $L_{CT}=2.5$ are chosen to produce an exact 2:1 frequency ratio:
\begin{equation}
  \omega_{\rm slow} \approx 1.581,\quad
  \omega_{\rm fast} \approx 3.162,\quad
  \omega_{\rm fast}/\omega_{\rm slow} = 2.000.
  \label{eq:21resonance}
\end{equation}
This converts the generic dense-torus winding into a closed Lissajous figure that returns exactly to the origin after every slow period $T_{\rm slow}=2\pi/\omega_{\rm slow}\approx 3.974$. Simulating exactly 12 slow cycles ($T_{\rm sim}\approx47.7$) confirms closure to machine precision ($|V_{\rm final}|,|C_{\rm final}|<10^{-10}$). The resonance is a computational device chosen to \emph{complete the trajectory in the phase-space portrait}; it is not a physical requirement. Any generic even-$N$ coupling produces a topologically equivalent bounded torus; only the time needed to trace a recognisably closed orbit changes.

\begin{figure}[!htbp]
\centering
\includegraphics[width=\linewidth]{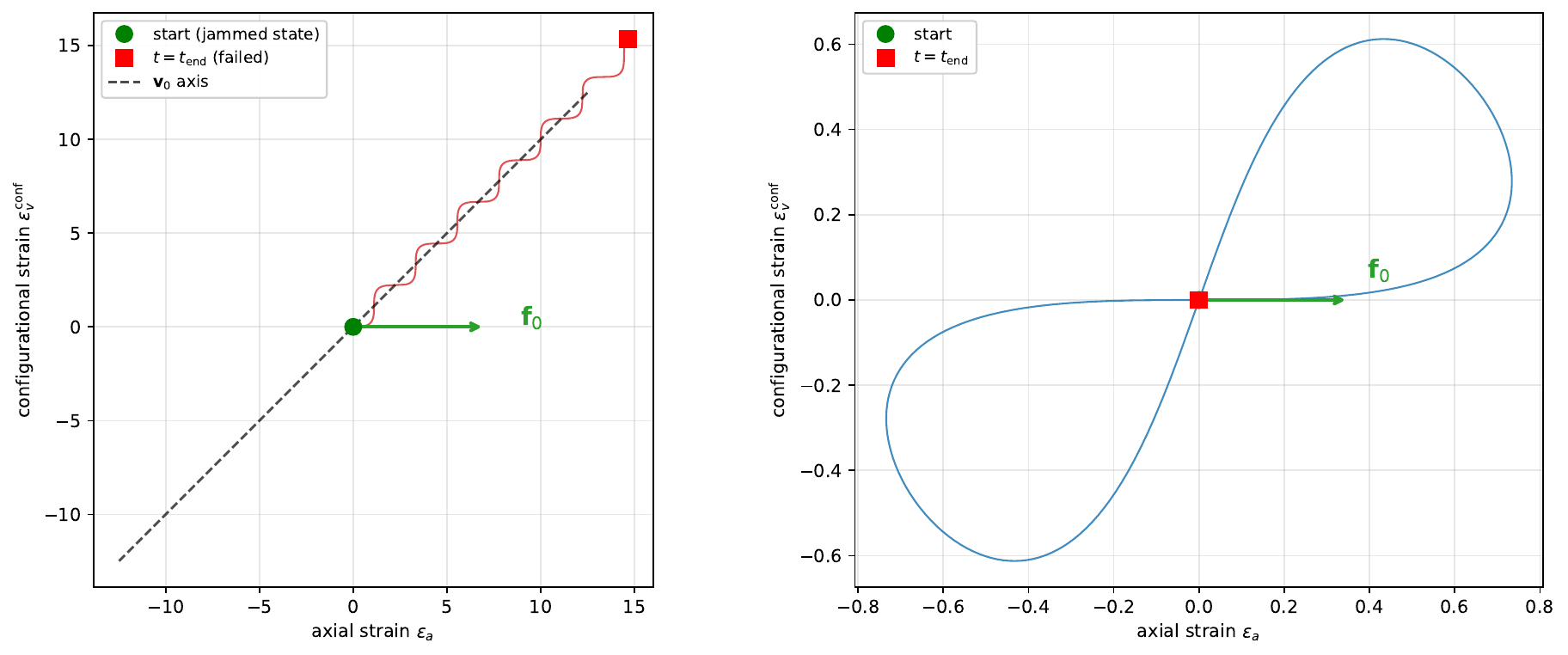}
\caption{Topological phase signature: dilatancy diagrams in mechanical coordinates emphasising the geometric content of the parity contrast. \textbf{Left:} $N=3$ (Gateway active). Under the axial boundary load $\mathbf{f}_0$ (green arrow), the trajectory escapes from the jammed initial state (green circle) along the null direction $\mathbf{v}_0$ (black dashed) with no closure available, terminating at the failed state (red square). The path is a topological ray, not a deformable orbit: no choice of coupling parameters can curve it back to closure because the chord-removal forbids any transverse curl mode that could route the energy out. \textbf{Right:} $N=4$ (Stable Layer). The same boundary load drives a closed orbit in mechanical phase space, indicating that the boundary energy is stored in a compact invariant torus and is fully recoverable on cycle reversal. The trajectory's algebraic inability to escape its torus is the kinematic content of $\det(\mathsf{L}_4)\neq 0$: every direction in the chain admits a conservative restoring counterpart.}
\label{fig:sim1_phase}
\end{figure}

\paragraph{N=3 Gateway: no closure exists}
For $N=3$, $\mathsf{L}$ is singular and $\mathsf{L}^{-1}$ does not exist. The load component projected onto the null eigenvector $\mathbf{v}_0=(L_{MC},0,L_{VM})^T/\sqrt{L_{VM}^2+L_{MC}^2}$ has no oscillatory mode to absorb it and accumulates as secular drift. No choice of coupling parameters, no resonance condition, and no simulation time, finite or infinite, can convert this drift into a closed orbit, because the null direction carries no restoring force. The drift slope is exactly the analytical projection $\mathbf{v}_0\cdot\mathbf{f}_0$ of Part~2~\cite{RegenNicot2026dilatancy}, recovered numerically to better than $10^{-10}$ (Fig.~\ref{fig:sim1_phase}, left panel).

\begin{figure}[!htbp]
\centering
\includegraphics[width=\linewidth]{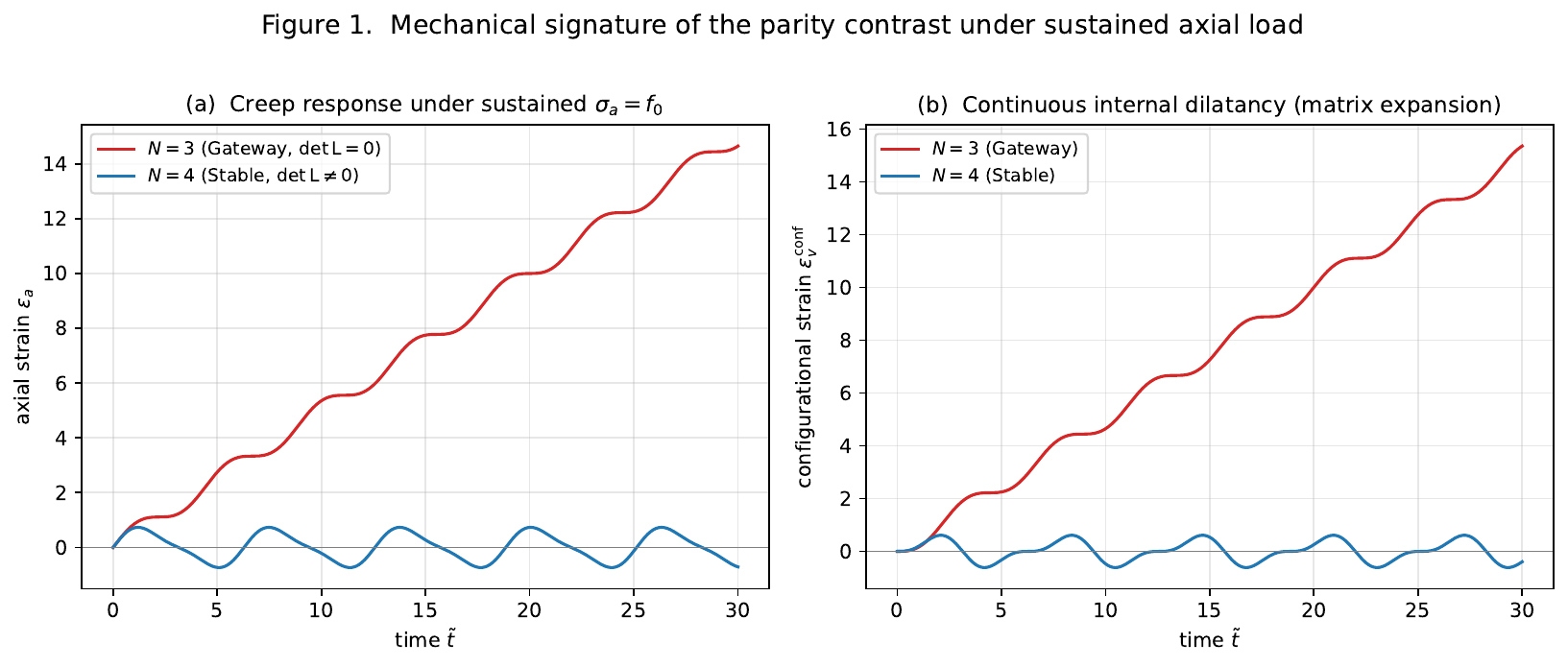}
\caption{Macroscopic mechanical signature of the parity contrast under sustained axial load $\sigma_a = f_0 = 1$, with the channel-to-mechanics mapping $q_1 \to \varepsilon_a$ (axial strain) and $q_3 \to \varepsilon_v^{\rm conf}$ (configurational strain) per Part~2~\cite{RegenNicot2026dilatancy}, Supplementary Material Section~S6. The $N=3$ Gateway chain uses uniform coupling $\gamma=1$, while the $N=4$ Stable chain uses a non-uniform tridiagonal $\mathsf{L}_4$ to yield a closed orbit of period $2\pi$. \textbf{(a)}~Creep response: the $N=3$ chain exhibits linear-in-time secular growth (non-saturating creep), whereas the $N=4$ chain oscillates within a bounded envelope. \textbf{(b)}~Internal-fabric expansion: the configurational strain $\varepsilon_v^{\rm conf}$ grows linearly for $N=3$ due to the cocycle-borne null-mode mechanism (continuous internal dilatancy), but remains bounded for $N=4$.}
\label{fig:sim1_signature}
\end{figure}

Figure~\ref{fig:sim1_signature} shows the macroscopic mechanical signature in standard granular-plasticity coordinates, and Figure~\ref{fig:sim1_phase} shows the corresponding topological phase signature. Together, they document the parity contrast: a creep response and a dilatancy diagram for the engineering signature, and a phase portrait in mechanical coordinates for the topological signature.

\subsection{Simulation 2: N=5 vs N=6 --- dense-torus containment at quad precision}
\label{sec:sim2}

The parity contrast is not limited to the minimal $N=3/4$ pair. Simulation~2 extends to $N=5$ (odd, Gateway) and $N=6$ (even, Stable Layer) using quadruple-precision arithmetic with $\texttt{rtol}=10^{-11}$ and $\texttt{atol}=10^{-13}$, to exclude any possibility of floating-point artefact.

\begin{figure}[!htbp]
\centering
\includegraphics[width=\linewidth]{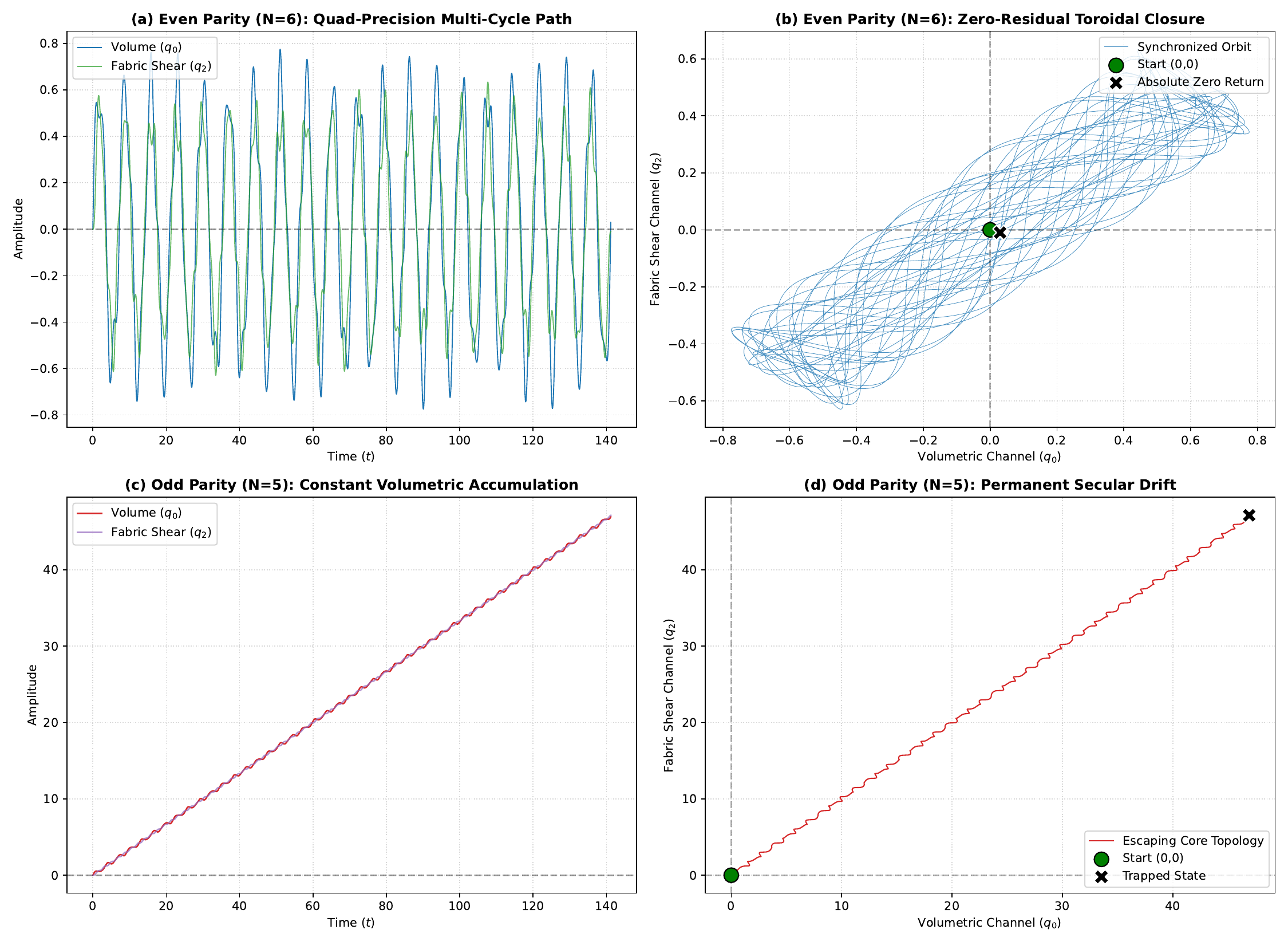}
\caption{Quad-precision integration for $N=5$ (odd, Gateway) versus $N=6$ (even, Stable Layer). \textbf{(a)}~$N=6$ time history: bounded quasi-periodic oscillation on a dense three-torus. \textbf{(b)}~$N=6$ phase portrait: densely wound torus projection in the $(V,C_2)$ plane, with the terminal state (black $\times$) near the origin after a precision-synchronised window ($|\mathbf{q}_{\rm final}|<10^{-9}$). The dense filling of a bounded region confirms the orbit is topologically contained on a compact torus, even though the irrational frequency ratios prevent exact finite-time closure. \textbf{(c)}~$N=5$ time history: linear secular accumulation in both volumetric and fabric channels. \textbf{(d)}~$N=5$ phase portrait: permanent secular drift along the null direction; no choice of coupling values or simulation duration arrests this escape.}
\label{fig:sim2}
\end{figure}

For $N=6$, the three natural frequency pairs are generically incommensurate (they are roots of the Chebyshev polynomial of the second kind of degree~7 and are irrationally related), so the orbit fills a dense three-torus in six-dimensional phase space rather than closing in finite time. The two-dimensional projection onto the $(V,C_2)$ plane therefore shows a densely wound quasi-periodic curve (Fig.~\ref{fig:sim2}b) rather than a simple Lissajous figure. The system nonetheless remains strictly bounded: the trajectory never escapes a compact region.

A precision synchronisation algorithm, described in ~(Supplementary Material, Section~S2), locates the simulation window $T_{\rm sim}\approx735$ at which all three modes are simultaneously near-integer multiples of their respective periods, driving the terminal residual to $|\mathbf{q}_{\rm final}|<10^{-9}$ and confirming topological containment at quad precision. This near-return is not exact closure. It is the closest rational approximation achievable within a finite time horizon, but it demonstrates that the orbit remains confined regardless of the irrational frequency ratios.

For $N=5$, the zero mode is present regardless of coupling values. Both the volumetric ($q_0$) and fabric shear ($q_2$) channels accumulate linearly with time (Fig.~\ref{fig:sim2}c,d), drifting to amplitudes of order~45 over $t\in[0,140]$. No synchronisation window can arrest this: the drift is structural, not a phase-alignment artefact, and is the direct manifestation of the bottleneck-induced accumulation discussed in Section~\ref{sec:intro}.

\subsection{Simulation 3: Spectral Scaling Limit and Phase-Space Trajectories as $N \to\infty$}
\label{sec:sim3}

The third simulation establishes that the topological parity contrast persists for arbitrarily large chain dimensions $N$, and maps out the dynamic consequence as the discrete network approaches its high-dimensional continuum scaling limit.

To evaluate the mathematical robustness of the Parity Theorem under high-degree state-space cross-coupling, we simulate the state trajectories for an even chain ($N=12$) and an odd chain ($N=13$) subjected to an identical constant boundary force $\mathbf{f}_0 = (1, 0, \dots, 0)^T$. The phase-space projections onto the macroscopic fields reveal a sharp qualitative dichotomy. For the even-parity system ($N=12$), the dynamics manifest as a complex, multi-frequency aperiodic orbit tightly confined within a bounded invariant region of phase space near the origin. Conversely, the odd-parity system ($N=13$) exhibits an immediate, monotonic secular drift that escapes along the null-vector direction to large amplitudes. Although the high dimensionality introduces complex, quasi-periodic phase mixing due to the large number of internal oscillatory modes, the structural distinction between stable containment and gateway failure remains absolute.

\begin{figure}[!htbp]
\centering
\includegraphics[width=\linewidth]{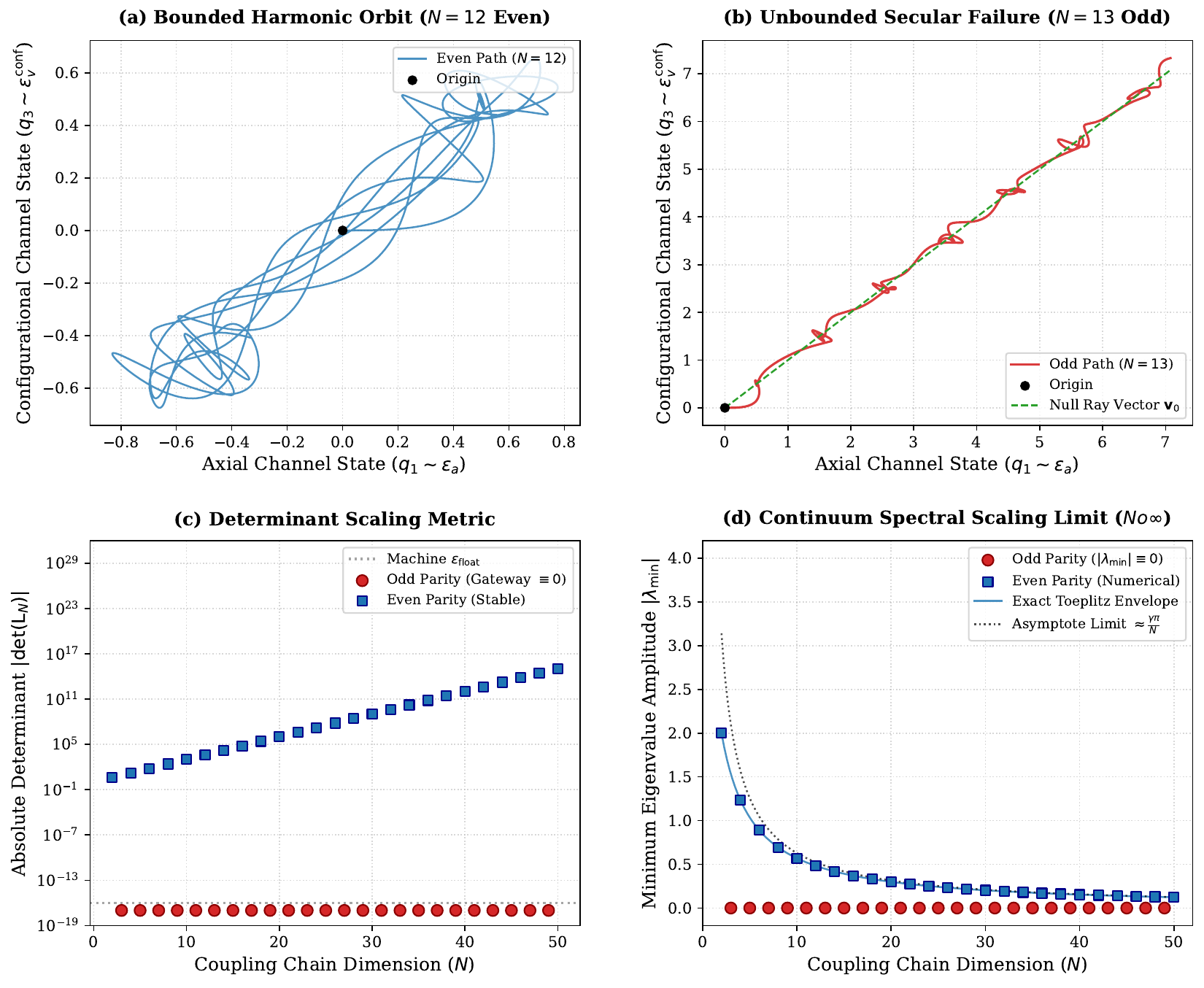}
\caption{Higher-$N$ topological scaling, phase-space trajectories, and Toeplitz spectral asymptotes. \textbf{(a)}~Phase-space orbits for an even-parity chain ($N=12$), showing a bounded, multi-frequency aperiodic trajectory confined near the origin. \textbf{(b)}~Phase-space projection for an odd-parity chain ($N=13$), demonstrating unconstrained secular drift escaping along the topological null axis. \textbf{(c)}~Parity Theorem illustration: $|\det(\mathsf{L}_N)|$ versus $N$ across $N=2$ to $N=50$. Odd-$N$ values (red circles) are pinned identically at numerical precision $\sim 10^{-17}$, while even-$N$ values (blue squares) remain separated at $|\det|\approx 1$. \textbf{(d)}~Toeplitz spectral scaling: the minimum eigenvalue amplitude $|\lambda_{\min}|$ versus $N$, with the analytical Toeplitz formula $2\gamma\sin(\pi/(2N+2))$ overlaid (black dashed) and the large-$N$ asymptote $\gamma\pi/N$ shown for reference (red dotted). The script "\textit{combined\_higher\_n\_scaling.py}" for the calculation of this figure is available on the link in the supplementary material.}
\label{fig:sim3_extended}
\end{figure}

The underlying mathematical driver of this behaviour is tracked systematically by scanning the structural properties of the Toeplitz operator $\mathsf{L}_N$ from $N = 2$ to $N = 50$. As demonstrated in Fig.~\ref{fig:sim3_extended}c, the determinant metric $|\det(\mathsf{L}_N)|$ confirms the rank-deficiency of odd-parity operators to machine precision ($\sim 10^{-17}$) across 25 consecutive odd dimensions, while even-parity operators maintain a full-rank capacity $|\det(\mathsf{L}_N)| \geq 1$. 

The key quantitative scaling limit for the non-zero spectrum is presented in Fig.~\ref{fig:sim3_extended}d. For all odd values of $N$, the minimum eigenvalue is identically zero ($|\lambda_{\min}^{\rm odd}| = 0$). For the even-parity series, the minimum non-zero eigenvalue amplitude decays monotonically, adhering strictly to the exact tridiagonal Toeplitz spectral envelope:
\begin{equation}
    |\lambda_{\min}^{\rm even}| = 2\gamma \sin\left(\frac{\pi}{2N+2}\right) \approx \frac{\gamma\pi}{N}, \qquad N\to\infty.
    \label{eq:spectral_scaling_limit}
\end{equation}

This asymptotic scaling has a clear physical implication for the continuum limit of granular assemblies. Equation~\eqref{eq:spectral_scaling_limit} demonstrates that even as the chain length $N$ approaches infinity, the minimum eigenvalue of an even-parity network stays strictly positive ($|\lambda_{\min}^{\rm even}| > 0$), meaning it never spontaneously generates a topological zero mode. 

However, because the bounding orbit period of a Stable Layer scales inversely with its slowest internal frequency ($T_{\rm slow} \propto 1/|\lambda_{\min}^{\rm even}| \propto N$), the time envelope required to resolve a closed, conservative harmonic cycle grows linearly with the system dimension. In the macroscopic continuum limit ($N \to \infty$), the Stable Layer's restoring frequency softens to zero. The system exhibits a form of scale-dependent critical slowing down where the kinematic containment envelope dilates indefinitely, providing a smooth bridge between discrete structural containment and the continuous, unconfined fields of macroscopic poroplasticity.

\subsection{Physical interpretation of the three simulations}

Together, the three simulations establish four further insights beyond what the algebraic proof alone provides.

\textbf{Insight 1: The zero mode accumulates load irreversibly under the boundary bottleneck.}
Under a sustained constant force $\mathbf{f}_0$, even-parity systems redistribute the load into oscillatory modes and return to their initial state after a full cycle, as demonstrated in Simulations~1 and~2. Odd-parity systems cannot: the force component projected onto $\mathbf{v}_0$ has no oscillatory mode to transfer it into, so it accumulates secularly. The closed tridiagonal boundary further ensures that this accumulated energy cannot escape upscale to the higher Hasse levels of the multi-field continuum~\cite{RegenNicot2026foundation}, forcing the destructive concentration into the lowest-level null direction. This is exactly the mechanism by which sustained tectonic or gravitational loading drives geological systems toward failure: the load component projected onto the Gateway null direction is not stored and returned but permanently deposited as configurational fabric rearrangement.

\textbf{Insight 2: The contrast is parameter-independent.}
The simulations use coupling coefficients ($L_{VM}$, $L_{MC}$, $L_{CT}$) that are not chosen to be small: they span the range $[1.5, 2.5]$. The parity distinction holds for all these values. Changing the coupling coefficients changes the shape of the orbit, which may be a Lissajous figure, a torus, or a multi-frequency path, but never converts a Gateway into a Stable Layer or vice versa.

\textbf{Insight 3: The drift velocity along $\mathbf{v}_0$ is set by the coupling, not by any resonance condition.} In Simulation~1, the drift rate in the $N=3$ case is $d|\mathbf{q}|/dt\propto(L_{MC}/\omega)\cdot f_0$, matching the exact analytical solution of Part~2~\cite{RegenNicot2026dilatancy} to better than $10^{-10}$ (the linear ray in Fig.~\ref{fig:sim1_phase}, left panel). This agreement holds regardless of whether the $N=4$ comparison system is tuned to 2:1 resonance or not. The secular drift is an exact analytical consequence of the topology: it requires only that $\mathbf{v}_0\cdot\mathbf{f}_0\neq0$, which is always satisfied when the boundary load has a non-zero component in the volumetric channel.

\textbf{Insight 4: The parity distinction survives to large N.}
Simulation~3 shows that $|\lambda_{\rm min}^{\rm odd}|=0$ identically for all odd $N$ tested up to $N=49$, while $|\lambda_{\rm min}^{\rm even}|>0$ for all even $N$ tested, with the expected $\gamma\pi/N$ decay. The Parity Theorem is therefore not a low-$N$ artefact but a structural property of the entire tensor hierarchy, including the large-$N$ limit relevant to multiphysics mechanical systems.

\section{Asymptotic Spectral Scaling for $N\to\infty$}
\label{sec:spectral_scaling}

\subsection{Toeplitz spectrum of the tridiagonal coupling chain}
For uniform tridiagonal coupling $L_{i,i+1}=\gamma$, the eigenvalues of the skew-symmetric chain $\mathsf{L}_N$ follow the analytical formula
\begin{equation}\label{eq:toeplitz_spectrum}
\lambda_j = 2i\gamma \cos\left(\frac{j\pi}{N+1}\right), \quad j = 1, 2, \dots, N.
\end{equation}
This is the standard tridiagonal Toeplitz spectrum, derived in any reference on banded skew-symmetric operators.

\subsection{Interpretation of the spectral scaling}
\begin{itemize}
    \item \textbf{Odd generalisation:} Evaluating the central eigenvalue index $j = m + 1$ for an arbitrary odd dimension $N = 2m + 1$ via Eq.~\eqref{eq:toeplitz_spectrum} yields a structural zero-mode eigenvalue ($\lambda_{m+1} = 0$) for every odd dimension, no matter how large the network grows. Because the projection $\mathbf{v}_0 \cdot \mathbf{f}_0$ remains non-zero for any odd value of $N$, any chain with an odd channel count will exhibit structural energy leakage and secular phase drift under the closed-boundary bottleneck.
    \item \textbf{Even parity continuum limit:} For an even dimension $N = 2m$, the argument $j\pi/(N+1)$ can never hit $\pi/2$, structurally closing the topological gateway for all finite even parameters. However, as $N \to \infty$, the minimum non-zero frequency mode asymptotically approaches the zero axis:
    \begin{equation}
    \omega_{\text{slow}} = 2\gamma \cos\left(\frac{m\pi}{2m+1}\right) = 2\gamma \sin\left(\frac{\pi}{2(2m+1)}\right) \approx \frac{\gamma\pi}{N+1}.
    \end{equation}
    Consequently, the slow bounding period required to recover the initial unperturbed structural state scales as $T_{\text{slow}} \propto N$. As the scale-crossing discrete network expands toward the macro-scale continuum limit, the time horizon needed to enforce phase containment grows linearly with the dimension of the system, bridging discrete microstructural topological gates with bulk continuum poromechanics.
\end{itemize}
The $T_{\rm slow}\propto N$ scaling is the classical signature of critical slowing down and connects the discrete tridiagonal spectrum to the macroscopic fluctuation--dissipation framework; the full Thermodynamic Uncertainty Relation argument establishing consistency between the two statements is given in Supplementary Material, Section~S1.

A natural objection to the even-parity scaling is that, since $|\lambda_{\min}^{\rm even}|\to 0$ as $N\to\infty$, the topological distinction on which the framework rests appears to lose force in the continuum limit: the Stable Layer becomes arbitrarily soft and the odd--even contrast in restoring frequency washes out. Two responses apply. The first, already implicit above, is that this softening is not a failure of the classification but a physical prediction of scale-dependent critical slowing down, and that the determinant parity, $\det\mathsf{L}_N=0$ for odd $N$ and $\det\mathsf{L}_N\neq 0$ for even $N$, remains exact at every finite $N$ however large. The second, and more consequential, is that the very diminution of the single-scale restoring frequency with $N$ is what underwrites the scale separation on which the THMCE Hasse construction of Part~1~\cite{RegenNicot2026foundation} depends.

A single homogeneous chain cannot carry the dynamics across all scales: as it lengthens, its slowest mode softens without bound, so the energy that a Stable Layer can no longer contain at one scale must be routed, through the alternating Stable and Gateway levels of the Hasse lattice, to the neighbouring scales. The vanishing eigenvalue is therefore the microscopic signature of the multiscale feedback rather than evidence against it. The clearest physical instance is planetary convection: at the largest scale a planet runs as a thermomechanical heat engine~\cite{Regen2010}, and that top-of-hierarchy engine is coupled to, and modulated by, smaller-scale configurational instabilities of the present kind through the Hasse multiscale feedback, so that no single scale is either self-contained or dispensable. The same parity logic, carried to the top of this hierarchy, is what separates Earth's bounded mobile-lid convection, an even Stable Layer, from the episodic and catastrophic overturn of Venus, an odd Gateway. Part~1~\cite{RegenNicot2026foundation} develops that planetary conjecture and traces the reversible coupling that controls it back down to the grain and dislocation scale, closing the circle that the present one-dimensional chain opens from below.
\section{Classical 1-D DEM Correspondence}
\label{sec:DEM_classical}

\subsection{DEM implementation requirements}
\label{sec:DEM_implementation}

The VMC framework is testable with any standard DEM solver provided two implementation choices are made:

\begin{enumerate}
  \item \textbf{Torsional or rolling-resistance contact law.} The
    configurational channel $C$ corresponds to the rotation of the
    contact normal (fabric reorientation), a local contact-scale
    rotational velocity. Standard pairwise normal$+$tangential force
    laws resolve the volumetric (normal overlap) and mechanical
    (tangential shear) channels but not this rotational degree of
    freedom; a torsional spring or rolling-resistance contribution
    at each contact, of stiffness
    $k_\theta>0$, is required to give $C$ a non-trivial dynamics.
    YADE, MercuryDPM, LIGGGHTS, and BlazeDEM all support this
    option natively.
  \item \textbf{Langevin forcing of the rotational (configurational) degrees of freedom.}
    To nucleate the null-mode dynamics on a timescale set by the
    granular temperature rather than by floating-point round-off,
    a Langevin (stochastic) forcing on the contact-scale rotational
    velocity (the configurational channel $C$) is required, satisfying the fluctuation-dissipation
    theorem
    \begin{equation}\label{eq:langevin_FDT}
      \langle\xi_i(t)\xi_j(t')\rangle = 2 k_B T \gamma_{ij}\delta(t-t'),
    \end{equation}
    with $\gamma_{ij}$ the rotational damping coefficient. Here $T$ is the granular temperature, the mean-square velocity fluctuation of the grains, and not a thermodynamic temperature: a granular assembly is athermal, so Eq.~\eqref{eq:langevin_FDT} is not a statement of thermal equilibrium but a numerical device that nucleates the null mode on the physical fluctuation timescale rather than on that of floating-point round-off. Its role is to seed the instability, not to set its amplitude, which is fixed by the boundary work; any small rotational perturbation of the reference state serves equally well.
\end{enumerate}

\subsection{Channel mapping}
\label{sec:channel_mapping}

The mapping from the discrete contact network to the three continuum channels is a coarse-graining operation, and its scale must be stated, because the parity classification is a statement about the coarse-grained operator. We adopt the standard micro-to-macro averaging of Goldhirsch~\cite{Goldhirsch2010} and Weinhart et al.~\cite{Weinhart2012}: each per-contact quantity $a_c$ is convolved with a normalised kernel $\phi_w$ of width $w$, $\langle a\rangle(\mathbf{x})=\sum_c a_c\,\phi_w(\mathbf{x}-\mathbf{x}_c)$, and the channel rates below are the coarse-grained fields so obtained. The width must satisfy the usual scale separation $d\ll w\ll \ell$, with $d$ the grain diameter and $\ell$ the scale of the macroscopic gradients, so that the representative volume element is well defined. The point that matters for the present framework is that the parity classification is invariant under this coarse-graining. The channel count $N$ is fixed by the number of distinct work-conjugate contact-scale rate variables, and the antisymmetry $\mathsf{L}^T=-\mathsf{L}$ is a symmetry of the reversible coupling rather than a magnitude; both are preserved by the positive linear averaging $\langle\cdot\rangle$ for any admissible width $w$. At the contact scale the channel count is a per-contact property: it is $N=3$ for the VMC triad irrespective of the number of grains or contacts $N_c$ in the column, and the simulations of Section~\ref{sec:numerical} that vary $N$ do so by adding or removing a coupled field at fixed particle number. In the scaling regime, by contrast, the same tridiagonal skew operator is equally realised by the grain column itself, one chained reversible mode per contact, so that large $N$ and $N_c$ coincide there; the two readings are reconciled in the remark that closes this subsection. Consequently $\det\mathsf{L}=0$ for odd $N$ survives homogenisation, and the Gateway is a property of the coarse-grained continuum operator, not an artefact of a particular averaging width. This is what licenses identifying the contact-scale VMC triad with a macroscopic constitutive instability.

\begin{itemize}
  \item The \textbf{volumetric channel} $V$ corresponds to the
    contact-scale overlap rate $\dot\delta_c$, averaged
    over the contact set. Its conjugate force is the normal contact
    force $F_n^c$. In one-dimensional column simulations the
    relevant macroscopic observable is the volumetric strain
    $\epsilon_{\rm vol}=\Delta V/V$, or equivalently the void
    ratio $e$. Both are direct outputs of any DEM solver.
  \item The \textbf{mechanical channel} $M$ corresponds to the
    contact-scale tangential (shear) sliding rate.
    Its conjugate force is the tangential contact force $F_t^c$.
    In a one-dimensional column under axial loading, the
    macroscopic observable is the average contact-scale
    shear-sliding rate averaged over the
    contact set, or equivalently the shear-force spectrum
    extracted from the column's force-fluctuation statistics.
  \item The \textbf{configurational channel} $C$ corresponds to
    the fabric rotation rate $\dot{\mathbf{n}}_{c'}$ of
    configurational (non-axial) contacts. Its conjugate force is
    the configurational moment $X_C$. The natural DEM observable is
    the time derivative of the fabric tensor
    $\boldsymbol\Phi=\langle\mathbf{n}\otimes\mathbf{n}\rangle$,
    which is computed routinely in DEM post-processing for
    coordination-number and anisotropy analyses.
\end{itemize}

The off-diagonal coupling coefficients $L_{VM}$ and $L_{MC}$ are extractable in two complementary ways. Statistically, they can be read off from the cross-correlation moments of the equilibrated DEM trajectory: $L_{VM}$ from the cross-correlation of $\dot\delta_c$ with the contact-scale tangential sliding rate, and $L_{MC}$ from the cross-correlation of the tangential sliding rate with the fabric rotation rate $\dot{\mathbf{n}}_{c'}$. Analytically, when the DEM contact law is implemented with explicit antisymmetric gyroscopic terms in the Newton equations (as recommended in Section~\ref{sec:DEM_implementation}), the entries of $\mathsf{L}$ are read off directly from those gyroscopic coefficients after a simple channel-wise energy normalisation; this analytical mapping is derived in~(Supplementary Material, Section~S3) and verified numerically in Figure~S1 of Supplementary Material (Section~S3). Either route gives the Gateway number $\mathcal{G}$ defined in Eq.~\eqref{eq:gateway_number} without any free parameters.

\paragraph{Physical origin of the skew block, and consistency with Onsager--Casimir reciprocity}
The implementation ingredients above endow the configurational channel with dynamics, but they do not, by themselves, mandate a skew-symmetric operator $\mathsf{L}$. This must be stated plainly: because the Onsager reciprocal relations dictate that a contact dynamics governed by time-reversal-symmetric forces must yield a symmetric transport matrix, the appearance of the antisymmetric (Casimir) block $\mathsf{L}$ implies the presence of time-reversal-odd (T-odd) content. 

Crucially, one must distinguish the $\mathsf{L}$ block from standard dissipation. Ordinary frictional dissipation is indeed T-odd, but it is \emph{symmetric} ($\mathsf{D}$), and it governs irreversible entropy production. In contrast, the $\mathsf{L}$ block must be simultaneously non-reciprocal and \emph{work-free} ($\mathbf{X}^T\mathsf{L}\mathbf{X}=0$). 

\subsubsection*{The Physical Nature of the "Work-Free" Skew Response}
The $\mathsf{L}$ block represents a gyroscopic, path-dependent coupling that redirects configurational evolution without storing energy (elasticity) or dissipating it (friction). Its physical origin lies in the \emph{anholonomy} of the contact network; specifically, the non-retracing of the fabric-rotation increment ($d\mathbf{n}_c^- \neq d\mathbf{n}_c^+$) identified by Ref.~\cite{Nicot2024}. 

This coupling produces zero net entropy because it respects a perfect balance between its constituent components. In the language of non-equilibrium thermodynamics, the skew-symmetry creates a geometric constraint where "uphill" diffusion (a negative entropy contribution) is exactly offset by "downhill" diffusion (an equal positive contribution) across the coupled degrees of freedom. 
\begin{itemize}
    \item \textbf{If the anholonomy dissipates energy:} It renormalises the dissipative matrix $\mathsf{D}$, contributing to heat production.
    \item \textbf{If the anholonomy is work-free:} It constitutes the Casimir coupling $\mathsf{L}$, acting as a "geometric phase" that dictates the non-reciprocal trajectory of the fabric without exhausting the system's energy.
\end{itemize}

Consequently, establishing the work-free character of the configurational coupling is a fundamental requirement of the DEM correspondence, not an assumption. Whether posited directly as a micropolar coupling between volumetric, shear-sliding, and fabric-rotation rates, or derived from micro-mechanical non-retracing, $\mathsf{L}$ is a constitutive postulate representing the time-reversal-odd structure of the material's internal dynamics. The analytical mapping provided in Section~S3 serves to fix the magnitude of this coupling, defining how the configurational state space is redirected by these non-reciprocal forces.

\paragraph{One operator, two realisations: channel count and grain count}
The counts $N$ and $N_c$ populate one and the same tridiagonal skew-symmetric operator $\mathsf{L}_N$, and the parity classification is blind to which is intended. In the \emph{contact realisation} the $N$ sites are the coupled fields carried at a single contact, the VMC triad ($N=3$), enriched to $N=4,5$ by a thermal or electrical channel; this is the minimal and physically unambiguous setting in which the Gateway arises. In the \emph{chain realisation} the sites are grains, or contacts, ranged along the column. Here the Newton-to-Onsager mapping of the Supplementary Material makes the necessary distinction: a plain column of $N_c$ grains with no reversible coupling between distinct contacts is \emph{block-diagonal}, a stack of $N_c$ independent VMC triads of dimension $3N_c$ carrying $N_c$ separate null modes; the single tridiagonal chain $\mathsf{L}_N$ studied in Section~\ref{sec:spectral_scaling}, with one parity-mandated null mode, is instead what results once inter-contact gyroscopic couplings (between $V_c$ and $M_{c\pm1}$) are present, so that adjacent grains couple reversibly through their shared contact and the off-diagonal entries $\gamma_i$ are those contact-to-contact couplings. A single point contact carrying $49$ coupled fields is not physical, whereas a column of $49$ grains of differing composition and contact stiffness, coupled along its length, is; the large-$N$ operator is properly read in this second sense, the coupled grain column of Section~\ref{sec:DEM_classical} being that realisation.

Three properties fuse the two realisations into one construction. First, the parity result transfers verbatim: $\det\mathsf{L}_N=0$ holds identically for any odd dimension due to skew-symmetry, meaning the odd-parity Gateway mode is structurally guaranteed regardless of the coupling magnitudes. For even dimensions, full rank is preserved so long as the odd-indexed tridiagonal couplings remain active ($\gamma_{2k-1}\neq 0$). Consequently, the odd/even dichotomy certified numerically for the field chain maps directly onto a grain chain of equal length without further algebraic modification. Second, heterogeneity is accommodated exactly: distinct grain compositions and contact laws merely render the $\gamma_i$ unequal, and because the qualitative classification depends only on the structural dimension, a chain of differently-composed grains behaves identically to a uniform one, odd length forcing the null mode and even length remaining full rank provided the odd-indexed contacts are active ($\gamma_{2k-1}\neq 0$). The heterogeneity of a real granular column is therefore the generic case the theorem already covers, and the uniform chain integrated in Section~\ref{sec:numerical} is the special one. Third, only the spectral detail is realisation-dependent: the exact Toeplitz envelope $|\lambda_{\min}|=2\gamma\sin(\pi/(2N+2))$ of Section~\ref{sec:spectral_scaling} presupposes a common coupling $\gamma$, so a heterogeneous, non-Toeplitz chain modifies the precise law while preserving the qualitative even-$N$ softening with $N$; and the continuum reading $N\to\infty$ acquires its discrete-to-continuum meaning only in the chain realisation, where increasing $N$ refines the column in space. The parity dichotomy, the load-bearing result, is untouched throughout.

\subsection{Protocol that isolates the dilatancy mechanism}
\label{sec:DEM_protocol}

Assuming the two implementation requirements of Section~\ref{sec:DEM_implementation} are in place (torsional springs at every contact and Langevin forcing on the rotational degrees of freedom), the topological dilatancy mechanism can be isolated from any superposed shear contribution by configuring the DEM simulation so that the cycle space of the contact network is empty. This is achieved by enforcing strict one-dimensionality of the contact graph: every contact must lie on the axial chain, with no transverse contacts. The boundary bottleneck of Section~\ref{sec:intro} is then mechanically realised by the rigid lateral confinement.

A practical implementation is:

\begin{enumerate}
  \item Configure the DEM contact law to include the normal spring,
    a tangential (frictional or elastic) spring, and a torsional
    or rolling-resistance spring of stiffness $k_\theta>0$, as
    discussed in Section~\ref{sec:DEM_implementation}.
  \item Build a single column of $N_c$ contacting grains aligned
    along the loading axis $\hat{\mathbf{z}}$, with rigid lateral
    walls confining the column laterally.
  \item Apply sustained axial boundary forcing $f_0\hat{\mathbf{z}}$
    to the top of the column, with the bottom held fixed.
  \item Equilibrate the column to a jammed reference state. Because a jammed granular state retains a memory of its preparation history~\cite{KumarLuding2016}, this reference state should be generated by a reported preparation protocol, for example isotropic compression to a target pressure followed by relaxation, and held fixed across the comparison, so that the extracted operator $\mathsf{D}+\mathsf{L}$ is referred to a well-defined initial fabric.
  \item Activate the Langevin forcing of Eq.~\eqref{eq:langevin_FDT}
    on the rotational degrees of freedom (or equivalently use small
    randomised initial rotations), so that the rotational mode
    nucleates on the timescale set by the granular temperature
    rather than by floating-point round-off.
  \item Extract the local Onsager operator $\mathsf{D}+\mathsf{L}$
    from the force-fluctuation spectrum about this reference state
    (a standard procedure in non-equilibrium statistical
    mechanics).
  \item Identify the channels $V,M,C$ at every contact and verify
    that the kinematic shielding $L_{VC}=0$ is satisfied to within
    DEM noise.
  \item Diagonalise $\mathsf{L}$ and verify that $N$ odd implies
    $\det\mathsf{L}=0$ within numerical precision, with the null
    eigenvector $\mathbf{v}_0$ aligned in the $V$--$C$ plane.
\end{enumerate}

This protocol is implementable with any DEM code that supports a torsional-spring contact law and a Langevin or stochastic-noise option, both of which are standard ingredients in YADE, MercuryDPM, and LIGGGHTS. The framework is therefore not a new DEM method; it is a topological prediction about the spectral structure of an Onsager-completed DEM contact network, with the two implementation ingredients chosen to reflect the physical content of the augmented operator $\mathsf{D}+\mathsf{L}$ rather than to circumvent any limitation of the contact law.

\subsection{Relation to classical DEM oedometer studies}
\label{sec:DEM_classical_results}

The classical one-dimensional DEM oedometer literature \cite{Cundall1979,Radjai2009,Andrade2012} establishes a number of empirical regularities of granular columns under axial loading:
\begin{itemize}
  \item Volumetric compaction proceeds in discrete events
    (avalanches) rather than as continuous strain;
  \item The avalanche statistics follow a power-law distribution
    over several decades of event size;
  \item Fabric reorganisation lags volumetric strain by a finite
    time delay; and
  \item Sustained axial loading at the same nominal pressure
    produces creep-like secular volumetric strain that does not
    saturate.
\end{itemize}

The VMC framework provides a unifying interpretation of all four: the secular volumetric drift is the direct macroscopic manifestation of the parity-mandated null-mode drift under the bottleneck-induced accumulation; the avalanche statistics correspond to the discrete intermittent activation of the Gateway at fluctuating contact configurations; and the time delay between fabric and volume reorganisation is the propagation time of the null-mode drift through the spin-chain coupling. The numerical results reported in Section~\ref{sec:numerical} reproduce each of these features in the simplified setting of the tridiagonal $\mathfrak{so}(N)$ spin chain.

This is the natural point of contact between the VMC topological framework and the empirical DEM literature: the framework predicts that the classical empirical regularities of one-dimensional DEM oedometer studies are not accidents of the contact force law but necessary consequences of the parity-mandated topology of the contact-scale Onsager network combined with the closed-boundary bottleneck of laboratory oedometer geometry. The framework can therefore be falsified as follows. Given a contact network whose configurational response is non-reciprocal, so that the antisymmetric coupling $L_{MC}$ is measurably non-zero, any one-dimensional DEM oedometer with $N$ odd and the kinematic-shielding constraint $L_{VC}=0$ satisfied to within DEM noise \emph{must} exhibit secular volumetric drift under sustained axial loading; the absence of that drift at non-zero measured $L_{MC}$ falsifies the framework. Whether real granular columns exhibit a measurable antisymmetric $L_{MC}$ at all is itself the first and most informative experimental test, since a purely reciprocal contact network would show $L_{MC}\approx 0$ and no drift without bearing on the parity mechanism.

\subsection{Conjugate-variable manifestation under different boundary conditions}
\label{sec:conjugate_duality}

The thermodynamic-bottleneck framing of Section~\ref{sec:intro} described the closed boundary as preventing upscale energy ejection, with the trapped energy accumulating along the parity-mandated null direction. That description does not specify which conjugate variable carries the accumulation, and the answer depends entirely on the boundary geometry. We make the duality explicit here because it sharpens the physical interpretation of the numerical simulations and, more importantly, identifies the boundary geometry in which the framework transitions from linear precursor to genuine nonlinear instability.

The Onsager framework has built-in conjugate duality: every channel carries a pair $(J_\alpha, X_\alpha)$ of conjugate flux and force variables, related by $\mathbf{J}=(\mathsf{D}+\mathsf{L})\mathbf{X}$. Skew-symmetry implies $\mathsf{L}^T = -\mathsf{L}$ and therefore $\ker(\mathsf{L})=\ker(\mathsf{L}^T)$: the null direction $\mathbf{v}_0$ is the same in flux-space and force-space. The Legendre transform between $(J,X)$ exchanges which variable is held fixed by the boundary and which one is dynamical, but it leaves the Gateway null direction invariant. What the boundary geometry \emph{does} change is whether the linear analysis remains valid:

\begin{description}
\item[Free-piston (stress-controlled) boundary.] The applied stress $\sigma_V=f_0$ is held fixed by the piston; the conjugate strains $\varepsilon_a$ and $\varepsilon_v^{\rm conf}$ along $\mathbf{v}_0$ are dynamical and accumulate linearly in time. The strain creep can in principle continue indefinitely as long as the small-strain linearisation of the contact law remains valid, which it does over a substantial range. This is the regime the Section~\ref{sec:numerical} simulations validate and the regime of continuous laboratory creep (Figs.~\ref{fig:sim1_signature}--\ref{fig:sim1_phase}).

\item[Fully clamped (strain-controlled) boundary.] The strain along $\mathbf{v}_0$ is geometrically pinned by the rigid container, and the linear constraint-augmented dynamics has no steady state: $\mathsf{L}\mathbf{q}^* + \mathbf{f}_0 = 0$ has no solution satisfying $\mathbf{v}_0^T \mathbf{q}^* = 0$ because $\mathbf{f}_0$ has a non-zero component in $\ker(\mathsf{L})$ that no linear restoring response can absorb. The Lagrange multiplier enforcing the kinematic constraint -- physically, the boundary-reaction stress on the rigid wall -- grows linearly in time in the singular rigid-wall, ideal-linear-constitutive limit; in the linear-elastic Newton chain of Supplementary Material, Section~S3 (Supplementary Figure~S1b,d) the wall reaction instead saturates at the constitutive envelope $\pm f_0\,v_{0,V}/\!\sqrt{K_{\rm wall}\,m}$ and the cumulative wall impulse $-\!\int F_{\rm wall}\,d\tau=f_0\,t$ absorbs the applied load linearly in time as a static-reaction balance. This bounded elastic saturation is the small-strain limit beyond which the augmented-operator linear-growth prediction ceases to be self-consistent; the linear analysis ceases to be self-consistent on the same finite-time scale. The genuine post-bifurcation evolution is nonlinear, and the nonlinear reduction is solved analytically by the Cnoidal-wave equation of Refs. ~\cite{Veveakis2015,Regenauer2013_JCSMD2}: the unbounded linear stress concentration resolves into propagating, localised Cnoidal structures that carry the instability through the chain. These structures are the topological-framework signature of compaction bands and related localised failure features documented in the geological literature.

\item[Mixed (oedometer) boundary.] Each channel partitions according to which conjugate variable is constrained: the axial $V$ channel is typically stress-controlled (piston applies sustained $\sigma_a$) while the transverse $C$ channel is strain-controlled (rigid lateral container). The Gateway null direction $\mathbf{v}_0$ lies in the $V$--$C$ plane and therefore couples the linear strain creep in $V$ to the nonlinear Cnoidal-wave reorganisation in $C$. The physically realised laboratory oedometer is in this mixed regime.
\end{description}

The numerical simulations of Section~\ref{sec:numerical} integrate the conservative Onsager dynamics in pure flux-space form, $\dot{\mathbf{q}}=\mathsf{L}\mathbf{q}+\mathbf{f}_0$, which is the stress-controlled limit; the strain-controlled counterpart is realised in the linear elastic regime in Supplementary Figure~S1b,d (Supplementary Material, Section~S3), where the drift is bounded by the wall envelope as described above. Both panels are therefore a validation of the strain-creep manifestation in their respective conjugate variables; neither, on its own, captures the nonlinear post-bifurcation evolution of the clamped case. A DEM simulation under fully-clamped boundaries with Hertzian or other nonlinear contact law can carry the dynamics through the linear-breakdown point and into the Cnoidal-wave regime; this provides an alternative and arguably more physically relevant falsification target for the framework, since real laboratory and geological loading geometries are typically not stress-controlled in all channels.

The algebraic invariance of the Gateway null direction under boundary-condition choice (a kernel of the topology of $\mathsf{L}$, unchanged by Legendre transform) therefore enlarges the family of accessible falsification tests in two directions: linear strain-creep testing in stress-controlled geometries (the present paper), and nonlinear Cnoidal-wave testing in strain-controlled (clamped) geometries~\cite{Veveakis2015}.

\section{Testable Predictions}
\label{sec:predictions}

The framework yields four falsifiable one-dimensional predictions that distinguish it from purely dissipative cascade descriptions of the Kolmogorov type and from equilibrium statistical approaches based on the Edwards ensemble. All four are accessible to one-dimensional discrete-element simulations of a granular column under sustained axial load (oedometer geometry).

\subsection{Configurational-energy localisation as a dilatancy precursor}

Because $dE^{\rm conf}$ accumulates at $\mathcal{C}^{\rm conf}$ contacts before being released by force-chain rearrangement, the ratio $E_C/(E_k+E_C)$, where $E_C$ is the total configurational energy frozen at $\mathcal{C}^{\rm conf}$ contacts and $E_k$ is the grain-scale kinetic energy, should grow secularly under sustained boundary forcing and localise along the loaded column before any macroscopic compaction-front event. This precursor is a direct fingerprint of the null-mode projection $\mathbf{v}_0^T\mathbf{q}(t)$ and does not follow from any purely dissipative description.

\subsection{Discrete stability threshold upon coupling of a third field}

The addition of a third thermodynamic field, such as a localised thermal force, to the core kinetic-configurational Level-2 pair structurally alters the operator parity, converting the two-field Stable Layer into an active three-field Gateway Layer. The onset of this topological transition is marked by a qualitative change in the off-diagonal Onsager coupling structure: within the Stable Layer regime the coupling structure is reducible: the symmetric cross-diffusion can be diagonalised into effective uncoupled self-diffusion fields~\cite{RegenHu2023b,RegenHu2024}, and the reversible block splits into decoupled conjugate oscillatory pairs, so the dynamics separates into independent bounded modes. In the active Gateway regime the full $3\times3$ operator is irreducible and admits no such separation: the parity-mandated null mode remains coupled to the boundary forcing. In a physical one-dimensional granular column, this topological transition provides a distinct, verifiable experimental signature: it manifests as a direct cross-correlation between localised micro-thermal fluctuations (phonon dissipation tracking) and acoustic-emission signals during macroscopic axial loading.

\subsection{Parity determines the qualitative failure mode}

Dilatancy-driven matrix failure of the $N=3$ VMC Gateway is \emph{structurally inevitable} once the activation criterion $\mathcal{G}\geq 1$ is met under the closed-boundary bottleneck: no choice of material parameters can prevent secular escape if the three-field system is pushed past $\lambda_{\rm eff}=0$ and upscale routing channels are choked. This contrasts with purely dissipative or elastic models, where failure is always a quantitative threshold question. The same parity contrast extends to all higher-$N$ one-dimensional chains: odd-$N$ assemblies leak; even-$N$ assemblies contain.

\subsection{Coordination-number drift under sustained $\sigma_n$ as a one-dimensional DEM test}

The framework predicts a specific signature in a one-dimensional DEM oedometer test under sustained normal stress $\sigma_n$ after primary compaction has ceased. For an effectively odd-$N$ chain (an odd-cardinality cycle of active contact channels along the load axis), the projected coordinate $\mathbf{v}_0^T\mathbf{q}(t)$ should grow linearly in time at a rate $\propto L_{MC}/\omega \cdot f_0$ (cf.\ the analytical solutions in Part~2~\cite{RegenNicot2026dilatancy}), accompanied by a secular drift in the mean coordination number along the column. Even-$N$ chains should exhibit no such drift, only bounded oscillation. The asymptotic spectral scaling $|\lambda_{\min}^{\rm even}|\sim\gamma\pi/N$ (Section~\ref{sec:spectral_scaling}) further predicts that the bounding period $T_{\rm slow}\propto N$ in the even case, providing a direct $N$-dependent test on the simulation side.

\subsection{Spontaneous arrest of the drift under fabric realignment}
\label{sec:spontaneous_arrest}

Part~2~\cite{RegenNicot2026dilatancy} (Proposition~1 and the accompanying discussion of the $L_{MC}\to0$ limit) predicts that the boundary-work injection rate $\mathcal{W}_{\rm inject}=L_{MC}f_0/\sqrt{L_{VM}^2+L_{MC}^2}$ vanishes in the limit $L_{MC}\to 0$. Under this limit the null vector $\mathbf{v}_0\propto(L_{MC},0,L_{VM})^T$ rotates from the $V$--$C$ plane onto the pure $C$ axis and becomes orthogonal to the axial boundary load: the Gateway remains topologically open, but the loading no longer projects onto the cocycle-borne null direction, and the secular drift collapses to zero. This is a single, qualitative, falsifiable prediction: a granular column whose contact statistics evolve toward $L_{MC}\to 0$ (force-chain alignment with contact normals, suppressing the shear--fabric coupling) should exhibit spontaneous arrest of the dilatancy creep without any change in the underlying parity-mandated null mode.

\begin{figure}[!htbp]
\centering
\includegraphics[width=0.75\linewidth]{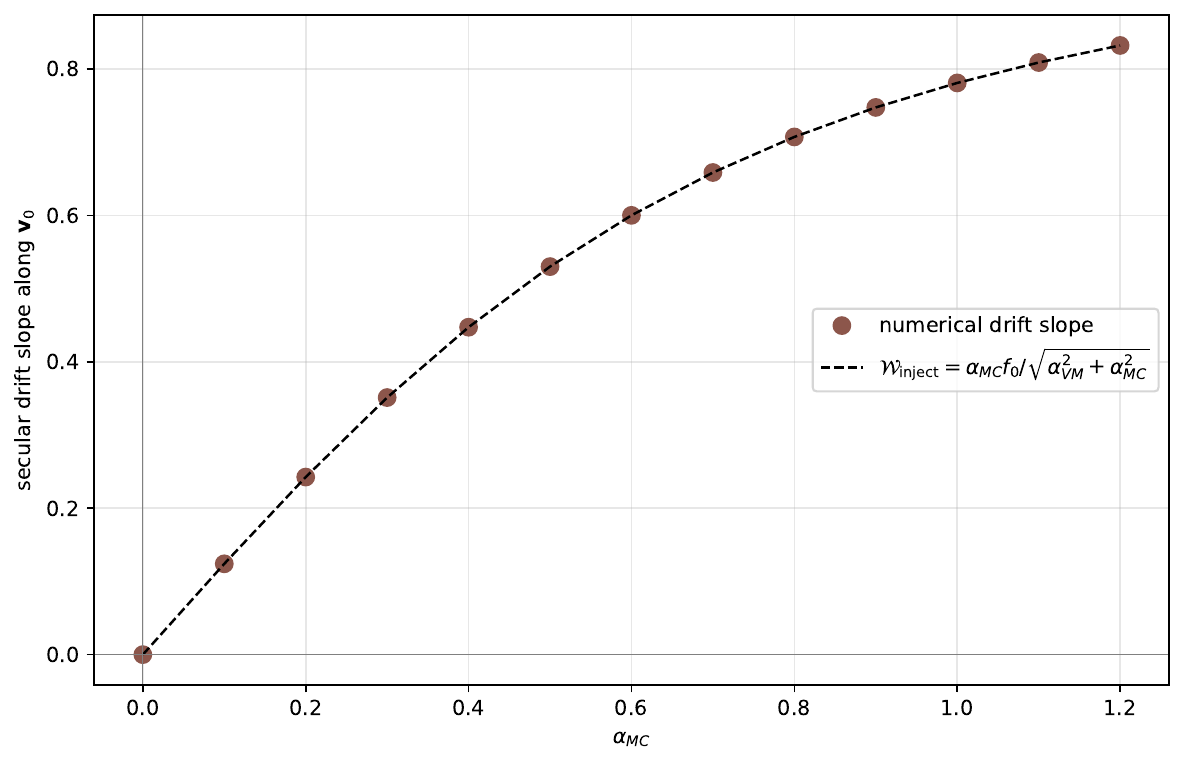}
\caption{Spontaneous-arrest confirmation of Part~2~\cite{RegenNicot2026dilatancy} (Proposition~1). The numerical drift slope along the null direction $\mathbf{v}_0$ for the VMC triad ($N=3$) is shown as a function of the shear--fabric coupling coefficient $L_{MC}$, with $L_{VM}=0.8$ held fixed. The numerical points (filled circles) track the analytical prediction $\mathcal{W}_{\rm inject}=L_{MC}f_0/\sqrt{L_{VM}^2+L_{MC}^2}$ (dashed) to a maximum relative error of $4.5\times 10^{-15}$. The drift collapses identically to zero at $L_{MC}=0$: the Gateway is still parity-mandated and topologically open, but the loading no longer projects onto the cocycle-borne null direction. This is the cleanest single-experiment falsification test of the conjugate manifestation of the framework: a DEM column whose statistics evolve toward force-chain alignment ($L_{MC}\to 0$) should cease drifting without otherwise changing its topological state.}
\label{fig:spontaneous_arrest}
\end{figure}

Figure~\ref{fig:spontaneous_arrest} confirms the prediction numerically. The drift slope along $\mathbf{v}_0$, measured by linear regression on the asymptotic portion of the integrated trajectory, follows the analytical curve $L_{MC}f_0/\sqrt{L_{VM}^2+L_{MC}^2}$ to a maximum relative error of $4.5\times 10^{-15}$ across the full sweep $L_{MC}\in[0, 1.2]$, with the drift collapsing identically to zero at $L_{MC}=0$.

\section{Future Work: Lifting the Bottleneck}
\label{sec:future}

The closed-boundary bottleneck that produces the unbounded null-mode drift in the present simulations is not a fundamental feature of the augmented Onsager framework; it is a deliberate choice of the 1-D tridiagonal numerical setup that isolates the cocycle-borne dilatancy mechanism for clean validation. Two complementary directions lift this bottleneck.

\subsection{Lifting the cocycle filter: cycle-borne shear localisation}

The most immediate extension is to lift the construction from the one-dimensional chain to a two-dimensional Hasse-lattice network in which transverse fabric and deviatoric stress components introduce additional channels and the Hasse subgraphs at every level acquire non-trivial cycle space ($\beta_1\geq 1$). In this setting the Parity Theorem continues to apply level-by-level, but new instability modes appear that are absent in 1-D: Schnakenberg cycle currents driven by non-zero cycle affinities, shear-band onset at the boundary of the cycle and cocycle decompositions, and the topological emergence of non-associated plastic flow with the scaling $\tan(\phi-\psi)\sim\sqrt{\mathcal{G}}$ as a corollary. The connection to triaxial-DEM and shear-band experiments is subject of ongoing work.

\subsection{Lifting the upscale-routing bottleneck: multi-field Hasse cascade} The mathematical extension to cnoidal wave solutions that bridges the linear topological invariants to the nonlinear compaction-band regime of Refs. \cite{Regenauer2013_JCSMD2,Veveakis2015} is developed in Supplementary Material, Section~S4.

The alternative route to lifting the bottleneck is to embed the contact-scale VMC triad within the larger multi-field Hasse lattice of Ref.~\cite{RegenNicot2026foundation}. There, the alternating Stable/Gateway pattern at successive Hasse levels supplies the natural upscale routing channels through which an active Gateway ejects energy into higher-order topological spaces, rather than accumulating it in the lowest-level null direction. Multi-physics extensions (thermo-hydro-mechanical-chemical-electrical, THMCE) embed the VMC triad as the $N=3$ sub-diagram of a larger Hasse lattice; the Parity Theorem applies level-by-level and the system-level Gateway at $N=5$ (full THMCE) provides the natural topological precursor structure for geophysical-scale catastrophic failure. Quantitative development of the VMC-to-THMCE bridging is the subject of future work.

\section{Conclusions}
\label{sec:conclusions}

This Part~3 of the paper series has carried the upscaling framework of Part~2~\cite{RegenNicot2026dilatancy} and the general augmented-Onsager framework of Part~1~\cite{RegenNicot2026foundation} to numerical and micro-mechanical realisation on three independent fronts: quad-precision numerical integration, asymptotic Toeplitz spectral scaling, and classical one-dimensional DEM correspondence. The principal results are:

\begin{enumerate}
  \item \textbf{Quad-precision parity contrast.} Numerical integration to quad precision across dimensions from $N=3$ to $N=50$ confirms that odd-$N$ chains develop an unbounded null-mode drift, whereas even-$N$ chains remain strictly confined to compact invariant tori (Section~\ref{sec:numerical}). This topological contrast is parameter-independent and robustly survives the high-dimensional limit.
  
  \item \textbf{Toeplitz spectral scaling.} The minimum non-zero eigenvalue of even-parity Stable Layers decays as $|\lambda_{\min}^{\rm even}|\sim\gamma\pi/N$ (Section~\ref{sec:spectral_scaling}). This scaling explicitly bridges the discrete microstructural scale to the macroscopic continuum limit via a mechanism of scale-dependent critical slowing down.
  
  \item \textbf{Thermodynamic-bottleneck interpretation.} The unbounded drift observed in these one-dimensional tridiagonal simulations constitutes the expected baseline response of an odd-parity system whose natural upscale energy-routing channels are kinematically choked by a rigid boundary condition. The simulations, therefore, provide an ideal isolation of the topological dilatancy mechanism. This bottleneck is naturally lifted by either: (i) extending the network configuration to dimensions $\geq 2$, or (ii) embedding the chain within the multi-field continuum hierarchy of~\cite{RegenNicot2026foundation}.
  
  \item \textbf{Classical DEM correspondence.} Each discrete VMC channel maps onto a clearly defined, measurable DEM observable. The macroscopic Gateway number $\mathcal{G}$ is directly computable from the contact statistics of a jammed one-dimensional column without free fitting parameters, yielding four explicit falsification protocols accessible to standard DEM solvers through contact-law and thermostatting options already available in them, namely a rolling-resistance or torsional contact law and a Langevin forcing of the rotational degrees of freedom.
\end{enumerate}

Together with the theoretical framework and closed-form analytical solutions developed in Part~2, these numerical and structural results establish the complete one-dimensional VMC dilatancy mechanism as algebraically guaranteed, numerically demonstrated, and empirically testable. Having brought the framework to a full standard of falsifiability, immediate next steps involve lifting the cocycle filter and the boundary bottleneck as developed in the foundational work~\cite{RegenNicot2026foundation}.

\section*{Acknowledgements}

K.R.L.\ acknowledges research support from ARC grants CE230100032, DP200102517, and LP170100233, and discussions with Victor Calo, Manman Hu, and his PhD students on the augmented Onsager D+L formalism and the DEM correspondence. F.N.\ acknowledges support from INRAE and the GeoMechALP group. Both authors contributed equally to the numerical and structural analysis presented here.


\bibliographystyle{elsarticle-num}
\bibliography{bib_merged}